\documentclass[openany,oneside,12pt,a4paper]{book}
\usepackage[english]{babel}
\usepackage[utf8]{inputenc}
\usepackage[symbols,nogroupskip]{glossaries-extra}
\usepackage{amsmath}
\usepackage{mathtools}
\usepackage{tikz}
\usepackage{tikz-cd}
\usepackage[RPvoltages]{circuitikz}
\usepackage{amsfonts}
\usepackage{graphicx}
\usepackage{epstopdf}
\usepackage{amssymb}
\usepackage{makeidx}
\usepackage{graphicx}
\usepackage{enumerate}
\usepackage{blindtext}
\usepackage{lipsum}
\usepackage{braket}
\usepackage{tabularx}
\usepackage[]{algorithm2e}
\usepackage{fancyhdr}
\usepackage{titlesec}
\usepackage{capt-of}
\usepackage{adjustbox}
\usepackage{etoolbox}
\usepackage{siunitx}
\usepackage{slantsc}
\usepackage{qcircuit}
\usepackage{csquotes}
\usepackage{float}
\usepackage{tabularx}
\usepackage{appendix}
\usepackage{afterpage}
\usepackage{array,booktabs}
\usepackage{pdflscape}
\usepackage{rotating}
\usepackage{hvfloat}
\usepackage{placeins}
\usepackage{bm}
\usepackage{acronym} 
\usepackage[backend=biber,backref,url=true,eprint=false,doi=false,isbn=false,citestyle=authoryear,sorting=nyt,style=alphabetic,maxbibnames=1,arxiv=true]{biblatex}
\usepackage[colorlinks=true, urlcolor=blue,linkcolor=blue,citecolor=blue,breaklinks=true,pdfauthor={Hannes Lagemann}]{hyperref}

\usetikzlibrary{matrix}
\usetikzlibrary{decorations.markings}
\usetikzlibrary{shapes.geometric}
\pgfdeclarelayer{edgelayer}
\pgfdeclarelayer{nodelayer}
\pgfsetlayers{edgelayer,nodelayer,main}

\tikzstyle{none}=[inner sep=0pt]

\tikzstyle{rn}=[circle,fill=Red,draw=Black,line width=0.8 pt]
\tikzstyle{gn}=[circle,fill=Lime,draw=Black,line width=0.8 pt]
\tikzstyle{yn}=[circle,fill=Yellow,draw=Black,line width=0.8 pt]

\tikzstyle{simple}=[-,draw=Black,line width=2.000]
\tikzstyle{arrow}=[-,draw=Black,postaction={decorate},decoration={markings,mark=at position .5 with {\arrow{>}}},line width=2.000]
\tikzstyle{tick}=[-,draw=Black,postaction={decorate},decoration={markings,mark=at position .5 with {\draw (0,-0.1) -- (0,0.1);}},line width=2.000]

\tikzstyle{newstyle}=[
  -,draw=Green,line width=2.000]

  \newcommand{\idxFFT}{\text{Fix.}}
\newcommand{\idxFTT}{\text{Tun.}}
\newcommand{\idxRES}{\text{Res.}}
\newcommand{\idxTLS}{\text{TLS}}
\newcommand{\idxINT}{\text{Int.}}

\newcommand{\TB}{transmon basis}

\newcommand{\width}{0.95}

\newcommand{\const}{\text{const.}}
\newcommand{\appref}[1]{Appendix~\ref{#1}}
\newcommand{\chapref}[1]{Chap.~\ref{#1}}
\newcommand{\chaapref}[2]{Chap.~\ref{#1} and \ref{#2}}

\newcommand{\chapsref}[2]{Chap.~\ref{#1}-\ref{#2}}
\newcommand{\secref}[1]{Sec.~\ref{#1}}
\newcommand{\secaref}[2]{Sec.~\ref{#1} and \ref{#2}}
\newcommand{\secaaref}[3]{Sec.~\ref{#1}, \ref{#2} and \ref{#3}}
\newcommand{\secsref}[2]{Secs.~\ref{#1}-\ref{#2}}

\newcommand{\figref}[1]{Fig.~\ref{#1}}
\newcommand{\figsref}[1]{Figs.~\ref{#1}}
\newcommand{\tabref}[1]{Table~\ref{#1}}
\newcommand{\tabsref}[2]{Tables~\ref{#1}-\ref{#2}}
\newcommand{\tabaref}[2]{Table~\ref{#1} and \ref{#2}}
\newcommand{\equref}[1]{Eq.~\eqref{#1}}
\newcommand{\equaref}[2]{Eqs.~\eqref{#1} and \eqref{#2}}
\newcommand{\equsref}[2]{Eqs.~\eqref{#1}--\eqref{#2}}
\newcommand{\sequref}[1]{Equation~\eqref{#1}}
\newcommand{\PANC}[1]{Panel(#1)}
\newcommand{\PANL}[1]{panel(#1)}
\newcommand{\PANSC}[1]{Panels(#1)}
\newcommand{\PANSL}[1]{panels(#1)}

\newcommand{\innerproduct}[2]{\braket{#1|#2}}
\newcommand{\ketbra}[2]{\ket{#1}\!\!\bra{#2}}
\newcommand{\norm}[1]{\left\lVert#1\right\rVert}
\newcommand{\BRR}[1]{\left( #1 \right)}
\newcommand{\brr}[1]{\left\{ #1 \right \}}
\newcommand{\OP}[1]{\hat{#1}}
\newcommand{\SUPOP}[1]{\check{#1}}
\newcommand{\tens}[1]{\mathbin{\mathop{\otimes}\limits_{#1}}}
\newcommand{\tensupper}[2]{\mathbin{\mathop{\otimes}\limits_{#1}^{#2}}}
\newcommand{\produpper}[2]{\mathbin{\mathop{\prod}\limits_{#1}^{#2}}}

\newcommand{\dx}[1]{d\!#1}

\newcommand{\mdot}{\mathord{\cdot}}

\newcommand{\der}[2]{\frac{d#1}{d#2}}
\newcommand{\derp}[2]{\frac{\partial#1}{\partial#2}}

\newcommand{\ie}{ i.e.~}
\newcommand{\eg}{ e.g.~}
\newcommand{\REF}{Ref.~}
\newcommand{\REFS}{Refs.~}

\newcommand{\nth}{$n^{\text{th}}$}
\newcommand{\trace}{\text{Tr}}
\newcommand{\hold}{test}

\newcommand{\lw}{0.30mm}
\newcommand{\Quote}[1]{“#1”}

\renewcommand{\tilde}[1]{\widetilde{#1}}
\renewcommand{\bar}[1]{\overline{#1}}

\def\getfirst#1.#2\relax{#1}

\mathchardef\mhyphen="2D 
\bibliography{./11_References/ref.bib}
\allowdisplaybreaks
\makeatletter
\patchcmd{\backmatter}
{\@mainmatterfalse}
{\@mainmatterfalse\pagenumbering{roman}}
{}
{}
\makeatother
\titleformat{\chapter}[display]
{\normalfont\huge\bfseries}{}{0pt}{\Huge}
\titleformat{name=\chapter,numberless}[display]
{\normalfont\huge\bfseries}{}{0pt}{\Huge}
\titlespacing*{\chapter}{0pt}{50pt}{*2}
\begin{document}

\let\oldUrl\url
\renewcommand{\url}[1]{\href{#1}{Link}}



\newcommand*{\plogo}{\fbox{$\mathcal{PL}$}} 
\newcommand*{\titleGP}{\begingroup 

	\centering 
	\vspace*{\baselineskip} 

	\rule{\textwidth}{1.6pt}\vspace*{-\baselineskip}\vspace*{2pt} 
	\rule{\textwidth}{0.4pt}\\[\baselineskip] 

	{\LARGE {R}eal-time simulations of transmon systems\\with\\time-dependent Hamiltonian models \\[0.2\baselineskip]} 

	\rule{\textwidth}{0.4pt}\vspace*{-\baselineskip}\vspace{3.2pt} 
	\rule{\textwidth}{1.6pt}\\[\baselineskip] 

	\scshape{{\footnotesize \mbox{Von der fakultät für mathematik, informatik und naturwissenschaften} \\ der rwth aachen university zur erlangung des akademischen  \\ grades eines doktors der naturwissenschaften  \\ genehmigte dissertation \\[\baselineskip]}} 

\vspace*{1.0\baselineskip} 

vorgelegt von \\[\baselineskip]
{\Large \href{mailto:hannes.a.l@me.com}{\color{black}{Hannes Alfred Lagemann, M. Sc.}} \\[\baselineskip]\par}  
aus \\[\baselineskip]
{\Large Emsbüren \\[\baselineskip]\par} 

\vspace*{1\baselineskip} 

Berichter \\[\baselineskip]
{\Large Prof.~Dr.~Kristel Francine Michielsen \par} 

\vspace*{1\baselineskip} 

{\Large Prof.~Dr.~David Peter DiVincenzo  \\[0.1\baselineskip]\par} 

\vspace*{1\baselineskip} 

{Tag der mündlichen prüfung: 6. März 2023} \\[0.1\baselineskip] 

\vspace*{1\baselineskip} 

{Diese dissertation ist auf den internetseiten der universitätsbibliothek verfügbar} \\[0.1\baselineskip] 

\endgroup}
\thispagestyle{empty}
\setcounter{page}{0}
\titleGP 
\clearpage

\clearpage

\clearpage
\frontmatter

\chapter{Zusammenfassung}

In dieser Dissertation untersuchen wir die Aspekte von zeitabhängigen Hamil-tonoperatoren (Modellen), die die zeitliche Entwicklung von Transmon Systemen beeinflussen können. Wir modellieren die Zeitentwicklung verschiedener Systeme als unitäre Echtzeitprozesse. Dazu lösen wir die zeitabhängige Schrödingergleichung (ZSG) numerisch. Die resultierenden Computermodelle bezeichnen wir als nicht-ideale-gatterbasierte Quantencomputer (NIGQC) Modelle, da Transmons in der Regel als Transmon Qubits in supraleitenden prototyp-gatterbasierten Quantencomputern (PGQC) verwendet werden.

Wir diskutieren zunächst das Modell des idealen-gatterbasierten Quantencomputers (IGQCs) und differenzieren zwischen den Begriffen IGQC, PG-QC und NIGQC, welche wir in dieser Arbeit betrachten. Im Anschluss leiten wir die Schaltkreis Hamiltonoperatoren ab, welche die Zeitentwicklung von frequenzfesten und stimmbaren Transmons generieren. Darüber hinaus diskutieren wir kurze und prägnante Ableitungen für effektive Hamiltonoperatoren (für beide Arten von Transmons). Wir verwenden die Schaltkreis Hamiltonoperatoren und die effektiven Hamiltonoperatoren, um zwei Mehr-teilchen Hamiltonoperatoren zu definieren. Die Wechselwirkungen zwischen den verschiedenen Teilsystemen werden als Dipol-Dipol Wechselwirkungen modelliert. Im Anschluss entwickeln wir zwei Produkt-Formel Algorithmen, welche es uns erlauben die ZSG für die von uns definierten Mehrteilchen Hamiltonoperatoren numerisch zu lösen.

Anschließend verwenden wir diese Algorithmen, um zu untersuchen wie sich verschiedene Näherungen (Annahmen) auf die Zeitentwicklung von Transmon Systemen auswirken. Hier betrachten wir den effektiven Mehrteilchen Hamiltonoperator und nutzen den Schaltkreis Mehrteilchen Hamiltonoperator als einen Referenzpunkt (im übertragenen Sinne). Wir betrachten ausschließlich Szenarien bei denen ein Steuerungssignal genutzt wird. Wir finden, dass die von uns untersuchten Näherungen die Zeitentwicklung der von uns modellierten Wahrscheinlichkeitsamplituden erheblich beeinflussen können. Des Weiteren untersuchen wir, wie anfällig die numerischen Werte verschiedener Gatter-Fehler Quantifikatoren gegenüber Näherungen (Annahmen) sind. Wir stellen fest, dass die von uns betrachteten Näherungen (Annahmen) eindeutig Gatter-Fehler Quantifikatoren wie die Diamant-Distanz und die durchschnittliche Fi­de­li­tät beeinflussen. Des Weiteren sind wir oft in der Lage klare und prägnante theoretische Erklärungen für die Ergebnisse, die wir in dieser Arbeit diskutieren, zu präsentieren.

\chapter{Abstract}

In this thesis we study aspects of Hamiltonian models which can affect the time evolution of transmon systems. We model the time evolution of various systems as a unitary real-time process by numerically solving the time-dependent Schrödinger equation (TDSE). We denote the corresponding computer models as non-ideal gate-based quantum computer (NIGQC) models since transmons are usually used as transmon qubits in superconducting prototype gate-based quantum computers (PGQCs).

We first review the ideal gate-based quantum computer (IGQC) model and provide a distinction between the IGQC, PGQCs and the NIGQC models we consider in this thesis. Then, we derive the circuit Hamiltonians which generate the dynamics of fixed-frequency and flux-tunable transmons. Furthermore, we also provide clear and concise derivations of effective Hamiltonians for both types of transmons. We use the circuit and effective Hamiltonians we derived to define two many-particle Hamiltonians, namely a circuit and an associated effective Hamiltonian. The interactions between the different subsystems are modelled as dipole-dipole interactions. Next, we develop two product-formula algorithms which solve the TDSE for the many-particle Hamiltonians we defined.

Afterwards, we use these algorithms to investigate how various frequently applied approximations (assumptions) affect the time evolution of transmon systems modelled with the many-particle effective Hamiltonian when a control pulse is applied. Here we also compare the time evolutions generated by the effective and circuit Hamiltonian. We find that the approximations we investigate can substantially affect the time evolution of the probability amplitudes we model. Next, we investigate how susceptible gate-error quantifiers are to approximations which make up the NIGQC model. We find that the approximations (assumptions) we consider clearly affect gate-error quantifiers like the diamond distance and the average infidelity. Furthermore, we provide clear and concise theoretical explanations for many of the findings we present in this thesis.

\clearpage
\mainmatter
{
	\setcounter{tocdepth}{4}
	\setcounter{secnumdepth}{4}
	\hypersetup{linkcolor=black}
	\pagestyle{empty}
	\pagenumbering{gobble}
	\tableofcontents
	\newpage
}

	\pagenumbering{arabic}
	\setcounter{page}{1}

\chapter{Introduction}\label{chap:I}

We begin this thesis with a fundamental distinction, namely we differentiate between a computer model and the corresponding computer hardware. Broadly speaking, a computer model consists of a collection of definitions and mathematical proofs which state that once we are provided with an ideal computing machine which is able to execute a certain number of elementary operations, we can solve a broad range of computational problems by executing sequences of these elementary operations. The computer model provides the knowledge of how to construct the sequence for a given computational problem. A famous example for such a computer model is the Turing machine, see \REF\cite{Turing36}. The computer hardware is tasked with the problem of realising the elementary operations from the abstract computer model in the real world and providing the necessary infrastructure for the computational process,\eg long-term and short-term memory storage. Note that the computer model does not discriminate between different technological solutions to the problem. For example, at one time, digital computer hardware was based on vacuum tubes which were later replaced by transistors.

The ideal gate-based quantum computer (IGQC) as discussed in \REF\cite{Nielsen:2011:QCQ:1972505} is an algebraic computer model that consists of a collection of definitions and mathematical proofs, see \secsref{sec:MathematicalFramework}{sec:Algorithms}. However, the construction of a fully functioning prototype gate-based quantum computer (PGQC) is an unsolved engineering task which requires a tremendous amount of precision in terms of control over a quantum system, see \REF\cite{QEC13}. Many studies with a focus on PGQCs (sometimes referred to as noisy intermediate-scale quantum devices, see \REF\cite{Preskill2018}) investigate a certain type of superconducting hardware, see for example \REFS\cite{Gu21,Roth19,Rol19,Arute19,Lacroix2020}, so-called transmon systems. In this thesis, we model the time evolution of transmon systems as a unitary real-time process. Therefore, throughout this work we assume that the dynamics of transmon systems is governed by the time-dependent Schrödinger equation (TDSE)
\begin{equation}\label{eq:TDSE_intro}
  i \partial_{t}\ket{\Psi(t)}=\OP{H}(t)\ket{\Psi(t)},
\end{equation}
where $\OP{H}(t)$ is the model Hamiltonian that generates the dynamics and $\ket{\Psi(t)}$ is the state vector which, by assumption, completely describes the state of the system. Note that we use $\hbar=1$ throughout this work. Unfortunately, analytical solutions of the TDSE are rare, especially for cases where the model Hamiltonian $\OP{H}(t)$ is explicitly time dependent. Hence, we use the product formula approach, see \REFS\cite{DeRaedt87,Huyghebaert90}, to construct two algorithms that can solve the TDSE numerically, for two many-particle Hamiltonians. We use these algorithms to construct non-ideal gate-based quantum computer (NIGQC) models which describe the time evolution of superconducting PGQCs used in experiments. In this thesis, we focus on certain aspects of the model which can substantially affect the state of a NIGQC,\ie the time evolution of transmon systems.

This thesis is structured as follows. In \chapref{chap:II}, we review the model of the IGQC and provide a distinction between the IGQC, PGQCs and NIGQCs. Next, in \chapref{chap:III}, we derive circuit and effective Hamiltonians for LC resonators, fixed-frequency and flux-tunable transmons. We use these subsystems to define two many-particle Hamiltonians, a circuit Hamiltonian and an associated effective Hamiltonian. The interactions between the different subsystems are modelled as dipole-dipole interactions. The effective Hamiltonian is supposed to mimic the dynamics of the circuit Hamiltonian. Then, in \chapref{chap:IV}, we review the fundamental problems we face if we numerically solve the TDSE. Furthermore, we discus the product-formula algorithms we use to mitigate these problems and to solve the TDSE for the Hamiltonians we define in \chapref{chap:III}. In \chapref{chap:NA}, we study how various approximations affect the time evolution of transmon systems modelled by the effective Hamiltonian. Here, we consider scenarios where control pulses are applied to the transmon systems. Additionally, we compare the time evolution of the circuit Hamiltonian with the one of the effective Hamiltonian. Note that in this chapter we only consider the probability amplitudes relevant to the transitions we study. Next, in \chapref{chap:GET}, we investigate how susceptible gate-error quantifiers are to the approximations which make up the NIGQC model. Gate-error quantifiers are usually used to quantify how successful a computation could be realised with a PGQC and/or with a NIGQC model. Finally, in \chapref{chap:End}, we summarise all the findings, provide concluding remarks and give an outlook for future research.


\chapter[An introduction to gate-based quantum computing]{An introduction to gate-based \\quantum computing}\label{chap:II}

In this chapter we provide a discussion of gate-based quantum computing. In \secref{sec:MathematicalFramework}, we introduce the mathematical framework we use to develop the model of the IGQC. Section \ref{sec:The single-qubit space} contains a discussion of the single-qubit Hilbert space. In \secref{sec:TheMultiQubitSpace}, we define the multi-qubit Hilbert space, which is constructed by combining single-qubit Hilbert spaces. The corresponding model constitutes the IGQC. Section \ref{sec:SimulationOfTheIdealGateBasedQuantumcomputer} contains a discussion of the simulation methods we use to numerically model the IGQC. In \secref{sec:Algorithms}, we discuss a handful of algorithms which can be used to solve different computational problems with an IGQC. We discuss algorithms which are of very different nature. In \secref{sec:Prototype gate-based quantum computers} we discuss PGQCs,\ie experimental setups which aim to realise the IGQC. The subject of \secref{sec:FromStaticsToDynamics} are NIGQC models,\ie models that can be used to describe the time evolution of the PGQCs. Such a model can potentially be used to explain and/or describe some of the basic trends we find in experimental data. The first four sections of this chapter are inspired by \REFS\cite{Nielsen:2011:QCQ:1972505,LagemannMSCThesis}.

\section{Mathematical framework}\label{sec:MathematicalFramework}
Quantum computing, on an IGQC, takes place in a finite-dimensional vector space $\mathcal{H}$ over the field of complex numbers $\mathbb{C}$. The vector space is also equipped with an inner product $\braket{\mdot|\mdot}\colon \mathcal{H} \times \mathcal{H} \rightarrow \mathbb{C}$. Such a space is usually referred to as finite-dimensional Hilbert space. In the following we omit the adjective finite-dimensional but add a label $D$ to every Hilbert space $\mathcal{H}^{D}$, where $D=\text{dim}(\mathcal{H}^{D})$ denotes the dimensionality of the vector space.

A computation on an IGQC might be understood as follows. If $\ket{\psi}\in \mathcal{H}^{D}\colon \innerproduct{\psi}{\psi}=1$ is some arbitrary state in some Hilbert space $\mathcal{H}^{D}$, we can perform a computation by mapping this state $\ket{\psi} \mapsto \ket{\psi^{\prime}}$ to another state $\ket{\psi^{\prime}}\in \mathcal{H}^{D}\colon \innerproduct{\psi^{\prime}}{\psi^{\prime}}=1$. We can generalise this idea by defining a function $\OP{U}\colon \mathcal{H}^{D} \rightarrow \mathcal{H}^{D}$ which preserves the norm for all $\ket{\psi} \in \mathcal{H}^{D}$. We also require $\OP{U}$ to be linear, which means we should speak of our function as an operator $\OP{U} \in \mathbb{L}(\mathcal{H}^{D})$. Here $\mathbb{L}(\mathcal{H}^{D})$ denotes the set of all linear functions,\ie operators, in $\mathcal{H}^{D}$. Additionally, operators $\OP{U}\in \mathbb{U}(\mathcal{H}^{D})$ which are norm preserving are referred to as unitary operators. Here $\mathbb{U}(\mathcal{H}^{D})\subseteq \mathbb{L}(\mathcal{H}^{D})$ denotes the set of all unitary operators in $\mathcal{H}^{D}$. All unitary operators $\OP{U}$ satisfy the equation
\begin{equation}
  \OP{U}^{\dagger}\OP{U}=\OP{I},
\end{equation}
where $\OP{U}^{\dagger}$ denotes the so-called adjoint operator. If $\OP{L} \in \mathbb{L}(\mathcal{H}^{D})$, we denote $\OP{L}^{\dagger}$ as the adjoint operator if for all $\ket{\psi},\ket{\psi^{\prime}} \in \mathcal{H}^{D}$ the relation
\begin{equation}
  \braket{\psi^{\prime}|\OP{L}^{\dagger}|\psi}=\braket{\psi|\OP{L}|\psi^{\prime}}^{*},
\end{equation}
holds true. The asterisk $^{*}$ denotes the complex conjugate of the complex number $\braket{\psi|\OP{L}|\psi^{\prime}}$.

The question we have to ask ourselves is: what is the relation between the algebraic formalism described above and quantum physics? The relation becomes clearer once we consider the postulates which often, see for example \REFS\cite{Weinberg2015,Nielsen:2011:QCQ:1972505,GustafsonSigal2011}, provide the foundation of the theoretical framework of quantum theory. In this thesis, we order the postulates in the same way as the authors of \REF\cite{Nielsen:2011:QCQ:1972505}. The first postulate states that an isolated physical system is completely described by a state vector $\ket{\psi}\in\mathcal{H}\colon\innerproduct{\psi}{\psi}=1$ in a Hilbert space $\mathcal{H}$. Note that the postulate does not specify the dimensionality of the Hilbert space. However, for simplicity and notational convenience we assume that $\mathcal{H}$ is finite-dimensional,\ie$\mathcal{H} \rightarrow \mathcal{H}^{D}$. A motivation for this assumption is provided in \secaref{sec:The single-qubit space}{sec:TheMultiQubitSpace}. The second postulate states that the evolution of a closed system is described by a unitary operator $\OP{U}$ which maps an initial state $\ket{\psi}$ to a final state $\ket{\psi^{\prime}}$ at some point later in time. In \secref{sec:FromStaticsToDynamics} we discuss this postulate in more detail. Both postulates together provide us with a conceptual bridge to the computational processes described above.

Another question we find ourselves confronted with is: how can we extract information from the abstract state vectors? In the following, we assume that this extraction of information is governed by the third postulate. Here we assume that the measurement, of some unspecified discrete physical observable, is governed by a set $\{\OP{E}_{i}\}\subseteq \mathbb{L}(\mathcal{H}^{D})$ of operators, where $i \in I \subseteq \mathbb{N}^{0}$ denotes a discrete variable and/or label which allows us to distinguish between different measurement outcomes,\ie discrete events. Here $I$ denotes an index set which contains the discrete indices $i$. In particular, if the system is in some arbitrary state $\ket{\psi}\in \mathcal{H}^{D}\colon \innerproduct{\psi}{\psi}=1$, we assume that the probability for measuring an event $i$ is given by the expression
\begin{equation}\label{eq:probabilty}
  p(i)=\braket{\psi|\OP{E}_{i}^{\dagger}\OP{E}_{i}|\psi}.
\end{equation}
The third postulate also determines the state vector which results from the measurement procedure
\begin{equation}
  \ket{\psi^{\prime}}=\frac{1}{\sqrt{p(i)}}\OP{E}_{i}\ket{\psi},
\end{equation}
where $\ket{\psi}$ is the state vector before the measurement and $\ket{\psi^{\prime}}$ is the state vector after the measurement. Here we assume that event $i$ has taken place. Since the probabilities $p(i)$ have to form a proper distribution, we require
\begin{equation}
  \sum_{i} \OP{E}_{i}^{\dagger}\OP{E}_{i} = \OP{I},
\end{equation}
as a normalisation condition.

Postulate one and two, refer to closed or isolated systems but often we are interested in physical systems which consist of two or more subsystems. In the following, we assume that the composition of two or more subsystems is governed by the fourth postulate. This postulate states that the Hilbert space $\mathcal{H}^{\bar{D}}$ of the composed system has to obey the algebraic rules of a specific tensor product structure. If we assume that $\mathcal{H}^{D}$ and $\mathcal{H}^{\tilde{D}}$ are different Hilbert spaces in accordance with the first postulate, the fourth postulate states that the state vectors of the composed system are elements of the space
\begin{equation}
  \mathcal{H}^{\bar{D}}=\mathcal{H}^{D} \tens{} \mathcal{H}^{\tilde{D}},
\end{equation}
where $\otimes$ is the so-called tensor product symbol. We define the tensor product structure axiomatically,\ie we assume that the equations
\begin{subequations}\label{eq:deftens}
	\begin{align}
	\left( \ket{\psi}_{D} + \ket{\bar{\psi}}_{D} \right) \otimes \ket{\psi}_{\tilde{D}} &= \ket{\psi}_{D} \otimes \ket{\psi}_{\tilde{D}} + \ket{\bar{\psi}}_{D} \otimes \ket{\psi}_{\tilde{D}}, \\
	 \ket{\psi}_{D}  \otimes	\left( \ket{\psi}_{\tilde{D}} + \ket{\bar{\psi}}_{\tilde{D}} \right) &= \ket{\psi}_{D} \otimes \ket{\psi}_{\tilde{D}} + \ket{\psi}_{D} \otimes \ket{\bar{\psi}}_{\tilde{D}}, \\
	 c \left(\ket{\psi}_{D} \otimes \ket{\psi}_{\tilde{D}}\right) &= \left(c \ket{\psi}_{D} \right) \otimes  \left(\ket{\psi}_{\tilde{D}} \right),\\
	 c \left(\ket{\psi}_{D} \otimes \ket{\psi}_{\tilde{D}}\right) &= \left( \ket{\psi}_{D} \right) \otimes  \left(c \ket{\psi}_{\tilde{D}} \right).
	\end{align}
\end{subequations}
are true, by definition, for all $c\in \mathbb{C}$, $\ket{\psi}_{D},\ket{\bar{\psi}}_{D} \in \mathcal{H}^{D} $ and $\ket{\psi}_{\tilde{D}},\ket{\bar{\psi}}_{\tilde{D}} \in \mathcal{H}^{\tilde{D}}$. Every Hilbert space $\mathcal{H}^{\bar{D}}$ also needs an inner product $\braket{\mdot|\mdot}\colon \mathcal{H}^{\bar{D}} \times \mathcal{H}^{\bar{D}} \rightarrow \mathbb{C}$. This function can be defined in terms of the inner products which are defined on the spaces $\mathcal{H}^{D}$ and $\mathcal{H}^{\tilde{D}}$. We define the inner product on $\mathcal{H}^{\bar{D}}$ as
\begin{equation}
  \braket{\bar{\psi}_{D} \otimes \bar{\psi}_{\tilde{D}}|\psi_{D} \otimes \psi_{\tilde{D}}} =  {\braket{\bar{\psi}|\psi}}_{D}  \,   \braket{\bar{\psi}|\psi}_{\tilde{D}},
\end{equation}
for all $\ket{\psi}_{D},\ket{\bar{\psi}}_{D} \in \mathcal{H}^{D} $ and $\ket{\psi}_{\tilde{D}},\ket{\bar{\psi}}_{\tilde{D}} \in \mathcal{H}^{\tilde{D}}$. Additionally, one can show, see \REF\cite{Nielsen:2011:QCQ:1972505}, that the dimensionality of the tensor product space $\mathcal{H}^{\bar{D}}$ is given by the relation
\begin{equation}
  \text{dim}(\mathcal{H}^{\bar{D}})=\text{dim}(\mathcal{H}^{D}) \text{dim}(\mathcal{H}^{\tilde{D}}).
\end{equation}

The algebraic relations stated above enable us to consider operators,\ie linear functions, which act on the states of the tensor-product space $\mathcal{H}^{\bar{D}}$. For example, if $\OP{A}\in \mathbb{L}(\mathcal{H}^{D})$, we can define the operator $\OP{A}\otimes\OP{I}$ such that
\begin{equation}
    \OP{A}\tens{}\OP{I} \left(\ket{\psi}_{D} \otimes \ket{\psi}_{\tilde{D}}\right)=\left(\OP{A}\ket{\psi}_{D} \otimes \OP{I}\ket{\psi}_{\tilde{D}}\right),
\end{equation}
for all $\ket{\psi}_{D} \in \mathcal{H}^{D} $ and $\ket{\psi}_{\tilde{D}} \in \mathcal{H}^{\tilde{D}}$. Additionally, we can define operators, in a similar manner, for all the other subspaces we model. A useful identity relation which relates such operators,\eg $\OP{A}\otimes \OP{I}$ and $\OP{B}\otimes\OP{I}$, reads
\begin{equation}\label{eq:productoperator}
  \BRR{\OP{A}\tens{}\OP{I}}\BRR{\OP{I}\tens{}\OP{B}}=\OP{A}\tens{}\OP{B}.
\end{equation}
We also find that all operators of the form $\OP{A}\otimes \OP{I}$ and $\OP{I}\otimes\OP{B}$ commute,\ie if we change the order of the product on the left-hand side of \equref{eq:productoperator}, we see that the result is still given by the right-hand side of \equref{eq:productoperator}. Note that in the following sections we also encounter operators which cannot be decomposed in the sense of \equref{eq:productoperator}.

In the end, we emphasise that so far we mainly provided definitions and stated results without mathematical prove. We refer the reader to \REFS\cite{Nielsen:2011:QCQ:1972505,Watrous2018} for more detailed discussions of the subjects we covered in this section. Furthermore, throughout this section we assumed that the Hilbert spaces are finite dimensional and the measurement events correspond to discrete observables. We can also describe observables which are continuous, in a more general mathematical framework. Discussions of the framework which cover both cases,\ie the discrete and the continuous case, can be found in \REFS\cite{Weinberg2015,GustafsonSigal2011}.

\section{The single-qubit Hilbert space}\label{sec:The single-qubit space}
In the previous section we introduced the algebraic calculus which is being used to model the IGQC. In this section we continue the review of the model by defining the single-qubit Hilbert space. We begin the construction process with a discussion of the Hilbert space $\mathcal{H}^{2}\subseteq \mathbb{C}^{2}$. This space is two dimensional. Consequently, we can express all state vectors $\ket{\psi}\in \mathcal{H}^{2}\colon \innerproduct{\psi}{\psi}=1$ in this space as
\begin{equation}\label{eq:qubit}
  \ket{\psi}= c_{0} \ket{0}+c_{1} \ket{1},
\end{equation}
where $\ket{0}$ and $\ket{1}$ are the basis states, whose existence is guaranteed by an existence theorem, see \REF\cite{HoffmanKunze71}. The coefficients have to satisfy the relation $|c_{0}|^{2} + |c_{1}|^{2} =1$ so that the state vectors are normalised. In the following, we assume that the states $\ket{0}$ and $\ket{1}$ are given by the eigenstates of the Pauli $\OP{\sigma}^{(z)}$ operator. We define this operator in the following manner
\begin{equation}
\OP{\sigma}^{(z)} \ket{z} = \begin{cases}
 +1 \ket{z} &\text{ iff } z=0\\
 -1 \ket{z} &\text{ iff } z=1.
\end{cases}
\end{equation}
In addition, we can define two other Pauli operators. The Pauli $\OP{\sigma}^{(y)}$ operator is defined as
\begin{equation}
\OP{\sigma}^{(y)} \ket{z} = \begin{cases}
 +i \ket{\neg z} &\text{ iff } z=0\\
 -i \ket{\neg z} &\text{ iff } z=1,
\end{cases}
\end{equation}
where $\neg\colon\{0,1\}\rightarrow \{0,1\}$ is the so-called bit-flip operator which maps a zero to a one and vice versa. Similarly, the Pauli $\OP{\sigma}^{(x)}$ operator is defined as
\begin{equation}
\OP{\sigma}^{(x)} \ket{z} = \begin{cases}
 1 \ket{\neg z} &\text{ iff } z=0\\
 1 \ket{\neg z} &\text{ iff } z=1.
\end{cases}
\end{equation}

A qubit is, by definition, a state $\ket{\psi}\in \mathcal{H}^{2}\colon \innerproduct{\psi}{\psi}=1$ in the sense of \equref{eq:qubit},\ie a normalised state in the Hilbert space $\mathcal{H}^{2}$. It is possible to represent all physically distinctive states in a graphical way. Here we map the different states to the points of a three-dimensional unit sphere in $\mathbb{R}^{3}$. To this end, we recast \equref{eq:qubit} by making use of the normalisation condition $\innerproduct{\psi}{\psi}=1$. This means we introduce the angles and/or variables $\vartheta\in[0,\pi]$ and $\theta\in [0,2\pi)$ such that the state vector reads
\begin{equation}
  \ket{\psi}= \cos\BRR{\frac{\vartheta}{2}}\ket{0}+ \sin\BRR{\frac{\vartheta}{2}}e^{i\theta}\ket{1}.
\end{equation}
If we compute the vector
\begin{equation}
  \mathbf{r}=\BRR{\braket{\psi|\OP{\sigma}^{(x)}|\psi},\braket{\psi|\OP{\sigma}^{(y)}|\psi},\braket{\psi|\OP{\sigma}^{(z)}|\psi}},
\end{equation}
we find
\begin{equation}
  \mathbf{r}=\BRR{\cos\BRR{\theta}\sin\BRR{\vartheta}, \sin\BRR{\theta}\sin\BRR{\vartheta}, \cos\BRR{\vartheta}}.
\end{equation}
Consequently, we see that the vector $\mathbf{r} \in \mathbb{R}^{3}$ is given by the well known expression for a cartesian vector of length one in spherical coordinates, where $\vartheta$ denotes the azimuthal angle and $\theta$ is referred to as the polar angle. In this picture, we might understand an arbitrary computation as an operation which is composed of rotations around the x-, y- and z-axes. We can express these rotations as
\begin{subequations}\label{eq:single_qubit_rot}
  \begin{align}
    \OP{R}^{(x)}\BRR{\gamma}=e^{-i\frac{\gamma}{2}\OP{\sigma}^{(x)}},\\
    \OP{R}^{(y)}\BRR{\gamma}=e^{-i\frac{\gamma}{2}\OP{\sigma}^{(y)}},\\
    \OP{R}^{(z)}\BRR{\gamma}=e^{-i\frac{\gamma}{2}\OP{\sigma}^{(z)}},
  \end{align}
\end{subequations}
where $\gamma\in[0,\pi)$ denotes the rotation angle and the orientation of the rotation is given by the right-hand rule. We find, see \REF\cite{Nielsen:2011:QCQ:1972505}, that all unitary operators $\OP{U}\in \mathbb{U}(\mathcal{H}^{2})$ can be expressed as
\begin{equation}\label{eq:rotation}
  \OP{U}=e^{i a}R^{(\mathbf{n})}(\gamma),
\end{equation}
where $a \in \mathbb{R}$ and the operator $R^{(\mathbf{n})}(\gamma)$ is given by the expression
\begin{equation}
  \OP{R}^{(\mathbf{n})}(\gamma)=e^{-i \frac{\gamma}{2} \mathbf{n} \cdot \OP{\bm{\sigma}}}.
\end{equation}
Here the vector $\mathbf{n}\in \mathbb{R}^{3}$ denotes a unit vector $\mathbf{n}=(n^{(x)},n^{(y)},n^{(z)})$ and $\OP{\bm{\sigma}}=(\OP{\sigma}^{(x)},\OP{\sigma}^{(y)},\OP{\sigma}^{(z)})$ is defined as a vector of Pauli operators. The operation in \equref{eq:rotation} can be visualised as a rotation around the vector $\mathbf{n}$. It is also possible to express $\OP{U}$ in terms of rotations around two axes only. For example, we can use the x-y-x representation
\begin{equation}
  \OP{U}=e^{i a} \OP{R}^{(x)}\BRR{\alpha}\OP{R}^{(y)}\BRR{\beta}\OP{R}^{(x)}\BRR{\gamma},
\end{equation}
where $\alpha,\beta,\gamma \in \mathbb{R}$. Note that the phase factors $e^{i a}$ have no measurable effect on the state vector.

Such decompositions might be used to implement arbitrary single-qubit operations $\OP{U}$ on PGQC or NIGQC. In later chapters, we use such a decomposition to implement an NIGQC model.

\section{The multi-qubit Hilbert space}\label{sec:TheMultiQubitSpace}

As mentioned before, we can understand a computation on an IGQC as a mapping between two states which are elements of a Hilbert space $\mathcal{H}^{D}$. In this section we construct this Hilbert space from smaller single-qubit Hilbert spaces by invoking the fourth postulate repeatedly, see \secref{sec:MathematicalFramework}. We define the Hilbert space of an $N$-qubit IGQC as the tensor-product space
\begin{equation}
  \mathcal{H}^{2^{N}}=\tensupper{n=0}{N-1}\mathcal{H}_{n}^{2},
\end{equation}
where $\mathcal{H}_{n}^{2}$ denotes the two-dimensional Hilbert space we discuss in \secref{sec:The single-qubit space}. The index $n\in \{0, ..., N-1\}$ allows us to express various mathematical objects in terms of entities which are only defined for the single-qubit spaces $\mathcal{H}_{n}^{2}$, see examples below. We begin with the set
\begin{equation}\label{eq:CBQ}
\mathcal{C}^{N} = \{\ket{\mathbf{z}} \in   \mathcal{H}^{2^{N}} | \exists! \mathbf{z} \in \mathcal{B}_{2}^{N} \colon \ket{\mathbf{z}}= \tensupper{n=0}{N-1} \ket{z_{n}} \},
\end{equation}
where $\mathcal{B}_{2}^{N}=\{0,1\}^{N}$ denotes a set of N-tuples and $\ket{z_{n}} \in \mathcal{H}_{n}^{2}$ are the Pauli $\OP{\sigma}^{(z)}$ operator eigenstates we discuss in \secref{sec:The single-qubit space}. One can show, see \REF\cite{Nielsen:2011:QCQ:1972505}, that this set constitutes a basis for the space $\mathcal{H}^{2^{N}}$. In the following, we denote $\mathcal{C}^{N}$ as the computational basis. An arbitrary $N$-qubit state $\ket{\psi}\in\mathcal{H}^{2^{N}}$ can be expressed as
\begin{equation}\label{eq:statevector}
  \ket{\psi}=\sum_{\mathbf{z}\in \mathcal{B}_{2}^{N}} c_{\mathbf{z}} \ket{\mathbf{z}},
\end{equation}
where the coefficients of the state vector have to satisfy the relation
\begin{equation}\label{eq:coeffconst}
  \sum_{\mathbf{z}\in \mathcal{B}_{2}^{N}} |c_{\mathbf{z}}|^{2} = 1.
\end{equation}
\sequref{eq:coeffconst} provides us with an intuitive way of looking at the states $\ket{\psi}\in\mathcal{H}^{2^{N}}$. We might interpret the different states as points on a high-dimensional hypersphere with radius one.

Addressing the different coefficients $c_{\mathbf{z}}\in \mathbb{C}$ by means of the bit strings $\mathbf{z}\in\mathcal{B}_{2}^{N}$ is often very convenient for analytical and/or numerical work. For example, we can define an arbitrary single-qubit operator $\OP{\mathcal{O}}_{n^{\prime}} \in \mathbb{U}(\mathcal{H}^{2^{N}})$ for an arbitrary operator $\OP{\mathcal{O}} \in \mathbb{U}(\mathcal{H}^{2})$ by means of the expression
\begin{equation}\label{eq:singlequbitoperatordef}
  \forall \mathbf{z}\in\mathcal{B}_{2}^{N}\colon \OP{\mathcal{O}}_{n^{\prime}} \ket{\mathbf{z}}=\tensupper{n=0}{N-1} \OP{\mathcal{O}}^{\delta_{n,n^{\prime}}} \ket{z_{n}},
\end{equation}
where $\delta_{n,n^{\prime}}$ denotes the Kronecker delta. This operator can also be expressed as
\begin{equation}
  \OP{\mathcal{O}}_{n^{\prime}}=\tensupper{n=0}{N-1} \OP{\mathcal{O}}^{\delta_{n,n^{\prime}}},
\end{equation}
see \equref{eq:productoperator}. The index $n^{\prime}$ determines which state $\ket{z_{n}}$ in \equref{eq:singlequbitoperatordef} is affected by the operator $\OP{\mathcal{O}}$. Therefore, we refer to $n^{\prime}$ as the target and/or qubit index. Note that an operator is fully defined once its action on a basis is determined.

In the model of the IGQC the operators $\OP{\mathcal{O}}_{n^{\prime}}$ are usually referred to as single-qubit gates and not operators. In the following, we follow this convention. One usually differentiates between single-qubit gates and the so-called multi-qubit gates. As the name indicates, multi-qubit gates affect more than one qubit index at a time. We can further distinguish between multi-qubit gates which can be expressed as a composition of single-qubit gates in the sense of \equref{eq:productoperator},\eg $\OP{\mathcal{O}}_{n^{\prime}} \otimes \OP{\mathcal{O}}_{\bar{n}}$ and multi-qubit gates which cannot be expressed in terms of a composition of single-qubit gates in the sense of \equref{eq:productoperator}. An example for the latter type of gate is given by the controlled gate $C\OP{\mathcal{O}}_{n^{\prime},\bar{n}}$, where $n^{\prime}$ is the so-called control index and $\bar{n}$ is again referred to as the target index. We define this gate as
\begin{equation}\label{eq:controlledgatedef}
  \forall \mathbf{z}\in\mathcal{B}_{2}^{N}\colon C\OP{\mathcal{O}}_{n^{\prime},\bar{n}} \ket{\mathbf{z}}= \OP{\mathcal{O}}_{\bar{n}}^{z_{n^{\prime}}} \ket{\mathbf{z}},
\end{equation}
where $z_{n^{\prime}}\in \{0,1\}$ is the $n^{\prime}$-th bit string component of the vector $\ket{\mathbf{z}}$ on the right-hand side of \equref{eq:controlledgatedef}. Consequently, we find that this operator only acts on the basis state $\ket{\mathbf{z}}$ for bit strings where $z_{n^{\prime}}=1$.

In \secref{sec:The single-qubit space} we discuss the procedure of decomposing an arbitrary operator $\OP{U}\in \mathbb{U}(\mathcal{H}^{2})$ into a composition of rotations around various axes. It is possible to show that something similar can be achieved in the multi-qubit space $\mathcal{H}^{2^{N}}$, with the operators $\OP{U}\in \mathbb{U}(\mathcal{H}^{2^{N}})$. We may express this result as follows: all unitary operators $\OP{U}$ can be approximated, with arbitrary precision $\epsilon>0$, by means of a composition
\begin{equation}
  \OP{T}=\produpper{i=0}{|I|-1} \OP{\mathcal{O}}_{i},
\end{equation}
of a finite sequence $(\OP{\mathcal{O}}_{i})_{i\in I\subseteq \mathbb{N}^{0}}$ of single-qubit and two-qubit gates $\OP{\mathcal{O}} \in \mathbb{U}(\mathcal{H}^{2^{N}})$. The finite length $|I|$ of the sequence is determined by the precision $\epsilon$ we require. Additionally, one can show that this finite sequence $(\OP{\mathcal{O}}_{i})_{i\in I\subseteq \mathbb{N}^{0}}$ can be constructed with only a handful of elementary gates. This set of elementary gates is usually referred to as a universal gate set. Note that there exists more than one such universal gate set. One of these sets, the so-called standard set, reads
\begin{equation}
  G=\{ C\OP{X}, \OP{H}, \OP{T}, \OP{S}\},
\end{equation}
where $\OP{X}, \OP{H}, \OP{T}, \OP{S} \in \mathbb{U}(\mathcal{H}^{2})$ are defined as
\begin{subequations}\label{eq:UniversalGateSet}
  \begin{align}
    \OP{X}\ket{z}&=\ket{\neg z},\\
    \OP{H}\ket{z}&=\frac{1}{\sqrt{2}}\BRR{(-1)^{z}\ket{z}+\ket{\neg z}},\\
    \OP{T}\ket{z}&=e^{z \frac{\pi}{4} i}\ket{z},\\
    \OP{S}\ket{z}&=e^{z \frac{\pi}{2} i}\ket{z}.
  \end{align}
\end{subequations}
Note that $C\OP{X}$ is often referred to as $\text{CNOT}$ gate and $\OP{X}=\OP{\sigma}^{(x)}$. Here, we simply use the letter X to address the corresponding Pauli operator, it is common practice to do the same with the $\OP{Y}=\OP{\sigma}^{(y)}$ and $\OP{Z}=\OP{\sigma}^{(z)}$ operators. The author would like to stress that the discussion presented in this section is far from exhaustive. The interested reader might consider \REFS\cite{Nielsen:2011:QCQ:1972505,Dawson2006,Barenco95} for more detailed information regarding this subject.

\section{Simulation of the ideal gate-based quantum computer}\label{sec:SimulationOfTheIdealGateBasedQuantumcomputer}
\newcommand{\spacelabel}{\text{Space}}
\newcommand{\timelabel}{\text{Time}}
In this section, we consider the subject of numerically modelling an IGQC. To this end, we intend to discuss the question, how can we simulate the execution of multiple computations $\OP{U}_{i} \in \mathbb{U}(\mathcal{H}^{2^{N}})$ in a computationally friendly way? Here, $i \in \{0, ..., I-1\}$ is an index variable and $I\in \mathbb{N}$ denotes the number of gates we intend to implement. There exist two main types of algorithms for this kind of problem, namely the Schrödinger algorithm and the Feynman algorithm, see \REFS\cite{Willsch20NIC,LagemannMSCThesis}. We discuss both algorithms and begin with the former. However, before we come to the algorithms, we first discuss the subject of computational complexity. Comparing both algorithms in terms of complexity is probably the fastest way to understand the differences between both methods.

Finding the complexity of an algorithm amounts to analytically determining bounds for the memory and the runtime requirements that a code instance is bounded by. Such an analysis is usually based on several assumptions,\ie we usually make assumptions about the algorithm and/or program parameters. In this work, we are only interested in the worst case running time complexity and the worst case memory requirements. To this end, we make use of so-called asymptotic notation. The symbols $\mathcal{O}^{\spacelabel}(g(n))$ and $\mathcal{O}^{\timelabel}(f(n))$ denote the space and time complexity, respectively. These symbols can be understood as follows, for some $n_{0}\in \mathbb{N}$, we find that the runtime $r(n)$ and the memory requirement $m(n)$ obey the inequalities $0 \leq m(n) \leq c_{\spacelabel} g(n)$ and $0 \leq r(n) \leq c_{\timelabel} f(n)$, where $c_{\spacelabel},c_{\timelabel} \in \mathbb{R}^{+}$ and $n\geq n_{0}$. Here $n\in \mathbb{N}$ denotes a discrete variable, the reader may consider $n$ to be the only relevant variable for the algorithm we investigate,\ie the fictitious algorithm we talk about while introducing the notation. For an extensive discussion of this subject we refer the reader to \REFS\cite{Cormen09,AB06}.

If we use an implementation of the Schrödinger algorithm to simulate the IGQC, we are required to store all state vector coefficients $c_{\mathbf{z}}$, see \equref{eq:statevector}, during the execution of the program. We often use double precision for our computations, in such a case we find that the data structure for the state vector has the size $2^{N+4}$ Byte, where $N$ denotes the number of qubits we simulate. Note that it is also possible to use encoding schemes which require a little bit less memory, see \REF\cite{DERAEDT201947}. However, the memory requirements of the Schrödinger algorithm still scale exponentially with the number of qubits. In the following, we assume that the state vector data structure dominates the memory requirements completely, therefore we can write $\mathcal{O}^{\spacelabel}(2^{N})$ for the space complexity. The memory requirements of the Schrödinger algorithm are a consequence of the tensor-product structure, see \secref{sec:MathematicalFramework}, we use to model multi-qubit systems.

We now turn our attention to the time complexity. Here we assume that the different $\OP{U}_{i}$ are either single-qubit gates or two-qubit gates of the controlled type. Two-qubit or multi-qubit gates which can be expressed as compositions of single-qubit gates are also simulated as a series of single-qubit gates, see \secref{sec:TheMultiQubitSpace} for more details. Single-qubit gates require at most updates of two state vector coefficients $c_{\mathbf{z}}$ at the same time. Two-qubit gates of the controlled type on the other hand might require updates of four state vector components $c_{\mathbf{z}}$ at the same time. This means we can partition the state vector data structure into groups of two and four coefficients $c_{\mathbf{z}}$ which we update together, sequentially. Consequently, we have to perform $2^{N}/2$ and $2^{N}/4$ updates in total and each update rule comes with a fixed number of floating-point operations. Therefore, we find that the total number of operations scales with $2^{N}$ and therefore we write $\mathcal{O}^{\timelabel}(2^{N})$. If we make use of parallel computing, we can reduce the run time for the simulation of a single gate to a certain extent. However, ultimately we cannot compensate for the exponential scaling of the runtime with $N$,\ie this would require nearly unlimited ideal hardware resources.

If we use an implementation of the Feynman algorithm to simulate the IGQC, we are only required to store a fraction of all state vector coefficients $c_{\mathbf{z}}$, see \equref{eq:statevector}, during the execution of the program. Additionally, since the computations of the different coefficients $c_{\mathbf{z}}$ are completely independent, we usually compute the coefficients one at a time.

The idea of the Feynman algorithm is to decompose certain two-qubit gates of the controlled type, which are part of the sequence of $\OP{U}_{i}$, in such a way that we can compute the different coefficients $c_{\mathbf{z}}$ by simulating a series of smaller qubit systems with sizes $P=\{P_{0},...,P_{|P|-1}\}\subseteq \mathbb{N}$. Here $P \subseteq \mathbb{N}$ denotes the set of subsystem sizes and $|P|$ refers to the cardinality of $P$. The simulations of the smaller qubit systems are performed with the Schrödinger algorithm and since we are in control of the sizes $P$, we find that we can overcome the memory bottleneck of a pure Schrödinger simulation. In the end, we combine the simulation results of the different subsystems and compute a single coefficient $c_{\mathbf{z}}$. However, we also pay a price for the reduced memory requirements. The amount of computations required to perform the Feynman simulation increases exponentially with the number of decomposed gates $S\in\mathbb{N}$ in the sequence $\OP{U}_{i}$.

If we consider the Feynman algorithm discussed in \REF\cite{LagemannMSCThesis}, we find that the space complexity of the algorithm is given by $\mathcal{O}^{\spacelabel}(2^{\max(P)})$, where $\max(P)$ denotes the maximum over all subsystem qubit sizes. The corresponding time complexity reads $\mathcal{O}^{\timelabel}(2^{S} |P| 2^{\max(P)})$. A thorough discussion of this algorithm, where the complexity results are derived properly, can be found in \REF\cite{LagemannMSCThesis}, which contains previous academic work by the author. This work is concerned with the development and implementation of a simulator for the IGQC. Here, a Feynman algorithm was used.

For this thesis the simulation code discussed in \REF\cite{LagemannMSCThesis} was extended so that one has the choice between an implementation of the Schrödinger and the Feynman algorithm. The IGQC simulator is integrated into a larger simulation software framework, see \secref{sec:StructureOfTheSimulationSoftware}.

\section{Algorithms}\label{sec:Algorithms}
\newcommand{\QFT}{\text{QFT}}

So far, we mainly considered computations in terms of abstract mappings $\OP{U} \in \mathbb{U}(\mathcal{H}^{2^{N}})$ between two states of the Hilbert space $\mathcal{H}^{2^{N}}$, for some fixed number of qubits $N$. In this section, we extend the picture of the computational process. To this end, we discuss various algorithms which at least require the implementation of a mapping $\OP{U}$ as part of a larger computational process. Note that this discussion is non-exhaustive, we simply intend to provide the reader with a better understanding of how the computation in terms of $\OP{U}$ fits into the bigger picture of things. We try to focus on algorithms and/or subjects which are relevant for the results of this thesis. Furthermore, the following discussion is rather general in the sense that we avoid a detailed mathematical analysis of the different algorithms. This lack of detail allows us to cover more ground. We refer the reader to specific references if necessary.

Broadly speaking, we might define, see \REF\cite{Cormen09}, algorithms as sequences of computational rules and/or steps which take well-defined data structures as inputs in order to produce well-defined output data structures. Furthermore, an algorithm is usually associated with a specific computational problem so that the input data structure represents a problem instance and the output data structure is considered to be a potential solution. With this in mind, we might classify algorithms as follows. An algorithm belongs to the class of correct algorithms if and only if for every input, the algorithm stops and the output is the correct solution to the computational problem, with absolute certainty. Algorithms which belong to the class of incorrect algorithms possibly do not terminate,\ie the output is simply not guaranteed to be the correct solution for the given problem after a finite number of computational steps. Obviously, we can classify algorithms by means of different criteria. For example, a classification system which takes into account the complexity, see \secref{sec:SimulationOfTheIdealGateBasedQuantumcomputer}, of algorithms is a very useful tool to better understand the relations between different classes of algorithms and/or problems, see \REFS\cite{Cormen09,AB06} for more details. In the following we discuss three different algorithms: the quantum Fourier transformation algorithm, Shor's factoring algorithm and variational hybrid algorithms.

We begin with the quantum Fourier transformation (QFT) algorithm. The input for this algorithm is some state vector $\ket{\psi} \in \mathcal{H}^{2^{N}}\colon \innerproduct{\psi}{\psi}=1$. For the moment we ignore the input state so that we can focus on the action of the QFT on the basis states $\ket{\mathbf{z}} \in \mathcal{C}^{N}$. The action of the QFT on these states reads
\begin{equation}
  \OP{\QFT}\ket{\mathbf{z}}=\frac{1}{\sqrt{2^{N}}} \sum_{\mathbf{z^{\prime}} \in \mathcal{B}_{2}^{N}} e^{2\pi i  \frac{z z^{\prime}}{2^{N}}} \ket{\mathbf{z^{\prime}}},
\end{equation}
where $z \in \mathbb{N}$ denotes the integer $z=\sum_{n=0}^{N-1} z_{n} 2^{n}$. Consequently, the action of the operator $\OP{\QFT}$ on an arbitrary state vector $\ket{\psi}$ can be expressed as
\begin{equation*}
  \sum_{\mathbf{z}\in \mathcal{B}_{2}^{N}} c_{\mathbf{z}} \ket{\mathbf{z}} \mapsto \sum_{\mathbf{z}^{\prime} \in \mathcal{B}_{2}^{N}} d_{\mathbf{z}^{\prime}} \ket{\mathbf{z}^{\prime}},
\end{equation*}
where the state vector coefficients $d_{\mathbf{z}^{\prime}}$ are given by
\begin{equation}\label{eq:DFT}
  d_{\mathbf{z}^{\prime}}=\frac{1}{\sqrt{2^{N}}} \sum_{\mathbf{z}\in \mathcal{B}_{2}^{N}} c_{\mathbf{z}} e^{2\pi i  \frac{z z^{\prime}}{2^{N}}}.
\end{equation}
We can use \equref{eq:DFT} as the definition for the discrete Fourier transformation,\ie by definition the discrete Fourier transformation maps the components $c_{\mathbf{z}}$ to the components $d_{\mathbf{z}^{\prime}}$. While in theory the quantum Fourier transformation algorithm exhibits an advantage in terms of time complexity, over classical Fourier transformation algorithms, we find that extracting the complex coefficients $d_{\mathbf{z}^{\prime}}$ and preparing the state $\ket{\psi}$ is a major practical problem, see \REF\cite{Nielsen:2011:QCQ:1972505}. In principle, information extraction and state preparation are governed by postulate three, see \secref{sec:MathematicalFramework}, which means we have no direct access to the complex coefficients $c_{\mathbf{z}}$ and $d_{\mathbf{z}^{\prime}}$. However, we can use this algorithm as an intermediate computational step as part of another algorithm, namely Shor's factoring algorithm, see \REFS\cite{shor1994factoring,shor1997algorithm}. This is the algorithm we discuss next.

The computational problem Shor's algorithm aims to solve can be expressed as follows. Assume we are given an integer $M=p q$, where $p,q \in \mathbb{N}$ are prime numbers. The question then is what are the prime factors $p,q$ of the integer $M$. The algorithm can be split into two parts. The first part can be executed on a classical computer. Here we transform the factoring problem into an order-finding problem. If $m \in \mathbb{N}\colon m < M$ and $m$ and $M$ have no common factors, we say that the order of $m$ modulo $M$ is the smallest integer $r\in \mathbb{N}$ which satisfies $m^{r}=1$ modulo $M$. The second part is meant to be executed on a quantum computer. Here we solve the order-finding problem by applying another algorithm which in turn relies on the QFT algorithm. The crux of Shor's algorithm is that the factoring problem can potentially be solved in polynomial time,\ie with a time complexity which is given by a polynomial. To the best knowledge of the author, no classical state-of-the-art implementable algorithm can solve this problem in polynomial time. This is why many cryptography systems are built on the assumption that, in the foreseeable future, no algorithm can be implemented which solves this problem in polynomial time. Shor's algorithm provides an example where many algorithms are concatenated in the sense that the computational rules are algorithms themselves. Therefore, Shor's algorithm provides us with a natural example of a problem where the original computational problem gets transformed, subproblems are solved with various algorithms and in the end the processes are aligned in the correct order. This is probably the best way to think about computing in general. We are allowed to creatively combine all resources at our disposal, the only important criterion is that the end result is the correct one.

The next algorithm or rather the next class of algorithms we discuss combines the generation of a specific state vector $\ket{\psi(\mathbf{x})} \in \mathcal{H}^{2^{N}}\colon \innerproduct{\psi}{\psi}=1$ with classical optimisation algorithms. Here $\mathbf{x}\in \mathbb{R}^{D^{\prime}}$ denotes a parameter vector and for large $N$ we assume that $D^{\prime}$ is substantially smaller than $D=2^{N}$. The class of variational hybrid algorithms may be characterised as follows. Assume we are given the following inputs: an initial state $\ket{\mathbf{0}} \in \mathcal{H}^{2^{N}}$, a set of gates which form the unitary operator $\OP{U}(\mathbf{x}) \in \mathbb{U}(\mathcal{H}^{2^{N}})$ and a Hermitian operator $\OP{H} \in \mathbb{H}(\mathcal{H}^{2^{N}})$. Here $\mathbb{H}(\mathcal{H}^{2^{N}}) \subseteq \mathbb{L}(\mathcal{H}^{2^{N}})$ denotes the set of all Hermitian operators,\ie $\OP{H} \in \mathbb{H}(\mathcal{H}^{2^{N}})$ implies $\OP{H}=\OP{H}^{\dagger}$. We can then use $\OP{U}(\mathbf{x})$ to compute the parameterised state vector
\begin{equation}
  \ket{\psi(\mathbf{x})}=\OP{U}(\mathbf{x})\ket{\mathbf{0}},
\end{equation}
on an IGQC. Then we use this state to determine the function value
\begin{equation}
  C(\mathbf{x})=\braket{\psi(\mathbf{x})|\OP{H}|\psi(\mathbf{x})},
\end{equation}
for a fixed $\mathbf{x}$. The function $C\colon \mathbb{R}^{D^{\prime}} \rightarrow \mathbb{R}$ is the so-called cost function. Note that the cost-function values are always real, this is a consequence of the fact that $\OP{H}$ is Hermitian. We might think about this function as some kind of subroutine which runs on an IGQC. A classical optimisation algorithm can call this subroutine in order to minimise the cost function $C$.

The procedure described above is motivated by the variational principle. This principle states that
\begin{equation}
   C(\mathbf{x})\geq E_{0},
\end{equation}
is true for all $\mathbf{x} \in \mathbb{R}^{D^{\prime}}$, see \REF\cite{Weinberg2015}. Here $E_{0}$ denotes the minimum of the eigenvalues of $\OP{H}$. Since $\OP{H}$ is defined on a finite-dimensional Hilbert space, we find that the existence of the minimum is guaranteed. Note that the states $\ket{\psi(\mathbf{x})}$ are normalised. The variational principle guarantees that we cannot surpass $E_{0}$ while minimising $C(\mathbf{x})$. Therefore, we might understand this class of algorithms as a method to find the ground state,\ie the eigenstate which corresponds to the eigenvalue $E_{0}$ of a Hermitian operator $\OP{H}$. For a general $\OP{U}(\mathbf{x})$ there is no guarantee that this method converges to $E_{0}$. Therefore, we denote this method as a heuristic method to find the ground state of a specific Hermitian operator $\OP{H}$.

We can define a generic operator $\OP{H}$ in terms of the Pauli operators we discuss in \secref{sec:The single-qubit space}. The full operator reads
\begin{equation}\label{eq:VQEHamiltonian}
  \OP{H}=\sum_{m=0}^{M-1} J_{m} \tensupper{n=0}{N-1} \OP{O}_{m,n},
\end{equation}
where $\OP{O}_{m,n} \in \{\OP{I},\OP{\sigma^{(x)}},\OP{\sigma^{(y)}},\OP{\sigma^{(z)}}\}$. This operator is sufficiently general to cover a variety of interesting cases,\eg the Ising model or the Heisenberg model. In principle, we can choose $\OP{U}(\mathbf{x})$ at random. However, we prefer to generate parameterised states $\ket{\psi(\mathbf{x})}$ which, for some reason, are likely to converge to the ground state for a given $\OP{H}$. It is sometimes the case that different states $\ket{\psi(\mathbf{x})}$ are associated with different algorithm names, see for example \REF\cite{farhi2014quantum} for the quantum approximate optimisation algorithm.

\section{Prototype gate-based quantum computers}\label{sec:Prototype gate-based quantum computers}

With this section we have reached a crucial juncture in this thesis. In the previous sections we talked about a static computer model,\ie all changes which affect the state vector $\ket{\psi}\in \mathcal{H}^{2^{N}}$ are in principle instantaneous. The IGQC is then simply an algebraic mathematical model and we connect this model to the physical theory and/or the experiment by means of the postulates we discussed in \secref{sec:MathematicalFramework}. In principle, we could formulate this computer model without considering the physical postulates.

\begin{figure}[!tbp]
    \centering
    \begin{minipage}{1.0\textwidth}
        \centering
        \includegraphics[width=\textwidth]{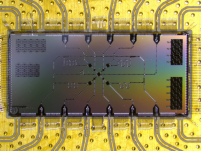}
    \end{minipage}\hfill
    \caption[Photograph of a chip which contains structures such that five superconducting qubits might be realised in an experiment.]{Photograph of a chip which contains structures such that five superconducting qubits might be realised in an experiment. The five dominant,\ie larger, circles in the centre of the photograph are supposed to be the qubits. The four small squares which connect the different qubits are the so-called coupler elements. The coupler elements are supposed to provide an indirect coupling between the different qubits. We can also see wave-like structures which are connected to all five qubits. These are used to read out the qubit state. The remaining structures are used to apply control pulses in terms of currents and voltages. With permission of Jonas Bylander from Chalmers University of Technology.}\label{fig:ChalmersChipImage}
\end{figure}

In this section, we discuss what we call are prototype gate-based quantum computers (PGQCs). There exist a variety of different PGQCs which aim to realise an IGQC, see \REF\cite{DiVincenzo13}. We focus our discussion on a specific type of PGQC: low-temperature superconducting integrated circuits and the experimental setups which operate these circuits. References \cite{Roth19,Krinner2020,Arute19,Bengtsson2020} discuss various experiments which make use of three different types of circuit architectures. Figure \ref{fig:ChalmersChipImage} shows a photograph of a superconducting integrated circuit,\ie a chip, which is supposed to form a five-qubit system. The five large circles in the middle of the photograph are meant to manifest the qubits. The four squares which connect the different qubits, via straight lines which form the letter X, are the so-called coupler elements. These elements provide an indirect coupling between the qubits. We can also see wave-like structures which are connected to the five qubits. These structures are used to read out the qubit state. The remaining structures on the chip can be used to apply control pulses in terms of currents and voltages, we denote these as feed lines. Therefore, we find that we might have to consider more system components than just the qubits themselves. In this case, we should probably at least take into account the couplers, read-out elements and feed lines.

\graphicspath{{./FiguresWithPermission/}}

Experiments, see \REF\cite{Naghiloo19}, show that the qubits on chips like the one in \figref{fig:ChalmersChipImage} are different from the qubits we define for the IGQC model. For example, if we initialise the qubit in state $\ket{1}$ at time $t_{0}$ and wait some time $t>t_{0}$, we usually find $p_{1}(t)<1$ for various times $t$. Here $p_{1}(t)$ denotes the probability for finding the qubit in the first-excited state. Additionally, if we consider the limit $t \gg t_{0}$, we often find that the qubit mainly resides in the ground state $\ket{0}$. The relaxation process can be described by an exponential function of the form $p_{1}(t)=a+b e^{-t/T_{1}}$, where $a$ is a real-valued constant which describes the population that resides in the state $\ket{1}$ after $ t \rightarrow \infty$. Furthermore, in this model the real-valued constants $b$ and $T_{1}$ describe the decay into the state $\ket{0}$. We denote the time $T_{1}$ as the relaxation time. State-of-the-art $T_{1}$ times are of the order of microseconds. Experimental qubits are usually characterised by more than just the relaxation law parameters. One of the most important parameters is the real-valued qubit frequency $\omega^{(01)}$. This parameter is determined by means of a spectroscopy experiment. Broadly speaking, we use one of the feed lines mentioned in our discussion of \figref{fig:ChalmersChipImage} to send a harmonic signal of frequency $\omega^{(D)}$ to the qubit. Then, the qubit frequency $\omega^{(01)}$ is the drive frequency $\omega^{(D)}$ which optimally excites the qubit from the ground $\ket{0}$ to the excited state $\ket{1}$ and vice versa. Another parameter of interest is the so-called dephasing time $T_{2}$. However, we do not intend to discuss this parameter here. Reference \cite{Naghiloo19} provides an introduction to the topic of qubit characterisation and measurement, including the relaxation time $T_{1}$, the qubit frequency $\omega^{(01)}$ and the dephasing time $T_{2}$.

\graphicspath{{./FiguresWithPermission/}}
\begin{figure}[!tbp]
    \centering
    \begin{minipage}{0.75\textwidth}
        \centering
        \includegraphics[scale=0.0727]{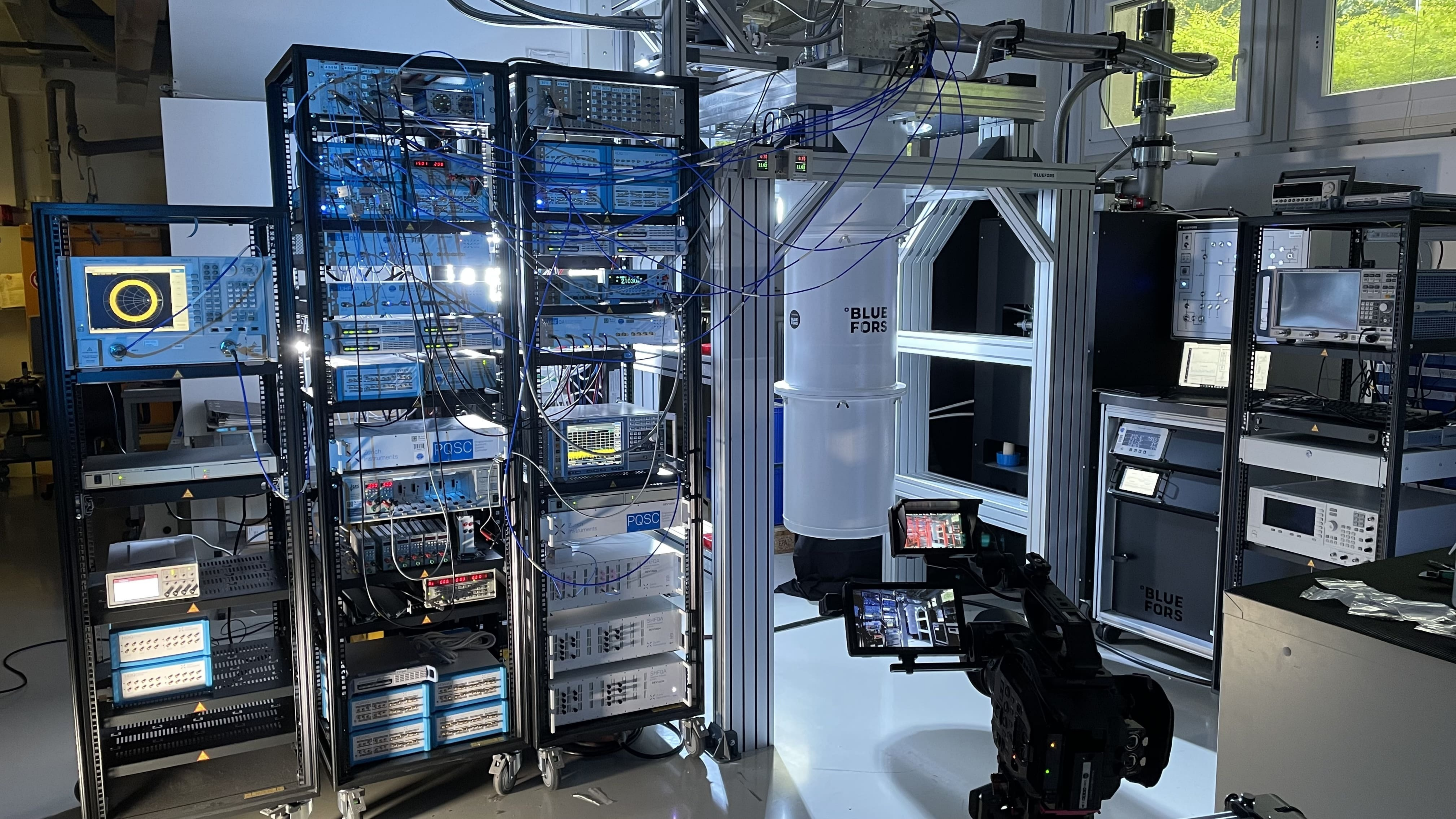}\\
        (a)
    \end{minipage}\hfill
    \begin{minipage}{0.25\textwidth}
        \centering
        \includegraphics[scale=0.0409]{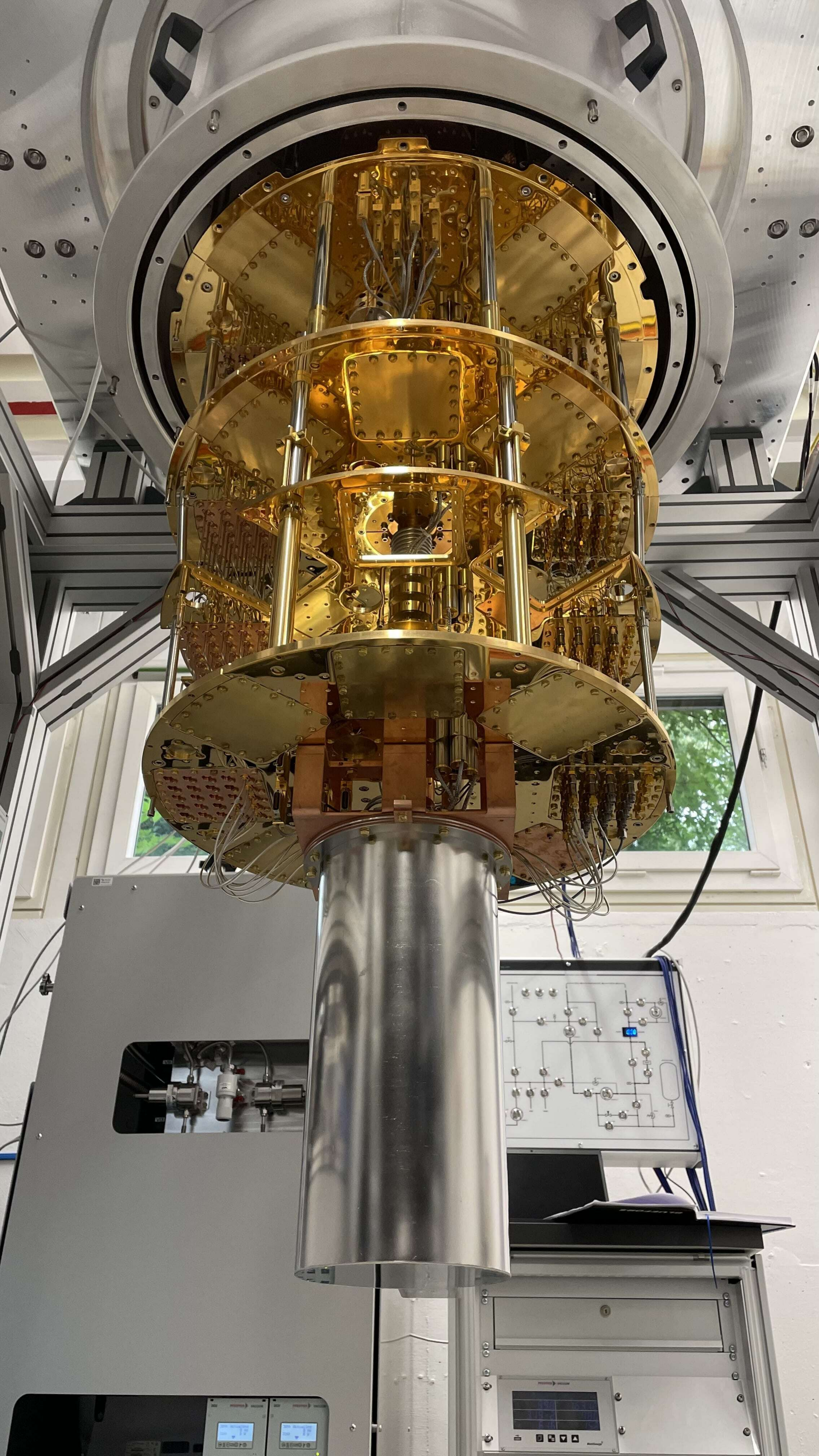}\\
        (b)
    \end{minipage}
    \caption[Photographs of an experimental setup to operate a superconducting integrated circuit at around 10 mK. ]{Photographs of an experimental setup to operate a superconducting integrated circuit at around 10 mK. Note that \figref{fig:ChalmersChipImage} shows a photograph of such a superconducting integrated circuit. The images show the same setup from different perspectives. The left photograph (a) shows a variety of control electronics which is connected to a closed cryostat, see white cylinder with the name tag Bluefors. The right photograph (b) shows the open cryostat from the bottom. We can see that the cryostat consists of different layers. Each layer usually corresponds to its own temperature zone. Additionally, at the bottom we can see a silver cylinder. This part of the setup is used to shield the chip from electromagnetic fields which might disturb the chip operation. See \REF\cite{Krinner2019} for more detailed information about the setup. With permission of Pavel Bushev from Forschungszentrum Jülich.}\label{fig:labsetup}
\end{figure}

Superconducting integrated circuits are usually part of a much larger experimental setup. Figure \ref{fig:labsetup} shows two photographs of an experimental setup to operate a chip at around 10 mK. Both images show the same setup from different perspectives. The left photograph in \figref{fig:labsetup}(a) shows different electronic devices, see the boxes in the rolling shelf carts. These devices are connected to a closed cryostat, see the white cylinder with the name tag Bluefors. The right photograph in \figref{fig:labsetup}(b) shows the open cryostat from the bottom. One can see different horizontal layers at the top of the image and a silver cylinder at the bottom of the image. The different layers correspond to different temperature zones within the setup, assuming the device is in operation. The silver cylinder is used as a magnetic shield. Note that such shields are usually made of superconducting materials. A much more detailed discussion of cryogenic setups is provided by \REF\cite{Krinner2019}.

The discussion so far should make one thing clear, namely that a complete, detailed, mathematical description of such a system is most likely an impossible task. The vast number of interdependent variables we have to consider is simply too great. For example, we might consider the variables material, see \REF\cite{Murray2021,Martinis14}, control signals, see \REF\cite{Rol19,Werninghaus2021}, temperatures at various points in the experimental setup, see \REF\cite{Krinner2019}, temporal stability of experimental parameters like $T_{1}$, $\omega^{(01)}$ and $T_{2}$, see \REF\cite{Burnett2019} and maybe even cosmic radiation, see \REF\cite{McEwen22}. Consequently, it seems to be impossible that we can formulate a model which describes all these aspects in one combined mathematical framework. However, we might be able to describe and/or explain some trends we see in the experimental data and gain some insight into the dynamics of certain system variables of importance.

Additionally, so far we have ignored another important issue, namely how we extract the qubit state data from the experimental setup. If we perform measurements with a PGQC, we find that we have to map the measurement data to discrete events, see \REF\cite{Naghiloo19}. These discrete events correspond to measuring the qubit states $\ket{0}$ and $\ket{1}$ and sometimes even other higher excited states like $\ket{2}$, see \REF\cite{Krinner21}. Note that these states are not part of the IGQC model. This is a consequence of the fact that the measurement protocols, see \REFS\cite{Naghiloo19}, do not naturally lead to perfectly differentiable discrete event data. This might be explained as follows, we repeat every experiment several times in the hope that we generate the exact same state vector or system state every time we perform the experiment. This is the state we intend to sample from, in the sense of the IGQC model $\ket{\psi}\in\mathcal{H}^{2^{N}}$. However, our discussion of the experimental setup suggests that this might not be the case. It is probably more realistic to assume that we sample from state vectors which are similar but not equal, due to all the environmental factors which play a role in the measurement process. Assuming, that this qualitative assessment is correct, we might even expect that the experimental data scatters to a certain extent and we have to employ data processing protocols which generate discrete event data. Note that not all experiments, see for example \REF\cite{Krinner21} and \REFS\cite{Andersen2020,Heinsoo18}, use the same data processing protocols, this makes a comparison of experiments and/or devices more difficult. Furthermore, the coupler states are usually not measured or only considered during the calibration of the control pulses, see \REF\cite{Bengtsson2020}. Consequently, we have to take into account the specific measurement protocol used if we compare the PGQC data to some mathematical model.

\section{Non-ideal gate-based quantum computers}\label{sec:FromStaticsToDynamics}
\newcommand{\parvec}{\mathbf{R}}
In \secsref{sec:MathematicalFramework}{sec:Algorithms} we discuss the IGQC model and in \secref{sec:Prototype gate-based quantum computers} we discuss superconducting PGQC. In this section we discuss another type of model: non-ideal gate-based quantum computers (NIGQCs). These models might be used to describe certain aspects of the PGQC. As discussed in \secref{sec:Prototype gate-based quantum computers}, superconducting PGQC are very complex devices and we should not be ignorant and attempt to describe the full device with one model. A NIGQC model should be used to study certain aspects of the PGQC in question.

We begin our discussion of NIGQC with the second postulate,\ie the postulate we discussed in \secref{sec:MathematicalFramework}. From here on, we assume that the time evolution of a closed physical system is governed by the time-dependent Schrödinger equation (TDSE)
\begin{equation}
  i\partial_{t}\ket{\Psi(t)}=\OP{H}(t)\ket{\Psi(t)},
\end{equation}
where $\OP{H}(t)=\OP{H}^{\dagger}(t)$ for all $t\in \mathbb{R}^{+}$,\ie $\OP{H}(t)$ is a Hermitian operator for every point in time. This operator generates the dynamics of the system and in the following we denote the operator $\OP{H}(t)$ as a Hamiltonian and the model,\ie the Hamiltonian in combination with the TDSE, as a Hamiltonian model. The formal solution of the TDSE can be expressed as
\begin{equation}
  \OP{\mathcal{U}}(t,t_{0})=\mathcal{T} \exp\BRR{-i \int_{t_{0}}^{t} \OP{H}(t^{\prime}) d t^{\prime}},
\end{equation}
where $\mathcal{T}$ denotes the time-ordering symbol. Since $\OP{H}(t)$ is Hermitian as defined above, we find that $\OP{\mathcal{U}}$ is unitary. Therefore, the assumption that the system dynamics is governed by a Hamiltonian model provides us with a way to determine an explicit operator $\OP{\mathcal{U}}$, once $\OP{H}(t)$ is fixed. Here we assume that we can determine $\OP{\mathcal{U}}$ analytically and/or numerically.

The science of physics, see for example \REFS\cite{Weinberg2015,Dirac1925,Blais2020circuit}, might provide us with systematic rules for determining operators $\OP{H}(t)$ whose predictions agree, to some extent, with most experiments performed in laboratories all over the world. However, these rules are not necessarily perfect and/or the corresponding models are so complex that we cannot solve the TDSE for the $\OP{H}(t)$ we derive. Therefore, we might consider simply guessing an operator $\OP{H}(t)$ or rather its form based on experience. Alternatively, we can also apply approximations to Hamiltonian models,\ie to the models which are too complex. The problem with this approach is that usually we do not know to what extent the results of the original and the approximate model deviate. This may lead to some kind of false confidence in one or both models.

The physical systems and/or experiments we intend to model are usually characterised by a set of parameters $\mathbf{R} \in \mathbb{R}^{D^{*}}$, where $D^{*} \in \mathbb{N}$, such that $\OP{H}(t)$ actually reads $\OP{H}(\parvec,t)$. Here $\parvec$ contains the parameters which characterise the static properties of a device as well as the dynamic features like the explicit form of $\OP{H}(t)$ at time $t > t_{0}$. Consequently, the solution of the TDSE should be expressed as $\ket{\Psi(\parvec,t)}$. It might be the case that the solutions $\ket{\Psi(\parvec,t)}$ and $\ket{\Psi(\parvec^{\prime},t)}$ for two close by parameter vectors $\parvec$ and $\parvec^{\prime}$ deviate quite strongly, at least after some time $t>t_{0}$. Therefore, we make this distinction here. We can make the scenario even more complex, by assuming that $\parvec \rightarrow \parvec(t)$ the parameter vectors are time dependent too. Experiments show that several quantities which are usually assumed to be constant actually fluctuate over time, see for example \REF\cite{Burnett2019}. Whether or not such a time dependence is relevant for an accurate description of the experiment and/or device has to be verified by considering the experimental data,\ie we have to study how the experimental data changes on different time scales.

This way of modelling computations $\OP{\mathcal{U}}$ in a finite-dimensional Hilbert space is quite different from the one we discuss in \secsref{sec:MathematicalFramework}{sec:Algorithms}. First, if we use a Hamiltonian model to implement $\OP{\mathcal{U}}$, we model a dynamic process which transforms the state vector continuously in case of an analytical model or as a real-time process in case of a numerical model. In the IGQC model the variable time does not play an explicit role,\ie here changes to the state vector occur instantaneously. Second, the qubit states we use for the Hamiltonian model are not necessarily decoupled from the extended state space,\ie the model we use is not necessarily a natural two-level model. If this is the case, we have to define a projection $\OP{P}$ to the computational subspace which is formed by the different qubit states. Note that we ourselves define what is considered to be a qubit state in the NIGQC model. We can then determine the operator $\OP{M}=\OP{P}\OP{\mathcal{U}}\OP{P}$ and by definition, consider the operator $\OP{M}$ to be our computational map, which we might compare to computational maps $\OP{U} \in \mathbb{U}(\mathcal{H}^{2^{N}})$ which are defined for the IGQC model. Third, we usually do not know how to choose the system parameters $\parvec$ such that the projection $\OP{M}$ perfectly fits a desired unitary operator $\OP{U}$.

In the following chapters we intend to make these issues more concrete. First we have to define a Hamiltonian which generates the dynamics of the system, see \chapref{chap:III}. Then we can build a numerical NIGQC model, see \chapref{chap:IV} and determine the parameters $\parvec$ which realise the computational map $\OP{M}$, see \chapref{chap:GET}. Additionally, we intend to study how the NIGQC models affect the implementation of simple algorithms, in comparison to the IGQC model, see \chapref{chap:GET}. Note that in principle we can relax the assumption that the dynamics is generated by a Hamiltonian model and consider other models, for example a Lindblad master equation model, see \REFS\cite{Lindblad76,GKS76}.

\section{Summary and conclusions}

In \secsref{sec:MathematicalFramework}{sec:Algorithms} we developed the model of the ideal gate-based quantum computer (IGQC) and discussed how to simulate the model. In this model, we describe the state of the IGQC by means of a time-independent state vector in a high-dimensional Hilbert space.

In \secref{sec:Prototype gate-based quantum computers} we discussed a superconducting prototype gate-based quantum computer (PGQC) and some problems which arise once we attempt to transfer the idea of a time-independent IGQC into the real world which is inherently dynamic. For example, we reported on experiments which show that even subtle issues like cosmic radiation affect state-of-the-art superconducting PGQCs. Note that there are more issues which need to be considered.

In \secref{sec:FromStaticsToDynamics} we introduced non-ideal gate-based quantum computer (NIGQC) models. With these models we attempt to describe different aspects of PGQCs. The state of a NIGQC is described by a time-dependent state vector or a time-dependent density matrix depending on which equation generates the dynamics of the system. In the former case we use the TDSE to generate a unitary time evolution. Similarly, in the latter case we use the Lindblad master equation to generate a non-unitary time evolution. Furthermore, we discussed some general issues regarding NIGQC models.

In conclusion, if we speak of gate-based quantum computers, we should at least clearly separate between the three different terms IGQCs, NIGQCs and PGQCs introduced in this chapter. Obviously, they are not one and the same.


\chapter{Hamiltonian models for lumped-element circuits}\label{chap:III}
In this chapter, we introduce the Hamiltonians we use to model NIGQCs, see \secref{sec:FromStaticsToDynamics}. In \secref{sec:TheLumpedElementApproximation}, we discuss the lumped-element circuit model we use to describe different electromagnetic systems. Next, in \secref{sec:CircuitQuantisationFormalism}, we review how one can derive a Hamiltonian for a given lumped-element circuit. In \secref{sec:ResAndTLS}, we apply this procedure to a so-called LC circuit without any power sources. This circuit is the electromagnetic equivalent of the harmonic oscillator. Furthermore, in this section we also discuss a bath model which can be formulated in terms of the LC circuit Hamiltonian. Then, in \secref{sec:Transmons}, we derive and discuss two circuit Hamiltonians, one for the fixed-frequency and one for the flux-tunable transmon. Later on, we use the bare eigenstates of both systems to model qubits in a NIGQC model. Furthermore, here we also discuss effective Hamiltonians which might be used to describe fixed-frequency and flux-tunable transmons, instead of the circuit Hamiltonians. In \secref{sec:TheQuantumComputerCircuitHamiltonianModel}, we define a many-particle circuit Hamiltonian which can be used to model a complete NIGQC. Similarly, in \secref{sec:TheQuantumComputerEffectiveHamiltonianModel}, we define an associated many-particle effective Hamiltonian which can be used to achieve the same goal. Note that throughout this chapter we use $\hbar=1$.

\section{The lumped-element approximation}\label{sec:TheLumpedElementApproximation}

\newcommand{\R}{(\mathbf{r})}
\newcommand{\RR}[1]{(\mathbf{#1})}
\newcommand{\VEC}{(\mathbf{r},t)}
\newcommand{\E}{\mathbf{E}\R}
\newcommand{\B}{\mathbf{B}\R}
\newcommand{\ET}{\mathbf{E}\VEC}
\newcommand{\BT}{\mathbf{B}\VEC}
\newcommand{\SIGN}[1]{\text{Sign}(#1)}

In \secref{sec:Prototype gate-based quantum computers}, we discussed superconducting PGQCs and we found that a complete description of such devices might be impossible,\ie there are simply too many variables which are relevant to the problem at hand. Consequently, a description in terms of Hamiltonian or NIGQC models, see \secref{sec:FromStaticsToDynamics}, forces us to make some simplifying approximations. In this section, we discuss the so-called lumped-element approximation. Ultimately, we intend to describe multi-qubit systems as electromagnetic systems, in a lumped-element model formulation.

Broadly speaking, we may describe the lumped-element approximation as follows. A circuit consists of different conductive (superconductive) electromagnetic structures and strip conductors (superconductors). The function of the strip conductors is to connect the different electromagnetic structures. In our model, we lump the more or less complex electromagnetic structures into fictitious two-terminal elements and the strip conductors become ideal wires. The two-terminal elements and ideal wires then form a network,\ie the different terminals are the nodes $n$ of the network, we connect the nodes via ideal wires and the two-terminal elements become branches $b$. Additionally, we assume that all elements are fully characterised by a constitutive relation. The corresponding constitutive relations connect the currents $I_{b}(t)$ following through the different two-terminal elements (branches) with the voltage differences $V_{b}(t)=V_{n}(t)-V_{n^{\prime}}(t)$ between the two nodes $n$ and $n^{\prime}$. Here $V_{n}(t)$ denotes the voltage at node $n$. In our network model, currents and voltages become directed edges of the network,\ie a change in direction corresponds to a sign flip in terms of currents and voltages. In the following, we assume that the constitutive relations for linear capacitive and inductive elements are given by the equations
\begin{equation}\label{eq:CapacitorCR}
  I_{b}(t)=C \dot{V}_{b}(t)
\end{equation}
and
\begin{equation}\label{eq:InductorCR}
  \dot{I}_{b}(t)=\frac{1}{L} V_{b}(t)
\end{equation}
respectively. Here $C$ denotes the capacitance and $L$ is the so-called inductance. Furthermore, we assume that the so-called Josephson junctions are described by a constitutive relation of the form
\begin{equation}\label{eq:Jeffect}
  I_{b}(t)=I_{c} \sin\BRR{\frac{2\pi}{\Phi_{0}} \Phi_{b}(t)},
\end{equation}
where the constant $I_{c}$ denotes the critical current and $\Phi_{0}=h/2e$ is the so-called magnetic flux quantum, which is of the order of $10^{-15}$ Weber. With \equref{eq:Jeffect} we also introduced the branch flux variable
\begin{equation}
  \Phi_{b}(t)=\int_{-\infty}^{t} V_{b}(t^{\prime}) dt^{\prime},
\end{equation}
where we assume that $V_{b}(t) \rightarrow 0$ for $t \rightarrow - \infty$. Josephson junctions can be understood as some kind of non-linear inductors. The capacitance $C$, the inductance $L$ and the critical current $I_{c}$ are the parameters we use to characterise the different elements in our model.

So far, we can only describe individual elements in terms of currents and voltages. However, we are interested in the case where the different elements interact with one another. To this end, we assume that the different elements or rather the currents and voltages of the different elements are connected by Kirchhoff's laws. The first law or the current law reads
\begin{equation}\label{eq:KirchI}
  \sum_{b \in \mathcal{N}_{i}} I_{b}(t)=0,
\end{equation}
for all nodes $i \in \mathbb{N}^{0}$ in the network. Here $\mathcal{N}_{i}\subseteq\mathbb{N}^{0}$ denotes an index set and this set contains all the branches $b$ which are connected to the node $i$. The second law or Kirchhoff's voltage law states that
\begin{equation}\label{eq:KirchII}
  \sum_{b \in \mathcal{L}_{j}} V_{b}(t)=\sum_{e \in \mathcal{E}_{j}} \mathcal{EMF}_{e}(t),
\end{equation}
for all loops $j$ in the network. Here $\mathcal{L}_{j} \subseteq \mathbb{N}^{0}$ denotes the index set of all two-terminal element branches $b$ which are part of the loop $j$. The functions $\mathcal{EMF}_{e}(t)$ denote so-called electromotive forces (EMFs), we consider EMFs to be model functions which allow us to describe idealised power sources. The index set $\mathcal{E}_{j} \subseteq \mathbb{N}^{0}$ contains all EMF branches $e$ which are part of the loop $j$.

Up to this point, we formulated our model in terms of assumptions, without actually having considered the physical scenario in question. There are two reasons for this approach. First, it is beyond the scope of this thesis to properly discuss all the relevant issues. Second, there are some questions which cannot be resolved completely. The remainder of this section is devoted to the issues just mentioned.

With our model we a priori assume that the two quantities, voltage $V_{b}(t)$ and current $I_{b}(t)$, are well defined for all branches $b$ in the network. From the point of view of Maxwell's theory of electromagnetism, see \REFS\cite{Maxwell10,Zangwill13,Wendt58}, we define currents as
\begin{equation}\label{eq:CurrentDef}
  I(t)= \int_{S} \mathbf{j}\VEC d\mathbf{S},
\end{equation}
where $\mathbf{j}\colon \mathbb{R}^{4} \rightarrow \mathbb{R}^{3}$ denotes the so-called current density and $S$ is a finite surface in three-dimensional space $\mathbb{R}^{3}$. This leads to the question, how can we determine the density $\mathbf{j}\VEC$ and the obvious answer would be to solve Maxwell's equations in combination with the Coulomb-Lorentz force for the physical system in question. However, it is unlikely that this approach yields actual results in terms of an analytical function $\mathbf{j}\VEC$,\ie exact solutions to this mathematical problem are the exception. Furthermore, once we consider $I_{b}(t)$, we find that there is no reference to the finite surface $S$ which determines the right-hand side of \equref{eq:CurrentDef}. Therefore, we might assume that $I_{b}(t)$ is constant throughout the two-terminal element we consider. A more detailed discussion of this assumption is provided by \REF\cite[Chapter II]{Wendt58}.

The voltage $V(t)$ is a quantity which is more difficult to grasp. Voltages are usually introduced in the context of electrostatics, where the electric field $\ET\colon \mathbb{R}^{4} \rightarrow \mathbb{R}^{3}$ and the magnetic field $\BT\colon \mathbb{R}^{4} \rightarrow \mathbb{R}^{3}$ decouple. This case occurs once both fields are constant for all times $t$. Here we find that the electric field $\E=-\nabla \varphi\R$ can be expressed in terms of a scalar potential $\varphi\R \colon \mathbb{R}^{3} \rightarrow \mathbb{R}$ alone. Therefore, in electrostatics we can define voltages as differences
\begin{equation}\label{eq:VoltageDef}
  V=\int_{\mathbf{r_{n}}}^{\mathbf{r_{n^{\prime}}}} \E d\mathbf{l}=\varphi\RR{r_{n}}- \varphi\RR{r_{n^{\prime}}},
\end{equation}
in terms of the scalar potential $\varphi\R$. Here we make use of the fact that the line integral in \equref{eq:VoltageDef} is path independent, see \REF\cite{Zangwill13}. This definition does not naturally translate to all non-static cases,\ie in some cases it is not possible to express the electric field with such ease. However, for quasi-static scenarios, where changes in the state of the system occur instantaneously, we might still be able to add some meaning to the right-hand side of \equref{eq:VoltageDef}, see \REF\cite[Chapter II]{Wendt58}.

Some authors, see \REF\cite[Section 6.10]{FANO60}, promote the view that voltages are not defined outside of electrostatics,\ie if the path independence in the computation of the voltage is not given. If this is the case, we have to ask ourself whether or not the constitutive relations and Kirchhoff's laws have a well-defined physical meaning in the context of Maxwell's theory of electromagnetism. The author of \REF\cite{Mcdonald12voltagedrop} advocates the use of so-called retarded electric scalar potentials in the Lorenz gauge. Another author, see \REF\cite[Section 1.4]{Balanis12}, provides a detailed discussion of the relation between Maxwell's field equations and the circuit equations introduced in this section, see \equsref{eq:CapacitorCR}{eq:KirchII}. Additionally, \REF\cite[Section 1.4]{Balanis12} also provides a detailed discussion of the relation between field quantities and the corresponding circuit quantities,\eg the electric field intensity $\ET$ and the voltage $V(t)$.

In conclusion, to the best knowledge of the author, there exists no standard definition of the quantity voltage $V(t)$ in time-dependent electromagnetism. Therefore, we might take the stand, in a qualitative manner, that time-dependent voltages can be associated with quasi-static potential differences $\varphi\RR{r_{n},t}- \varphi\RR{r_{n^{\prime}},t}$ in the sense of \equref{eq:VoltageDef}. The condition for the quasi-static or quasi-stationary state is often, see \REFS\cite{Wendt58,Zangwill13}, expressed as
\begin{equation}
    \omega s \ll c,
\end{equation}
where $\omega \in \mathbb{R}$ denotes the frequency of the oscillating currents and/or voltages in the circuit, $s \in \mathbb{R}$ refers to the distance which characterises the circuit and $c \in \mathbb{R}$ is the speed of light, which is of the order $10^{8}$ meter per second.

In this thesis, we prefer the axiomatic view presented at the beginning of this section. To this end, we treat EMFs as model functions,\ie EMFs allow us to describe idealised power sources. Furthermore, the constitutive relations in \equsref{eq:CapacitorCR}{eq:Jeffect} allow us to include energy conserving two-terminal elements. In the end, we can construct arbitrary networks and compare the predictions made by the circuit theory with the pointer readings, on different meters, we find in the laboratory, this view is motivated by \REF\cite{Wendt58}.

\section{The circuit quantisation formalism}\label{sec:CircuitQuantisationFormalism}

In this section, we discuss how one can derive so-called circuit Hamiltonians. Circuit Hamiltonians, sometimes in combination with the TDSE, provide us with quantum theoretical models, which allow us to make predictions for electromagnetic circuits in the quantum regime,\ie in the regime where quantum effects dominate the physical scenario in question. In \secref{sec:TheQuantumComputerCircuitHamiltonianModel}, we use different circuit Hamiltonians to define a many-particle Hamiltonian, which might describe some aspects of superconducting PGQCs, see \secref{sec:Prototype gate-based quantum computers}, we are interested in. The derivation procedure for circuit Hamiltonians is usually referred to as circuit quantisation.

The first step in the quantisation of an electromagnetic circuit corresponds to choosing a set of independent variables such that all voltages $V_{b}(t)$ and currents $I_{b}(t)$ of the circuit can be expressed in terms of these variables and the corresponding derivatives. Furthermore, we also require that the Euler-Lagrange equations, which are associated with a certain Lagrangian function $\mathcal{L}\colon\mathbb{R}^{N}\rightarrow\mathbb{R}$, enforce Kirchhoff's current and voltage law.

To the best knowledge of the author, there exist two well-studied methods to systematically quantise a non-dissipative electromagnetic circuit. The method of nodes and the method of loops, see \REFS\cite{Burkard04,Yurke84,DV97,Vool17,Ulrich16}. Broadly speaking, both methods deviate in their choice of variables. The method of nodes makes use of the fact that if we use the generalised branch fluxes
\begin{equation}
  \Phi_{b}(t)=\int_{-\infty}^{t} V_{b}(t^{\prime}) dt^{\prime},
\end{equation}
as our dynamic variables, where $V_{b}(t) \rightarrow 0$ for $t\rightarrow -\infty$, we find that Kirchhoff's voltage law can be satisfied without further ado by expressing $\Phi_{b}(t)$ in terms of independent generalised node fluxes $\Phi_{n}(t)$. Similarly, the method of loops makes use of the fact that if we use the generalised branch charges
\begin{equation}
  Q_{b}(t)=\int_{-\infty}^{t} I_{b}(t^{\prime}) dt^{\prime},
\end{equation}
as our dynamic variables, where $I_{b}(t) \rightarrow 0$ for $t\rightarrow -\infty$, Kirchhoff's current law can be satisfied by choosing another set of independent variables, see \REF\cite{Ulrich16}. Which method to choose depends on the problem at hand. For an extensive discussion of this issue see \REF\cite{Ulrich17}. In the end, we have to construct the Lagrangian $\mathcal{L}$ such that the remaining second law is satisfied too. In this thesis, we are interested in circuits, which contain linear capacitors, linear inductors, Josephson junctions,\ie non-linear inductors, voltage sources and time-dependent external fluxes. For this case,\ie the case where no non-linear capacitors are present, the method of nodes is well suited.

We can determine the node fluxes $\Phi_{n}$ with the following procedure. First, we choose a designated node, the so-called ground node $\Phi_{g}=0$. Then, we choose a spanning tree $\mathcal{S}$,\ie an undirected subgraph, which contains all nodes of the network but no loops. We use the discrete variable $j \in \mathbb{N}^{0}$ to refer to the different loops of the network. Branches, which close the loops are denoted as chords or closure branches. Similarly, the branches of the spanning tree are simply called tree branches. Next, we give the tree branches an orientation, they should point away from the ground node. In addition, we give each loop $j$ an orientation, this orientation is chosen randomly. In the end, we align the orientation of the closure branches with the ones of the loops. Consequently, at this point all branches possess an orientation. In the end, we define the node fluxes
\begin{equation}
  \Phi_{n}=\sum_{b\in \mathcal{S}} S_{n,b} \Phi_{b},
\end{equation}
where $S_{n,b}=1$ if branch $b$ is part of the path $\mathcal{P}(0,n)$ form the ground node to node $n$. If branch $b$ is not part of path $\mathcal{P}(0,n)$, we set $S_{n,b}=0$. Note that all branches on the paths $\mathcal{P}(0,n)$ have the same orientation because we defined it this way,\ie this is a convenient choice. Finally, we can use the node fluxes $\Phi_{n}$ to write
\begin{equation}
  \Phi_{b}=\Phi_{n}-\Phi_{n^{\prime}},
\end{equation}
for every branch $b$. The branch $b$ points from node $n^{\prime}$ to node $n$. This choice of variables ensures that if we form the sum of all branch fluxes $\Phi_{b}$, for an arbitrary loop $j$, the sum equals zero. Consequently, the time derivatives of these sums also add up to zero and this in turn means that Kirchhoff's voltage law is satisfied. A more detailed discussion of this topic with emphasis on graph theory, which goes beyond what we discuss in this thesis, can be found in \REF\cite{Burkard04}.

So far, we have ignored the presence of voltage sources $V_{g}(t)$ and external fluxes $\Phi_{\text{e}}(t)$. If we model both as EMFs, we can include $V_{g}(t)$ and $\Phi_{\text{e}}(t)$ by means of Kirchhoff's voltage law. Assuming, the external flux $\Phi_{\text{e}}$ present in loop $j$ is time-independent, we add a corresponding term
\begin{equation}
  \Phi_{b}=\Phi_{n}-\Phi_{n^{\prime}}\pm\Phi_{\text{e}},
\end{equation}
to the closure branch $b$ of this loop. The sign $\pm 1$ is chosen according to the right-hand rule. The case of time-dependent external fluxes is subject of ongoing research, see \REFS\cite{You,Riwar21}. We discuss this issue in \secref{sec:Transmons} in more detail. If a voltage source $V_{g}(t)$ is part of a branch $b$ which contains another circuit element whose constitutive relation is expressed in terms of the variable $\Phi_{e}$, we write
\begin{equation}
  \Phi_{b}=\Phi_{e} \pm \int_{-\infty}^{t} V_{g}(t^{\prime}) dt^{\prime}.
\end{equation}
The time derivative of this relation states that the voltage drop across the whole branch $V_{b}$ is given by the sum of the voltage drop across the element $V_{e}(t)=\dot{\Phi}_{e}(t)$ and the voltage source $V_{g}(t)$. We can solve this equation for $\Phi_{e}$ and add the corresponding term for the element $e$ to the Lagrangian, see \secref{sec:Transmons}.

Kirchhoff's current law can be enforced by constructing the Lagrangian $\mathcal{L}$ accordingly,\ie such that the Euler-Lagrange equations yield this law. The first construction rule states that every branch $b$ with an inductive element, which is characterised by a constitutive relation of the form $I_{b}(t)=g(\Phi_{b}(t))$, results in an additional potential term
\begin{equation}\label{eq:PotTerm}
  U(\Phi_{b})=\int_{-\infty}^{t} \dot{\Phi}_{b}(t^{\prime}) g(\Phi_{b}(t^{\prime})) dt^{\prime},
\end{equation}
of the Lagrangian $\mathcal{L}=T-U$. The second construction rule states that every branch $b$ with a capacitive element, which is characterised by a constitutive relation of the form $Q_{b}(t)=f(V_{b}(t))$, results in an additional kinetic term
\begin{equation}\label{eq:KinTerm}
  T(\dot{\Phi}_{b})=\int_{-\infty}^{t} \dot{\Phi}_{b}(t^{\prime}) \dot{f}(\dot{\Phi}_{b}(t^{\prime})) dt^{\prime},
\end{equation}
of the same Lagrangian $\mathcal{L}$. We have to perform variable substitutions in \equaref{eq:PotTerm}{eq:KinTerm} to see that the energies $U$ and $T$ are functions of $\Phi_{b}$ and $\dot{\Phi}_{b}$, respectively. The results read
\begin{equation}
  U(\Phi_{b})=\int_{\lim_{t\rightarrow -\infty} \Phi_{b}(t)}^{\Phi_{b}(t)} g(\Phi_{b}) d\Phi_{b},
\end{equation}
and
\begin{equation}
  T(\dot{\Phi}_{b})=f(\dot{\Phi}_{b})\dot{\Phi}_{b} - \int_{\lim_{t\rightarrow -\infty} \dot{\Phi}_{b}(t)}^{\dot{\Phi}_{b}(t)} f(\dot{\Phi}_{b}) d\dot{\Phi}_{b}.
\end{equation}
For the case we are interested in, we can verify that the Euler-Lagrange equations
\begin{equation}\label{eq:Euler-Lagrange equations}
  \der{}{t} \derp{\mathcal{L}}{\dot{\Phi}_{n}}-\derp{\mathcal{L}}{\Phi_{n}}=0,
\end{equation}
for the different node fluxes $\Phi_{n}$ yield Kirchhoff's current law. If we have a linear capacitive element, we find a contribution $\pm C\ddot{\Phi}_{b}$ to the sum of currents, see first term on the left-hand side of \equref{eq:Euler-Lagrange equations}. Similarly, an inductive element contributes the term $\pm g(\Phi_{b})$ to the sum of currents, see second term on the left-hand side of \equref{eq:Euler-Lagrange equations}. The sign is determined by the orientation of the branch $\Phi_{b}=\pm(\Phi_{n}-\Phi_{n^{\prime}})$. Therefore, changing the orientation results in a sign flip of both currents. This ensures that the in and outgoing currents carry the correct sign with regard to Kirchhoff's current law.

In a next step, we perform a Legendre transformation, see \REF\cite{Fenchel1949}, to derive the Hamiltonian function
\begin{equation}
 H= \sum_{n=0}^{N-1} Q_{n}\dot{\Phi}_{n}  - \mathcal{L},
\end{equation}
where
\begin{equation}
  Q_{n}=\derp{\mathcal{L}}{\dot{\Phi}_{n}},
\end{equation}
is the so-called \nth{} conjugate variable. Here we assume that $\mathcal{L}$ is a differential, convex function.

In the end, we perform the canonical quantisation. This means we make the same fundamental assumptions as the author of \REF\cite{Dirac1925}, namely we replace the variables $\Phi_{n}$ and $Q_{n}$ with the operators $\OP{\Phi}_{n}$ and $\OP{Q}_{n}$ and map the Poisson brackets $\{\Phi_{n},Q_{n^{\prime}}\}=\delta_{n,n^{\prime}}$ to the commutators $[\OP{\Phi}_{n},\OP{Q}_{n^{\prime}}]=i \hbar \delta_{n,n^{\prime}}\OP{I}$.

\section{Resonators and two-level systems}\label{sec:ResAndTLS}
\newcommand{\HBS}{\ket{\psi^{(z)}}}
\newcommand{\HBSV}[2]{\ket{\psi^{(z_{#2})}#1}}
In circuit quantum electrodynamics there exists a circuit equivalent to the well-studied model of a mechanical harmonic oscillator. This is the first system we quantise with the formalism discussed in \secref{sec:CircuitQuantisationFormalism}.
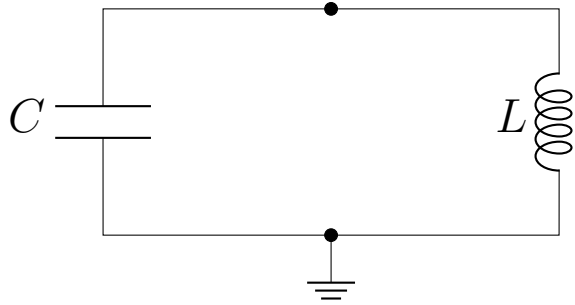
\begin{figure}[!tbp]
  \centering
  \begin{circuitikz}[scale = 1.5, transform shape]
      \node[ground] at (0,0) {};
      \node[circ] at (0,0) {};
      \node[circ] at (0,2) {};
      \draw (0,0) to[short] (-2,0)
      to[C,l=$C$] (-2,2)
      to[short] (0,2);
      \draw (0,0) to[short] (2,0)
      to[L,l=$L$] (2,2)
      to[short] (0,2);
  \end{circuitikz}
  \caption[Circuit diagram of a network containing a linear capacitor with capacitance $C$ and a linear inductor with inductance $L$. ]{Circuit diagram of a network containing a linear capacitor with capacitance $C$ and a linear inductor with inductance $L$. The ground node is marked by a dashed triangle.}\label{fig:LC_circuit}
\end{figure}
Figure \ref{fig:LC_circuit} shows a circuit diagram of exactly this system. As one can see, the left branch contains a linear capacitor with capacitance $C$ and the right branch contains a linear inductor with inductance $L$. The ground node is marked by a dashed triangle. Here we choose $\Phi_{g}=0$ such that the left branch flux is given by $\Phi_{C}=\Phi_{n}$ and the right one reads $\Phi_{L}=-\Phi_{n}$. Consequently, the Lagrangian has two terms, the kinetic energy term of the capacitor
\begin{equation}
  T(\dot{\Phi}_{n})= \frac{C}{2} \dot{\Phi}_{n}^{2}
\end{equation}
and the potential energy term
\begin{equation}
  U(\Phi_{n})= \frac{1}{2 L} \Phi_{n}^{2}.
\end{equation}
The Lagrangian then reads
\begin{equation}
  \mathcal{L}=\frac{C}{2} \dot{\Phi}_{n}^{2}-\frac{1}{2 L} \Phi_{n}^{2},
\end{equation}
and the conjugate variable can be expressed as
\begin{equation}
  Q=\derp{\mathcal{L}}{\dot{\Phi}_{n}}= C \dot{\Phi}_{n}.
\end{equation}
If we perform the Legendre transformation as well as the canonical quantisation, we find the Hamiltonian
\begin{equation}
  \OP{H}=\frac{\OP{Q}^{2}}{2 C}+\frac{\OP{\Phi}^{2}}{2 L}.
\end{equation}
Note that in this last step we removed the node label. We can express this Hamiltonian as
\begin{equation}\label{eq:Harmonic}
  \OP{H}=E_{C}\OP{n}^2+\frac{E_{L}}{2}\OP{\varphi}^{2},
\end{equation}
where
\begin{align}
  \OP{n}&=\frac{1}{2 e}\OP{Q},\\
  \OP{\varphi}&=\frac{2\pi}{\Phi_{0}}\OP{\Phi},\\
  E_{C}&=\frac{(2 e)^{2}}{2 C},\\
  E_{L}&=\left( \frac{\Phi_{0}}{2\pi} \right)^{2}\frac{1}{L}.
\end{align}
Here $E_{C}$ and $E_{L}$ denote the capacitive and inductive energy of the systems, respectively. The operator given by \equref{eq:Harmonic} and the corresponding eigenvalues and eigenstates are discussed in almost all quantum theory textbooks, see for example \REF\cite[Section 2.5]{Weinberg2015}. The eigenstates $\HBS$ in $\varphi$-space read
\begin{equation}\label{eq:HarmoicBasisWaveFunction}
    \psi^{(z)}(\varphi)= \frac{1}{\sqrt{2^{z} z!}}\sqrt[4]{\frac{\xi}{\pi}} e^{\frac{-(\sqrt{\xi}\varphi)^{2}}{2}} H_{z}(\sqrt{\xi}\varphi),
\end{equation}
where $z \in \mathbb{N}^{0}$, $\xi=\sqrt{E_{L}/(2 E_{C})}$ and $H_{z}(\sqrt{\xi}\varphi)$ denotes the Hermite polynomial of order $z$. We can express the Hamiltonian in its eigenbasis as
\begin{equation}\label{eq:resonator}
  \OP{H}_{\idxRES}= \omega^{(R)} \BRR{\OP{a}^{\dagger}\OP{a}+\frac{\OP{I}}{2}},
\end{equation}
where the $\omega^{(R)}$ is the resonance frequency
\begin{equation}\label{eq:res_freq}
  \omega^{(R)} = \sqrt{2 E_{C} E_{L}},
\end{equation}
and $\OP{a}$ and $\OP{a}^{\dagger}$ are the bosonic number operators which can be defined in terms of their action
\begin{align}
  \OP{a}^{\dagger} \ket{ \psi^{(z)} }&=\sqrt{z+1}\ket{\psi^{(z+1)}},\\
    \OP{a} \ket{ \psi^{(z)} } &= \sqrt{z} \ket{\psi^{(z-1)}},
\end{align}
on the discrete eigenstates $\HBS$. In analogy with the classical AC circuit theory we call this system simply a resonator. This is why we changed the label $\OP{H} \rightarrow \OP{H}_{\idxRES}$ of the Hamiltonian. Furthermore, from here on we omit the term $\OP{I}/2$ which only contributes a non-measurable phase to the dynamics of the system.

The resonator model in its formulation in terms of bosonic operators, see \equref{eq:resonator}, is the basis of many quantum theoretical models, see for example \REFS\cite{Ashcroft76,Balian07}. In the context of superconducting PGQCs we are interested in one particular model namely the two-level system (TLS) model, see \REF\cite{Mueller2019} for a review of this subject. We define the generic Hamiltonian
\begin{equation}\label{eq:TLS_def}
  \OP{H}=\OP{H}_{Sys.}+\OP{H}_{\idxTLS,\Sigma}+\OP{W}_{\idxINT},
\end{equation}
where $\OP{W}_{\idxINT}$ is some arbitrary interaction and $\OP{H}_{Sys.}$ is a Hamiltonian which describes some arbitrary system. Furthermore, the Hamiltonian
\begin{equation}
  \OP{H}_{\idxTLS,\Sigma}= \sum_{l \in L} \omega_{l}^{(T)} \OP{b}_{l}^{\dagger} \OP{b}_{l},
\end{equation}
describes a collection of non-interacting TLSs. Here we simply changed the notation,\ie $\omega^{(R)} \rightarrow \omega^{(T)}$, $\OP{a} \rightarrow \OP{b}$, $\OP{a}^{\dagger} \rightarrow \OP{b}^{\dagger}$ and $L \subseteq \mathbb{N}^{0}$ denotes an index set. The reason for this change in notation is that in the TLS model we only consider the lowest two levels of $\OP{H}_{\idxRES}$, for every TLS. We can look at this model from the following point of view, $\OP{H}_{\idxTLS}$ constitutes a bath for the system $\OP{H}_{Sys.}$ and the interactions between bath and system are governed by the operator $\OP{W}_{\idxINT}$. Reference \cite{Willsch2020FluxQubitsQuantumAnnealing} contains a numerical study where TLSs are used to include temperature effects into a model. Obviously, once we consider a specific physical scenario, we have to fix the TLS parameters $\omega_{l}^{(T)}$ and the operator $\OP{W}_{\idxINT}$.

TLS models are used to describe and/or explain various phenomena. For example, in experiments we find that relaxation times $T_{1}$, dephasing times $T_{2}$ and qubit frequencies $\omega^{(01)}$, see \secref{sec:Prototype gate-based quantum computers} for a definition of these parameters, fluctuate around a mean value on different time scales. Such fluctuations might be explained by the presence of TLSs, see \REF\cite{Burnett2019}. TLSs are also used to describe other types of solid state phenomena, see \REF\cite{Esquinazi98}.

\section{Fixed-frequency and flux-tunable transmons}\label{sec:Transmons}

In this section, we quantise and analyse two circuits. These systems are usually referred to as fixed-frequency and flux-tunable transmons. The eigenstates of these systems are often used as qubit states in NIGQC models. The so-called transmission line shunted plasma oscillation qubit or transmon for short was first introduced by the authors of \REF\cite{Koch}. The circuit Hamiltonian model for NIGQC we introduce in \secref{sec:TheQuantumComputerCircuitHamiltonianModel} makes use of two types of transmons, namely fixed-frequency and flux-tunable transmons. Note that the term transmon is actually reserved for a specific parameter regime in which both systems operate. We first quantise both circuits and then discuss the corresponding eigenproblems together. Additionally, we introduce effective Hamiltonians for both systems.

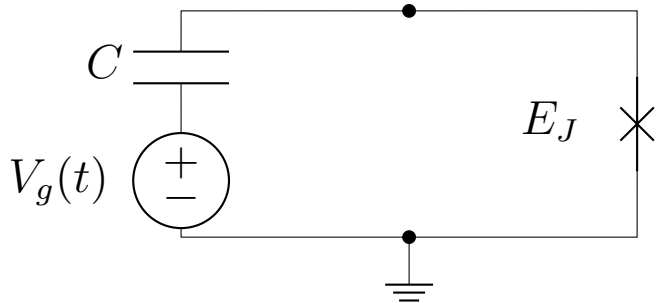
\begin{figure}[!tb]
  \centering
  \begin{circuitikz}[scale = 1.5, transform shape,american]
      \node[circ] at (0,2) {};
      \node[circ] at (0,0) {};
      \node[ground] at (0,0) {};
      \draw (0,0) to[short] (-2,0)
      to[V,v=$V_{g}(t)$] (-2,1)
      to[capacitor,l=$C$] (-2,2)
      to[short] (0,2);
      \draw (0,0) to[short] (2,0)
      to[barrier,l=$E_{J}$] (2,2)
      to[short] (0,2);
  \end{circuitikz}
    \caption[Circuit diagram of a network containing a linear capacitor with capacitance $C$, a Josephson junction with Josephson energy $E_{J}$ and a voltage source $V_{g}(t)$.]{Circuit diagram of a network containing a linear capacitor with capacitance $C$, a Josephson junction with Josephson energy $E_{J}$ and a voltage source $V_{g}(t)$. The ground node is marked by a dashed triangle.}\label{fig:FFT_circuit}
\end{figure}

Figure \ref{fig:FFT_circuit} shows the circuit diagram of a fixed-frequency transmon system. The left branch contains a linear capacitor with capacitance $C$ as well as a voltage source $V_{g}(t)$. The right branch contains a Josephson junction with Josephson energy $E_{J}$.
\newcommand{\var}{\frac{2\pi}{\Phi_{0}}\Phi_{n}}
We choose $\Phi_{g}=0$ such that the left branch flux is given by
\begin{equation}
  \Phi_{n}=\Phi_{C}+\int_{-\infty}^{t} V_{g}(t^{\prime}) dt^{\prime} ,
\end{equation}
where $\Phi_{C}$ is the branch flux variable associated with the capacitor. Therefore, the right branch flux reads $\Phi_{E_{J}}=-\Phi_{n}$. We find that the kinetic energy of the Lagrangian is given by
\begin{equation}
  T(\dot{\Phi}_{n})= \frac{C}{2} \BRR{\dot{\Phi}_{n}-V_{g}(t)}^{2},
\end{equation}
and the potential energy can be expressed as
\begin{equation}
  U(\Phi_{n})= -E_{J} \cos\BRR{\var}.
\end{equation}
The full Lagrangian then reads
\begin{equation}
  \mathcal{L}=\frac{C}{2} (\dot{\Phi}_{n}-V_{g}(t))^{2} + E_{J} \cos\BRR{\var}.
\end{equation}
Therefore the conjugate variable is given by the expression
\begin{equation}
  Q=\derp{\mathcal{L}}{\dot{\Phi}_{n}}= C  \BRR{\dot{\Phi}_{n}-V_{g}(t)}.
\end{equation}
After performing the Legendre transformation and the canonical quantisation, we find the Hamilton operator
\renewcommand{\var}{\frac{2\pi}{\Phi_{0}}\OP{\Phi}}
\begin{equation}
  \OP{H}= \frac{\OP{Q}^{2}}{2 C} + V_{g}(t) \OP{Q} - E_{J} \cos\BRR{\var}.
\end{equation}
Note that here we removed the node label $n$ from the operator $\OP{\Phi}_{n}$. After completing the square and removing some terms which only contribute non-measurable phases to the dynamics of the system, the Hamilton operator reads
\begin{equation}
  \OP{H}= \frac{\BRR{2e}^{2}}{2 C}\BRR{\frac{\OP{Q}}{2 e} + \frac{C V_{g}(t)}{2 e} }^{2} - E_{J} \cos\BRR{\var}.
\end{equation}
If we perform the variable substitutions
\begin{subequations}
  \begin{align}
     \OP{n}       &=\frac{\OP{Q}}{2 e},\\
     \OP{\varphi} &=\var,\\
     n_{g}(t)        &=-\frac{C V_{g}(t)}{2 e},
  \end{align}
\end{subequations}
we arrive at the final Hamilton operator for a fixed-frequency transmon
\begin{equation}\label{eq:fixed-frequency transmon}
 \OP{H}_{\idxFFT} = E_{C} (\OP{n}-n_{g}(t))^2 - E_{J} \cos(\OP{\varphi}),
\end{equation}
where $E_{C}=\BRR{2e}^{2}/(2 C)$ denotes the capacitive energy of the system.

\begin{figure}[!tb]
  \centering
  \begin{circuitikz}[scale = 1.4, transform shape,american]
      \node[circ] at (0,2) {};
      \node[circ] at (0,0) {};
      \node[] at (0,1) {$\Phi_{\text{e}}(t)$};
      \node[ground] at (0,0) {};
      \draw (0,0) to[short] (-4,0)
      to[capacitor,l=$C_{l}$] (-4,2)
      to[short] (0,2);
      \draw (0,0) to[short] (-2,0)
      to[barrier,l=$E_{J,l}$] (-2,2)
      to[short] (-2,2)
      to[short] (0,2);
      \draw (0,0) to[short] (2,0)
      to[barrier,l=$E_{J,r}$] (2,2)
      to[short] (2,2)
      to[short] (0,2);
      \draw (0,0) to[short] (4,0)
      to[capacitor,l=$C_{r}$] (4,2)
      to[short] (4,2)
      to[short] (0,2);
  \end{circuitikz}
    \caption[Circuit diagram of a network containing linear capacitors with capacitances $C_{l}$ (left) and $C_{r}$ (right), two Josephson junctions with Josephson energies $E_{J,l}$ (left) and $E_{J,r}$ (right) as well as an external flux $\Phi_{\text{e}}(t)$ threading through the center loop. ]{Circuit diagram of a network containing linear capacitors with capacitances $C_{l}$ (left) and $C_{r}$ (right), two Josephson junctions with Josephson energies $E_{J,l}$ (left) and $E_{J,r}$ (right) as well as an external flux $\Phi_{\text{e}}(t)$ threading through the center loop. The ground node is marked by a dashed triangle.}\label{fig:FTT_circuit}
\end{figure}
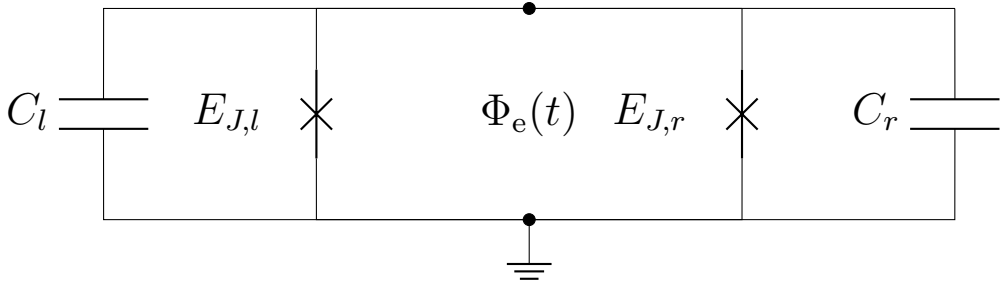
Figure \ref{fig:FTT_circuit} shows a circuit diagram of a flux-tunable transmon system. The network contains linear capacitors with capacitances $C_{l}$ (left) and $C_{r}$ (right), two Josephson junctions with Josephson energies $E_{J,l}$ (left) and $E_{J,r}$ (right) as well as an external flux $\Phi_{\text{e}}(t)$ threading through the center loop. We model the external flux as an EMF and not as an actual physical implementation in terms of additional circuit elements. Therefore, we include the external flux by means of Kirchhoff's voltage law. The following derivation is motivated by the work in \REF\cite{You}. The authors performed a critical analysis of the standard quantisation formalism which is often used in the literature. Note that in the meantime the work in \REF\cite{You} was extended by the authors of \REF\cite{Riwar21}.

\renewcommand{\brr}[1]{\left( #1 \right)}
\newcommand{\PL}{\Phi_{l}}
\newcommand{\PLD}{\dot{\Phi}_{l}}
\newcommand{\ML}{m_{l}}
\newcommand{\PR}{\Phi_{r}}
\newcommand{\MR}{m_{r}}
\newcommand{\PE}{\Phi_{e}(t)}
\newcommand{\PED}{\dot{\Phi}_{e}(t)}
\newcommand{\PV}{\Phi_{n}}
\newcommand{\PVD}{\dot{\Phi}_{n}}
\newcommand{\PVO}{\OP{\Phi}}
\newcommand{\CV}{Q}
\newcommand{\CVO}{\OP{Q}}
\newcommand{\MD}{m_{\Delta}}
\newcommand{\CS}{C_{\Sigma}}
\newcommand{\CL}{C_{l}}
\newcommand{\CR}{C_{r}}
\newcommand{\VG}{V_{g}(t)}
\newcommand{\GP}{\beta}

Kirchhoff's voltage law for the center loop yields the time-dependent constraint
\begin{equation}\label{eq:HolonomiConstraint}
  \PL+\PR=\PE.
\end{equation}
We can define a new degree of freedom
\begin{equation}
  \PV=\ML\PL+\MR\PR,
\end{equation}
such that the left
\begin{eqnarray}
  \PL=\frac{\brr{\PV-\MR\PE}}{\MD},
\end{eqnarray}
and the right
\begin{eqnarray}
  \PR=-\frac{\brr{\PV-\ML\PE}}{\MD},
\end{eqnarray}
branch flux satisfy \equref{eq:HolonomiConstraint} for all $\ML,\MR \in \mathbb{R}$. Here $\MD=\ML-\MR$. We use the two relations $V_{l}=V_{C_{l}}=V_{E_{J_{l}}}$ and $V_{r}=V_{C_{r}}=V_{E_{J_{r}}}$ given by Kirchhoff's voltage law for the left and right loop to express the Lagrangian
\begin{equation}
    \mathcal{L}=\frac{C_{l}}{2} \brr{\frac{\brr{\PVD-\MR\PED}}{\MD}}^{2} + \frac{C_{r}}{2} \brr{\frac{\brr{\PVD-\ML\PED}}{\MD}}^{2} - U\brr{\PV},
\end{equation}
solely in terms of the variable $\PV$. Here the potential energy is given by
\begin{equation}
  \begin{split}
      U\brr{\PV}=&-E_{J_{l}} \cos\brr{\frac{2 \pi}{\Phi_{0}}\frac{\brr{\PV-\MR\PE}}{\MD}} \\
                &-E_{J_{r}} \cos\brr{\frac{2 \pi}{\Phi_{0}}\frac{\brr{\PV-\ML\PE}}{\MD}}.
  \end{split}
\end{equation}
After evaluating the squares and neglecting all factors proportional to $\PED^{2}$, which in the end only contribute a non-measurable phase to the dynamics of the system, the Lagrangian reads
\begin{equation}
    \mathcal{L}=\frac{\CS}{2\MD^{2}} \PVD^{2} - \frac{\brr{\CL\MR+\CR\ML}}{\MD^{2}}  \PED \PVD - U\brr{\PV}.
\end{equation}
The conjugate variable $Q$ can be expressed as
\begin{equation}
  Q=\frac{\CS}{\MD^{2}} \PVD - \frac{\brr{\CL\MR+\CR\ML}}{\MD^{2}}\PED,
\end{equation}
such that the Hamiltonian\renewcommand{\PV}{\Phi}\renewcommand{\PVD}{\dot{\Phi}}
\begin{equation}
  H=\frac{\MD^{2}}{2\CS} \CV^{2} + \frac{\brr{\CL\MR+\CR\ML}}{\CS}\PED \CV + U\brr{\PV},
\end{equation}
can be determined by means of the Legendre transformation. Note that here we removed the node label $n$ from the variable $\Phi_{n}$. If we promote the conjugate variables $\Phi$ and $Q$ to the conjugate operators $\PVO$ and $\CVO$ and perform the substitutions
\begin{subequations}\label{eq:substitutions}
  \begin{align}
     \OP{\varphi}&=\frac{2 \pi}{\Phi_{0}} \PVO,\\
     \OP{n}&=\frac{1}{2 e} \CVO,\\
     \varphi(t)&=\frac{2 \pi}{\Phi_{0}} \Phi_{e}(t),
  \end{align}
\end{subequations}
we find the Hamilton operator
\renewcommand{\PVO}{\OP{\varphi}}
\renewcommand{\CVO}{\OP{n}}
\renewcommand{\PE}{\varphi(t)}
\renewcommand{\PED}{\dot{\varphi}(t)}
\begin{equation}\label{eq:flux-tunable transmon}
  \OP{H}=E_{C_{\Sigma}} \MD^{2} \CVO^{2} + \frac{\brr{\CL\MR+\CR\ML}}{\CS} \PED \CVO + U\brr{\PVO},
\end{equation}
where the second term on the right-hand side contains the factor $\hbar=1$. We now simplify the parameterisation by assuming $\MD=1$ and $\CL=\CR$. Additionally, we define $\MR=-\beta$ and $\CS=C$ such that the Hamiltonian for a flux-tunable transmon can be expressed as
\begin{equation}\label{eq:flux-tunable transmon_beta}
  \begin{split}
    \OP{H}_{\idxFTT}&=E_{C} \CVO^{2} + \brr{\frac{1}{2}-\beta} \PED \CVO - E_{J_{l}} \cos\brr{\OP{\varphi}+\beta\PE}\\
    &- E_{J_{r}} \cos\brr{\OP{\varphi}+(\beta-1)\PE}.
  \end{split}
\end{equation}
In the following we impose the so-called irrotational constraint, see \REF\cite{You}, onto the variable $\PV$. This means we use $\beta=1/2$ as a specific choice for the variable $\PV$.

The flux-tunable transmon Hamiltonian for $\GP=1/2$ can be expressed as
\begin{equation}\label{eq:flux-tunable transmon recast}\renewcommand{\brr}[1]{\left\{ #1 \right\}}
 \OP{H}_{\idxFTT} = E_{C} \OP{n}^{2} -  E_{J,\text{eff.}}(t) \cos(\OP{\varphi}-\varphi_{\text{eff.}}(t)),
\end{equation}
where
\newcommand{\EJEFF}{E_{J,\text{eff.}}(t)}
\newcommand{\EJEFFVAR}[1]{E_{J_{#1},\text{eff.}}(t)}
\begin{equation}\label{eq: eff Josephson energy}
  \EJEFF=E_{\Sigma} \sqrt{\cos\left(\frac{\varphi(t)}{2}\right)^{2}+d^{2} \sin\left(\frac{\varphi(t)}{2}\right)^{2}},
\end{equation}
denotes the effective Josephson energy and
\begin{equation}\label{eq: eff flux}
  \varphi_{\text{eff.}}(t)=\arctan\left(d \tan\left(\frac{\varphi(t)}{2}\right)\right),
\end{equation}
refers to the effective external flux. Here we also define the auxiliary system parameters
\begin{equation}
  E_{\Sigma}=(E_{J,l}+E_{J,r}),
\end{equation}
and
\begin{equation}\label{eq:asymmetry_factor}
   d=(E_{J_{r}}-E_{J_{l}})/(E_{J_{l}}+E_{J_{r}}).
\end{equation}
The latter is usually referred to as the asymmetry factor. The obvious structural similarity with Hamiltonian \equref{eq:fixed-frequency transmon} gives this system its name: the flux-tunable transmon. Both systems, the fixed-frequency and flux-tunable transmon are usually operated in the regime $E_{J}/E_{C} \gg 1$ or $\EJEFF/E_{C} \gg 1$, see \REF\cite{Koch}.

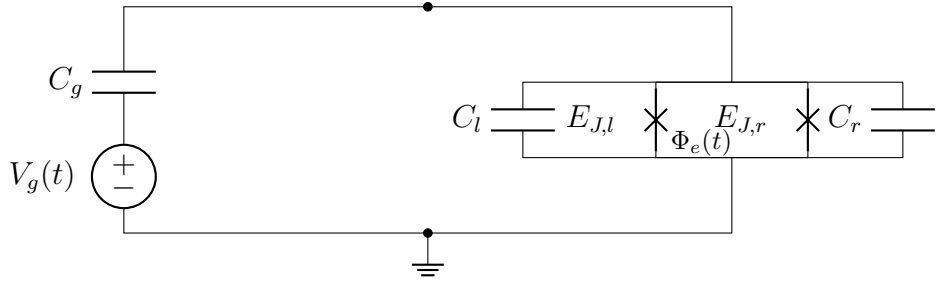
\begin{figure}[!tb]
  \centering
  \begin{circuitikz}[scale = 1.0, transform shape,american]
      \node[] at (3.6,1.2) {\footnotesize $\Phi_{e}(t)$};
      \node[ground] at (0,0) {};
      \node[circ] at (0,0) {};
      \draw (0,0) to[short] (-4,0)
      to[V,v=$V_{g}(t)$] (-4,1.5)
      to[capacitor,l=$C_{g}$] (-4,2.5)
      to[short] (-4,3)
      to[short] (0,3);
      \node[circ] at (0,3) {};
      \draw (0,0) to[short] (4,0)
      to[short] (4,1);
      \draw (4,1) to[short] (1.25,1)
      to[capacitor,l=$C_{l}$] (1.25,2)
      to[short] (4,2);
      \draw (4,1) to[short] (3,1)
      to[barrier,l=$E_{J,l}$] (3,2)
      to[short] (3,2)
      to[short] (4,2);
      \draw (4,1) to[short] (5,1)
      to[barrier,l=$E_{J,r}$] (5,2)
      to[short] (5,2)
      to[short] (4,2);
      \draw (4,1) to[short] (6.25,1)
      to[capacitor,l=$C_{r}$] (6.25,2)
      to[short] (6.25,2)
      to[short] (4,2);
      \draw (4,2) to[short] (4,3)
      to[short] (0,3);
  \end{circuitikz}
    \caption[Circuit diagram of a network. The right branch contains linear capacitors with capacitances $C_{l}$ (left) and $C_{r}$ (right), two Josephson junctions with Josephson energies $E_{J,l}$ (left) and $E_{J,r}$ (right) as well as an external flux $\Phi_{e}(t)$ threading through the loop which contains the two Josephson junctions.]{Circuit diagram of a network. The right branch contains linear capacitors with capacitances $C_{l}$ (left) and $C_{r}$ (right), two Josephson junctions with Josephson energies $E_{J,l}$ (left) and $E_{J,r}$ (right) as well as an external flux $\Phi_{e}(t)$ threading through the loop which contains the two Josephson junctions. The left branch contains a linear capacitor $C_{g}$ and a voltage source $V_{g}(t)$. The ground node is marked by a dashed triangle.}\label{fig:FTT_D_circuit}
\end{figure}\renewcommand{\PV}{\Phi_{n}}\renewcommand{\PVD}{\dot{\Phi}_{n}}
The next aim is to add a driving term of the form $n_{g}(t) \OP{n}$ to the model of a flux-tunable transmon. Figure \ref{fig:FTT_D_circuit} shows the circuit of a flux-tunable transmon, see \figref{fig:FTT_circuit}, with an additional branch which contains a voltage source characterised by the real-valued function $V_{g}(t)$ and an additional capacitor with the capacitance $C_{g}$. In the following we make use of Kirchhoff's voltage law for the center loop
\begin{equation}
  V_{C_{g}}+V_{g}(t)=V_{C_{l}},
\end{equation}
to derive the circuit Hamiltonian $\OP{H}^{*}$ for this system. The Lagrangian for this system reads
\renewcommand{\PE}{\Phi_{e}(t)}
\renewcommand{\PED}{\dot{\Phi}_{e}(t)}
\begin{equation}
  \mathcal{L}^{*}=\mathcal{L}+\frac{C_{g}}{2} \brr{\PLD-\VG}^{2}.
\end{equation}
Consequently, we can express $\mathcal{L}^{*}$ in terms of $\PV$
\begin{equation}
  \mathcal{L}^{*}=\mathcal{L}+\frac{C_{g}}{2\MD^{2}}\PVD^{2} - \frac{C_{g}}{\MD^{2}}\PVD\brr{\MR \PED+\MD \VG},
\end{equation}
where we dropped all terms proportional to $\PED^{2}$, $\VG^{2}$ and $\PED \VG$. In the end, these terms only contribute a non-measurable phase to the dynamics of the system. Again, we simplify the parametrisation by assuming $\MD=1$, $\MR=-\GP$, $\CL=\CR$ and $C=\CS$, the result reads
\begin{equation}
  \begin{split}
    \mathcal{L}^{*}&=\frac{\brr{C + C_{g}}}{2} \PVD^{2} - \brr{\frac{C}{2} - \GP \brr{C + C_{g}}} \PED \PVD\\
    &- C_{g} \VG \PVD - U\brr{\PV},
  \end{split}
\end{equation}
where as before we neglect all terms proportional to $\PED^{2}$. In an ad hoc manner, we assume that $C+C_{g} \rightarrow C $ such that the system's dynamic behaviour is described by the Lagrangian
\begin{equation}
  \mathcal{L}^{*}=\frac{C}{2} \PVD^{2} - C \brr{\frac{1}{2}-\GP} \PED \PVD - C_{g} \VG \PVD - U\brr{\PV}.
\end{equation}
Therefore, the conjugate variable reads
\begin{equation}
  Q=C \PVD - C \brr{\frac{1}{2}-\GP} \PED  - C_{g} \VG,
\end{equation}
and the first term of the Hamiltonian function can be expressed as
\begin{equation}
  Q\PV= \frac{Q^{2}}{C} + \brr{\frac{1}{2}-\GP} \PED Q + \frac{C_{g}}{C } \VG Q .
\end{equation}
After expressing the Lagrangian in terms of the variables $\PV$ and $Q$, the function reads
\begin{equation}
  \mathcal{L}^{*}=\frac{Q^{2}}{2C} - U(\PV),
\end{equation}
where we neglect all terms which only contribute non-measurable phases to the dynamics of the system. Consequently, the Hamiltonian can be expressed as
\begin{equation}
  H^{*}=E_{C} \brr{\frac{\CV}{2 e}}^{2} + \brr{\frac{1}{2}-\GP} (2 e) \PED \brr{\frac{\CV}{2 e}}  - 2 E_{C} n_{g}(t) \brr{\frac{\CV}{2 e}} + U\brr{\PV},
\end{equation}
where the real-valued function $n_{g}(t)$ is defined as
\begin{equation}
  n_{g}(t)=- \frac{C_{g} V_{g}(t)}{2 e}.
\end{equation}
After performing the canonical quantisation, completing the square with regard to the term $n_{g}(t) \CVO$, using the substitutions in \equref{eq:substitutions}(a-c) with $\hbar=1$ and dropping some terms which only contribute non-measurable phases to the dynamics of the system, we find the Hamilton operator\renewcommand{\PED}{\dot{\varphi}(t)}\renewcommand{\PE}{\varphi(t)}
\begin{equation}\label{eq:flux-tunable transmon with charge drive}
  \begin{split}
    \OP{H}_{\idxFTT}^{*}&=E_{C} \brr{\CVO -n_{g}(t) }^{2} + \brr{\frac{1}{2}-\GP} \PED \CVO - E_{J,l} \cos\brr{\PVO+\GP \PE}\\
    &- E_{J,r} \cos\brr{\PVO+ (\GP-1) \PE}.
  \end{split}
\end{equation}
Note that here we removed the node label $n$ from the operator $\OP{\Phi}_{n}$. Furthermore, as before the second term on the right-hand side of \equref{eq:flux-tunable transmon with charge drive} contains the factor $\hbar=1$. We use the Hamiltonian in \equref{eq:flux-tunable transmon with charge drive} to describe flux-tunable transmons with an additional charge drive component.

\renewcommand{\PV}{\Phi(t)}
\renewcommand{\PVD}{\dot{\Phi}(t)}
\renewcommand{\brr}[1]{\left\{ #1 \}\right}

In this thesis, almost all results are obtained numerically. Therefore, we have to express the Hamiltonians in \equaref{eq:fixed-frequency transmon}{eq:flux-tunable transmon} in some discrete basis. The basis we use is the so-called charge basis $\{\ket{n}\}_{n \in \mathbb{Z}}$. We define the basis states by
\begin{equation}\label{eq:charge_basis_states}
  \ket{n}=\frac{1}{2\pi} \int_{0}^{2\pi} e^{-i n \varphi} \ket{\varphi} d\varphi,
\end{equation}
where $n \in \mathbb{Z}$, such that the charge operator can be expressed as
\begin{equation}
  \OP{n}= \sum_{n=-\infty}^{\infty} n \ketbra{n}{n}.
\end{equation}
The states $\ket{\varphi}$ are the eigenstates of the operator $\OP{\varphi}$. Here $\varphi \in [0,2\pi)$. Both operators are Hermitian and Fourier transform duals of each other. The charge operator $\OP{n}$ can be expressed as a differential operator
\begin{equation}
  \OP{n}=- i\partial_{\varphi},
\end{equation}
in the flux basis. Furthermore, the flux basis states read
\begin{equation}
  \ket{\varphi}=\sum_{n=-\infty}^{\infty} e^{i n \varphi} \ket{n}.
\end{equation}
If we consider the action of the operator
\begin{equation}
  e^{i\OP{\varphi}} \ket{n}=\frac{1}{2\pi} \int_{0}^{2\pi} \braket{\varphi|n} e^{i\varphi} \ket{\varphi} d\varphi,
\end{equation}
on the charge states $\ket{n}$, we find
\begin{equation}
  e^{i\OP{\varphi}} \ket{n}=\ket{n-1}.
\end{equation}
Consequently, we can express the cosine operator as
\begin{equation}
  \cos(\OP{\varphi})=\frac{1}{2} \sum_{n=-\infty}^{\infty} \BRR{\ketbra{n}{n+1} + \ketbra{n+1}{n}}.
\end{equation}
The statements made regarding the charge basis states are not as accurate as required. However, a detailed discussion of the mathematical subtleties would require us to extend the mathematical framework considerably. Therefore, we abstain from doing so and refer the interested reader to \REF\cite[Section 2.1.3]{Willsch2016Master}, which provides a summary of mathematical issues that appear upon closer inspection. For the simulations in this thesis, we always use a finite number $2 N_{c} +1$, where $N_{c} \in \mathbb{N}^{0}$, of basis states $\{\ket{-N_{c}}, ...,\ket{0}, ..., \ket{+N_{c}}\}$ to express the Hamiltonians \equaref{eq:fixed-frequency transmon}{eq:flux-tunable transmon} as matrices in a computer program.

If we consider qubits, which are abstract states by definition, see \secref{sec:The single-qubit space}, as our object of study, we usually assume that the dynamics of the system can, in some basis, be confined to a small subspace. An even smaller subspace is used as the computational basis, see \equref{eq:CBQ}, for NIGQC models. Since this is a rather subtle issue, it is best to further clarify this statement. The aim is to control the system such that the state vector can be expressed in terms of the computational basis at certain points in time,\ie the moments in time where a measurement would take place. The qubit states themselves are usually associated with the eigenstates $\ket{\phi^{(z)}(t)}$ and eigenvalues $E^{(z)}(t)$ of some time-dependent Hamiltonian $\OP{H}(t)$, given by
\begin{equation}\label{eq: stationary Schroedinger equation}
  \OP{H}(t)\ket{\phi^{(z)}(t)}=E^{(z)}(t)\ket{\phi^{(z)}(t)},
\end{equation}
where $z\in \mathbb{N}^{0}$. Note that we usually use the states at $t=0$ as qubit states.

In this thesis, we use the eigenstates of the fixed-frequency and flux-tunable transmons as qubit states. In this context, we define the circuit Hamiltonian NIGQC model qubit frequency $\omega^{(Q_{0})}=E^{(1)}(0)-E^{(0)}(0)$ and anharmonicity $\alpha^{(Q_{0})}=E^{(2)}(0)-E^{(0)}(0)- 2\omega^{(Q_{0})}$ in terms of the eigenvalues $E^{(z)}(t)$ of the corresponding systems. Additionally, in both cases we can express $\ket{\phi^{(z)}(t)}$ and $E^{(z)}(t)$ in terms of the Mathieu functions $\mathcal{M}_{A}$, $\mathcal{M}_{C}$ and $\mathcal{M}_{S}$. Here we use the same notation as the author of \REF\cite{Cottet2002}. This is a consequence of the fact that the Hamiltonian in \equref{eq:fixed-frequency transmon} and the Hamiltonian in \equref{eq:flux-tunable transmon recast} for $\beta=1/2$ are structurally equivalent,\ie the only difference between both Hamiltonians is that they show different time dependencies, with regard to the variables $n_{g}(t)$ and $\varphi(t)$. Despite the fact that both systems share this particular kind of equivalence, we find that their behaviour, with respect to the variables $n_{g}(t)$ and $\varphi(t)$ is rather different. Therefore, we discuss and compare both systems, with respect to their corresponding dynamic variables, by analysing the eigenstates and eigenvalues.

\begin{table}[!tbp]
\caption[Parameters for a fixed-frequency (row $i=0$) and a flux-tunable transmon (row $i=1$). The first column contains indices $i$ for the different parameter sets. The second and third column show the qubit frequency $\omega^{(Q_{0})}$ and anharmonicity $\alpha^{(Q_{0})}$. ]{Parameters for a fixed-frequency (row $i=0$) and a flux-tunable transmon (row $i=1$). The first column contains indices $i$ for the different parameter sets. The second and third column show the qubit frequency $\omega^{(Q_{0})}$ and anharmonicity $\alpha^{(Q_{0})}$. The fourth, fifth and sixth column show the capacitive energy $E_{C}$, the left Josephson energy $E_{J_{l}}$ and the right Josephson energies $E_{J_{r}}$, respectively. The seventh column shows the flux offset parameter $\varphi_{0}$ which is determined by $\varphi(0)$, see \equref{eq:flux-tunable transmon}. All units are in GHz except the flux offset parameter $\varphi_{0}$ which is without units. The parameters are motivated by \REFS\cite{Ganzhorn20,Lacroix2020}.}\label{tab:overlap}
\begin{tabularx}{\textwidth}{ X X X X X X X  }
\hline\hline
$i$ &$\omega_{i}^{(Q_{0})}/2 \pi$&$ \alpha_{i}^{(Q_{0})}/2\pi$ &$E_{C_{i}}/2\pi$&$E_{J_{l,i}}/2\pi$& $E_{J_{r,i}}/2\pi$ & $\varphi_{0,i}/2\pi$ \\
\hline
0 & 6.200 & -0.285 & 1.027 & 20.371  & n/a    & n/a \\
1 & 5.200 & -0.295 & 1.036 & 4.817  & 9.633 & 0 \\
\hline\hline
\end{tabularx}
\end{table}

The eigenstates of a fixed-frequency transmon in the $\varphi$-space read
\begin{equation}\label{eq:Mathieufunctions}
  \phi^{(z)}(\varphi,t)=\frac{e^{i n_{g}(t) \varphi}}{\sqrt{2 \pi}}\left(\mathcal{M}_{C}(a(t),b,\frac{\varphi}{2}) +i^{1+2(z+1)}\mathcal{M}_{S}(a(t),b,\frac{\varphi}{2})\right),
\end{equation}
where we introduced the auxiliary functions
\begin{subequations}
\begin{align}
  a(t)&= \frac{4 E^{(z)}(t)}{E_{C}}, \\
   b&= \frac{-2 E_{J}}{E_{C}},
\end{align}
\end{subequations}
and the eigenvalues
\begin{equation}
   E^{(z)}(t)=E_{C} \mathcal{M}_{A}( z+1-(z+1) [\mathrm{mod} 2] + 2 n_{g}(t)(-1)^{z},b)).
\end{equation}
The variable $n_{g}(t)$ enters the eigenstates twice:~explicitly through the global phase $e^{i n_{g}(t) \varphi}$ and implicitly through $E^{(z)}(t)$. We emphasise that eigenstates and eigenvalues are taken from \REF\cite{Cottet2002}.

\graphicspath{{./FiguresAndData/EigenvaluesAndEigenfunctions/}}
\begin{figure}[!tp]
    \centering
    \begin{minipage}{0.75\textwidth}
        \centering
        \includegraphics[width=\width\textwidth]{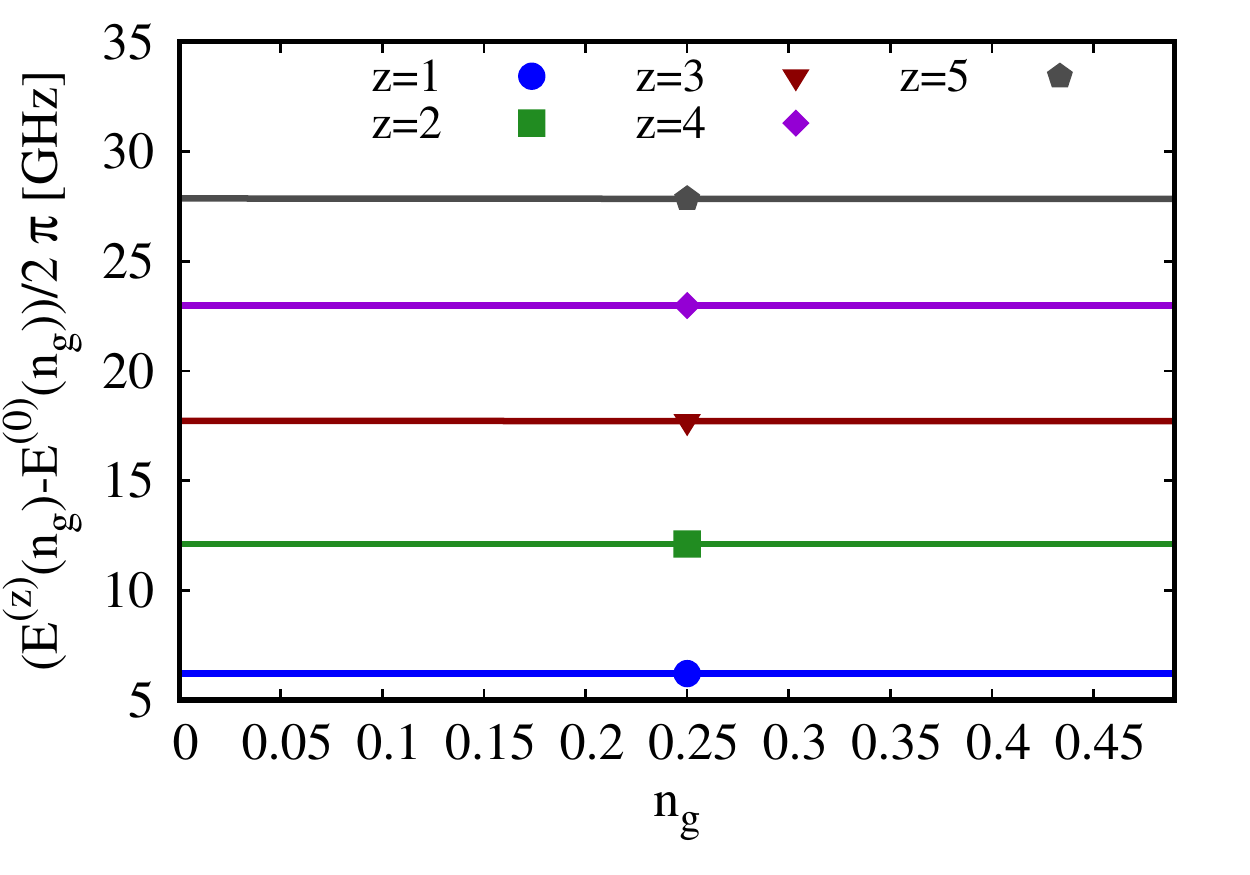}
    \end{minipage}\hfill
    \caption[The lowest five energy levels of a fixed-frequency transmon as a function of the effective offset charge $n_{g}$.]{The lowest five energy levels of a fixed-frequency transmon as a function of the effective offset charge $n_{g}$. We use the Hamiltonian \equref{eq:fixed-frequency transmon}, $N_{c}=50$ charge states and the parameters listed in \tabref{tab:overlap}, row $i=0$, to obtain the results.}\label{fig:spectrum_fft}
\end{figure}

Figure \ref{fig:spectrum_fft} shows the lowest five states of the energy spectrum of a fixed-frequency transmon as a function of effective offset charge $n_{g}$. Here we use the Hamiltonian given by \equref{eq:fixed-frequency transmon}, $N_{c}=50$ charge states and the parameters listed in \tabref{tab:overlap}, row $i=0$, to diagonalise the corresponding matrix representation. As one can see, \figref{fig:spectrum_fft} shows five quasi-constant energies. To the best knowledge of the author, the contributors of \REF\cite{Koch} were the first to notice that if the system is operated in the transmon limit $E_{J}/E_{C} \gg 1$, the eigenvalues $E^{(z)}(t)$ become quasi constant with regard to the variable $n_{g}$. For the remainder of this discussion, we assume that the eigenvalues are actually constant. Therefore, in the transmon limit the eigenstates only vary in time by means of a global phase. If we assume that $n_{g}(0)=0$, the eigenstates read
\begin{equation}
  \phi^{(z)}(\varphi,t)=e^{i n_{g}(t) \varphi} \phi^{(z)}(\varphi,0).
\end{equation}
We define that the \TB{} states are the eigenstates of $\OP{H}_{\idxFFT}(t)$ for $t=0$. The overlap between the \TB{} states and $\ket{\phi^{(z)}(t)}$ then reads
\begin{equation}
  \braket{ \phi^{(z)}(t)|\phi^{(z)}(0)}=\int_0^{2\pi} e^{-i n_{g}(t) \varphi} |\phi^{(z)}(\varphi,0)|^{2}\mathrm{d}\varphi.
\end{equation}
For practical purposes,\ie the implementation of quantum gates, it is fair to assume that $n_{g}(t) \leq 0.05$, see for example Refs.~\cite{Wi17,Willsch2020}. In such cases, we find that the argument $n_{g}(t) \varphi$ only varies slowly. Therefore, one may assume that $\braket{ \phi^{(z)}(t)|\phi^{(z)}(0)} \simeq \braket{ \phi^{(z)}(0)|\phi^{(z)}(0)}$ for all $t$. Note that the overlap cannot be fully time independent, otherwise we would not be able to implement any gates with a fixed-frequency transmon.

\begin{figure}[!tbp]
    \centering
    \begin{minipage}{0.49\textwidth}
        \centering
        \includegraphics[width=\width\textwidth]{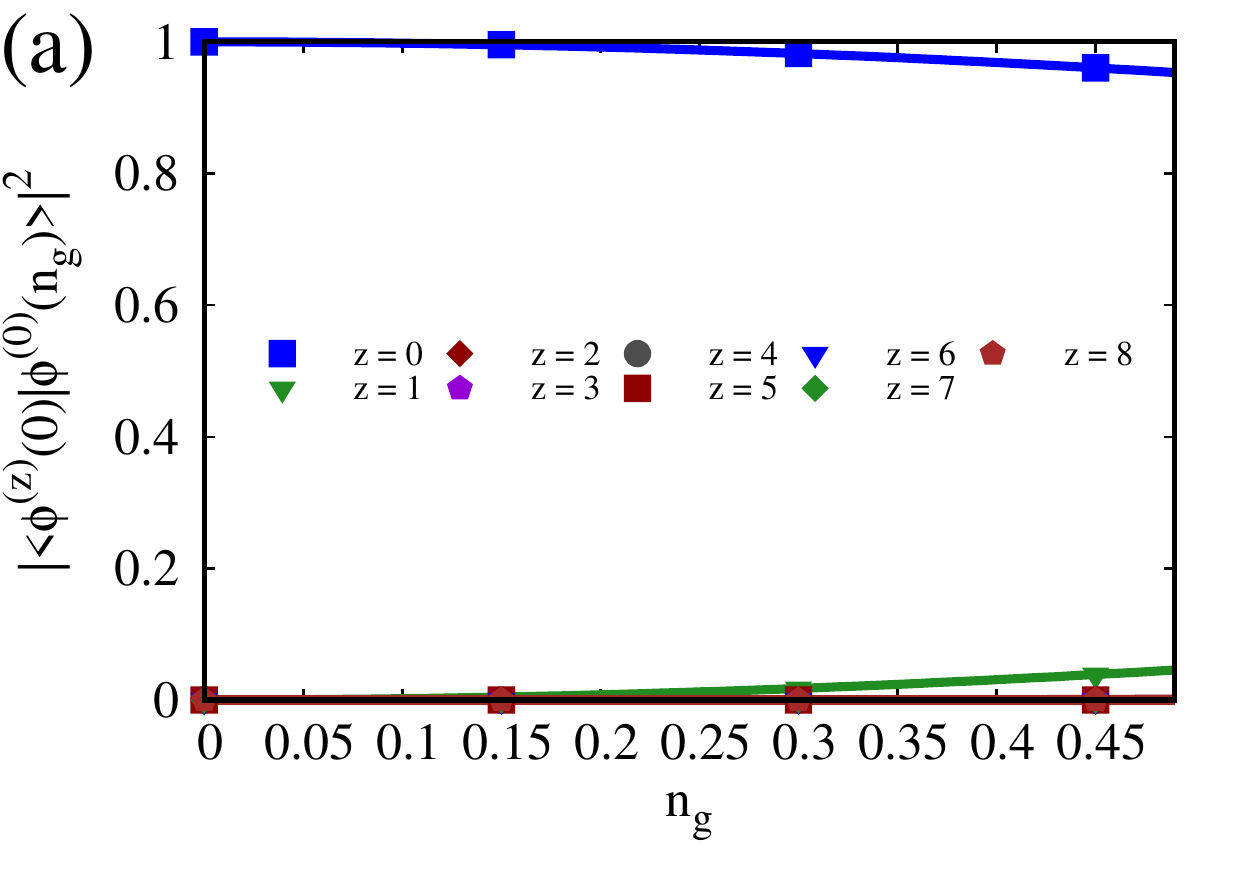} 
    \end{minipage}\hfill
    \begin{minipage}{0.49\textwidth}
        \centering
        \includegraphics[width=\width\textwidth]{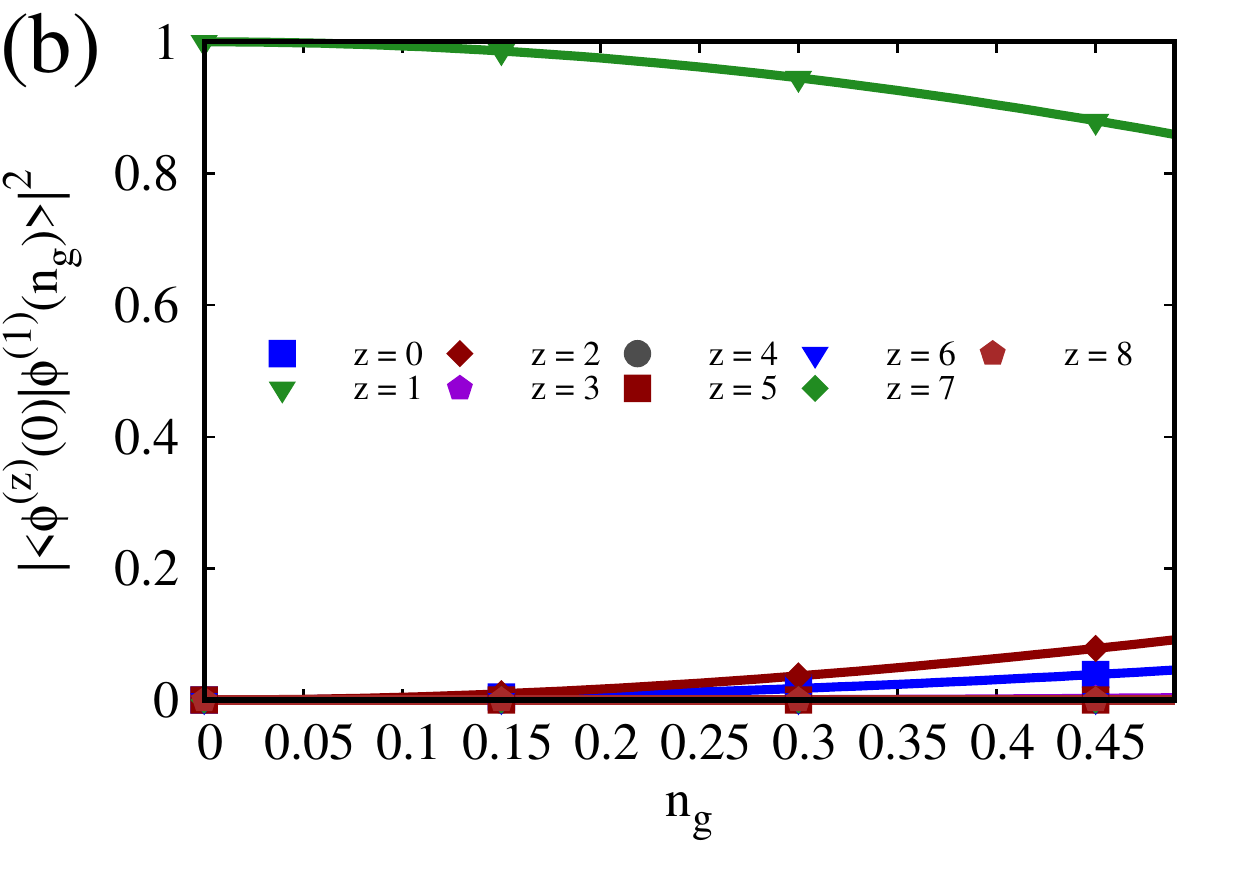} 
    \end{minipage}
    \caption[Overlaps $|\braket{\phi^{(z)}(0)|\phi^{(0)}(n_{g})}|^{2}$ (a) and $|\braket{\phi^{(z)}(0)|\phi^{(1)}(n_{g})}|^{2}$ (b) between the instantaneous eigenstates $\ket{\phi^{(0/1)}(\varphi)}$ and various measurement basis states $\ket{\phi^{(z)}(0)}$ as functions of the external charge $n_{g}$.]{Overlaps $|\braket{\phi^{(z)}(0)|\phi^{(0)}(n_{g})}|^{2}$ (a) and $|\braket{\phi^{(z)}(0)|\phi^{(1)}(n_{g})}|^{2}$ (b) between the instantaneous eigenstates $\ket{\phi^{(0/1)}(\varphi)}$ and various measurement basis states $\ket{\phi^{(z)}(0)}$, where $z \in \{0,1,2,3,4,5,6,7,8,9\}$, as functions of the external charge $n_{g}$. The basis states $\ket{\phi^{(z)}(\varphi)}$ are obtained by numerical diagonalisation. We use Hamiltonian \equref{eq:fixed-frequency transmon}, $N_{c}=50$ charge states and the parameters listed in \tabref{tab:overlap}, row $i=0$, to obtain the results.}\label{fig:overlap_charge}
\end{figure}

To study this relation for various values of $n_{g}$, \figref{fig:overlap_charge}(a-b) show the overlaps $|\braket{\phi^{(z)}(0)|\phi^{(0)}(n_{g})}|^{2}$ (a) and $|\braket{\phi^{(z)}(0)|\phi^{(1)}(n_{g})}|^{2}$ (b) between the instantaneous eigenstates $\ket{ \phi^{(0)}( n_{g} )}$ and $\ket{ \phi^{(1)}( n_{g} )}$ and various transmon basis states $\ket{\phi^{(z)}(0)}$, where $z \in \{0,1, ...,9\}$, as functions of the external charge $n_{g}$. We use Hamiltonian \equref{eq:fixed-frequency transmon}, $N_{c}=50$ charge states and the parameters listed in \tabref{tab:overlap}, see $i=0$, to compute the overlap.

\graphicspath{{./FiguresAndData/EigenvaluesAndEigenfunctions/}}

Figures~\ref{fig:overlap_charge}(a-b) show that for $n_{g} \leq 0.05$ the overlaps barely deviate from one or zero. However, these small deviations are sufficient to implement single-qubit gates. For $n_{g} > 0.05$, the state $\ket{ \phi^{(0)}( n_{g} )}$ can possibly be expressed as a superposition of the states $\ket{ \phi^{(0)}( 0 )}$ and $\ket{ \phi^{(1)}( 0 )}$. Similarly, the state $\ket{ \phi^{(1)}( n_{g} )}$ can possibly be expressed as a superposition of the lowest three \TB{} states.

As a consequence, we find that for small $n_{g}$ fixed-frequency transmons behave like artificial atoms,\ie like systems with eigenvalues and eigenstates which are approximately time independent. These properties allow us to activate transitions, by means of Rabi oscillations, see \REF\cite{CT10}, between different \TB{} states, potentially without leaving the lowest three or four states. The dynamics of certain circuit architectures, which solely rely on fixed-frequency transmons and resonators as coupler elements, might be modelled with four \TB{} states only, see Ref. \cite{Willsch2020}.

\newcommand{\idxfft}{\text{fix.}}
Fixed-frequency transmons are often described by effective models, see Refs.\cite{Roth19,McKay16,Koch,Blais2020circuit,Gu21} and many more. We can develop such an effective model in a step-wise manner. Note that for simplicity, we set $n_{g}(0)=0$ in the beginning. If we expand the cosine in \equref{eq:fixed-frequency transmon} to the second order, we find the Hamiltonian given by \equref{eq:Harmonic} with $E_{L}=E_{J}$ such that the harmonic oscillator frequency given by \equref{eq:res_freq} reads $\omega=\sqrt{2 E_{C} E_{J}}$. Note that here we neglect all terms which only contribute a non-measurable phase to the dynamics of the system. If we expand the cosine in \equref{eq:fixed-frequency transmon} to the quartic order, we find the Hamiltonian
\begin{equation}
  \OP{H}_{\text{quartic}}= \omega \OP{c}^{\dagger}\OP{c} + -\frac{E_{C}}{48} \left(\OP{c}^{\dagger} - \OP{c}\right)^{4},
\end{equation}
where $\OP{c}^{\dagger}$ and $\OP{c}$ are the bosonic operators, see \secref{sec:ResAndTLS}, which act on the harmonic basis states $\HBS$. We can split the operator
\begin{equation}\label{eq:quatric_term_decompo}
  \left(\OP{c}^{\dagger} - \OP{c}\right)^{4}= \OP{D}+\OP{V},
\end{equation}
into an operator $\OP{D}$ which is diagonal in the harmonic basis and one $\OP{V}$ which is non-diagonal in the same basis. We neglect the operator $\OP{V}$ and define another Hamiltonian
\begin{equation}\label{eq:fixed-frequency eff}
  \OP{H}_{\idxfft}= \omega^{(q_{0})} \OP{c}^{\dagger}\OP{c} + \frac{\alpha^{(q_{0})}}{2} \BRR{\OP{c}^{\dagger}\OP{c}\BRR{\OP{c}^{\dagger}\OP{c}-\OP{I}}},
\end{equation}
where $\omega^{(q_{0})}=\omega + \alpha^{(q_{0})}$ and $\alpha^{(q_{0})}=-E_{C}/4$. The model parameters $\omega^{(q_{0})}$ and $\alpha^{(q_{0})}$ determine the first and second energy levels of the system $\OP{H}_{\idxfft}$. The former $\omega^{(q_{0})}$ denotes the transmon qubit frequency and the latter $\alpha^{(q_{0})}$ is referred to as the transmon qubit anharmonicity. Note that the derivation of Hamiltonian \equref{eq:fixed-frequency eff} is similar but not equivalent to one which can be found in \REF\cite[Section B 4.1.3]{DiVincenzo13}.

We can also incorporate the driving term in \equref{eq:fixed-frequency transmon} by noting that we can factor out the term $-2E_{C}n_{g}(t)\hat{n}$. The corresponding driving term reads
\begin{equation}\label{eq:fixed-frequency drive}
   \OP{D}_{\text{Dri.}}(t)= \Omega(t) \BRR{\OP{c}^{\dagger} + \OP{c}},
\end{equation}
where $\Omega(t) \propto -2 E_{C}n_{g}(t)$. In the end, we define the effective Hamiltonian for a fixed-frequency transmon as
\begin{equation}
  \OP{H}_{\idxfft}(t)= \omega^{(q_{0})} \OP{c}^{\dagger}\OP{c} + \frac{\alpha^{(q_{0})}}{2} \BRR{\OP{c}^{\dagger}\OP{c}\BRR{\OP{c}^{\dagger}\OP{c}-\OP{I}}} + \Omega(t) \BRR{\OP{c}^{\dagger} + \OP{c}}.
\end{equation}
It is often the case that only the lowest two or three basis states are used to model these systems. If we use two basis states only, we can express the effective Hamiltonian in terms of Pauli spin operators as
\begin{equation}
  \OP{H}_{\idxfft,\idxTLS}(t) = -\frac{\omega}{2} \OP{\sigma}^{(z)} + \Omega(t) \OP{\sigma}^{(x)},
\end{equation}
where we omit the contribution of all higher-order terms in the cosine expansion, see $\omega^{(q_{0})}$.

We emphasise that one cannot claim that the circuit Hamiltonian given by \equref{eq:fixed-frequency transmon} and any of the effective Hamiltonians discussed so far generate the same dynamics in terms of the TDSE. Since we introduce more effective Hamiltonians in this section, we address this issue in a general manner at the end of this section.

Next, we turn our attention to the flux-tunable transmon. As mentioned before, the Mathieu functions are also the eigenstates of the flux-tunable transmon Hamiltonian given by \equref{eq:flux-tunable transmon recast}. The Hamiltonians in \equref{eq:fixed-frequency transmon} and \equref{eq:flux-tunable transmon recast} are structurally equivalent, only the parameter dependencies are different,\ie we have to perform the substitutions $E_{\mathrm{J}} \rightarrow E_{\mathrm{J, eff}}(t)$ and $\varphi \rightarrow \varphi-\varphi_{\mathrm{eff}}(t)$. However, this means that flux-tunable transmons have time-dependent eigenvalues and eigenstates.

\graphicspath{{./FiguresAndData/EigenvaluesAndEigenfunctions/}}
\begin{figure}[!tbp]
    \centering
    \begin{minipage}{0.75\textwidth}
        \centering
        \includegraphics[width=\width\textwidth]{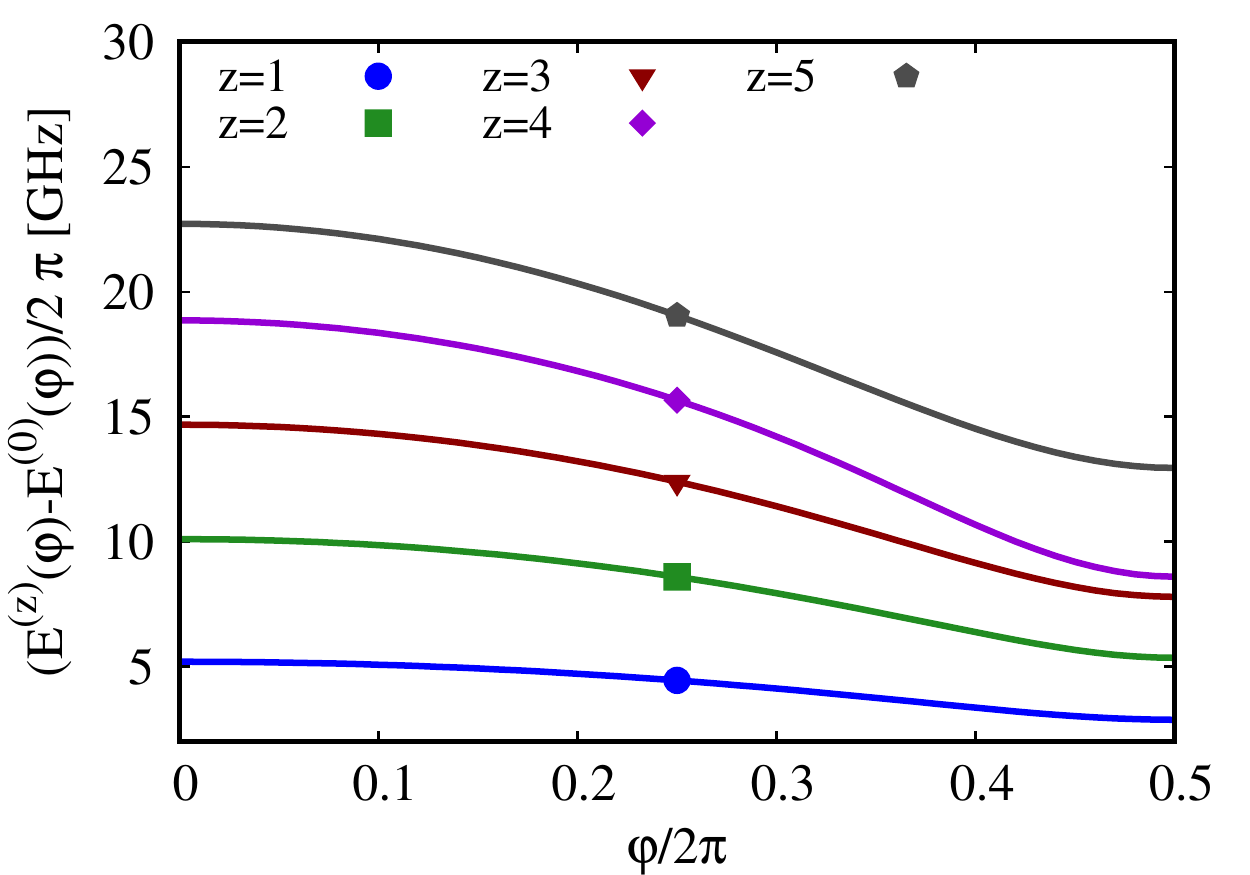}
    \end{minipage}
    \caption[The lowest five energy levels of a flux-tunable transmon as a function of the flux offset $\varphi$.]{The lowest five energy levels of a flux-tunable transmon as a function of the flux offset $\varphi$. We numerically diagonalise Hamiltonian \equref{eq:flux-tunable transmon} with $N_{c}=50$ charge states and the parameters listed in \tabref{tab:overlap}, row $i=1$.}\label{fig:spectrum_ftt}
\end{figure}
Figure \ref{fig:spectrum_ftt} shows the lowest five energy levels of a flux-tunable transmon as a function of the flux offset $\varphi$. Here we use the Hamiltonian given by \equref{eq:flux-tunable transmon}, $N_{c}=50$ charge states and the parameters listed in \tabref{tab:overlap}, row $i=1$, to obtain the results. As one can see, all energies decrease with increasing flux $\varphi$. If one uses the lowest two states as qubit states, one might associate the tunable energy levels with a tunable qubit frequency.

\graphicspath{{./FiguresAndData/EigenvaluesAndEigenfunctions/}}
\begin{figure}[!tbp]
    \centering
    \begin{minipage}{0.49\textwidth}
        \centering
        \includegraphics[width=\width\textwidth]{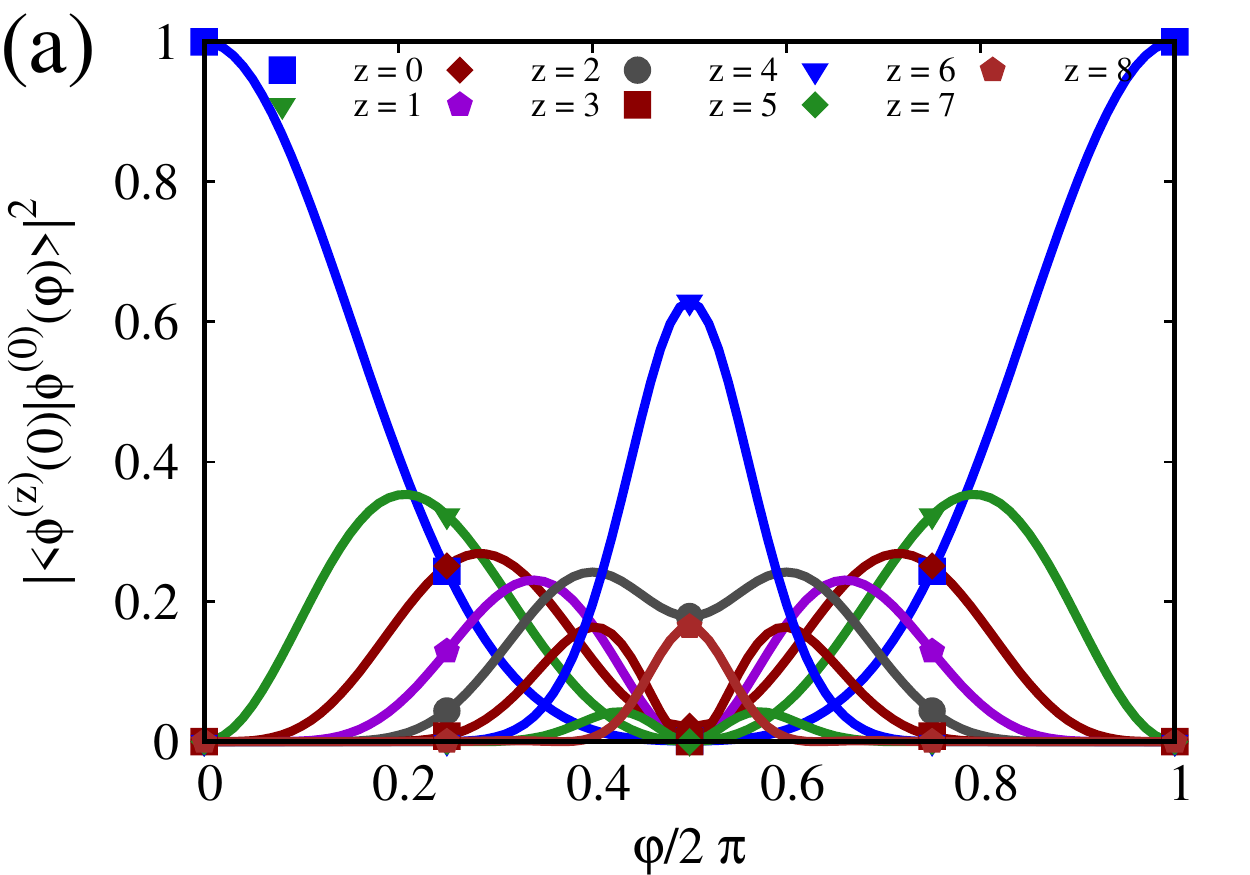} 
    \end{minipage}\hfill
    \begin{minipage}{0.49\textwidth}
        \centering
        \includegraphics[width=\width\textwidth]{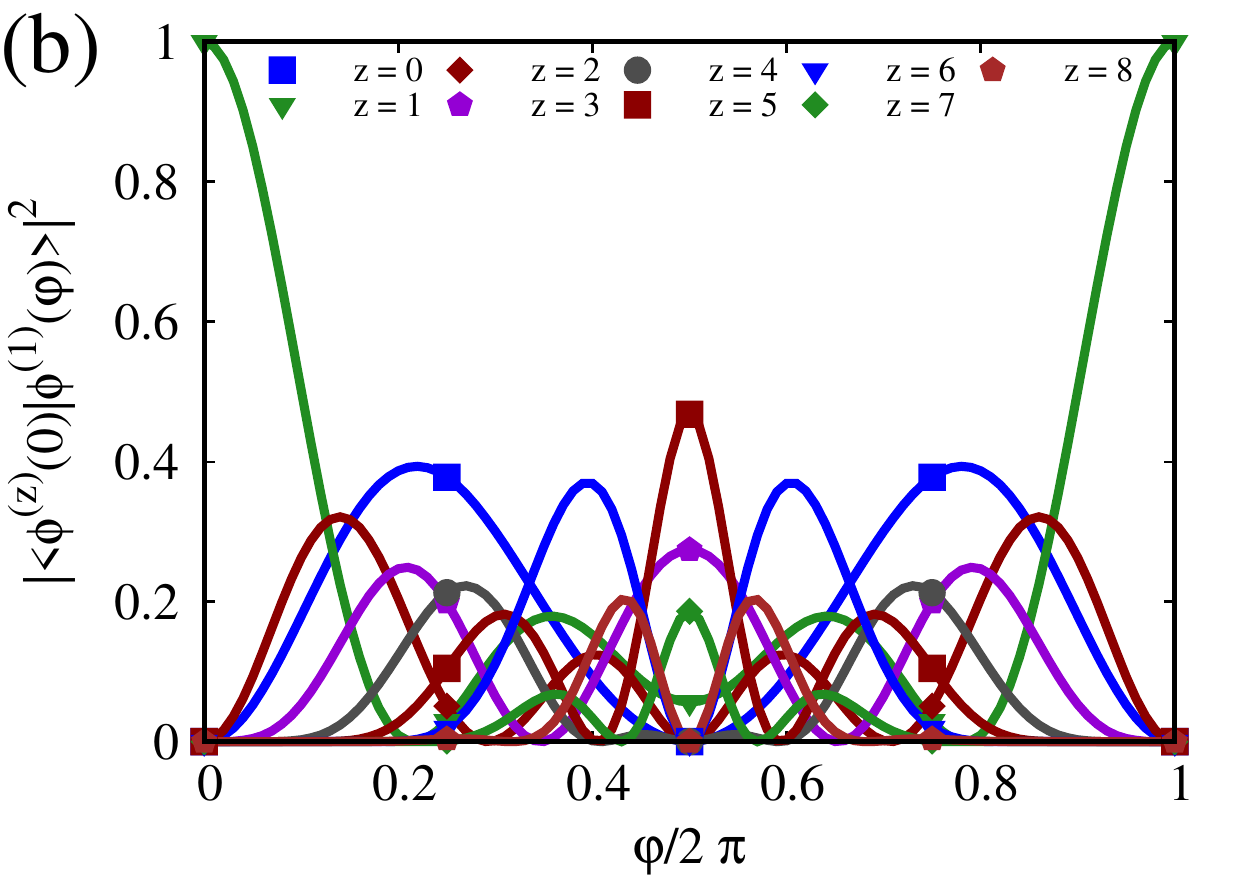} 
    \end{minipage}
    \caption[Overlaps $|\braket{\phi^{(z)}(0)|\phi^{(0)}(\varphi)}|^{2}$ (a) and $|\braket{\phi^{(z)}(0)|\phi^{(1)}(\varphi)}|^{2}$ (b) between the instantaneous eigenstates $\ket{\phi^{(0/1)}(\varphi)}$ and various measurement basis states $\ket{\phi^{(z)}(0)}$ as functions of the external flux $\varphi$.]{Overlaps $|\braket{\phi^{(z)}(0)|\phi^{(0)}(\varphi)}|^{2}$ (a) and $|\braket{\phi^{(z)}(0)|\phi^{(1)}(\varphi)}|^{2}$ (b) between the instantaneous eigenstates $\ket{\phi^{(0/1)}(\varphi)}$ and various measurement basis states $\ket{\phi^{(z)}(0)}$, where $z \in \{0,1,2,3,4,5,6,7,8,9\}$, as functions of the external flux $\varphi$. The basis states $\ket{\phi^{(z)}(\varphi)}$ are obtained by numerical diagonalisation of Hamiltonian \equref{eq:flux-tunable transmon}. We use $N_{c}=50$ charge states and the parameters listed in \tabref{tab:overlap}, row $i=1$, to obtain the results.}
    \label{fig:overlap_flux}
\end{figure}
Figures~\ref{fig:overlap_flux}(a-b) show the overlaps $|\braket{\phi^{(z)}(0)|\phi^{(0/1)}(\varphi)}|^2$ between the instantaneous eigenstates $\ket{ \phi^{(0)}( \varphi )}$(a) and $\ket{ \phi^{(1)}( \varphi )}$(b) and various \TB{} states $\ket{\phi^{(z)}(0)}$, where $z \in \{0,1, ...,9\}$, as functions of the external flux $\varphi/2\pi \in [0,1]$. Here we use the parameters listed in \tabref{tab:overlap}, row $i=1$, the Hamiltonian given by \equref{eq:flux-tunable transmon} and $N_{c}=50$ charge states to compute the overlaps between the different states. Figures \ref{fig:overlap_flux}(a-b) show that the overlaps between the instantaneous basis states and the \TB{} states $z=0$ and $z=1$ vanish as $\varphi/2 \pi$ approaches the point $0.5$, afterwards they return back to one. Hamiltonian \equref{eq:flux-tunable transmon recast} has two symmetry points, one at $\varphi/2 \pi=0.5$ and one at $\varphi/2 \pi=1.0$. The two symmetry points explain this behaviour.

In accordance with the adiabatic theorem, see \REFS\cite{Weinberg2015,Amin09}, the external flux might allow us to move the system from a state $\ket{\phi^{(z)}(\varphi_{0})}$, where $\varphi_{0}=\varphi(0)$ is an arbitrary flux offset which defines the \TB{}, with energy $E^{(z)}(\varphi_{0})$ to a state $\ket{\phi^{(z)}(\varphi(t))}$ with energy $E^{(z)}(\varphi(t))$, see \figref{fig:spectrum_ftt}, as long as we vary the external flux slowly enough. Since in general $\braket{\phi^{(z)}(\varphi(t))|\phi^{(z)}(\varphi_{0})}\neq 1$, we expect that the description of these systems in the \TB{} probably requires more than just two or three basis states. The instantaneous basis states are linear combinations of the lowest transmon basis states, see \figsref{fig:overlap_flux}(a-b) and the amplitude of $\varphi(t)$ determines the overlap or how many basis states are needed to describe the system.

\newcommand{\idxftt}{\text{tun.}}
As for fixed-frequency transmons, we find that flux-tunable transmons are often modelled by means of effective models, see for example \REFS\cite{McKay16,Roth19,Yan18,Didier,Rol19,Blais2020circuit,Gu21} and we can develop such an effective model in a step-wise manner. If we expand the Hamiltonian given by \equref{eq:flux-tunable transmon recast} to the second order in the cosine, we can define the Hamiltonian
\begin{equation}\label{eq:Harmonic_td}
  \OP{H}_{\text{second}}(t)= E_{C} \OP{n}^{2} + \frac{E_{J,\text{eff.}}(t)}{2} \left(\OP{\varphi}-\varphi_{\text{eff.}}(t)\right)^{2},
\end{equation}
such that the structural similarity to the harmonic oscillator Hamiltonian in \equref{eq:Harmonic} should be obvious. The difference between the Hamiltonians in \equaref{eq:Harmonic}{eq:Harmonic_td} is that the eigenstates and eigenvalues of the Hamiltonian in \equref{eq:Harmonic_td} are now time dependent. The time-dependent eigenstates in $\varphi$-space read
\begin{equation}\label{eq:HarmoicBasisWaveFunctionTimeDep}
    \psi^{(z)}(x(t))= \frac{1}{\sqrt{2^{z} z!}}\sqrt[4]{\frac{\xi(t)}{\pi}} e^{\frac{-x(t)^{2}}{2}} H_{z}(x(t)),
\end{equation}
where $z\in \mathbb{N}^{0}$, $\xi(t)=\sqrt{E_{J,\text{eff.}}(t)/(2 E_{C})}$, $x(t)=\sqrt{\xi(t)}(\varphi-\varphi_{\text{eff.}}(t))$ and $H_{z}(x(t))$ denotes the Hermite polynomial of order $z$. The eigenvalues are characterised by the time-dependent frequency
\begin{equation}
  \omega(t)=\sqrt{2 E_{C} E_{J,\text{eff.}}(t)}.
\end{equation}
We use the time-dependent basis states $\ket{\psi^{(z)}(t)}$ to model the dynamics of the system. Consequently, we work with the transformed state vector
\begin{equation}
  \ket{\Psi^{*}(t)}=\mathcal{W}(t)\ket{\Psi(t)},
\end{equation}
where $\mathcal{W}(t)$ is a unitary transformation which maps the basis states $\ket{\psi^{(z)}(0)}$ at time $t=0$ to the states $\ket{\psi^{(z)}(t)}$ at time $t$. The TDSE for the state vector $\ket{\Psi^{*}(t)}$ should retain its original form, see \REFS\cite{WillschM2020,Weinberg2015}. Therefore, we have to transform the Hamiltonian as
\begin{equation}\label{eq:HamiltonianTrafo}
  \OP{H}^{*}(t)= \mathcal{W}(t)\OP{H}(t)\mathcal{W}^{\dagger}(t) -i \mathcal{W}(t)\partial_{t}\mathcal{W}^{\dagger}(t).
\end{equation}
We denote the second term in \equref{eq:HamiltonianTrafo} as the driving term
\newcommand{\DTF}{\OP{\mathcal{D}}_{\text{Dri.}}(t)}
\begin{equation}
  \DTF= -i \mathcal{W}(t)\partial_{t}\mathcal{W}^{\dagger}(t),
\end{equation}
and we can determine this term to be
\begin{equation}\label{eq:drive_term_ftt}
  \DTF= - i \sqrt{\frac{\xi(t)}{2}} \dot{\varphi}_{\text{eff.}}(t) \BRR{\OP{c}^{\dagger}- \OP{c}} + \frac{i}{4} \frac{\dot{\xi}(t)}{\xi(t)} \BRR{\OP{c}^{\dagger} \OP{c}^{\dagger} - \OP{c} \OP{c}},
\end{equation}
where we assume that $\xi(t)\neq0$ for all times $t$. Here we adjusted the derivation in \REF\cite[Section 5.1.2]{WillschM2020} which considers a similar model. Additionally, we find
\begin{equation}\label{eq:flux_factor_one}
  \dot{\varphi_{\text{eff}}}(t)=\dot{\varphi}(t)\frac{d}{2 \left(\cos\left(\frac{\varphi(t)}{2}\right)^{2}+d^{2} \sin\left(\frac{\varphi(t)}{2}\right)^{2}\right)},
\end{equation}
and
\begin{equation}\label{eq:flux_factor_two}
  \frac{\dot{\xi}(t)}{\xi(t)}=\dot{\varphi}(t)\frac{(d^{2}-1) \sin(\varphi(t))}{8 \left(\cos\left(\frac{\varphi(t)}{2}\right)^{2}+d^{2} \sin\left(\frac{\varphi(t)}{2}\right)^{2}\right)}.
\end{equation}
The transformed Hamiltonian in the time-dependent harmonic basis can be expressed as
\begin{equation}\label{eq:Harmonic_td_trafo}
  \OP{H}_{\text{second}}^{*}(t)=\omega(t) \OP{c}^{\dagger}\OP{c} +  -i \sqrt{\frac{\xi(t)}{2}} \dot{\varphi}_{\text{eff.}}(t) \BRR{\OP{c}^{\dagger}- \OP{c}} + \frac{i}{4} \frac{\dot{\xi}(t)}{\xi(t)} \BRR{\OP{c}^{\dagger} \OP{c}^{\dagger} - \OP{c} \OP{c}},
\end{equation}
where we neglect all terms which only contribute a non-measurable phase to the dynamics of the system. Note that so far we only considered a time-dependent harmonic oscillator model.

If we use the lowest two eigenstates only, we can use the effective Hamiltonian
\begin{equation}
  \OP{H}_{\text{second},\idxTLS}^{*}(t)=-\frac{\omega(t)}{2} \OP{\sigma}^{(z)} + -\sqrt{\frac{\xi(t)}{2}} \dot{\varphi}_{\text{eff.}}(t) \OP{\sigma}^{(y)},
\end{equation}
to model the dynamics of the system. Some authors, see for example \REFS\cite{Roth19,McKay16,Yan18}, use the term $- (\omega(t)/2) \OP{\sigma}^{(z)}$ only to model flux-tunable transmons as two-level systems. If we consider the Hamiltonian given by \equref{eq:flux-tunable transmon recast} and compare it with the term $- (\omega(t)/2) \OP{\sigma}^{(z)}$, we find that with such an approximation all higher-order terms in the cosine expansion do not contribute to the dynamics of the system. Additionally, the driving term in \equref{eq:drive_term_ftt} and all higher states are ignored completely.

We can include some contributions of the higher-order terms in the cosine expansion as for the case of the fixed-frequency transmon, see \equref{eq:quatric_term_decompo}. First, we expand the cosine in \equref{eq:flux-tunable transmon recast} to the quartic order and define the Hamiltonian
\begin{equation}
  \OP{H}_{\text{quartic}}^{*}(t)=\omega(t) \OP{c}^{\dagger}\OP{c} + -\frac{E_{C}}{48} \BRR{\OP{c}^{\dagger} -\OP{c}}^{4}  + \DTF.
\end{equation}
Second, we split the operator
\begin{equation}
  \BRR{\OP{c}^{\dagger} -\OP{c}}^{4}=\OP{D}+\OP{V},
\end{equation}
into a diagonal $\OP{D}$ and non-diagonal $\OP{V}$ part, in the time-dependent harmonic basis. Third, we define the two effective Hamiltonians
\begin{equation}\label{eq:fft_eff_II}
  \OP{H}_{\idxftt,I}^{*}(t)= \omega^{(q)}(t) \OP{c}^{\dagger}\OP{c} + \frac{\alpha^{(q_{0})}}{2} \BRR{\OP{c}^{\dagger}\OP{c}\BRR{\OP{c}^{\dagger}\OP{c}-\OP{I}}}+ \DTF,
\end{equation}
and
\begin{equation}\label{eq:tunable-frequency eff}
  \OP{H}_{\idxftt,II}^{*}(t)= \omega^{(q)}(t) \OP{c}^{\dagger}\OP{c} + \frac{\alpha^{(q_{0})}}{2} \BRR{\OP{c}^{\dagger}\OP{c}\BRR{\OP{c}^{\dagger}\OP{c}-\OP{I}}},
\end{equation}
where $\omega^{(q)}(t)=\omega(t)+\alpha^{(q_{0})}$ and $\alpha^{(q_{0})}=-E_{C}/4$. With the definition of the Hamiltonian given by \equref{eq:fft_eff_II} we neglect all non-diagonal contributions $\OP{V}$. Similarly, with the definition of the Hamiltonian in \equref{eq:tunable-frequency eff} we neglect all non-diagonal contributions $\OP{V}$ and the driving term $\DTF$.

To the best knowledge of the author, one often finds that the tunable frequency $\omega^{(q)}(t)$ is modelled with the function
\begin{equation}\label{eq:tunable frequency}
  \omega^{(q)}(t)=\omega^{(q_{0})} \sqrt[4]{\cos\left(\frac{\varphi(t)}{2}\right)^{2} + d^{2} \sin\left(\frac{\varphi(t)}{2}\right)^{2}},
\end{equation}
only, where $\omega^{(q_{0})}=\const$ or $\omega^{(q_{0})}=\sqrt{2 E_{C} E_{\Sigma}}$ and flux-tunable transmons are modelled with the effective Hamiltonian given by \equref{eq:tunable-frequency eff}, see for example \REFS\cite{McKay16,Roth19,Ganzhorn20,Gu21}.

We now turn to the question how the dynamics generated by an original model Hamiltonian $\OP{H}_{M}(t)$ and the corresponding effective Hamiltonian $\OP{H}_{E}(t)$ deviate in time. The formal solutions of the TDSE for both Hamiltonians can be expressed as
\begin{equation}
  \hat{\mathcal{U}}_{\text{M}}(t,t_{0})=\mathcal{T} \exp\left( -i \int_{t_{0}}^{t} \hat{H}_{\text{M}}(t^{\prime}) dt^{\prime} \right)
\end{equation}
and
\begin{equation}
  \hat{\mathcal{U}}_{\text{E}}(t,t_{0})=\mathcal{T} \exp\left( -i \int_{t_{0}}^{t} \hat{H}_{\text{E}}(t^{\prime}) dt^{\prime} \right),
\end{equation}
where $\hat{\mathcal{U}}_{\text{M}}$ and $\hat{\mathcal{U}}_{\text{E}}$ denote the time-evolution operators for both Hamiltonians. Consequently, a quantitative assessment of the deviations between both models should be based on the time-evolution operators in combination with an appropriate operator norm. Unfortunately, we find that in almost all cases the solution of the TDSE is not known such that a detailed comparison between both solutions is a non-trivial problem in itself. A general discussion of this issues and explicit examples can be found in \REF\cite{Burgarth21}.

Comparing the time evolution of two Hamiltonian models can pose a rather difficult problem alone. However, we have to consider additional complications once we consider larger systems,\ie interacting many-particle systems. For example, by defining the effective Hamiltonian in \equref{eq:tunable-frequency eff}, we neglect the driving term given by \equref{eq:drive_term_ftt}. For simplicity, we consider the effective Hamiltonian in \equref{eq:fft_eff_II} to be the original model. This approximation constitutes some type of adiabatic approximation. The Hamiltonian in \equref{eq:tunable-frequency eff} is diagonal so that the formal solution of the TDSE can be obtained. The structure of this solution is similar but not equivalent to the one of a state vector under the adiabatic approximation, see \REF\cite{Weinberg2015}. In case of the Hamiltonian in \equref{eq:tunable-frequency eff}, we additionally neglect the so-called geometric phase. This well-studied approximation is based on a formulation of the problem in terms of the instantaneous eigenstates. Consequently, if we use the Hamiltonian in \equref{eq:tunable-frequency eff} to define an interacting many-particle system, we find that the corresponding error bounds are not valid any more. Furthermore, moving from the Hamiltonian given by \equref{eq:flux-tunable transmon recast} to the Hamiltonian in \equref{eq:tunable-frequency eff} requires a series of redefinitions of the model of a flux-tunable transmon. Additionally, it is usually the case that only the lowest states of the system are used to model the dynamics. This step constitutes an additional approximation. To the best knowledge of the author, none of these steps comes with an explicit analysis of the error in terms of the time-evolution operators involved in the process, see \REF\cite{Burgarth21} for explicit examples. Consequently, we should not claim that the different models generate the same dynamics.

\section{A circuit Hamiltonian model for non-ideal gate-based transmon quantum computers}\label{sec:TheQuantumComputerCircuitHamiltonianModel}
We can use fixed-frequency transmons, flux-tunable transmons, resonators and TLS, see \secaref{sec:ResAndTLS}{sec:Transmons}, to define a many-particle circuit Hamiltonian model. The generic model Hamiltonian reads
\begin{equation}\label{eq:CHM}
  \OP{H}(t)=\OP{H}_{\idxFFT,\Sigma}(t)+\OP{H}_{\idxFTT,\Sigma}(t)+\OP{H}_{\beta,\Sigma}(t)+\OP{H}_{\idxRES,\Sigma}+\OP{H}_{\idxTLS,\Sigma}+\OP{V}_{\idxINT},
\end{equation}
where the individual terms are defined as
\begin{subequations}\label{eq:CHMDEF}
  \begin{align}
      \OP{H}_{\idxFFT,\Sigma}(t) &=\sum_{i\in I} E_{C_{i}} \BRR{\OP{n}_{i}-n_{g,i}(t) }^{2} - E_{J_{i}} \cos\BRR{\OP{\varphi}_{i}}, \\
      \begin{split}
      \OP{H}_{\idxFTT,\Sigma}(t) &=\sum_{j\in J} \left( \right. E_{C_{j}} \BRR{\OP{n}_{j}-n_{g,j}(t) }^{2} + (\frac{1}{2}-\beta_{j}) \dot{\varphi}_{j}(t) \OP{n}_{j}\\
        &- E_{J_{l,j}} \cos\BRR{\OP{\varphi}_{j}+ \beta_{j} \varphi_{j}(t)} \\
        &- E_{J_{r,j}} \cos\BRR{\OP{\varphi}_{j}+(\beta_{j}-1)\varphi_{j}(t)} \left. \right),
      \end{split}\\
      \OP{H}_{\idxRES,\Sigma} &=\sum_{k\in K} \omega_{k}^{(R)} \OP{a}_{k}^{\dagger}\OP{a}_{k},\\
      \OP{H}_{\idxTLS,\Sigma} &=\sum_{l\in L} \omega_{l}^{(T)} \OP{b}_{l}^{\dagger}\OP{b}_{l},\\
      \begin{split}
      \OP{V}_{\idxINT}  &=\sum_{(i,i^{\prime})\in I \times I^{\prime}}  G_{i,i^{\prime}}^{(0)} \BRR{\OP{n}_{i} \tens{} \OP{n}_{i^{\prime}}}\\
                        &+\sum_{(j,i)\in J \times I}                    G_{j,i}^{(1)} \BRR{\OP{n}_{j} \tens{} \OP{n}_{i}}\\
                        &+\sum_{(j,j^{\prime})\in J \times J^{\prime}}  G_{j,j^{\prime}}^{(2)} \BRR{\OP{n}_{j} \tens{} \OP{n}_{j^{\prime}}}\\
                        &+\sum_{(k,i)\in K \times I}                    G_{k,i}^{(3)} \BRR{\OP{a}_{k}+\OP{a}_{k}^{\dagger}} \tens{} \OP{n}_{i} \\
                        &+\sum_{(k,j)\in K \times J}                    G_{k,j}^{(4)} \BRR{\OP{a}_{k}+\OP{a}_{k}^{\dagger}} \tens{} \OP{n}_{j} \\
                        &+\sum_{(l,i)\in L \times I}                    G_{l,i}^{(5)} \BRR{\OP{b}_{l}+\OP{b}_{l}^{\dagger}} \tens{} \OP{n}_{i} \\
                        &+\sum_{(l,j)\in L \times J}                    G_{l,j}^{(6)} \BRR{\OP{b}_{l}+\OP{b}_{l}^{\dagger}} \tens{} \OP{n}_{j} \\
                        &+\sum_{(k,k^{\prime})\in K \times K^{\prime}}  G_{k,k^{\prime}}^{(7)} \BRR{\OP{a}_{k}+\OP{a}_{k}^{\dagger}} \tens{} \BRR{\OP{a}_{k^{\prime}}+\OP{a}_{k^{\prime}}^{\dagger}}\\
                        &+\sum_{(l,k)\in L \times K}                    G_{l,k}^{(8)} \BRR{\OP{b}_{l}+\OP{b}_{l}^{\dagger}} \tens{} \BRR{\OP{a}_{k}+\OP{a}_{k}^{\dagger}}\\
                        &+\sum_{(l,l^{\prime})\in L \times L^{\prime}}  G_{l,l^{\prime}}^{(9)} \BRR{\OP{b}_{l}+\OP{b}_{l}^{\dagger}} \tens{} \BRR{\OP{b}_{l^{\prime}}+\OP{b}_{l^{\prime}}^{\dagger}}.
      \end{split}
  \end{align}
\end{subequations}
The full Hamiltonian $\OP{H}(t)$ describes a set of interacting fixed-frequency transmons, flux-tunable transmons, resonators and TLSs. Here $I\subseteq \mathbb{N}^{0}$ denotes the index set of the fixed-frequency transmons, $J\subseteq \mathbb{N}^{0}$ denotes the index set of the flux-tunable transmons, $K\subseteq \mathbb{N}^{0}$ denotes the index set of the resonators and $L\subseteq \mathbb{N}^{0}$ denotes the index set of the TLSs. We use the term $\OP{V}_{\idxINT}$ to model dipole-dipole interactions, see \REFS\cite{Zangwill13,CT10}, between the different subsystems. The interaction term $\OP{V}_{\idxINT}$ has ten terms in total,\ie each type of circuit element can interact with the remaining circuit element types. For simplicity and notational convenience we express the interaction operator as
\begin{equation}\label{eq:INTOPTABRV}
  \OP{V}_{\idxINT} = \sum_{n \in \{0,...,9\}} \sum_{(i,j) \in  I_{n} \times J_{n}} G_{i,j}^{(n)} \BRR{\OP{v}_{n,i} \tens{} \OP{v}_{n,j}},
\end{equation}
where the different $\OP{v}_{n,i}$, $\OP{v}_{n,j}$, $I_{n}$ and $J_{n}$ are defined in terms of the interaction operator given by \equref{eq:CHMDEF}(e). In the following, we use the interaction strength constants $G_{i,j}^{(0 \mhyphen 9)}$ to address the different terms.

The generic form of the Hamiltonian \equref{eq:CHM} allows us to model a wide range of physical scenarios. First, we can model different circuit architectures and quantum gates, see \REF\cite{Blais2020circuit} for a review of different circuit architectures and gate strategies. Second, we are able to describe small circuits in a thermal bath environment of TLSs, see \REF\cite{Willsch2020FluxQubitsQuantumAnnealing} and \secref{sec:ResAndTLS}.

In practice, we find that even if we are able to solve the TDSE for $\OP{H}(t)$, the work effort to set up simulations for these very different scenarios is immense,\ie each physical scenario comes with its own special needs. Furthermore, the capabilities of the simulation code in terms of the number of circuit elements which can be simulated, are determined by the simulation algorithm we introduce in \chapref{chap:IV}, the exact trajectories of $n_{g,i/j}(t)$ and $\varphi_{j}(t)$ and the system parameters we introduce later. This issue is further discussed in \chaapref{chap:NA}{chap:GET}.

\section{An effective Hamiltonian model for non-ideal gate-based transmon quantum computers}\label{sec:TheQuantumComputerEffectiveHamiltonianModel}
\newcommand{\idxtrans}{\text{Tra.}}
\newcommand{\idxinter}{\text{Int.}}
\newcommand{\idxres}{\text{Res.}}
\newcommand{\idxdri}{\text{Dri.}}
\newcommand{\idxtls}{\text{TLS}}
\newcommand{\EJ}[1]{ \tilde{E}_{J,#1}(t) }
\newcommand{\EC}[1]{ \tilde{E}_{C,#1} }
\newcommand{\DTFS}{\OP{\mathcal{D}}_{\text{Dri.},\Sigma}(t)}

We can use the resonator model defined in \secref{sec:ResAndTLS} and the effective models for fixed-frequency and flux-tunable transmons defined in \secref{sec:Transmons} to define a many-particle effective Hamiltonian model in the harmonic basis. Note that here we do not explicitly take into account the time-dependent harmonic basis which is used to formulate the effective flux-tunable transmon model. The generic model Hamiltonian reads
\begin{equation}\label{eq:EHM}
  \OP{H}(t)= \OP{H}_{\idxtrans,\Sigma}(t) + \OP{H}_{\idxres,\Sigma} + \OP{D}_{\idxdri,\Sigma}(t) + \DTFS + \OP{W}_{\idxinter}(t).
\end{equation}
The first term
\begin{equation}\label{eq:transeff}
  \OP{H}_{\idxtrans,\Sigma}(t)= \sum_{i \in I} \omega_{i}^{(q)}(t) \OP{c}_{i}^{\dagger}\OP{c}_{i} + \frac{\alpha_{i}^{(q_{0})}}{2} \BRR{\OP{c}_{i}^{\dagger}\OP{c}_{i}\BRR{\OP{c}_{i}^{\dagger}\OP{c}_{i}-\OP{I}}},
\end{equation}
describes a collection of non-interacting transmon qubits, fixed frequency and/or flux tunable. Here $I \subseteq \mathbb{N}^{0}$ denotes an index set for the different transmons. If we choose $\omega_{i}^{(q)}(t)=\const$ for all times $t$ and a certain transmon term $i\in I$, we describe a fixed-frequency transmon. Otherwise, we model a flux-tunable transmon, see \secref{sec:Transmons}.

As in \secref{sec:TheQuantumComputerCircuitHamiltonianModel}, we use the second term
\begin{equation}
  \OP{H}_{\idxres,\Sigma} = \sum_{j \in J} \omega_{j}^{(R)} \OP{a}_{j}^{\dagger}\OP{a}_{j},
\end{equation}
to model a set of non-interacting resonator elements. Here $J \subseteq \mathbb{N}^{0}$ denotes an index set for the different resonator elements.

The third term
\begin{equation}\label{eq:drive_term_charge}
  \OP{D}_{\idxdri,\Sigma}(t) = \sum_{i \in I} \Omega_{i}(t) \BRR{\OP{c}_{i}^{\dagger} + \OP{c}_{i}},
\end{equation}
is defined as the sum of individual charge driving terms which act on the different transmon subspaces. Here $\Omega_{i}(t) \propto n_{g,i}$(t) denotes a real-valued function which allows us to model quantum gates.

The fourth term
\begin{equation}\label{eq:drive_term_flux}
  \DTFS = \sum_{i \in I} - i \sqrt{\frac{\xi_{i}(t)}{2}} \dot{\varphi}_{\text{eff.},i}(t) \BRR{\OP{c}_{i}^{\dagger}- \OP{c}_{i}} + \frac{i}{4} \frac{\dot{\xi}_{i}(t)}{\xi_{i}(t)} \BRR{\OP{c}_{i}^{\dagger} \OP{c}_{i}^{\dagger} - \OP{c}_{i} \OP{c}_{i}},
\end{equation}
is another driving term which allows us to model flux-tunable transmons non adiabatically. The individual terms in \equref{eq:drive_term_flux} are proportional to the time derivative $\dot{\varphi}_{i}(t)$ of the external flux $\varphi_{i}(t)$ which one can use to control the system, see \equaref{eq:flux_factor_one}{eq:flux_factor_two}. In the following we refer to the Hamiltonian \equref{eq:EHM} as the adiabatic effective Hamiltonian model when the term in \equref{eq:drive_term_flux} is neglected. Similarly, if we include the term we refer to the Hamiltonian \equref{eq:EHM} as the non-adiabatic effective Hamiltonian model. Additionally, we set $\dot{\varphi}_{i}(t)=0$ for all fixed-frequency transmons $i$.

The fifth term describes the interactions between the different subsystems and reads
\begin{align}\label{eq:intereff}
\begin{split}
  \OP{W}_{\idxinter}(t) &=\sum_{(i,i^{\prime})\in I \times I^{\prime}} g_{i,i^{\prime}}^{(c, c)}(t) \BRR{\OP{c}_{i}^{\dagger} + \OP{c}_{i}} \tens{} \BRR{\OP{c}_{i^{\prime}}^{\dagger} + \OP{c}_{i^{\prime}}}\\
                     &+\sum_{(j,j^{\prime})\in J \times J^{\prime}} g_{j,j^{\prime}}^{(a, a)}(t) \BRR{\OP{a}_{j}^{\dagger} + \OP{a}_{j}} \tens{} \BRR{\OP{a}_{j^{\prime}}^{\dagger} + \OP{a}_{j^{\prime}}}\\
                     &+\sum_{(j,i)\in J \times I}                   g_{j,i}^{(a, c)}(t)          \BRR{\OP{a}_{j}^{\dagger} + \OP{a}_{j}} \tens{} \BRR{\OP{c}_{i}^{\dagger} + \OP{c}_{i}},
\end{split}
\end{align}
where $g_{i,j}^{(\mdot, \mdot)}(t)$ is a real-valued function which describes a time-dependent interaction strength. Here too, see \secref{sec:TheQuantumComputerCircuitHamiltonianModel}, we model dipole-dipole interactions. In principle, we can choose the function $g_{i,j}^{(\mdot, \mdot)}(t)$ as we see fit. However, in this thesis, we restrict ourselves to the following three cases. The first case is defined as
\begin{equation}\label{eq:eff_int_res_res}
  g_{j,j^{\prime}}^{(a, a)}(t)=G_{j,j^{\prime}},
\end{equation}
where $G_{j,j^{\prime}}$ is a constant real-valued parameter which determines the order of magnitude of the interaction strength. We use this constant function to model the coupling between two resonator elements $j$ and $j^{\prime}$. The second case is defined as
\begin{equation}\label{eq:eff_int_trans_res}
  g_{j,i}^{(a, c)}(t)=G_{j,i} \sqrt[4]{ \frac{\EJ{i}}{8 \EC{i}} },
\end{equation}
where $\EJ{i}$ denotes a Josephson energy and $\EC{i}$ refers to a capacitive energy. Here we model the coupling between a transmon $i$ and a resonator $j$. The third case is defined as
\begin{equation}\label{eq:eff_int_trans_trans}
  g_{i,i^{\prime}}^{(c, c)}(t)=G_{i,i^{\prime}} \sqrt[4]{ \frac{\EJ{i}}{8 \EC{i}} } \sqrt[4]{ \frac{\EJ{i^{\prime}}}{8 \EC{i^{\prime}}}}.
\end{equation}
Here we model the interactions between two transmons $i$ and $i^{\prime}$. The function $\EJ{i}$ and the parameter $\EC{i}$ are chosen as follows. Assuming we model a fixed-frequency transmon $i$ with Josephson energy $E_{J,i}$ and capacitive energy $E_{C,i}$, we then set $\EJ{i}=E_{J,i}$ and $\EC{i}=E_{C,i}$. If we model a flux-tunable transmon $i$ with effective Josephson energy $\EJEFFVAR{i}$ and capacitive energy $E_{C,i}$, we set $\EJ{i}=\EJEFFVAR{i}$ and $\EC{i}=E_{C,i}$. This effective interaction strength model is motivated by the work of \REF\cite{Koch}.

\section{Summary and conclusions}
In this chapter we discussed the fundamental mathematical relations we use to describe electromagnetic systems in this thesis, see \secref{sec:TheLumpedElementApproximation}. Furthermore, we reviewed the circuit quantisation formalism. This formalism allows us to determine Hamiltonians for different electromagnetic systems, see \secref{sec:CircuitQuantisationFormalism}. Additionally, we applied the formalism and derived circuit Hamiltonians for LC resonators as well as fixed-frequency and flux-tunable transmons, see \secaref{sec:ResAndTLS}{sec:Transmons}. In \secref{sec:ResAndTLS}, we also introduced a two-level system (TLS) model. Similarly, in \secref{sec:Transmons}, we also derived several effective Hamiltonians for fixed-frequency and flux-tunable transmons. These models can potentially mimic the dynamic behaviour of the associated circuit Hamiltonian models. Finally, see \secaref{sec:TheQuantumComputerCircuitHamiltonianModel}{sec:TheQuantumComputerEffectiveHamiltonianModel}, we defined two generic models for non-ideal gate-based transmon quantum computers,\ie these models are supposed to describe certain aspects of superconducting prototype gate-based quantum computers (PGQCs), see also \secaref{sec:FromStaticsToDynamics}{sec:Prototype gate-based quantum computers}. In \secref{sec:TheQuantumComputerCircuitHamiltonianModel} we define a generic circuit Hamiltonian model which describes a set of interacting fixed-frequency and flux-tunable transmons as well as LC resonators and TLSs. Here we model the interactions as dipole-dipole interactions. Similarly, in \secref{sec:TheQuantumComputerEffectiveHamiltonianModel} we define a generic effective Hamiltonian which describes a set of interacting fixed-frequency transmons, flux-tunable transmons and LC resonators. Here too, we model the interactions as dipole-dipole interactions.

The Hamiltonians we derive and/or define are a result of the assumptions we make, see \secsref{sec:TheLumpedElementApproximation}{sec:TheQuantumComputerEffectiveHamiltonianModel}. Consequently, the time evolution of a NIGQC modelled with the TDSE and a Hamiltonian is also a direct product of these assumptions. However, we can never be absolutely certain that the assumptions we make are sufficient or actually correct, when it comes to modelling a physical scenario. Consequently, we should always be sceptical about the models we use to describe the physical scenario of interest.


\chapter{Real-time simulation algorithms for time-dependent Hamiltonians}\label{chap:IV}
The aim of this chapter is to provide a description of the simulation algorithms we employ to solve the time-dependent Schrödinger equation (TDSE) for the Hamiltonians \equaref{eq:CHM}{eq:EHM} as well as the extend simulation software we use to obtain most results in this thesis.

This chapter is structured as follows. In \secref{sec:TDSEIntro}, we review the problems we face once we try to solve the TDSE numerically. Next, in \secref{sec:TheProductFormulaAlgorithm} we provide the reader with a general discussion of the so-called product-formula algorithm, this is the algorithm we mainly use to solve the TDSE in this work. In \secaref{sec:SOTEO_HB}{sec:SOTEO_MB} we discuss two second-order time-evolution operators which approximately solve the TDSE for the Hamiltonians \equaref{eq:CHM}{eq:EHM}. Next, in \secref{sec:SimulationsOfTheCircuitHamiltonianModelInAlternativeBases}, we discus why simulations of the circuit Hamiltonian model in two alternative computational bases were abandoned. In \secref{sec:ImplementationOfTheTime-evolutionOperator}, we discuss the update rules we use to implement the time-evolution operators we discuss in \secaref{sec:SOTEO_HB}{sec:SOTEO_MB}. In the end, in \secref{sec:StructureOfTheSimulationSoftware}, we provide a high-level overview of the complete simulation software.

Since we solve the TDSE numerically, we always work with finite-dimensional Hilbert spaces. Therefore, in this chapter we always implicitly assume that operators are linear maps which act on a finite-dimensional Hilbert space. Furthermore, we use $\hbar=1$ throughout this chapter.

\newcommand{\DIM}{\text{dim}}
\section{Solving the time-dependent Schrödinger equation numerically}\label{sec:TDSEIntro}
Unfortunately, we usually find that analytical solutions of the TDSE
\begin{equation}\label{eq:TDSE}
  i \partial_{t}\ket{\Psi(t)}=\OP{H}(t)\ket{\Psi(t)},
\end{equation}
where $\OP{H}(t)$ denotes an arbitrary model Hamiltonian which acts on the Hilbert space $\mathcal{H}\subseteq \mathbb{C}^{\DIM}$, are rare and can usually only be obtained for simple model Hamiltonians. Here we assume that no approximations are being made to simplify $\OP{H}(t)$. In particular, in most cases the problem becomes more difficult once we consider a composition of different subsystems which interact with one another. In this thesis, we are interested in exactly this case.

The formal solution of the TDSE reads
\begin{equation}\label{eq:TDSE_OP}
  \hat{\mathcal{U}}(t,t_{0}) = \mathcal{T} \exp\left( -i \int_{t_{0}}^{t} \hat{H}(t^{\prime}) dt^{\prime} \right),
\end{equation}
where $\mathcal{T}$ denotes the time-ordering symbol. Numerical solutions of the TDSE require a discretisation of the time domain,\ie $t \rightarrow t_{j}= \tau j + t_{0}$. If we assume that $\OP{H}(t)$ is piecewise constant between the two time steps $t_{j}$ and $t_{j+1}$, for a small enough time step difference $\tau$, we might write
\begin{equation}\label{eq:EVSTE}
   \hat{\mathcal{U}}(t,t_{0}) = \prod_{j=0}^{J-1} e^{-i \tau \hat{H}\BRR{t_{j}+\frac{\tau}{2}}}.
\end{equation}
This assumption enables us to reduce the problem to the implementation of a sequence of unitary operators
\begin{equation}\label{eq:TEDIS}
  \OP{\mathcal{U}}\BRR{t_{j+1},t_{j}}=e^{-i \tau \hat{H}\BRR{t_{j}+\frac{\tau}{2}}},
\end{equation}
such that
\begin{equation}
  \ket{\Psi(t_{j+1})} = \OP{\mathcal{U}}(t_{j+1},t_{j}) \ket{\Psi(t_{j})}.
\end{equation}
There exist different algorithms which allow us to implement $\OP{\mathcal{U}}$ in a computer program. One popular approach is named full diagonalisation. If we assume that $\OP{H}$ is represented by some matrix, in some unspecified basis, for some time step, we might use the QR algorithm, see \REFS\cite{GoluVanl96,FRAN61,FRAN62}, to diagonalise this matrix. The QR algorithm and others are implemented in various numerical libraries,\eg the Linear Algebra Package (LAPACK), see \REF\cite{PACK99} or Intel's Math Kernel Library (MKL), see \REF\cite{MKL09}. If we supply the library subroutines with a matrix representation $H$ of $\OP{H}$, we can potentially obtain matrices for the basis transformation $W$ and the diagonal matrix $\lambda$ such that $H=W\lambda W^{\dagger}$. Assuming we can repeatedly diagonalise $\OP{H}(t_{j}+\frac{\tau}{2})$ for every time step $t_{j}$, we can solve our problem by implementing the time-evolution operator
\begin{equation}\label{eq:full_diagonalisation}
   \OP{\mathcal{U}}\BRR{t_{j+1},t_{j}}=\OP{W}\BRR{t_{j}+\frac{\tau}{2}} e^{-i \tau \OP{\lambda}\BRR{t_{j}+\frac{\tau}{2}}} \OP{W}^{\dagger}\BRR{t_{j}+\frac{\tau}{2}}.
\end{equation}

The advantage of this method is that it is conceptually very simple and a great deal of complexity is hidden inside the library subroutines. The disadvantage of this method is that the computational work, in terms of time $\mathcal{O}^{\text{Time}}(\text{dim}^{3})$ and space $\mathcal{O}^{\text{Space}}(\text{dim}^{2})$ complexity, see \secref{sec:SimulationOfTheIdealGateBasedQuantumcomputer}, grows with the third and second power in terms of the dimensionality of the system, see \REF\cite{GoluVanl96}. Since the dimensionality itself grows exponentially with the number of subsystems added to the quantum-theoretical model, see \secref{sec:MathematicalFramework}, we find that this approach and other approaches whose time and space complexity show a non-linear scaling in terms of dimensionality, are not applicable to the problem at hand. Consequently, we would prefer to implement $\OP{\mathcal{U}}$ with an algorithm whose time and space complexity scale linearly in terms of dimensionality. Obviously, this does not solve the exponential scaling problem which is an intrinsic part of the quantum-theoretical formalism. However, assuming we can find an algorithm which exhibits sufficient concurrency,\ie independent computational tasks which can be parallelised, we might be able to remedy this issue by using more computational resources, see \REF\cite{Hager:2010}.

Methods which might allow us to solve the problem at hand in a reasonable time and with reasonable memory requirements are the Chebyshev polynomial algorithm, a Lanczos type algorithm or the product-formula algorithm. Reference \cite{HTCN06} provides an overview of all methods mentioned. The product-formula algorithm can often be implemented by only consuming computational resources which grow linearly with the dimension of the Hilbert space. Furthermore, usually we are able to create sufficient concurrency by making use of the freedoms the algorithm provides us with. In this thesis, we use the product-formula algorithm to solve the TDSE for the Hamiltonian models defined in \secaref{sec:TheQuantumComputerCircuitHamiltonianModel}{sec:TheQuantumComputerEffectiveHamiltonianModel}.

\section{The product-formula algorithm}\label{sec:TheProductFormulaAlgorithm}
In this section, we discuss the product-formula algorithm, see \REFS\cite{DeRaedt83,DeRaedt87}, in general terms. Therefore, we assume that the problem at hand can be described by an arbitrary model Hamiltonian $\OP{H}$,\ie we assume that this operator is Hermitian and the Hilbert space in question is $\mathcal{H}^{\DIM}\subseteq \mathbb{C}^{\DIM}$. Since we intend to implement $\OP{\mathcal{U}}(t_{j+1},t_{j})$ given by \equref{eq:TEDIS} for the discretised time steps $t_{j}$, we can assume that this operator is time independent,\ie this operator represents the piecewise constant Hamiltonian operator between the time steps $t_{j+1}$ and $t_{j}$.

In the beginning, we assume that there exists an additive decomposition
\begin{equation}
  \OP{H}=\sum_{m=0}^{M-1} \OP{A}_{m},
\end{equation}
of the model Hamiltonians. Here $ M \in \mathbb{N}^{0}$ and the operators $\OP{A}_{m}$ are assumed to be Hermitian. Note that we always implicitly assume that operators act on finite-dimensional Hilbert spaces. Consequently, we find that the different $\OP{A}_{m}$ are bounded operators. These assumptions are sufficient to make use of the so-called generalised (Trotter) decomposition formula
\begin{equation}\label{eq:Trotter}
  e^{\OP{H}}= \lim_{n \rightarrow \infty} \BRR{\prod_{m=0}^{M-1} e^{\frac{\OP{A}_{m}}{n}}}^{n},
\end{equation}
see \REFS\cite{Suzuki85,Suzuki77}. We have in mind to systematically apply this formula to the time-evolution operator $\OP{\mathcal{U}}$, for finite $n$ and a given operator decomposition $\OP{A}_{m}$. In this way, we can shift the problem to the implementation of a product of operators $e^{-i a \tau \OP{A}_{m}}$, where $a$ is some real-valued constant. Since the different $\OP{A}_{m}$ are Hermitian, we find that the resulting time evolution is unitary,\ie norm preserving. This is the general idea of the product-formula algorithm.

In practice, this raises two questions. First, in order to obtain efficient simulation code, how should we choose the operator decomposition $\OP{A}_{m}$? This question can only be answered properly once we have fixed the simulation basis and the model Hamiltonian,\ie this question is very problem specific. We discuss this issue in the following sections. The second question which we have to address is, how valid is the approximation once we consider a finite $n$,\ie what is the numerical precision of this approximation. This question can be answered for the local error if we construct the approximations in a specific way.

For the time-evolution operator
\begin{equation}
   \OP{\mathcal{U}}=e^{-i\tau \hat{H}},
\end{equation}
the first-order approximant reads
\begin{equation}
  \OP{\mathcal{U}_{1}}\BRR{\tau}=\prod_{m=0}^{M-1} e^{-i\tau \OP{A}_{m}}.
\end{equation}
We can use the first-order approximation $\OP{\mathcal{U}}_{1}$ to construct certain higher-order approximations, see \REF\cite{HTCN06}. The second-order approximation can be expressed as
\begin{equation}
  \OP{\mathcal{U}_{2}}\BRR{\tau}=\OP{\mathcal{U}}_{1}^{\dagger}\BRR{-\frac{\tau}{2}}\OP{\mathcal{U}}_{1}\BRR{\frac{\tau}{2}}.
\end{equation}
Similarly, the fourth-order approximation can be expressed in terms of the second-order approximation
\begin{equation}
  \OP{\mathcal{U}}_{4}\BRR{\tau}=\OP{\mathcal{U}}_{2}\BRR{a \tau} \OP{\mathcal{U}}_{2}\BRR{a\tau} \OP{\mathcal{U}}_{2}\BRR{\BRR{1- 4 a}\tau} \OP{\mathcal{U}}_{2}\BRR{a\tau} \OP{\mathcal{U}}_{2}\BRR{a\tau},
\end{equation}
where $a=1/(4-4^{\frac{1}{3}})$. For the first-, second- and fourth-order approximation $p \in \{1,2,4\}$ there exists a common bound
\begin{equation}\label{eq:PFA_local_error}
  \norm{\OP{\mathcal{U}}(\tau)-\OP{\mathcal{U}}_{p}(\tau)} \leq b_{p} \tau^{p+1},
\end{equation}
for the local error, see \REF\cite{DeRaedt87}. Here the constants $b_{p}$ are real valued and
\begin{equation}
  \norm{\OP{X}}=\sup_{\norm{\psi}_{2}=1}\BRR{\norm{\OP{X}\ket{\psi}}_{2}},
\end{equation}
denotes the operator norm for the operator $\OP{X}$ which is induced by the Euclidean norm $\norm{\psi}_{2}=\sqrt{\braket{\psi|\psi}}$.

Constructing a simulation algorithm in this way has several benefits. First, since we require a Hermitian operator decomposition $\OP{A}_{m}$, we find that the time evolution is norm preserving,\ie unconditionally stable. Consequently, we find that the approximation does not affect this general property of the model. Second, we find that some higher-order approximations can be implemented recursively. Third, the local error can be controlled by choosing an appropriate $\tau$ in combination with a particular higher-order approximation.  Last but not least, since the $\OP{A}_{m}$ are chosen by ourselves, we are often able to implement the operators $e^{-i\tau \OP{A}_{m}}$ with computational resources that only scale linearly with the dimension of the system. These features make the product-formula algorithm a suitable candidate for simulating the real-time dynamics of a quantum system.

\section{The time-evolution operator for the effective Hamiltonian model}\label{sec:SOTEO_HB}
\newcommand{\UHB}{\OP{\mathcal{U}}^{\text{HB}}}
In \secref{sec:TheProductFormulaAlgorithm}, we discussed the product-formula algorithm. This algorithm provides us with the recipe to determine a unitary operator $\OP{\mathcal{U}}$ that approximately solves the TDSE for a given Hamiltonian operator $\OP{H}$. In this section, we use this recipe to determine the operator $\UHB$ for the Hamiltonian \equref{eq:EHM}. For the remainder of this chapter we use the variable $t$ to denote the discretised time steps $t_{j}+\tau/2$. This is done for simplicity and notational convenience. In the following, we use the states
\begin{equation}\label{eq:basisstates_HB}
    \ket{\mathbf{z}_{J},\mathbf{z}_{I}}=\tens{j \in  J} \ket{\psi^{(z_{j})}}\tens{i \in I} \ket{\psi^{(z_{i})}},
\end{equation}
as basis states for the state vector $\ket{\Psi(t)}$, see \equref{eq:TDSE}. Here $\mathbf{z}_{J} \in \mathcal{I}_{n_{J}}^{|J|}$, $\mathbf{z}_{I} \in \mathcal{I}_{n_{I}}^{|I|}$ and $\mathcal{I}_{n}^{N}=\{0,1, ...,n-1\}^{N}$ denotes a set of N-tuples, where the tuple entries can take on values from $0$ to $n-1$. Furthermore, the state vectors $\HBSV{}{i/j}$ denote the basis states of the harmonic oscillator given by \equaref{eq:HarmoicBasisWaveFunction}{eq:HarmoicBasisWaveFunctionTimeDep}. We use the indices $i$ and $j$ to differentiate between transmons and resonators, respectively. The states in \equref{eq:basisstates_HB} form a basis
\begin{equation}
  \mathcal{J}_{\mathbf{n}}^{\prime}=\{\ket{\mathbf{z}} \in \mathcal{H}_{\text{Sim.}}^{\prime}|\exists! \mathbf{z}\in\mathcal{I}_{\mathbf{n}}^{\prime}\colon \ket{\mathbf{z}}=\ket{\mathbf{z}_{J},\mathbf{z}_{I}}\},
\end{equation}
for the Hilbert space
\begin{equation}
  \mathcal{H}_{\text{Sim.}}^{\prime}=\tens{j\in J}\mathcal{H}_{\idxres,j}\tens{i \in I}\mathcal{H}_{\idxtrans,i},
\end{equation}
and we find
\begin{equation}
  \DIM(\mathcal{H}_{\text{Sim.}}^{\prime}) = n_{J}^{|J|} n_{I}^{|I|},
\end{equation}
for the dimensionality of the space we use to perform the simulations. The index set $\mathcal{I}_{\mathbf{n}}^{\prime}$ is defined as
\begin{equation}
  \mathcal{I}_{\mathbf{n}}^{\prime}= \mathcal{I}_{n_{J}}^{|J|} \times \mathcal{I}_{n_{I}}^{|I|},
\end{equation}
where $\times$ denotes the Cartesian product and $\mathbf{n}=(n_{J},n_{I})$. Here $n_{J}$ and $n_{I}$ denote the number of basis states we use to model the resonators and transmons, respectively.

One can show, see \secref{sec:MathematicalFramework} and \REF\cite{Nielsen:2011:QCQ:1972505}, that the product states given by \equref{eq:basisstates_HB} form a basis of the Hilbert space $\mathcal{H}_{\text{Sim.}}^{\prime}$. We name this basis the bare harmonic basis or harmonic basis for short. Note that the notation we use for the simulation basis overloads the notational framework we use in this thesis. In \chapref{chap:I}, we used the states $\ket{\mathbf{z}}\in \mathcal{C}^{N}$ as the basis states of the state vector for the IGQC. In this section, we use $\ket{\mathbf{z}}\in \mathcal{J}_{\mathbf{n}}^{\prime}$ as the simulation basis states of the NIGQC. Both states are elements of the same mathematical structure, namely a finite-dimensional Hilbert space. However, we usually find that the dimensionality is different. Furthermore, in \secref{sec:SOTEO_MB}, we use the states $\ket{\mathbf{z}}$ as basis states of another Hilbert space. For the remainder of this thesis, we always make clear which basis is meant if we refer to the states $\ket{\mathbf{z}}$.

As discussed in \secref{sec:TheProductFormulaAlgorithm}, the first step in finding $\UHB$ is choosing a Hermitian decomposition of the model Hamiltonian $\OP{H}$. Our choice for this decomposition is
\begin{equation}\label{eq:decomposition}
  \OP{H}(t)=\OP{H}_{0}(t)+\OP{V}_{0}(t)+\OP{V}_{1}(t)+\OP{V}_{2}(t).
\end{equation}
The first term
\begin{equation}
    \OP{H}_{0}(t)=\OP{H}_{\idxtrans,\Sigma}(t)+\OP{H}_{\idxres,\Sigma},
\end{equation}
is defined as the sum of terms which describe non-interacting transmon and resonator elements. The second term
\begin{equation}
  \OP{V}_{0}(t)=\sum_{i \in  I} q_{i}(t) i\BRR{\OP{c}_{i}^{\dagger} - \OP{c}_{i}},
\end{equation}
where $q_{i}(t)= - \sqrt{\xi_{i}(t)} \dot{\varphi}_{\text{eff.},i}(t)/\sqrt{2}$ for all $i \in I$, is given by the sum of the first summand of the non-adiabatic driving term in \equref{eq:drive_term_ftt}. Similarly, the third term
\begin{equation}
  \OP{V}_{1}(t)=\sum_{i \in  I} s_{i}(t)i \BRR{\OP{c}_{i}^{\dagger}\OP{c}_{i}^{\dagger} - \OP{c}_{i}\OP{c}_{i}},
\end{equation}
where $s_{i}(t)=  \dot{\xi}_{i}(t)/(4\xi_{i}(t))$ for all $i \in I$, consists of the sum of the second summand in \equref{eq:drive_term_ftt}. The fourth term
\begin{equation}
  \OP{V}_{2}(t)=\OP{D}_{\idxdri,\Sigma}(t)+\OP{W}_{\idxinter}(t),
\end{equation}
is defined as the sum of the charge drive term $\OP{D}_{\idxdri,\Sigma}(t)$ and the dipole-dipole interaction term $\OP{W}_{\idxinter}(t)$. Note that this decomposition is motivated by the work presented in \REF\cite{Willsch2020}.

In the following, we express the different decomposition terms as
\begin{subequations}\label{eq:decomposition_term_1}
\begin{align}
  \OP{H}_{0}(t) &=\sum_{i \in  I} \OP{\lambda}_{\idxtrans,i}(t) + \sum_{j \in  J} \OP{\lambda}_{\idxRES,j},\\
  \OP{V}_{0}(t) &=\OP{Q}\OP{\lambda}_{Q}(t)\OP{Q}^{\dagger},\\
  \OP{V}_{1}(t) &=\OP{S}\OP{\lambda}_{S}(t)\OP{S}^{\dagger},\\
  \OP{V}_{2}(t) &=\OP{T}\OP{\lambda}_{T}(t)\OP{T}^{\dagger}.
\end{align}
\end{subequations}
The operator $\OP{H}_{0}(t)$ is diagonal in the basis $\mathcal{J}_{\mathbf{n}}$. The operators $\OP{\lambda}_{\idxtrans,i}(t)$ and $\OP{\lambda}_{\idxRES,j}$ simply contain the eigenvalues of the transmon and resonator terms, see \equaref{eq:tunable-frequency eff}{eq:resonator}, respectively. Furthermore, the operators $\OP{V}_{0}(t)$, $\OP{V}_{1}(t)$ and $\OP{V}_{2}(t)$ in \equref{eq:decomposition_term_1}(b-d) are defined in terms of the basis transformations
\begin{subequations}\label{eq:basistrafo_HB}
\begin{align}
  \OP{Q}&=\tens{i \in  I} \OP{Q}_{c_{i}},\\
  \OP{S}&=\tens{i \in  I} \OP{S}_{c_{i}},\\
  \OP{T}&=\tens{j \in  J} \OP{T}_{a_{j}} \tens{i \in  I} \OP{T}_{c_{i}},
\end{align}
\end{subequations}
and the operators $\OP{\lambda}_{Q}(t)$, $\OP{\lambda}_{S}(t)$ and $\OP{\lambda}_{T}(t)$ are diagonal in the transformed bases $\{\ket{\mathbf{w}_{Q/S/T}}\}$, which can be obtained by making use of $\OP{Q}$, $\OP{S}$ and $\OP{T}$. Note that $\OP{\lambda}_{Q}(t)$, $\OP{\lambda}_{S}(t)$ and $\OP{\lambda}_{T}(t)$ are usually non-diagonal in the basis $\mathcal{J}_{\mathbf{n}}^{\prime}$. The different basis transformations in \equref{eq:basistrafo_HB}(a-c) are themselves defined in terms of basis transformations $\OP{Q}_{c_{i}}$, $\OP{S}_{c_{i}}$, $\OP{T}_{c_{i}}$ and $\OP{T}_{a_{j}}$ which act on the individual transmon and resonator subspaces
\begin{subequations}\label{eq:basistrafo_HB_subspace}
\begin{align}
    \OP{Q}_{c_{i}}  \OP{\lambda}_{Q_{c_{i}}} \OP{Q}_{c_{i}}^{\dagger} &=i\BRR{\OP{c}_{i}^{\dagger} - \OP{c}_{i}},\\
    \OP{S}_{c_{i}}  \OP{\lambda}_{S_{c_{i}}} \OP{S}_{c_{i}}^{\dagger} &=i\BRR{\OP{c}_{i}^{\dagger}\OP{c}_{i}^{\dagger} - \OP{c}_{i}\OP{c}_{i}},\\
    \OP{T}_{c_{i}}  \OP{\lambda}_{T_{c_{i}}} \OP{T}_{c_{i}}^{\dagger} &=\BRR{\OP{c}_{i}^{\dagger} + \OP{c}_{i}},\\
    \OP{T}_{a_{j}}  \OP{\lambda}_{T_{a_{j}}} \OP{T}_{a_{j}}^{\dagger} &=\BRR{\OP{a}_{j}+\OP{a}_{j}^{\dagger}}.
\end{align}
\end{subequations}
We emphasise that if we represent the different operators $\OP{Q}_{c_{i}}$, $\OP{S}_{c_{i}}$, $\OP{T}_{c_{i}}$ and $\OP{T}_{a_{j}}$ numerically as matrices, the diagonalisation algorithm yields the same result for all $i$ and $j$. Consequently, if we implement the algorithm and/or the corresponding time-evolution operator in a program, we only have to make use of a very small amount of memory to store the three matrices. This is one of the reasons why we express the problem in this way. Another reason for choosing this decomposition follows right away.

In the transformed bases $\{\ket{\mathbf{w}_{Q/S/T}}\}$, see basis transformations $\OP{Q}$, $\OP{S}$ and $\OP{T}$, we can express the different diagonal operators as
\begin{align}
  \OP{\lambda}_{Q}(t) &=\sum_{j \in  J} q_{i}(t)\OP{\lambda}_{Q_{c_{i}}},\\
  \OP{\lambda}_{S}(t) &=\sum_{j \in  J} s_{i}(t)\OP{\lambda}_{S_{c_{i}}},\\
\begin{split}
  \OP{\lambda}_{T}(t) &=\sum_{i \in I} \Omega_{i}(t) \OP{\lambda}_{T_{c_{i}}} \\
                      &+ \sum_{(i,i^{\prime})\in I \times I^{\prime}} g_{i,i^{\prime}}^{(c, c)}(t) \BRR{\OP{\lambda}_{T_{c_{i}}} \tens{} \OP{\lambda}_{T_{c_{i^{\prime}}}}}\\
                      &+ \sum_{(j,j^{\prime})\in J \times J^{\prime}} g_{j,j^{\prime}}^{(a, a)}(t) \BRR{\OP{\lambda}_{T_{a_{j}}} \tens{} \OP{\lambda}_{T_{a_{j^{\prime}}}}}\\
                      &+ \sum_{(j,i) \in J \times I}                  g_{j,i}^{(a, c)}(t)          \BRR{\OP{\lambda}_{T_{a_{j}}} \tens{} \OP{\lambda}_{T_{c_{i}}}}.
\end{split}
\end{align}
Therefore, we can lump the time dependencies into the diagonal operators completely.

The resulting second-order time-evolution operator for the Hermitian decomposition given by \equref{eq:decomposition} reads
\begin{equation}
  \begin{split}
  \UHB=& e^{-\frac{\tau}{2} \OP{H}_{0}(t)} \OP{S}e^{-\frac{\tau}{2} \OP{\lambda}_{S}(t)}\OP{S}^{\dagger} \OP{Q} e^{-\frac{\tau}{2}\OP{\lambda}_{Q}(t)} \OP{Q}^{\dagger} \OP{T} e^{-i \tau \OP{\lambda}_{T}(t)} \OP{T}^{\dagger} \\ &\OP{Q}e^{-\frac{\tau}{2} \OP{\lambda}_{Q}(t)}\OP{Q}^{\dagger} \OP{S} e^{-\frac{\tau}{2} \OP{\lambda}_{S}(t)} \OP{S}^{\dagger} e^{-\frac{\tau}{2} \OP{H}_{0}(t)}.
  \end{split}
\end{equation}
In order to reduce the number of basis transformations which need to be implemented in a computer program, we introduce the operators $ \OP{R} = \OP{S}^{\dagger} \OP{Q} $ and $ \OP{U} =\OP{Q}^{\dagger} \OP{T}$ such that
\begin{equation}\label{eq:TEO HB}
  \begin{split}
  \UHB=& e^{-\frac{\tau}{2} \OP{H}_{0}(t)} \OP{S} e^{-\frac{\tau}{2} \OP{\lambda}_{S}(t)} \OP{R} e^{-\frac{\tau}{2}\OP{\lambda}_{Q}(t)}\OP{U} e^{-i \tau \OP{\lambda}_{T}(t)} \OP{U}^{\dagger}  \\ & e^{-\frac{\tau}{2} \OP{\lambda}_{Q}(t)} \OP{R}^{\dagger} e^{-\frac{\tau}{2} \OP{\lambda}_{S}(t)} \OP{S}^{\dagger} e^{-\frac{\tau}{2} \OP{H}_{0}(t)}.
  \end{split}
\end{equation}
For the adiabatic effective Hamiltonian \equref{eq:EHM} we can derive the much simpler time-evolution operator
\begin{equation}\label{eq:TEO HB adiabatic}
  \UHB_{\text{adia.}}= e^{-\frac{\tau}{2} \OP{H}_{0}(t)} \OP{T}e^{-\tau \OP{\lambda}_{T}(t)}\OP{T}^{\dagger} e^{-\frac{\tau}{2} \OP{H}_{0}(t)},
\end{equation}
where we assume $\OP{V}_{0}(t)=0$ and $\OP{V}_{1}(t)=0$ for all times $t$ or $\dot{\varphi}_{i}(t) \rightarrow 0 $ for all $i$, see \equaref{eq:flux_factor_one}{eq:flux_factor_two}.

Consequently, we find that we have to implement two types of diagonal operators and one type of non-diagonal operators for every iteration step $t$. The algorithmic update rules that allow us to implement these operators are subject of \secref{sec:ImplementationOfTheTime-evolutionOperator}.

\section{The time-evolution operator for the circuit Hamiltonian model}\label{sec:SOTEO_MB}
\newcommand{\UTB}{\OP{\mathcal{U}}^{\text{TB}}}

In \secref{sec:TheProductFormulaAlgorithm}, we discuss how to solve the TDSE numerically by implementing a unitary time-evolution operator $\OP{\mathcal{U}}$, for an arbitrary model Hamiltonian $\OP{H}$. In \secref{sec:SOTEO_HB}, we make use of this recipe and derive a time-evolution operator for the effective Hamiltonian \equref{eq:EHM}. In this section, we do the same for Hamiltonian \equref{eq:CHM}. Note that this model Hamiltonian is more complex than the one we discussed in \secref{sec:SOTEO_HB}. In the following, we restrict the discussion to the case where each type of subsystem,\ie transmons, resonators and TLSs, is modelled with a fixed number of basis states. The simulation code allows us more freedom,\ie we can choose two different values for the number of basis states we use to model the subsystem in question. A detailed discussion of this part of the simulation code goes beyond the scope of this section. Here, we intend to understand the structure of the second-order time-evolution operator as clearly as possible. Consequently, we omit some details regarding the actual simulation code in this section.

In the following, we use the product states
\begin{equation}\label{eq:basisstates}
    \ket{\mathbf{z}_{L},\mathbf{z}_{K},\mathbf{z}_{J},\mathbf{z}_{I}}=\tens{l\in  L} \ket{\psi^{(z_{l})}}\tens{k\in  K} \ket{\psi^{(z_{k})}}\tens{j\in  J} \ket{\phi^{(z_{j})}}\tens{i\in  I} \ket{\phi^{(z_{i})}},
\end{equation}
as basis states for the state $\ket{\Psi(t)}$, see \equref{eq:TDSE}. Here $\mathbf{z}_{L}\in \mathcal{I}_{n_{L}}^{|L|}$ , $\mathbf{z}_{K}\in \mathcal{I}_{n_{K}}^{|K|}$, $\mathbf{z}_{J}\in \mathcal{I}_{n_{J}}^{|J|}$, $\mathbf{z}_{I}\in \mathcal{I}_{n_{I}}^{|I|}$. Consequently, $n_{L}$, $n_{K}$, $n_{J}$ and $n_{I}$ denote the numbers of basis states we use to model the dynamics of the system. We discuss the systems associated with the harmonic basis states $\ket{\psi^{(z_{l/k})}}$ and the transmon basis states $\ket{\phi^{(z_{i/j})}}$ in \secaref{sec:ResAndTLS}{sec:Transmons}, respectively. Note that we differentiate between the fixed-frequency and flux-tunable transmon states by means of the indices $i$ and $j$. The simulation basis can be expressed as
\begin{equation}
  \mathcal{J}_{\mathbf{n}}=\{\ket{\mathbf{z}} \in \mathcal{H}_{\text{Sim.}}|\exists! \mathbf{z}\in\mathcal{I}_{\mathbf{n}}\colon \ket{\mathbf{z}}=\ket{\mathbf{z}_{L},\mathbf{z}_{K},\mathbf{z}_{J},\mathbf{z}_{I}}\},
\end{equation}
where
\begin{equation}
  \mathcal{H}_{\text{Sim.}}=\tens{l\in L}\mathcal{H}_{\idxTLS,l}\tens{k\in K}\mathcal{H}_{\idxRES,k}\tens{j\in J}\mathcal{H}_{\idxFTT,j}\tens{i \in I}\mathcal{H}_{\idxFFT,i},
\end{equation}
denotes a Hilbert space with dimensionality
\begin{equation}\label{eq:dimensionsimulationbasis}
  \DIM(\mathcal{H}_{\text{Sim.}})= n_{L}^{|L|} n_{K}^{|K|} n_{J}^{|J|} n_{I}^{|I|},
\end{equation}
and
\begin{equation}
  \mathcal{I}_{\mathbf{n}}= \mathcal{I}_{n_{L}}^{|L|} \times \mathcal{I}_{n_{K}}^{|K|} \times \mathcal{I}_{n_{J}}^{|J|} \times \mathcal{I}_{n_{I}}^{|I|},
\end{equation}
is another index set, which is specified by the four tuple $\mathbf{n}=(n_{K},n_{L},n_{J},n_{I})$. Here, we denote $\mathcal{J}_{\mathbf{n}}$ as the transmon bare basis or \TB{} for short. As discussed in \secref{sec:MathematicalFramework} and \REF\cite{Nielsen:2011:QCQ:1972505}, one can show that the product states given by \equref{eq:basisstates} form a basis of the space $\mathcal{H}_{\text{Sim.}}$. As mentioned before, in this thesis, see \chapref{chap:II} and \secref{sec:SOTEO_HB}, we use the ket vectors $\ket{\mathbf{z}}$ to refer to basis states which belong to different bases. This has actually an advantage once we compare different models with one another, see \chapref{chap:NA}.

Since all simulations in this thesis are performed by imposing the irrotational constraint $\beta_{j}=1/2$, see \REF\cite{You} and \equref{eq:flux-tunable transmon_beta}, on all flux-tunable transmons $j \in J$, we discus the second-order time-evolution operator for this special case only. Furthermore, the notation we use for the circuit Hamiltonian in \equref{eq:CHM} is already quite cumbersome and this choice simplifies the notation to some extent. We begin the discussion of the second-order time-evolution operator $\UTB$ by recasting Hamiltonian \equref{eq:CHM} into the form
\begin{equation}\label{eq:decomposition_MB}
  \OP{H}(t)=\OP{H}_{0}+\OP{W}_{0}(t)+\OP{W}_{1}(t)+\OP{W}_{2}(t),
\end{equation}
where the individual terms are defined as
\begin{subequations}
\begin{align}
    \OP{H}_{0} &=\OP{H}_{\idxFFT,\Sigma}(0)+\OP{H}_{\idxFTT,\Sigma}(0)+\OP{H}_{\idxRES,\Sigma}+\OP{H}_{\idxTLS,\Sigma},\\
    \OP{W}_{0}(t) &=\sum_{j \in J} \BRR{E_{J_{l,j}}+E_{J_{r,j}}}\BRR{\cos(\frac{\varphi_{0,j}}{2}) - \cos(\frac{\varphi_{j}(t)}{2})} \cos(\OP{\varphi}_{j}),\\
    \OP{W}_{1}(t) &=\sum_{j \in J} \BRR{E_{J_{l,j}}-E_{J_{r,j}}}\BRR{\sin(\frac{\varphi_{0,j}}{2}) - \sin(\frac{\varphi_{j}(t)}{2})} \sin(\OP{\varphi}_{j}),\\
    \OP{W}_{2}(t) &=\sum_{i \in I} -2 E_{C,i} n_{g,i}(t) \OP{n}_{i} +\sum_{i \in J} -2 E_{C,j} n_{g,j}(t) \OP{n}_{j} + \OP{V}_{\idxINT}.
\end{align}
\end{subequations}
The terms $\OP{H}_{\idxFFT,\Sigma}(0)$ and $\OP{H}_{\idxFTT,\Sigma}(0)$ are the fixed-frequency and flux-tunable transmon terms at time $t=0$, see \equref{eq:CHM}. These terms are specified by the charge-offset values $n_{0,i/j}=n_{g,i/j}(0)$ and the flux-offset values $\varphi_{0,j}=\varphi_{j}(0)$. Note that we neglect all terms which only contribute a global phase factor to the time evolution.

We use the form given by \equref{eq:decomposition_MB} as a product-formula decomposition, see \secref{sec:TheProductFormulaAlgorithm}. In the transmon basis, we express these terms as
\begin{subequations}\label{eq:CHMrecast}
\begin{align}
    \OP{H}_{0} &=\sum_{i \in  I} \OP{\lambda}_{\idxFFT,i}+\sum_{j \in  J} \OP{\lambda}_{\idxFTT,j}+\sum_{k \in  K} \OP{\lambda}_{\idxRES,k}+\sum_{l \in  L} \OP{\lambda}_{\idxTLS,l},\\
    \OP{W}_{0}(t) &=\OP{Q}\OP{\lambda}_{Q}(t)\OP{Q}^{\dagger},\\
    \OP{W}_{1}(t) &=\OP{S}\OP{\lambda}_{S}(t)\OP{S}^{\dagger},\\
    \OP{W}_{2}(t) &=\OP{T}\OP{\lambda}_{T}(t)\OP{T}^{\dagger}.
\end{align}
\end{subequations}
The operators $\OP{\lambda}_{\idxFFT,i}$, $\OP{\lambda}_{\idxFTT,j}$, $\OP{\lambda}_{\idxRES,k}$ and $\OP{\lambda}_{\idxTLS,l}$ are diagonal in the \TB{} $\mathcal{J}_{\mathbf{n}}$. Furthermore, the operators $\OP{Q}$, $\OP{S}$ and $\OP{T}$ denote basis transformations
\begin{subequations}\label{eq:basistrafo}
\begin{align}
    \OP{Q}&=\tens{j\in  J}  \OP{Q}_{\varphi_{j}},\\
    \OP{S}&=\tens{j\in  J}  \OP{S}_{\varphi_{j}},\\
    \OP{T}&=\tens{l \in  L} \OP{T}_{b_{l}} \tens{k \in  K} \OP{T}_{a_{k}} \tens{j \in  J} \OP{T}_{n_{j}} \tens{i \in  I} \OP{T}_{n_{i}},
\end{align}
\end{subequations}
which are defined in terms of another set of basis transformations
\begin{subequations}
\begin{align}
    \OP{Q}_{ \varphi_{j} }    \OP{\lambda}_{Q_{\varphi_{j}}}     \OP{Q}_{\varphi_{j}}^{\dagger}   &=\cos{\OP{\varphi}_{j}},\\
    \OP{S}_{ \varphi_{j} }    \OP{\lambda}_{S_{\varphi_{j}}}     \OP{S}_{\varphi_{j}}^{\dagger}   &=\sin{\OP{\varphi}_{j}},\\
    \OP{T}_{n_{i/j}}  \OP{\lambda}_{T_{n_{i/j}}} \OP{T}_{n_{i/j}}^{\dagger} &=\OP{n}_{i/j},\\
    \OP{T}_{a_{k}}  \OP{\lambda}_{T_{a_{k}}} \OP{T}_{a_{k}}^{\dagger} &=\BRR{\OP{a}_{k}+\OP{a}_{k}^{\dagger}},\\
    \OP{T}_{b_{l}}  \OP{\lambda}_{T_{b_{l}}} \OP{T}_{b_{l}}^{\dagger} &=\BRR{\OP{b}_{l}+\OP{b}_{l}^{\dagger}},
\end{align}
\end{subequations}
which act on the subspaces of the individual subsystems. These transformations allow us to express the operators $\OP{\lambda}_{Q}(t)$, $\OP{\lambda}_{S}(t)$ and $\OP{\lambda}_{T}(t)$ as diagonal operators
\begin{subequations}\label{eq:diagonaloperators}
\begin{align}
    \OP{\lambda}_{Q}(t)&=\sum_{j\in J} \BRR{E_{J_{l,j}}+E_{J_{r,j}}}\BRR{\cos(\frac{\varphi_{0,j}}{2}) - \cos(\frac{\varphi_{j}(t)}{2})} \OP{\lambda}_{Q_{\varphi_{j}}},\\
    \OP{\lambda}_{S}(t)&=\sum_{j\in J} \BRR{E_{J_{l,j}}-E_{J_{r,j}}}\BRR{\sin(\frac{\varphi_{0,j}}{2}) - \sin(\frac{\varphi_{j}(t)}{2})} \OP{\lambda}_{S_{\varphi_{j}}},\\
    \begin{split}
    \OP{\lambda}_{T}(t) &=\sum_{i\in I} -2 E_{C_{i}} n_{g,i}(t) \OP{\lambda}_{T_{n_{i}}} + \sum_{j\in J} -2 E_{C_{j}} n_{g,j}(t) \OP{\lambda}_{T_{n_{j}}}\\
                      &+\sum_{n \in \{0,...,9\}} \sum_{(i,j) \in  I_{n} \times J_{n}} G_{i,j}^{(n)} \BRR{ \OP{\lambda}_{T_{v_{n,i}}}  \tens{} \OP{\lambda}_{T_{v_{n,j}}}},
    \end{split}
\end{align}
\end{subequations}
in their respective basis, see the labels $Q$, $S$ and $T$. Here, we implicitly defined a function which maps the different interaction operators $\OP{v}_{n,i}$, see \equref{eq:INTOPTABRV}, to the correct operators $\OP{\lambda}_{T_{v_{n,i}}}$. This formulation of the problem allows us to separate the time dependencies from the operators, in the individual summation terms.

We intend to implement a high-performance simulation code. Therefore we restrict the number of allowed trajectories of $\OP{H}(t)$. To this end, we require that in an arbitrary time interval $t\in[t_{0},T]$ either the flux variables $\varphi_{j}(t)$ or the charge variables $n_{g,i/j}(t)$ vary but not both. This assumption allows us to derive the time-evolution operator $\OP{\mathcal{U}}^{\text{TB}}(t)$ by considering two special cases. We label these cases with Roman numerals I and II.

If we consider time intervals $t\in[t_{0},T]$, where $\varphi_{j}(t)=\varphi_{0,j}$ for all $j \in  J$, we find $\OP{W}_{0}(t)=0$ and $\OP{W}_{1}(t)=0$. Consequently, the Hamiltonian \equref{eq:CHM} simplifies considerably. The corresponding second-order time-evolution operator reads
\begin{equation}\label{eq:UTBI}
  \OP{\mathcal{U}}^{\text{I}}(t)=e^{-i\frac{\tau}{2}\OP{H}_{0}} \OP{T} e^{-i \tau \OP{\lambda}_{T}(t)} \OP{T}^{\dagger} e^{-i \frac{\tau}{2} \OP{H}_{0}}.
\end{equation}
The operator we implement here is structurally equivalent to the one discussed in \REF\cite{Willsch2020}.

If we consider time intervals $t\in[t_{0},T]$, where $\varphi_{j}(t)\neq\varphi(0)$ for some $j \in  J$ and $n_{g,i}(t)=0$ for all $i \in  I$ and $n_{g,j}(t)=0$ for all $j \in  J$, we find that the second-order time-evolution operator can be expressed as
\begin{equation}
  \OP{\mathcal{U}}^{\text{II}}(t)= \OP{Q} e^{-i \frac{\tau}{2} \OP{\lambda}_{Q}(t)} \OP{Q}^{\dagger} \OP{S} e^{-i \frac{\tau}{2} \OP{\lambda}_{S}(t)}  \OP{S}^{\dagger} \OP{\mathcal{U}}^{\text{I}}(t_{0}) \OP{S} e^{-i \frac{\tau}{2} \OP{\lambda}_{S}(t)}  \OP{S}^{\dagger}  \OP{Q} e^{-i \frac{\tau}{2} \OP{\lambda}_{Q}(t)} \OP{Q}^{\dagger}.
\end{equation}
We can simplify $\OP{\mathcal{U}}^{\text{II}}(t)$, in the sense that we aim to reduce the number of operators which have to be implemented in the computer program. To this end, we define an additional unitary operator $\OP{R}=\OP{Q}^{\dagger} \OP{S}$ such that the operator $\OP{\mathcal{U}}^{\text{II}}(t)$ reads
\begin{equation}\label{eq:UTBII}
  \OP{\mathcal{U}}^{\text{II}}(t)= \OP{Q} e^{-i \frac{\tau}{2} \OP{\lambda}_{Q}(t)} \OP{R} e^{-i \frac{\tau}{2} \OP{\lambda}_{S}(t)} \OP{S}^{\dagger} \OP{\mathcal{U}}^{\text{I}}(t_{0}) \OP{S} e^{-i \frac{\tau}{2} \OP{\lambda}_{S}(t)}  \OP{R}^{\dagger} e^{-i \frac{\tau}{2} \OP{\lambda}_{Q}(t)} \OP{Q}^{\dagger},
\end{equation}
where we made use of the fact that $\OP{R}^{\dagger}=\OP{S}^{\dagger} \OP{Q}$. Consequently, we can express the operator $\OP{\mathcal{U}}^{\text{TB}}(t)$ on the interval $t\in[t_{0},T]$ as
\begin{equation}\label{eq:TEO MB}
\OP{\mathcal{U}}^{\text{TB}}(t)=
\begin{cases}
\begin{aligned}
\OP{\mathcal{U}}^{\text{I}}(t)    &\text{ if Case I }\\
\OP{\mathcal{U}}^{\text{II}}(t)   &\text{ if Case II}.
\end{aligned}
\end{cases}
\end{equation}
An implementation of this operator in a computer program allows us to further reduce the number of operators which have to be applied to the state vector $\ket{\Psi(t)}$ given by \equref{eq:TDSE}. We can see the simplification of the operator structure once we consider the product of two operators $\OP{\mathcal{U}}^{\text{II}}(t)$, for two different time steps $t^{\prime \prime} > t^{\prime}$. The corresponding product reads
\begin{equation}\label{eq:UTBproduct}
\begin{split}
\begin{aligned}
  \OP{\mathcal{U}}^{\text{II}}(t^{\prime \prime}) \OP{\mathcal{U}}^{\text{II}}(t^{\prime}) &= \OP{Q} e^{-i \frac{\tau}{2} \OP{\lambda}_{Q}(t^{\prime \prime})} \OP{R} e^{-i \frac{\tau}{2} \OP{\lambda}_{S}(t^{\prime \prime})} \OP{S}^{\dagger} \OP{\mathcal{U}}^{\text{I}}(t^{\prime \prime}) \OP{S} e^{-i \frac{\tau}{2} \OP{\lambda}_{S}(t^{\prime \prime})}  \OP{R}^{\dagger} e^{-i \frac{\tau}{2} \OP{\lambda}_{Q}(t^{\prime \prime})} \\
  &  e^{-i \frac{\tau}{2} \OP{\lambda}_{Q}(t^{\prime})} \OP{R} e^{-i \frac{\tau}{2} \OP{\lambda}_{S}(t^{\prime})} \OP{S}^{\dagger} \OP{\mathcal{U}}^{\text{I}}(t^{\prime}) \OP{S} e^{-i \frac{\tau}{2} \OP{\lambda}_{S}(t^{\prime})}  \OP{R}^{\dagger} e^{-i \frac{\tau}{2} \OP{\lambda}_{Q}(t^{\prime})} \OP{Q}^{\dagger},
\end{aligned}
\end{split}
\end{equation}
as one can see, in this case we can avoid implementing the two operators $\OP{Q}^{\dagger}$ and $\OP{Q}$ due to the fact that the product of both yields the identity operator. However, if we implement the time-evolution operator in this way, we have to take into account that between two time steps $t^{\prime \prime}, t^{\prime}$ we cannot sample from $\ket{\Psi(t)}$ in the \TB{},\ie in this case we have to apply the transformations $\OP{Q}^{\dagger}$ and $\OP{Q}$ once again. If we simulate complete quantum circuits, we rarely sample from $\ket{\Psi(t)}$. In such cases this approach is preferable.

\section{Simulations of the circuit Hamiltonian model in alternative bases}\label{sec:SimulationsOfTheCircuitHamiltonianModelInAlternativeBases}
In the previous section we discussed a product-formula time-evolution operator $\UTB$ for a specific Hermitian decomposition of the circuit Hamiltonian given by \equref{eq:CHM}. Here we use the transmon basis $\{\ket{\phi^{(z)}(t=0)}\}_{z \in \mathbb{N}^{0}}$, see \equaref{eq:basisstates}{eq: stationary Schroedinger equation}, as the computational basis to simulate the state vector $\ket{\Psi(t)}$ in \equref{eq:TDSE}. Note that in this context the computational basis is simply the basis we use to represent the abstract state vector $\ket{\Psi(t)}$ in a computer program,\ie we should not confuse this computational basis with the one of the IGQC, see \chapref{chap:II}. The first question which comes to mind is: why do we use this basis to solve the TDSE? For the work in this thesis we implemented several product-formula time-evolution operators which make use of different computational bases. In this section we briefly summarise the results of this work.

If we use the charge basis $\{\ket{n}\}_{n \in \mathbb{Z}}$, see \equref{eq:charge_basis_states}, as the computational basis, we find (data not shown) that a quasi approximation-free solution of the TDSE can be obtained by simulating flux-tunable transmons with up to fifty one charge basis states. This means adding an additional flux-tunable transmon to the model increases the state vector size by a factor of fifty one. This number was obtained by performing the simulations presented in \chapref{chap:NA} with the charge basis simulator. Note that the corresponding simulations in the transmon basis require at most sixteen basis states for the flux-tunable transmon. Therefore, using the charge basis simulator for large-scale simulations of flux-tunable transmons is not feasible. The core simulation algorithm was devised by the author of \REF\cite{Willsch2016Master}. The author of this thesis simply modified the algorithm presented in \REF\cite{Willsch2016Master} such that the changes in the circuit Hamiltonian given by \equref{eq:flux-tunable transmon_beta} are accounted for, see also \REF\cite{You}.

If we use the instantaneous basis $\{\ket{\phi^{(z)}(t)}\}_{z \in \mathbb{N}^{0}}$, see \equaref{eq:basisstates}{eq: stationary Schroedinger equation}, to model the different flux-tunable transmons in the system,\ie the fixed-frequency transmons and resonators are still modelled with the basis states $\{\ket{\phi^{(z)}(t=0)}\}_{z \in \mathbb{N}^{0}}$ and $\{\ket{\psi^{(z)}}\}_{z \in \mathbb{N}^{0}}$, we find that we need to modify the circuit Hamiltonian in \equref{eq:CHM} such that the TDSE stays form invariant. For simplicity, we consider a single flux-tunable transmon whose dynamics is modelled with the circuit Hamiltonian in \equref{eq:flux-tunable transmon recast}. The state vector which describes the time-evolution of the system can be expressed as
\begin{equation}
  \ket{\Psi^{*}(t)}=\mathcal{Y}(t)\ket{\Psi(t)},
\end{equation}
where $\mathcal{Y}(t)$ is a unitary transformation which maps the basis states $\ket{\phi^{(z)}(0)}$ at time $t=0$ to the states $\ket{\phi^{(z)}(t)}$ at time $t$. We require the TDSE to retain its original form, see \REF\cite{Weinberg2015}. Therefore, we transform the Hamiltonian as
\begin{equation}
  \OP{H}^{*}(t)= \mathcal{Y}(t)\OP{H}(t)\mathcal{Y}^{\dagger}(t) -i \mathcal{Y}(t)\partial_{t}\mathcal{Y}^{\dagger}(t).
\end{equation}
In order to proceed, we have to determine an analytical expressions for the inner product $\braket{\phi^{(z^{\prime})}(t)|\partial_{t}|\phi^{(z)}(t)}$ such that the term
\begin{equation}\label{eq:BT}
  \OP{\mathcal{D}}(t)= -i \mathcal{Y}(t)\partial_{t}\mathcal{Y}^{\dagger}(t),
\end{equation}
can be implemented. However, to the best knowledge of the author, this expression is not known to the research community for the eigenfunctions in \equref{eq:Mathieufunctions}, see also \REF\cite{Cottet2002}. Note that the Josephson energy and the flux variables in \equref{eq:Mathieufunctions} need to be substituted if Hamiltonian \equref{eq:flux-tunable transmon recast} is considered, see \secref{sec:Transmons} for more details. Therefore, this approach was abandoned too. Furthermore, all attempts to implement the operator in \equref{eq:BT} numerically failed.

\section{Implementation of the second-order time-evolution operator}\label{sec:ImplementationOfTheTime-evolutionOperator}

In \secaref{sec:SOTEO_HB}{sec:SOTEO_MB}, we discuss two second-order time-evolution operators $\OP{\mathcal{U}}^{\text{HB}}$ and $\OP{\mathcal{U}}^{\text{TB}}$ which allow us to solve the TDSE for Hamiltonians \equaref{eq:CHM}{eq:EHM} numerically,\ie we can obtain an approximate solution. In this section, we discuss how to implement the operator $\OP{\mathcal{U}}^{\text{TB}}$ with respect to the state vector
\begin{equation}\label{eq:state_vector}
  \ket{\Psi(t)}=\sum_{\mathbf{z} \in  \mathcal{I}_{\mathbf{n}}} c_{\mathbf{z}}(t) \ket{\mathbf{z}}.
\end{equation}
Since we can use the same algorithms to implement the time-evolution operators $\OP{\mathcal{U}}^{\text{HB}}$ and $\OP{\mathcal{U}}^{\text{TB}}$ given by \equaref{eq:TEO HB}{eq:TEO MB}, respectively, we only discuss the implementation of $\OP{\mathcal{U}}^{\text{TB}}$ in detail. Transferring the results of the following discussion to the case of $\OP{\mathcal{U}}^{\text{HB}}$ should come at ease.

\renewcommand{\OP}[1]{#1}

Broadly speaking, the time-evolution matrix $\OP{\mathcal{U}}^{\text{TB}}$ contains two types of matrices, namely diagonal and non-diagonal ones. Note that the diagonal matrices are not all diagonal in the same basis. In fact, this is the reason why we find the non-diagonal matrices in $\OP{\mathcal{U}}^{\text{TB}}$,\ie these matrices transform the basis for every term in the product-formula decomposition.

Since both types of matrices are defined in terms of tensor products, see \equref{eq:basistrafo} and the sums of tensor products, see \equref{eq:diagonaloperators} and the product-formula decomposition allows us to treat the terms $\OP{H}_{0}$, $\OP{W}_{0}(t)$ and $\OP{W}_{1}(t)$ and $\OP{W}_{2}(t)$ individually, we find that we can evaluate the action of all operators either by considering their action with respect to the individual basis states $\ket{\mathbf{z}}$ or by considering a small set $\{\ket{\mathbf{z}}\}$ of interdependent basis states. In the following, we derive a set of update rules which allow us to implement these operators.

We begin with the update rules for the diagonal matrices. Here we can further differentiate. The approximant  $\mathcal{U}^{\text{TB}}$ contains diagonal matrices which exclusively act on individual subspaces, see $\OP{H}_{0}$, $\OP{W}_{0}(t)$ and $\OP{W}_{1}$(t) as well as diagonal matrices which act on two distinct subspaces, see $\OP{W}_{2}(t)$,\ie the matrices which describe the interaction between two subsystems.

Without loss of generality, we consider a diagonal matrix $\OP{\lambda}_{\idxFTT,j}$ which is defined with respect to the subspace $\mathcal{H}_{\idxFTT,j}$ of some flux-tunable transmon. The action of this matrix reads
\begin{equation}
  e^{-i \tau \OP{\lambda}_{\idxFTT,j}} \ket{\mathbf{z}}= e^{-i \tau \lambda_{\idxFTT}^{(z_{j})}} \ket{\mathbf{z}},
\end{equation}
where $z_{j}\in\{0,...,n_{J}-1\}$ is the jth tuple entry of $\mathbf{z}_{J}=\BRR{z_{|J|-1}, ...,z_{0}}$ in $\ket{\mathbf{z}}$, see \equref{eq:basisstates}. Consequently, $\lambda_{\idxFTT}^{(z_{j})}$ is the eigenvalue which corresponds to the eigenvector $\ket{\phi^{(z_{j})}}$. In the end, we simply have to implement the update rule
\begin{equation}
  c_{\mathbf{z}}^{\prime}= c_{\mathbf{z}} e^{-i \tau \lambda_{\idxFTT}^{(z_{j})}},
\end{equation}
where $c_{\mathbf{z}}^{\prime}$ denotes the updated state vector coefficient.

Similarly, without loss of generality, we can consider a diagonal matrix  product of the form
\begin{equation}
  \OP{\lambda}_{\idxRES,k}\tens{}\OP{\lambda}_{\idxFTT,j},
\end{equation}
where the matrix $\OP{\lambda}_{\idxRES,k}$ acts on the subspace $\mathcal{H}_{\idxRES,k}$. The matrix  $\OP{\lambda}_{\idxFTT,j}$ is defined as before. The action of the tensor product can be expressed as
\begin{equation}
  e^{-i \tau \OP{\lambda}_{\idxRES,k}\tens{}\OP{\lambda}_{\idxFTT,j}} \ket{\mathbf{z}}= e^{-i \tau \lambda_{\idxRES}^{(z_{k})}\lambda_{\idxFTT}^{(z_{j})}} \ket{\mathbf{z}},
\end{equation}
where $z_{k}\in\{0,...,n_{K}-1\}$ is the kth tuple entry of $\mathbf{z}_{K}=\BRR{z_{|K|-1}, ...,z_{0}}$ in $\ket{\mathbf{z}}$ and $\lambda_{\idxRES}^{(z_{k})}$ is the eigenvalue which corresponds to the eigenvector $\ket{\psi^{(z_{k})}}$. The update rule for this case reads
\begin{equation}
  c_{\mathbf{z}}^{\prime}= c_{\mathbf{z}} e^{-i \tau \lambda_{\idxRES}^{(z_{k})}\lambda_{\idxFTT}^{(z_{j})}}.
\end{equation}
In principle, we can extend the procedure to products of more than two diagonal matrices. The argumentation is straightforward.

If we implement the matrix products in $\UTB$, we actually work with the transformed basis states $\ket{\mathbf{w}_{Q}}$, $\ket{\mathbf{w}_{S}}$ and $\ket{\mathbf{w}_{T}}$, which are related to the basis states $\ket{\mathbf{z}}$ by means of the transformations $\OP{Q}$, $\OP{S}$ and $\OP{T}$, see \equref{eq:basistrafo}. However, in order to provide the reader with a discussion which makes use of the new basis states, we would need to introduce new notation. We circumvent this inconvenience by providing a discussion which considers matrix products that are diagonal in the basis $\mathcal{J}_{\mathbf{n}}$. The arguments we make can be transferred to any basis, assuming that the matrices in question are diagonal in the corresponding bases.

So far, we only discussed how to implement diagonal matrices which exclusively act on specific subspaces or products of diagonal matrices which are diagonal in one shared basis. However, the approximant $\mathcal{U}^{\text{TB}}$ contains the sums in \equref{eq:diagonaloperators} of those types of matrices. Nonetheless, this does not lead to further complications since the individual summation terms commute. Consequently, we can replace the exponential functions that contain the sums by products of exponential functions, which contain the individual summation terms. This means we can implement the individual terms with the update rules we have discussed so far.

The non-diagonal basis transformations $\OP{Q}$, $\OP{S}$ and $\OP{T}$ all have the same form,\ie we can express them as tensor products of matrices which act on the different subspaces. Therefore, the tensor algebra calculus, see \secref{sec:MathematicalFramework}, allows us to update the state vector by applying the matrices which act on the different subspaces. We simply make repeated use of the identity given by \equref{eq:productoperator}. In the end, we only have to find one additional update rule for a full implementation of $\mathcal{U}^{\text{TB}}$.

As before, without loss of generality, we consider the matrix $\OP{T}_{n_{j}}$ which acts on the subspace $\mathcal{H}_{\idxFTT,j}$. Evaluating this matrix with respect to the basis vector $\ket{\mathbf{z}^{(z_{j}^{\prime})}}$ and the index $z_{j}^{\prime}$ yields
\begin{equation}\label{eq:basistrafo_state}
  \OP{T}_{n_{j}}\ket{\mathbf{z}^{(z_{j}^{\prime})}}= \sum_{z_{j} \in \{0,...,n_{J}-1\}} T_{n_{j}}^{(z_{j}, z_{j}^{\prime})} \ket{\mathbf{z}^{(z_{j})}}.
\end{equation}
Here, we introduced the label $z_{j}$ to the notation. This index simply highlights the corresponding tuple entry in $\mathbf{z}_{J}$. We might call the label $j$ the target index. If we apply $\OP{T}_{n_{j}}$ to the state vector in \equref{eq:state_vector}, we can make use of \equref{eq:basistrafo_state} and rearrange the summation terms to find an update rule.

Assuming we fix all the remaining tuple entries in the basis state $\ket{\mathbf{z}^{(z_{j}^{\prime})}}$,\ie all tuple entries, which are not $z_{j}^{\prime}$, we find that the matrix $\OP{T}_{n_{j}}$ relates all basis states $\ket{\mathbf{z}^{(z_{j}^{\prime})}}$, where $z_{j}^{\prime} \in \{0, ..., n_{J}-1\}$. The corresponding update rule reads
\begin{equation}\label{eq:nondiagonalupdaterule}
  c_{\mathbf{z}^{(z_{j}^{\prime})}}^{\prime}=\sum_{z_{j} \in \{0, ..., n_{J}-1\}} T_{n_{j}}^{(z_{j}^{\prime}, z_{j})} c_{\mathbf{z}^{(z_{j})}}.
\end{equation}
Consequently, if we model the subspace $\mathcal{H}_{\idxFTT,j}$ with $n_{J}$ basis states, we find that $n_{J}$ different state vector coefficients $c_{\mathbf{z}}$ are related by \equref{eq:nondiagonalupdaterule}. This allows us to partition the full state vector into groups of size $n_{J}$. We can then update these groups independently,\ie the groups which are built by fixing the tuple entries that are not the target tuple entry $z_{j}^{\prime}$. Note that this reasoning can be applied to all subspaces $\mathcal{H}_{\idxTLS,l}$, $\mathcal{H}_{\idxRES,k}$, $\mathcal{H}_{\idxFTT,j}$ and $\mathcal{H}_{\idxFFT,i}$. This update rule can be implemented in many different ways and the space of possible implementations is quite large. For example, two important aspects of the particular implementation instance are the way we address the different state vector coefficients $c_{\mathbf{z}}$ and the programming paradigm itself.

A detailed discussion of all the subroutines we use to perform the simulations and the extended simulation code is beyond the scope of this chapter. However, in \secref{sec:StructureOfTheSimulationSoftware} we provide the reader with an overview of the simulation software.

\renewcommand{\OP}[1]{\hat{#1}}

\section{An overview of the simulation software}\label{sec:StructureOfTheSimulationSoftware}
\graphicspath{{./Diagram/}}
\begin{figure}[!tbp]
    \centering
    \includegraphics[width=\width\textwidth]{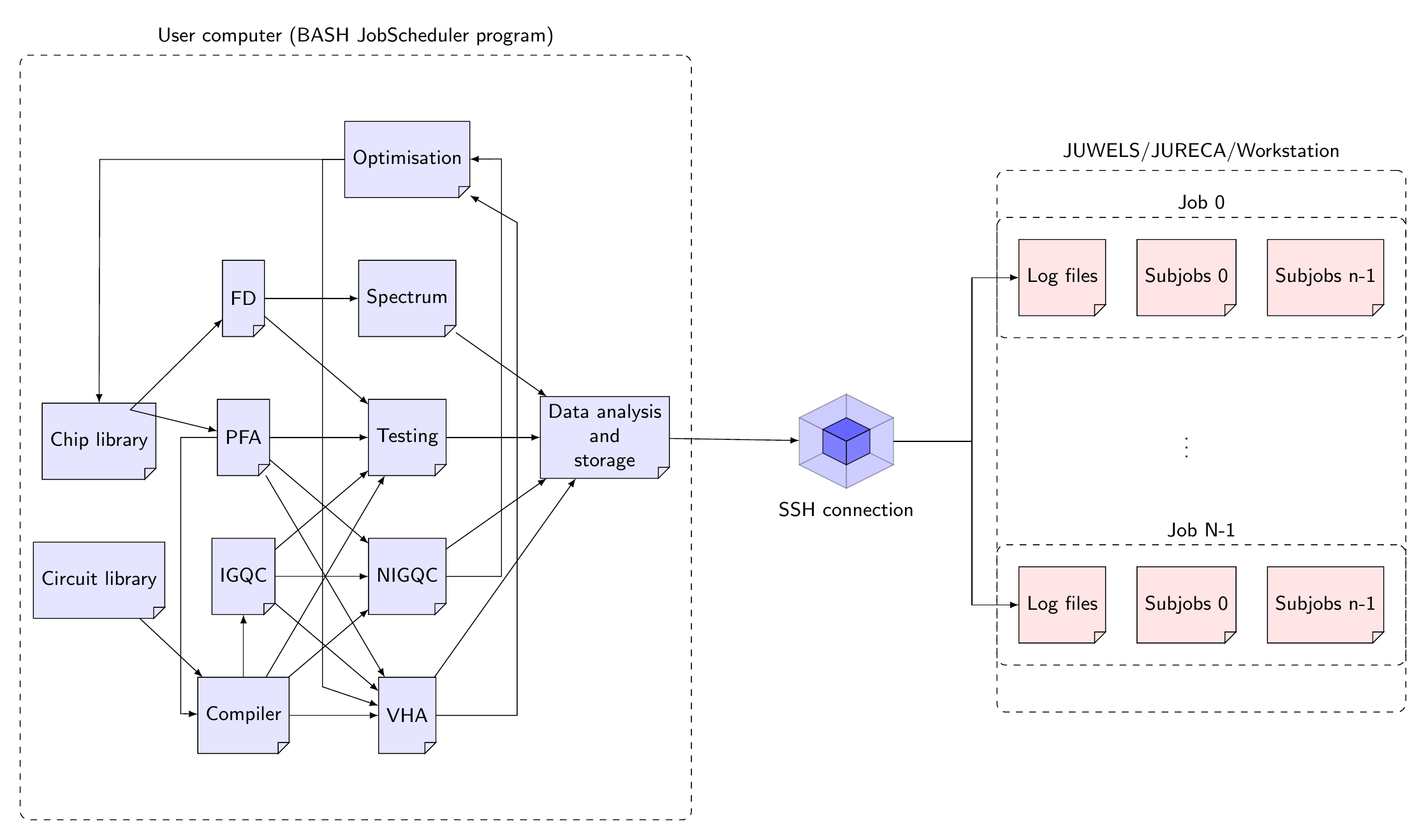}
    \caption[High-level sketch of the simulation software and the surrounding compute infrastructure.]{High-level sketch of the simulation software and the surrounding compute infrastructure. We use the blue rectangles to lump C++ source code into modules. The dashed lines surrounding the modules indicate the borders of the C++ simulation software but also the domain of a BASH program called JobScheduler. We use the program JobScheduler in combination with the MAKE build automation tool to control the compilation, execution, logging and data storage on various computer systems, including the supercomputers JUWELS and JURECA, see \REFS\cite{JUWELS,JURECA}. The modules which are named by means of abbreviations are full diagonalisation (FD), product-formula algorithm (PFA), ideal gate-based quantum computer (IGQC), non-ideal gate-based quantum computer (NIGQC) and variational hybrid algorithm (VHA). The remaining module names are self explanatory. The functionality of all modules is explained in the main text.} \label{fig:SoftwareDiagram}
\end{figure}

In this section, we provide a high-level overview of the simulation software. So far, we discussed some of the components or modules of this simulation software. For example, in \secref{sec:SimulationOfTheIdealGateBasedQuantumcomputer} we discussed how to simulate the IGQC and in \secsref{sec:TDSEIntro}{sec:ImplementationOfTheTime-evolutionOperator} we discussed how to solve the TDSE with full diagonalisation and the product-formula algorithm. The algorithms can be used to implement a numerical NIGQC model. Note that in the following we capitalise almost all of the computer science terminology.

Figure \ref{fig:SoftwareDiagram} shows a high-level sketch of the simulation software. Here we lump different parts of the simulation code into software modules, see blue rectangles and show the surrounding compute infrastructure, see dashed lines, red rectangles and other symbols. The dashed rectangle on the right side of \figref{fig:SoftwareDiagram} shows the boundaries of the source code. We use C++ as the basic programming language. In addition, we use Open Multi Processing (OpenMP), Message Passing Interface (MPI) and Compute Unified Device Architecture (CUDA) to parallelise the C++ simulation code.

Since we execute tasks on different computer systems like the supercomputers JUWELS and JURECA, see \REFS\cite{JUWELS,JURECA}, we use a homemade BASH program named JobScheduler to control compilation, execution, logging and data storage. If the program JobScheduler executes a job, the following steps are performed. First, we update the source code by means of an SSH connection. Second, we execute a MAKE program to recompile some parts of the source code if changes have been made. Third, we create a directory structure which contains the result data, log files, SLURM job scripts and executables. Fourth, if the task is executed on a supercomputer system, we write a SLURM job script for every task or subjob. Fifth, we move executables and SLURM job scripts to the correct directories. In the end, if we use a supercomputer system, we submit the SLURM job scripts to the program SLURM and the program JobScheduler terminates. Note that every step in this sequence relies on the successful execution of the preceding step. In the following we briefly summarise the functionality of each module and sometimes discuss the relation between different modules.

The module named chip library, see first column on the left side, contains all NIGQC chip parameters. Here we manage the parameters which specify the Hamiltonians \equaref{eq:CHM}{eq:EHM} and the pulse parameters which determine the values of the time dependencies that can be found in the model Hamiltonians. All these parameters specify a NIGQC model, see \chapsref{chap:NA}{chap:GET}.

The module named circuit library, see first column on the left side, has a similar functionality as the module chip library. Here we manage all the quantum circuits we use to perform simulations.

The module named FD, see second column on the left side, allows us to perform full diagonalisation of the Hamiltonians \equaref{eq:CHM}{eq:EHM}. Here, we use the software packages LAPACK and MKL, see \REFS\cite{PACK99,MKL09}, to compute the eigenvalues and eigenvectors for a given Hamiltonian. The module is split into two parts. The first part consists of a class which organises all the data and the second part allows us to simulate the time evolution of a system by implementing \equref{eq:EVSTE} for various time steps.

The module named PFA, see second column on the left side, allows us to solve the TDSE with various product-formula algorithms. This module is also split into two parts,\ie a part with classes which organise the simulation data and a part with subroutines which update the state vector data structure by means of the update rules we discuss in \secref{sec:ImplementationOfTheTime-evolutionOperator}. We can determine an approximate solution of the TDSE for the Hamiltonian \equref{eq:EHM} in the harmonic basis, see \secref{sec:SOTEO_HB} and the Hamiltonian \equref{eq:CHM} in the transmon basis, see \secref{sec:SOTEO_MB}. Furthermore, the user can choose between serial execution, parallel execution with OpenMP and parallel execution on a GPU, where we use CUDA to parallelise the source code.

The module named IGQC, see second column on the left side, enables us to simulate the IGQC. Here we update the state vector of a $N$-qubit system for a given quantum circuit, see \chapref{chap:II}. We can perform the simulations with the Feynman or the Schrödinger algorithm, see \secref{sec:SimulationOfTheIdealGateBasedQuantumcomputer}. The user has the choice between serial execution, parallel execution with OpenMP and parallel execution with OpenMP and MPI.

The module named Compiler, see second column on the left side, takes a quantum circuit and optionally a PFA class, with information about the chip architecture and the pulse parameters, as an input and creates control data structures. We can use the control data structures in combination with the PFA and IGQC modules to simulate complete quantum circuits, see modules NIGQC and VHAs. The data processing steps for the PFA module are quite complex and the computational effort grows with the number of qubits and the quantum circuit length. We discuss some of the details of these computations in \chapref{chap:GET}. The corresponding data processing steps for the IGQC module are relatively simple and there is quasi no overhead in time and space.

The module named Optimisation, see third column on the left side, is a module to perform various types of optimisation. Here we use the library NLopt, see \REF\cite{NLopt}, to solve the optimisation problems we encounter. The library provides us with more than a dozen optimisation algorithms.

The module named Spectrum, see third column on the left side, enables us to compute the spectrum of a system for different model parameters in parallel. Here we use the packages LAPACK and MKL, see \REFS\cite{PACK99,MKL09}, to obtain the eigenvalues for a given Hamiltonian. We parallelised this procedure with MPI. This means every MPI process computes the eigenvalues for one subset of all model parameters. Furthermore, we can make use of the fact that the packages LAPACK and MKL are parallelised with OpenMP.

The module named Testing, see third column on the left side, allows us to systematically test critical parts of the software. We automatically test the modules FD, PFA, IGQC and Compiler every time after changes have been made to the corresponding source code files.

The module named NIGQC, see third column on the left side, enables us to compute various gate-error measures for a given quantum circuit. This means we can study how gate errors dynamically evolve over time,\ie for consecutive gates in the quantum circuit. In \secref{sec:GET_errormeasures} we discuss all gate-error measures we can compute with this module. We use MPI to parallelise this part of the simulation code.

The module named VHA, see third column on the left side, enables us to simulate variational hybrid algorithms with different optimisation algorithms and quantum computer models,\ie IGQC or NIGQC. The user can specify the cost Hamiltonian given by \equref{eq:VQEHamiltonian} in combination with a suitable optimisation algorithm, see module Optimisation. This module takes a quantum circuit as an input. However, the quantum circuit is first preprocessed by the Compiler module. We use MPI to parallelise the simulation code. This means the user can run several problem instances in parallel.

The module named Data analysis and storage, see fourth column on the left side, manages data analysis and storage, as the name suggests. In almost all cases we can store data in binary or text format. The module also enables us to automatically store data in LATEX tables, see \appref{app:ControlPulseParameters} and \appref{app:TablesWithGateErrorMetrics}. Here a subroutine generates the corresponding LATEX source code.

\section{Summary, conclusions and outlook}\label{sec:ConclusionsAlgorithms}
In this chapter we reviewed the fundamental computational problems we face when modelling time-evolution processes with the time-dependent Schrödinger equation (TDSE), see \secref{sec:TDSEIntro}. We provided a general discussion of the product-formula algorithm which can potentially remedy some of computational issues we face, see \secref{sec:TheProductFormulaAlgorithm}. Furthermore, in \secref{sec:SOTEO_HB} we derived a product-formula time-evolution operator which allows us to approximately solve the TDSE for the effective Hamiltonian \equref{eq:EHM}. Similarly, in \secref{sec:SOTEO_MB} we derived a product-formula time-evolution operator which allows us to approximately solve the TDSE for the circuit Hamiltonian \equref{eq:CHM}. Additionally, we reported on attempts to model the time evolution of the circuit Hamiltonian \equref{eq:CHM} in the charge basis and the instantaneous flux-tunable transmon basis, see \secref{sec:SimulationsOfTheCircuitHamiltonianModelInAlternativeBases}. We also discussed the algorithmic update rules which allow us to implement the time-evolution of the state vector for the effective and the circuit Hamiltonian in a computer program, see \secref{sec:ImplementationOfTheTime-evolutionOperator}. Finally, we provided an overview of the simulation software we use to obtain the results in this thesis, see \secref{sec:StructureOfTheSimulationSoftware}.

In order to obtain the results in this thesis, we implemented several time-evolution operators which approximately solve the TDSE for the circuit Hamiltonian \equref{eq:CHM}. However, we were not able to model various systems, see \chapref{chap:NA}, with less than sixteen basis states per flux-tunable transmon. Therefore, in practice it is not possible to model more than eight interacting flux-tunable transmons with the circuit model. Here we work under the constraints that we have to simulate the original circuit Hamiltonian \eqref{eq:CHM} and that the computer hardware can store up to $2^{32}$ state vector coefficients $c_{\mathbf{z}}$.

For future work, it might be interesting to look into the question whether or not there exists a computational basis which allows us to simulate more than eight interacting flux-tunable transmons. Here one might consider the time-dependent harmonic basis, see \equref{eq:HarmoicBasisWaveFunctionTimeDep} and \REF\cite{WillschM2020}. Alternatively, one might solve the analytical problem discussed in the latter part of  \secref{sec:SimulationsOfTheCircuitHamiltonianModelInAlternativeBases}.


\chapter{Analysis of effective models for flux-tunable transmon systems}\label{chap:NA}
\newcommand{\ISWAP}{\text{ISWAP}}
\newcommand{\CZ}{\text{CZ}}
\newcommand{\DF}{\omega^{(D)}}
\newcommand{\DFTP}{\omega^{(D)}/2\pi}
\newcommand{\PA}{\delta}
\newcommand{\PATP}{\delta/2\pi}
\newcommand{\TRF}{T_{r/f}}
\newcommand{\TD}{T_{d}}
\newcommand{\FQ}{\Phi_{0}}
\newcommand{\FQTP}{\phi_{0}/2\pi}
\newcommand{\FP}{\varphi(t)}
\newcommand{\FPTP}{\varphi(t)/2\pi}

Many studies with a focus on flux-tunable transmons are based on effective models and not the associated lumped-element circuit Hamiltonian models introduced in \chapref{chap:III}. Unfortunately, usually it is not known to what extend the predictions made by the effective and the circuit model deviate when a control pulse is applied,\ie when the Hamiltonians are time-dependent. In this chapter, we compare the time evolution of state vectors which are obtained by solving the TDSE for both the effective and the circuit Hamiltonians, for microwave and unimodal control pulses. Here, we consider single-qubit (X) and two-qubit (\ISWAP{} and \CZ{}) gate type transitions. Note that we consider these transitions only in terms of the probability amplitudes and all transitions are activated by means of an external flux. Furthermore, in order to obtain a quasi approximation-free solution of the TDSE, we increase the number of basis states up to the point where the probability amplitudes have converged up to the third decimal.

The work in this chapter is structured as follows. In \secref{sec:NA_model_parameters} we discuss the three different systems we study and the control pulse we use to activate the different transitions. Here, we also discuss the approximations which give rise to different effective models. Next, in \secref{sec:NA_single_flux_tunable_transmon} we study the pulse response of a single flux-tunable transmon for microwave and unimodal pulse forms. Here, we consider two effective models and the associated circuit model. In \secaref{sec:NA_E_suppressed}{sec:NA_unsuppressed_arch_I} we investigate various single-qubit (X) and two-qubit (\ISWAP{} and \CZ{}) transitions for a first two-qubit NIGQC model, see \figref{fig:arch_sketch}(a). Here we model the different transitions with a microwave pulse for various effective models and the associated circuit model. Similarly, in \secaref{sec:NA_E_suppressed_AII}{sec:NA_unsuppressed_arch_II}, we study excitations of the coupler element and two-qubit (\ISWAP{} and \CZ{}) transitions for a second two-qubit NIGQC model, see \figref{fig:arch_sketch}(b). Here we activate the different transitions with a microwave and a unimodal pulse for various effective models and the associated circuit model. Note that we use $\hbar=1$ throughout this chapter.

\section{System specification and simulation parameters}\label{sec:NA_model_parameters}

\renewcommand{\width}{0.95}
\newcommand{\scale}{0.22}
\begin{figure}[!tbp]
  \begin{minipage}{0.5\textwidth}
    \centering
    \begin{tikzpicture}[thick,scale=\scale, every node/.style={transform shape}]
	     \begin{pgfonlayer}{nodelayer}
		\node [style=none] (8) at (6, 7) {};
		\node [style=none] (9) at (10, 7) {};
		\node [style=none] (10) at (8, 9) {};
		\node [style=none] (11) at (8, 5) {};
		\node [style=none] (16) at (-2, 0) {};
		\node [style=none] (17) at (2, 0) {};
		\node [style=none] (18) at (0, 2) {};
		\node [style=none] (19) at (0, -2) {};
		\node [style=none] (20) at (14, 0) {};
		\node [style=none] (21) at (18, 0) {};
		\node [style=none] (22) at (16, 2) {};
		\node [style=none] (23) at (16, -2) {};
		\node [style=none] (26) at (8, 7) {\Huge $\omega_{2}^{(q)}(t)$};
		\node [style=none] (29) at (0, 0) {\Huge $\omega_{0}^{(q_{0})}$};
		\node [style=none] (30) at (16, 0) {};
		\node [style=none] (32) at (16, 0) {\Huge $\omega_{1}^{(q_{0})}$};
		\node [style=none] (46) at (8, 10) {};
		\node [style=none] (47) at (8, 10) {\Huge \textbf{Flux-tunable transmon} $i=2$};
		\node [style=none] (49) at (0, -3) {\Huge \textbf{Fixed-frequency transmon} $i=0$};
		\node [style=none] (51) at (16, -3) {\Huge \textbf{Fixed-frequency transmon} $i=1$};
		\node [style=none] (52) at (8, 12) {};
		\node [style=none] (53) at (8, 12) {\Huge \textbf{\underline{Architecture I}}};
		\node [style=none] (54) at (8, -5) {};
		\node [style=none] (60) at (4, 3) {};
		\node [style=none] (61) at (3, 3) {\Huge $g_{2,0}^{(c,c)}(t)$};
		\node [style=none] (62) at (12, 3) {};
		\node [style=none] (63) at (13, 3) {\Huge $g_{2,1}^{(c,c)}(t)$};
	\end{pgfonlayer}
	     \begin{pgfonlayer}{edgelayer}
          \draw [bend left=45,line width=\lw] (8.center) to (10.center);
    		  \draw [bend left=45,line width=\lw] (11.center) to (8.center);
      		\draw [bend left=45,line width=\lw] (10.center) to (9.center);
      		\draw [bend left=45,line width=\lw] (9.center) to (11.center);
      		\draw [bend left=45,line width=\lw] (16.center) to (18.center);
      		\draw [bend left=45,line width=\lw] (19.center) to (16.center);
      		\draw [bend left=45,line width=\lw] (18.center) to (17.center);
      		\draw [bend left=45,line width=\lw] (17.center) to (19.center);
      		\draw [bend left=45,line width=\lw] (20.center) to (22.center);
      		\draw [bend left=45,line width=\lw] (23.center) to (20.center);
      		\draw [bend left=45,line width=\lw] (22.center) to (21.center);
      		\draw [bend left=45,line width=\lw] (21.center) to (23.center);
      		\draw [line width=\lw]              (11.center) to (17.center);
      		\draw [line width=\lw]              (11.center) to (20.center);
    	\end{pgfonlayer}
  \end{tikzpicture}\\
    (a)
  \end{minipage}
  \begin{minipage}{0.5\textwidth}
    \centering
    \begin{tikzpicture}[thick,scale=\scale, every node/.style={transform shape}]
	     \begin{pgfonlayer}{nodelayer}
		\node [style=none] (8) at (6, 7) {};
		\node [style=none] (9) at (10, 7) {};
		\node [style=none] (10) at (8, 9) {};
		\node [style=none] (11) at (8, 5) {};
		\node [style=none] (16) at (-2, 0) {};
		\node [style=none] (17) at (2, 0) {};
		\node [style=none] (18) at (0, 2) {};
		\node [style=none] (19) at (0, -2) {};
		\node [style=none] (20) at (14, 0) {};
		\node [style=none] (21) at (18, 0) {};
		\node [style=none] (22) at (16, 2) {};
		\node [style=none] (23) at (16, -2) {};
		\node [style=none] (26) at (8, 7) {\Huge $\omega_{2}^{(R)}$};
		\node [style=none] (29) at (0, 0) {\Huge $\omega_{0}^{(q)}(t)$};
		\node [style=none] (30) at (16, 0) {};
		\node [style=none] (32) at (16, 0) {\Huge $\omega_{1}^{(q)}(t)$};
		\node [style=none] (46) at (8, 10) {};
		\node [style=none] (47) at (8, 10) {\Huge \textbf{Resonator} $i=2$};
		\node [style=none] (49) at (0, -3) {\Huge \textbf{Flux-tunable transmon} $i=0$};
		\node [style=none] (51) at (16, -3) {\Huge \textbf{Flux-tunable transmon} $i=1$};
		\node [style=none] (52) at (8, 12) {};
		\node [style=none] (53) at (8, 12) {\Huge \textbf{\underline{Architecture II}}};
		\node [style=none] (54) at (8, -5) {};
		\node [style=none] (60) at (4, 3) {};
		\node [style=none] (61) at (3, 3) {\Huge $\bar{g}_{2,0}^{(a,c)}(t)$};
		\node [style=none] (62) at (12, 3) {};
		\node [style=none] (63) at (13, 3) {\Huge $\bar{g}_{2,1}^{(a,c)}(t)$};
	\end{pgfonlayer}
	     \begin{pgfonlayer}{edgelayer}
		\draw [bend left=45,line width=\lw] (8.center) to (10.center);
		\draw [bend left=45,line width=\lw] (11.center) to (8.center);
		\draw [bend left=45,line width=\lw] (10.center) to (9.center);
		\draw [bend left=45,line width=\lw] (9.center) to (11.center);
		\draw [bend left=45,line width=\lw] (16.center) to (18.center);
		\draw [bend left=45,line width=\lw] (19.center) to (16.center);
		\draw [bend left=45,line width=\lw] (18.center) to (17.center);
		\draw [bend left=45,line width=\lw] (17.center) to (19.center);
		\draw [bend left=45,line width=\lw] (20.center) to (22.center);
		\draw [bend left=45,line width=\lw] (23.center) to (20.center);
		\draw [bend left=45,line width=\lw] (22.center) to (21.center);
		\draw [bend left=45,line width=\lw] (21.center) to (23.center);
		\draw [line width=\lw]              (11.center) to (17.center);
		\draw [line width=\lw]              (11.center) to (20.center);
	\end{pgfonlayer}
    \end{tikzpicture}\\
    (b)
  \end{minipage}
  \caption[Sketches of two different device architectures I(a) and II(b) to realise NIGQCs, see \secref{sec:FromStaticsToDynamics}.]{Sketches of two different device architectures I(a) and II(b) to realise NIGQCs, see \secref{sec:FromStaticsToDynamics}. \PANC{a} shows architecture I. Here we couple two fixed-frequency transmons characterised by the qubit frequencies $\omega_{0}^{(q_{0})}$ and $\omega_{1}^{(q_{0})}$ to a flux-tunable transmon characterised by the tunable frequency $\omega_{2}^{(q)}(t)$. We use the device parameters listed in \tabref{tab:device_parameter_flux_tunable_coupler_chip} and \tabref{tab:device_parameter_flux_tunable_coupler_chip_effective} for our simulations of this system. \PANC{b} shows architecture II. Here we couple two flux-tunable transmons characterised by the qubit frequencies $\omega_{0}^{(q)}(t)$ and $\omega_{1}^{(q)}(t)$ to a resonator characterised by the resonance frequency $\omega_{2}^{(R)}$. We use the device parameters listed in \tabref{tab:device_parameter_resonator_coupler_chip} and \tabref{tab:device_parameter_resonator_coupler_chip_effective} for our simulations of this system. The dynamics is determined by solving the TDSE for the circuit Hamiltonian \equref{eq:CHM} and the effective Hamiltonian \equref{eq:EHM}.}\label{fig:arch_sketch}
\end{figure}

Figures \ref{fig:arch_sketch}(a-b) show sketches of two different transmon systems or device architectures which potentially allow us to realise NIGQCs, see \secref{sec:FromStaticsToDynamics}.

Architecture I consists of two fixed-frequency transmons $\omega_{0}^{(q_{0})}$ and $\omega_{1}^{(q_{0})}$ coupled to a flux-tunable transmon $\omega_{2}^{(q)}(t)$. The flux-tunable transmon functions as a coupler element. We use the device parameters listed in \tabref{tab:device_parameter_flux_tunable_coupler_chip} and \tabref{tab:device_parameter_flux_tunable_coupler_chip_effective} for our simulations of this system. This architecture is subject of several experimental and theoretical studies, see \REFS\cite{McKay16,Roth19,Ganzhorn20,Gu21,Blais2020circuit}.

Architecture II consists of two flux-tunable transmons $\omega_{0}^{(q)}(t)$ and $\omega_{1}^{(q)}(t)$ coupled to a coupling resonator $\omega_{2}^{(R)}$. We use the device parameters listed in \tabref{tab:device_parameter_resonator_coupler_chip} and \tabref{tab:device_parameter_resonator_coupler_chip_effective} for our simulations of this architecture. This device type is discussed in \REFS\cite{Rol19,Krinner2020,Lacroix2020,Blais2020circuit}.

The device parameters in \tabsref{tab:device_parameter_flux_tunable_coupler_chip}{tab:device_parameter_resonator_coupler_chip_effective} are used to fully specify the circuit Hamiltonian \equref{eq:CHM} and the effective Hamiltonian \equref{eq:EHM}. These are the Hamiltonians we use to model the NIGQCs. One can use both architectures to implement \ISWAP{} and \CZ{} gate transitions with an external flux $\FP$,\ie these transitions can potentially be used to implement the corresponding gate matrices we discuss in \chapref{chap:I}.

\begin{table}[!tbp]
\caption[Device parameters for the circuit Hamiltonian (Architecture I). ]{\label{tab:device_parameter_flux_tunable_coupler_chip} Device parameters for the circuit Hamiltonian \equref{eq:CHM}. We use these parameters to model a device of type architecture I. The first column shows the subsystem indices $i$. The second and third columns show the transmon qubit frequencies $\omega_{i}^{(Q_{0})}$ and anharmonicities $\alpha_{i}^{(Q_{0})}$, respectively. The fourth column shows the capacitive energies $E_{C_{i}}$. The fifth and sixth columns show the left $E_{J_{l,i}}$ and right $E_{J_{r,i}}$ Josephson energies. The seventh column shows the flux-offset values $\varphi_{0,i}$. The eighth column shows the interaction strength constants $G_{2,i}^{(1)}$. Here we model dipole-dipole interactions between fixed-frequency and flux-tunable transmons. We show all parameters except the ones in the first and seventh column in units of GHz. These parameters are motivated by experiments discussed in \REF\cite{Ganzhorn20}.}
\centering
{\small
\setlength{\tabcolsep}{4pt}
\begin{tabularx}{\textwidth}{ X X X X X X X X X }
\hline\hline
$i$ & $\omega_{i}^{(Q_{0})}/2\pi$ & $\alpha_{i}^{(Q_{0})}/2\pi$ & $E_{C_{i}}/2\pi$ & $E_{J_{l,i}}/2\pi$ & $E_{J_{r,i}}/2\pi$ & $\varphi_{0,i}/2\pi$ & $G_{2,i}^{(1)}/2\pi$\\
\hline
0 & 5.100 & -0.310 & 1.079 & 13.456  & n/a & n/a & 0.085\\
1 & 6.200 & -0.285 & 1.027 & 20.371  & n/a & n/a & 0.085\\
2 & 8.100 & -0.235 & 0.880 & 17.897  & 21.486 & 0.15 & n/a\\
\hline\hline
\end{tabularx}
}
\caption[Device parameters for the circuit Hamiltonian (Architecture II). ]{\label{tab:device_parameter_resonator_coupler_chip} Device parameters for the circuit Hamiltonian \equref{eq:CHM}. We use these parameters to model a device of type architecture II. The first column shows the subsystem indices $i$. The second and third columns show the transmon qubit frequencies $\omega_{i}^{(Q_{0})}$ and anharmonicities $\alpha_{i}^{(Q_{0})}$, respectively. The fourth column shows the capacitive energies $E_{C_{i}}$. The fifth and sixth columns show the left $E_{J_{l,i}}$ and right $E_{J_{r,i}}$ Josephson energies. The seventh column shows the flux-offset values $\varphi_{0,i}$. The eighth column shows the interaction strength constants $G_{2,i}^{(4)}$. Here we model dipole-dipole interactions between flux-tunable transmons and resonators. We show all parameters except the ones in the first and eighth column in units of GHz. These device parameters are motivated by experiments discussed in \REF\cite{Lacroix2020}.}
\centering
{\small
\setlength{\tabcolsep}{4pt}
\begin{tabularx}{\textwidth}{ X X X X X X X X X }
\hline\hline
$i$ & $\omega_{i}^{(R)}/2\pi$ & $\omega_{i}^{(Q_{0})}/2\pi$ & $\alpha_{i}^{(Q_{0})}/2\pi$ & $E_{C_{i}}/2\pi$ & $E_{J_{l,i}}/2\pi$ & $E_{J_{r,i}}/2\pi$ & $\varphi_{0,i}/2\pi$ & $G_{2,i}^{(4)}/2\pi$\\
\hline
0 & n/a & 4.200& -0.320 & 1.068 & 3.140  & 9.419 & 0 & 0.300\\
1 & n/a & 5.200& -0.295 & 1.036 & 4.817  & 9.633 & 0 & 0.300\\
2 & 45.000 & n/a & n/a & n/a & n/a & n/a & n/a & n/a\\
\hline\hline
\end{tabularx}
}
\end{table}

\begin{table}[!tbp]

\caption[Device parameters for the adiabatic effective Hamiltonian \equref{eq:EHM} where the interaction strength is constant.]{\label{tab:device_parameter_flux_tunable_coupler_chip_effective} Device parameters for the adiabatic effective Hamiltonian \equref{eq:EHM} where the interaction strength is constant. We use these parameters to model a device of type architecture I, see also \tabref{tab:device_parameter_flux_tunable_coupler_chip} for the corresponding circuit Hamiltonian device parameters. The first column shows the subsystem indices $i$. The second and third columns show the transmon qubit frequencies $\omega_{i}^{(q_{0})}$ and anharmonicities $\alpha_{i}^{(q_{0})}$, respectively. The fourth column shows the flux-offset values $\varphi_{0,i}$. The fifth column shows the effective interaction strength $g_{2,i}^{(c,c)}(\varphi_{0,2})$ at the operating point $\varphi_{0,2}$. Here we model dipole-dipole interactions between fixed-frequency and flux-tunable transmons. All device parameters except the ones in the first and fourth columns are given in units of GHz.}
\centering
{\small
\setlength{\tabcolsep}{4pt}
\begin{tabularx}{\textwidth}{ X X X X X }
\hline\hline
$i$ & $\omega_{i}^{(q_{0})}/2 \pi$&$\alpha_{i}^{(q_{0})}/2\pi$ & $\varphi_{0,i}/2\pi$ & $ g_{2,i}^{(c,c)}(\varphi_{0,2})/2\pi$\\
\hline
  0 & 5.100 & -0.310 &  0 & 0.146\\
  1 & 6.200 & -0.285 &  0 & 0.164\\
  2 & 8.100 & -0.235 &  0.15 & n/a\\
\hline\hline
\end{tabularx}
}

\caption[Device parameters for the adiabatic effective Hamiltonian \equref{eq:EHM} where the interaction strength is constant.]{\label{tab:device_parameter_resonator_coupler_chip_effective} Device parameters for the adiabatic effective Hamiltonian \equref{eq:EHM} where the interaction strength is constant. We use these parameters to model a device of architecture II, see also \tabref{tab:device_parameter_resonator_coupler_chip} for the corresponding circuit Hamiltonian device parameters. The first column shows the subsystem indices $i$. The second column shows the resonator frequencies $\omega_{i}^{(R)}$. The third and fourth columns show the transmon qubit frequencies $\omega_{i}^{(q_{0})}$ and anharmonicities $\alpha_{i}^{(q_{0})}$, respectively. The fifth column shows the flux-offset values $\varphi_{0,i}$. The sixth column shows the effective interaction strength $g_{2,i}^{(a,c)}(\varphi_{0,i})$ at the operating points $\varphi_{0,i}$. Here we model dipole-dipole interactions between flux-tunable transmons and resonators. All device parameters except the ones in the first and fifth columns are given in units of GHz.}
\centering
{\small
\setlength{\tabcolsep}{4pt}
\begin{tabularx}{\textwidth}{ X X X X X X }
\hline\hline
$i$ & $\omega_{i}^{(R)}/2 \pi$ &$\omega_{i}^{(q_{0})}/2 \pi$&$\alpha_{i}^{(q_{0})}/2\pi$ & $\varphi_{0,i}/2\pi$ & $ g_{2,i}^{(a,c)}(\varphi_{0,i})/2\pi$\\
\hline
  0 & n/a & 4.200& -0.320 & 0 &0.307\\
  1 & n/a & 5.200& -0.295 & 0 &0.344\\
  2 & 45.000 & n/a & n/a & n/a & n/a\\
\hline\hline
\end{tabularx}
}
\end{table}
For our simulations we use the external flux (control pulse)
\begin{equation}\label{eq:NA_control_pulse}
\varphi(t)=\varphi_{0}+\PA e(t) \cos(\DF t),
\end{equation}
where the real-valued pulse parameters $\varphi_{0}$, $\PA$ and $\DF$ denote the flux offset, the pulse amplitude and the drive frequency, respectively. Note that the external flux $\varphi(t)$ and consequently the flux offset $\varphi_{0}$ and the pulse amplitude $\PA$ are without units. Furthermore, the Hamiltonian \equref{eq:flux-tunable transmon recast} is $2\pi$ periodic. Therefore, in this chapter we give the external flux $\varphi(t)$ and the parameters $\varphi_{0}$, $\PA$ always in units of $2\pi$. The real-valued envelope function $e(t)$ is defined as
\begin{equation}
e(t) = \begin{cases}
\sin(\lambda t) &\text{if $0 \leq t < \TRF$}\\
1     &\text{if $\TRF \leq t \leq \Delta T $}\\
\sin(\frac{\pi}{2}+\lambda (t-\Delta T)) &\text{if $ \Delta T < t \leq \TD $,}
\end{cases}
\end{equation}
where the real-valued pulse parameters $\TRF$ and $\TD$ denote the rise and fall time and the pulse duration, respectively. Furthermore, $\lambda=\pi/2\TRF$ and $\Delta T=\TD-\TRF$.

\graphicspath{{./FiguresAndData/NAPaper/EffIntAndPulse/}}
\begin{figure}[!tbp]
    \centering
    \begin{minipage}{0.49\textwidth}
        \centering
        \includegraphics[width=\width\textwidth]{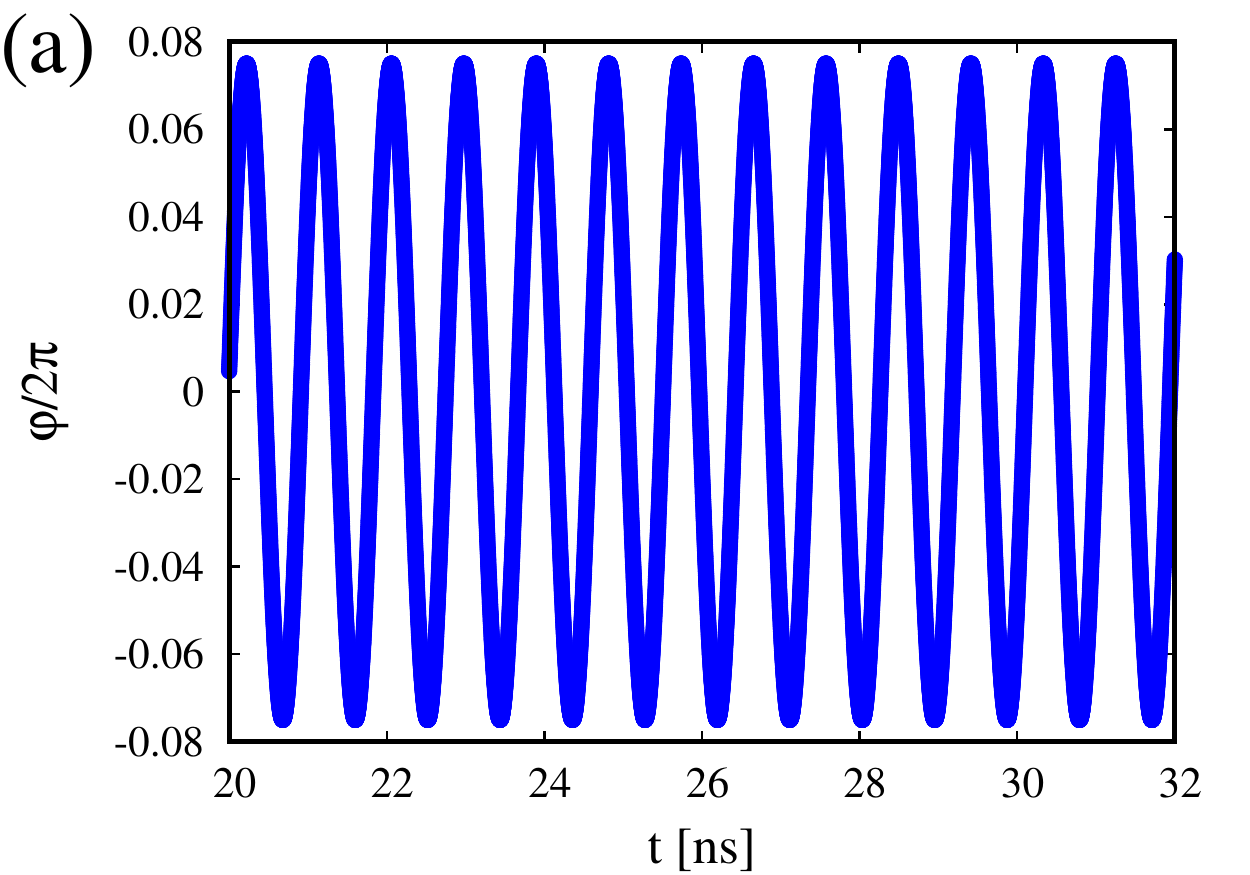}
    \end{minipage}\hfill
    \begin{minipage}{0.49\textwidth}
        \centering
        \includegraphics[width=\width\textwidth]{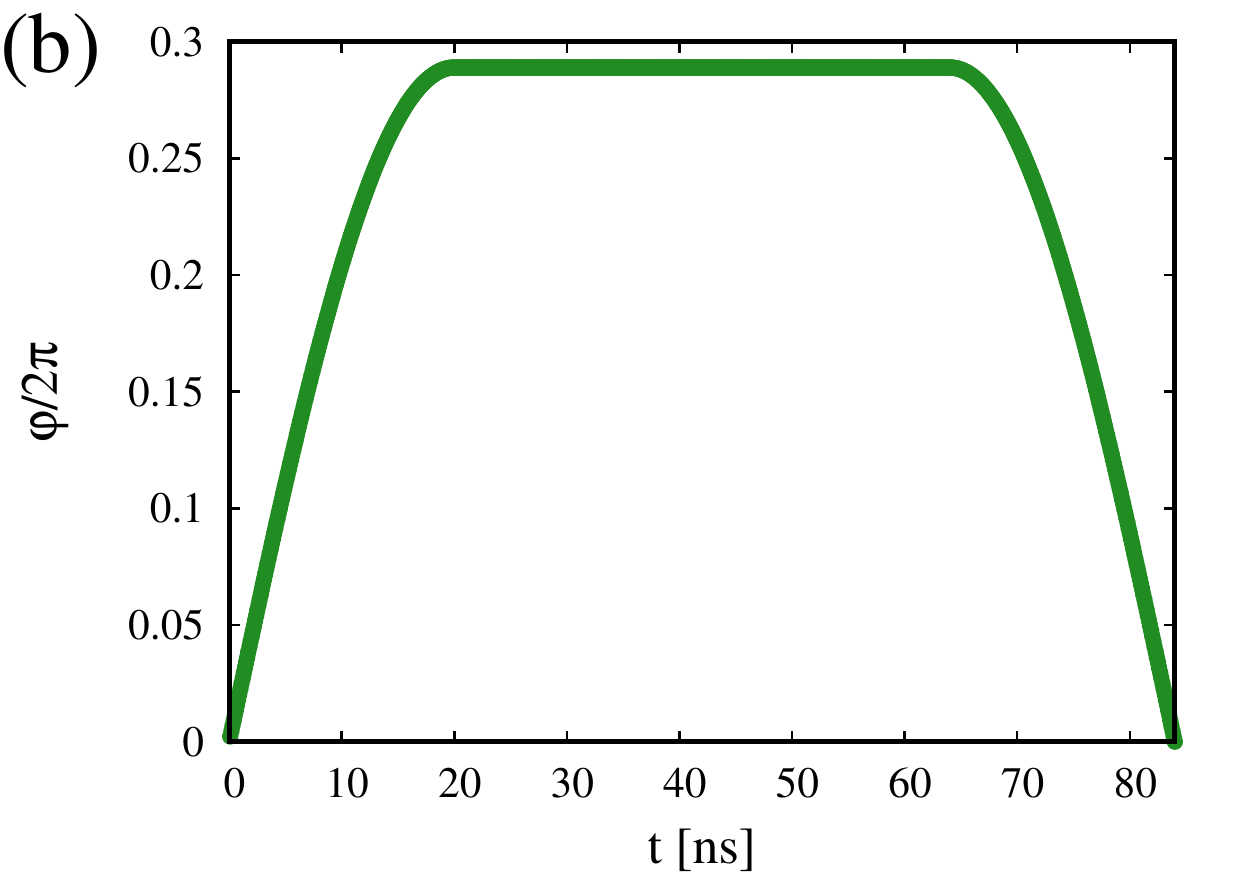}
    \end{minipage}
    \caption[Fluxes $\varphi/2\pi$ given by \equref{eq:NA_control_pulse} as a functions of time $t$ for two sets of pulse parameters. ]{Fluxes $\varphi/2\pi$ given by \equref{eq:NA_control_pulse} as a functions of time $t$ for two sets of pulse parameters. \PANC{a} shows a microwave pulse. Here we use the parameters $\DFTP=1.089$ GHz, $\PATP=0.075$, $\TRF=13$ ns and $\TD=205.4$ ns to obtain the results. The parameters are also listed in \tabref{tab:summary_effective_hamiltonian_results} row six. \PANC{b} shows a unimodal pulse. Here we use the parameters $\DFTP=0$ GHz, $\PATP=0.297$, $\TRF=20.0$ ns and $\TD=84.0$ ns to acquire the results. The parameters are also listed in \tabref{tab:summary_effective_hamiltonian_results} row nine.}\label{fig:NA_pulse_time_evo}
\end{figure}
Figures \ref{fig:NA_pulse_time_evo}(a-b) show the external flux $\FPTP$ as functions of time $t$ for two different sets of pulse parameters. In \PANL{a} we model a microwave pulse with the pulse amplitude $\PATP=0.075$, the drive frequency $\DFTP=1.089$ GHz, the rise and fall time $\TRF=13$ ns and the pulse duration $\TD=205.4$ ns. This microwave pulse can be used to activate \ISWAP{} transitions in architecture I if we drive the tunable coupler, see \figref{fig:arch_sketch}(a). In \PANL{b} we model a unimodal pulse with the pulse amplitude $\PATP=0.297$, the drive frequency $\DFTP=0$ GHz, the rise and fall time $\TRF=20$ ns and the pulse duration $\TD=84$ ns. This unimodal pulse can be used to activate \ISWAP{} transitions in architecture II if we drive the flux-tunable transmon $\omega_{1}^{(q)}(t)$, see \figref{fig:arch_sketch}(b).

Table \ref{tab:summary_circuit_hamiltonian_results} contains a summary of the pulse parameters we use to model transitions with the circuit Hamiltonian \equref{eq:CHM}. Similarly, in \tabref{tab:summary_effective_hamiltonian_results} we summarise the pulse parameters we use to model various transitions with the adiabatic effective Hamiltonian \equref{eq:EHM}.

The overall objective of this chapter is to analyse how well the circuit and the effective Hamiltonian models agree with one another if we model architectures I and II, see \figref{fig:arch_sketch}(a-b). To this end, we implemented two simulation codes. One allows us to simulate the effective Hamiltonian model and the other allows to do the same with the circuit Hamiltonian model, see \chapref{chap:IV}. The simulation code for the former model is equipped with the option to turn various approximations,\ie simplifications, on and off.

To the best knowledge of the author, we often find that flux-tunable transmons are modelled adiabatically, see \REFS\cite{McKay16,Roth19,Gu21,Yan18,Baker22}. This means that the non-adiabatic driving term
\begin{equation}\label{eq:drive_term_ftt_second_time}
    \DTF= - i \sqrt{\frac{\xi(t)}{2}} \dot{\varphi}_{\text{eff.}}(t) \BRR{\OP{c}^{\dagger}- \OP{c}} + \frac{i}{4} \frac{\dot{\xi}(t)}{\xi(t)} \BRR{\OP{c}^{\dagger} \OP{c}^{\dagger} - \OP{c} \OP{c}},
\end{equation}
first introduced in \equref{eq:drive_term_ftt} is neglected completely, see also \equref{eq:EHM}. We can control this assumption in the simulation code.

Furthermore, we can find instances where the interaction strength between flux-tunable transmons and other circuit elements is modelled as time independent, see \REFS\cite{Ganzhorn20,McKay16,Roth19,Gu21}. We model the interaction strength between a flux-tunable transmon $i$ and a fixed-frequency transmon $j$ with the function
\begin{equation}\label{eq:eff_int_trans_trans_second_time}
  g_{i,j}^{(c, c)}(t)=G_{i,j} \sqrt[4]{\frac{\EJ{i}}{8 \EC{i}}} \sqrt[4]{ \frac{E_{J_{j}}}{8 E_{C_{j}}} },
\end{equation}
where $G_{i,j}$ is a constant real-valued parameter, $\EJ{j}$ and $E_{J_{i}}$ denote Josephson energies and $\EC{j}$ and $E_{C_{i}}$ refer to capacitive energies. Similarly, we model the interaction strength between a resonator element $i$ and a flux-tunable transmon $j$ with the function
\begin{equation}\label{eq:eff_int_trans_res_second_time}
  g_{i,j}^{(a, c)}(t)=G_{i,j} \sqrt[4]{ \frac{\EJ{j}}{8 \EC{j}} }.
\end{equation}
This interaction strength model is taken from \REF\cite{Koch} and explained in more detail in \secref{sec:TheQuantumComputerEffectiveHamiltonianModel}. We can control the time dependence in the simulation code. Additionally, we can also control the accuracy of the spectrum of a single flux-tunable transmon, see \secref{sec:NA_single_flux_tunable_transmon} for more details.

Although, we are mainly interested in architecture I and II, it seems plausible to first investigate how well the effective models in \equaref{eq:fft_eff_II}{eq:tunable-frequency eff} for a single flux-tunable transmon can cover the dynamics of the corresponding circuit Hamiltonian \equref{eq:flux-tunable transmon recast}. This question is the subject of the next section.

\begin{table}[!tbp]
\caption[Summary of the pulse parameters we use to perform simulations with the circuit Hamiltonian model.]{\label{tab:summary_circuit_hamiltonian_results} Summary of the pulse parameters we use to perform simulations with the circuit Hamiltonian model, see \equref{eq:CHM} and \secref{sec:TheQuantumComputerCircuitHamiltonianModel}. Here, we use a pulse of the form \equref{eq:NA_control_pulse} to drive different systems. The first column contains table references. The corresponding tables contain the device parameters which specify the Hamiltonian. The second column shows figure references which refer the reader to the simulation results. The third column shows different gate types. We can potentially use the corresponding pulses to model these gates, assuming we can arrange the phases of the state vectors accordingly. If we cannot assign a gate to the transition, we denote the gate as not applicable (n/a).  The fourth column shows the two states which exchange population due to the driving terms. The fifth column contains the drive frequencies $\DF$ in units of GHz. The sixth column shows the pulse amplitudes $\PA$. The seventh and eighth columns show the rise and fall time $\TRF$ and the pulse duration $\TD$ in units of ns, respectively. The ninth column shows the minimum number of transmon basis states $n_{J}$ we recommend for modelling the dynamics of the system in the transmon basis.}
\centering
{\footnotesize
\setlength{\tabcolsep}{3pt}
\begin{tabularx}{\textwidth}{ c c c c c c c c c  }
\hline\hline
Parameters &  Figure & Gate &  States $z$  & $\DFTP$ & $\PATP$ & $\TRF$ & $\TD$ & $n_{J}$  \\
\hline

\tabref{tab:device_parameter_flux_tunable_coupler_chip} & \figref{fig:NA_RESMAPS_CUTS}(a) & $X$ &   $\{(0),(1)\}$                     & $7.636$ & $0.001$ & $100.0$ & $200.0$ & $3$ \\

\tabref{tab:device_parameter_flux_tunable_coupler_chip} & \figref{fig:NA_supressed_chevron_pattern}(a)  & $X$ &  $\{(0,0,0),(0,1,0)\}$             &$6.183$ & $0.045$  & $22.5$ & $45.0$ & $3$\\

\tabref{tab:device_parameter_flux_tunable_coupler_chip} & \figref{fig:NA_supressed_chevron_pattern}(b)  & $X$ &  $\{(0,0,0),(0,0,1)\}$             &$5.092$ & $0.085$  & $25.0$ & $50.0$ & $3$\\

\tabref{tab:device_parameter_flux_tunable_coupler_chip} & \figref{fig:NA_cir_ISWAP_cases_chalmers}(a-d)  & $\ISWAP{}$ &   $\{(0,1,0),(0,0,1)\}$ &$1.089$ & $0.075$ & $13.0$ & $209.40$  & $6$  \\

\tabref{tab:device_parameter_flux_tunable_coupler_chip} & \figref{fig:NA_cir_cz_cases_chalmers}(a-d)  & $\CZ{}$ &  $\{(0,1,1),(0,2,0)\}$     & $0.809$ & $0.085$ & $13.0$ & $297.55$ & $8$\\

\tabref{tab:device_parameter_resonator_coupler_chip}    & \figref{fig:NA_supressed_AII}(a-c)  & n/a &   $\{(0,0,0),(1,0,0)\}$ & $45.00$     & $0.020$ & $150.0$ & $300.0$ & $4$\\

\tabref{tab:device_parameter_resonator_coupler_chip}    & \figref{fig:NA_supressed_AII}(a-c)  & n/a &   $\{(0,0,0),(2,0,0)\}$    & $45.00$     & $0.020$ & $150.0$ & $300.0$ & $4$\\

\tabref{tab:device_parameter_resonator_coupler_chip}    & \figref{fig:NA_cir_ISWAP_cases_eth}(a-d)  & $\ISWAP{}$ &   $\{(0,1,0),(0,0,1)\}$ & $0$     & $0.289$ & $20.0$ & $100.0$ & $14$\\

\tabref{tab:device_parameter_resonator_coupler_chip}    & \figref{fig:NA_cir_cz_cases_eth}(a-d)  & $\CZ{}$ &   $\{(0,1,1),(0,0,2)\}$    & $0$     & $0.334$ & $20.0$ & $125.0$ & $16$\\
\hline\hline
\end{tabularx}}
\end{table}

\begin{table}[!tbp]
\caption[Summary of pulse parameters we use to perform simulations with the adiabatic effective Hamiltonian model.]{\label{tab:summary_effective_hamiltonian_results} Summary of pulse parameters we use to perform simulations with the adiabatic effective Hamiltonian model, see \equref{eq:EHM} and \secref{sec:TheQuantumComputerEffectiveHamiltonianModel}. Here, we use a pulse of the form \equref{eq:NA_control_pulse} to drive different systems. The first column contains table references. The corresponding tables contain the device parameters which specify the Hamiltonian. The second column shows figure references which refer the reader to the simulation results. The third column shows different gate types. We can potentially use the corresponding pulses to model these gates, assuming we can arrange the phases of the state vectors accordingly. The fourth column shows the two states which exchange population due to the driving terms. The fifth column contains the drive frequencies $\DF$ in units of GHz. The sixth column shows the pulse amplitudes $\PA$. The seventh and eighth columns show the rise and fall time $\TRF$ and the pulse duration $\TD$ in units of ns, respectively. If we cannot model the transition with the adiabatic effective Hamiltonian, we denote the parameters and gates as not applicable (n/a). The ninth column shows the different cases we model with the effective model Hamiltonian. In case A we model the system with a time-independent interaction strength and a non-adjusted spectrum. In case B we model the system with a time-dependent interaction strength and a non-adjusted spectrum. In case C we model the system with a time-dependent interaction strength and an adjusted spectrum.}
\centering
{\footnotesize
\setlength{\tabcolsep}{3pt}
\begin{tabularx}{\textwidth}{ c c c c c c c c c }
\hline\hline
Parameters & Figure & Gate & States $z$ & $\DFTP$ & $\PATP$ & $\TRF$ & $\TD$ & Case  \\

\hline

\tabref{tab:device_parameter_flux_tunable_coupler_chip_effective} & n/a &n/a            & $\{(0),(1)\}$           & n/a & n/a & n/a & n/a &  n/a  \\

\tabref{tab:device_parameter_flux_tunable_coupler_chip_effective} & n/a &n/a            & $\{(0,0,0),(0,1,0)\}$   & n/a & n/a  & n/a & n/a &  n/a  \\

\tabref{tab:device_parameter_flux_tunable_coupler_chip_effective} & n/a &n/a            & $\{(0,0,0),(0,0,1)\}$   & n/a & n/a  & n/a & n/a &  n/a  \\

\tabref{tab:device_parameter_flux_tunable_coupler_chip_effective} & \figref{fig:NA_eff_cz_ISWAP_cases_chalmers}(a) &$\ISWAP{}$ &  $\{(0,1,0),(0,0,1)\}$  &$1.088$ & $0.075$ & $13.0$ & $139.6$ & A  \\

\tabref{tab:device_parameter_flux_tunable_coupler_chip_effective} & \figref{fig:NA_eff_cz_ISWAP_cases_chalmers}(b) &$\ISWAP{}$ &  $\{(0,1,0),(0,0,1)\}$  &$1.089$ & $0.075$ & $13.0$ & $205.4$  & B \\

\tabref{tab:device_parameter_flux_tunable_coupler_chip_effective} & \figref{fig:NA_eff_cz_ISWAP_cases_chalmers}(c) &$\CZ{}$    & $\{(0,1,1),(0,2,0)\}$   & $0.807$ & $0.085$ & $13.0$ & $196.5$ &  A\\

\tabref{tab:device_parameter_flux_tunable_coupler_chip_effective} & \figref{fig:NA_eff_cz_ISWAP_cases_chalmers}(d) &$\CZ{}$    & $\{(0,1,1),(0,2,0)\}$   & $0.807$ & $0.085$ & $13.0$ & $272.0$ & B  \\

\tabref{tab:device_parameter_resonator_coupler_chip_effective} & n/a &n/a &             $\{(0,0,0),(1,0,0)\}$   & n/a & n/a & n/a & n/a &  n/a  \\

\tabref{tab:device_parameter_resonator_coupler_chip_effective} & n/a &n/a &             $\{(0,0,0),(2,0,0)\}$   & n/a & n/a & n/a & n/a &  n/a  \\

\tabref{tab:device_parameter_resonator_coupler_chip_effective} & \figref{fig:NA_eff_cz_ISWAP_cases_eth}(a)    &$\ISWAP{}$ & $\{(0,1,0),(0,0,1)\}$   & $0$ & $0.297$ & $20.0$ & $84.0$ &  A  \\

\tabref{tab:device_parameter_resonator_coupler_chip_effective} & \figref{fig:NA_eff_cz_ISWAP_cases_eth}(b)    &$\ISWAP{}$ & $\{(0,1,0),(0,0,1)\}$   & $0$ & $0.289$ & $20.0$ & $96.0$ & C  \\

\tabref{tab:device_parameter_resonator_coupler_chip_effective} & \figref{fig:NA_eff_cz_ISWAP_cases_eth}(c)    &$\CZ{}$    & $\{(0,1,1),(0,0,2)\}$   & $0$ & $0.343$ & $20.0$ & $105.0$ & A  \\

\tabref{tab:device_parameter_resonator_coupler_chip_effective} & \figref{fig:NA_eff_cz_ISWAP_cases_eth}(d)    &$\CZ{}$    & $\{(0,1,1),(0,0,2)\}$   & $0$ & $0.334$ & $20.0$ & $121.0$ & C   \\
\hline\hline
\end{tabularx}}
\end{table}

\section{Simulations of a single flux-tunable transmon}\label{sec:NA_single_flux_tunable_transmon}
In this section, we compare how the effective Hamiltonians \equref{eq:fft_eff_II} and \equref{eq:tunable-frequency eff} and the circuit Hamiltonian \equref{eq:flux-tunable transmon recast} react to different types of pulses $\FP$,\ie external fluxes. All these models are used to describe isolated flux-tunable transmons. We study how the different systems,\ie models, react to a resonant microwave pulse, see \figref{fig:NA_pulse_time_evo}(a) and a unimodal pulse, see \figref{fig:NA_pulse_time_evo}(b). If not stated otherwise, we use the system parameters listed in \tabref{tab:device_parameter_flux_tunable_coupler_chip} row $i=2$, the pulse given by \equref{eq:NA_control_pulse} and three basis states to perform the simulations.

The effective Hamiltonian \equref{eq:tunable-frequency eff} is so simple that we can solve the TDSE for an arbitrary pulse $\FP$. If we assume that the initial state of the system is given by
\begin{equation}
  \ket{\Psi(t_{0})}= \sum_{z \in \mathbb{N}^{0}} c_{z}(t_{0}) \ket{\psi^{(z)}(t_{0})},
\end{equation}
where $c_{z}(t) \in \mathbb{C}$ for all times $t$ and $\ket{\psi^{(z)}(t)}$ denotes the instantaneous harmonic basis states, we find that the state vector at time $t$ is given by the expression
\begin{equation}\label{eq:NA_eff_ham_solution_tdse}
  \ket{\Psi(t)}= \sum_{z \in \mathbb{N}^{0}} e^{-i \int_{t_{0}}^{t} E^{(z)}(t^{\prime}) dt^{\prime}}c_{z}(t_{0}) \ket{\psi^{(z)}(t)}.
\end{equation}
Consequently, the probability amplitudes $|\braket{z|\Psi(t)}|^{2}$ do not change over time, no matter how we modulate the energies $E^{(z)}(t)$.

\renewcommand{\width}{1.0}
\graphicspath{{./FiguresAndData/NAPaper/CircuitHamiltonianGaugeSimulations/SingleFTT/}{./FiguresAndData/NAPaper/EffectiveHamiltonianGaugeSimulations/SingleFTT/ResMap/}}
\begin{figure}[!tbp]
    \centering
    \begin{minipage}{0.49\textwidth}
        \centering
        \includegraphics[width=\width\textwidth]{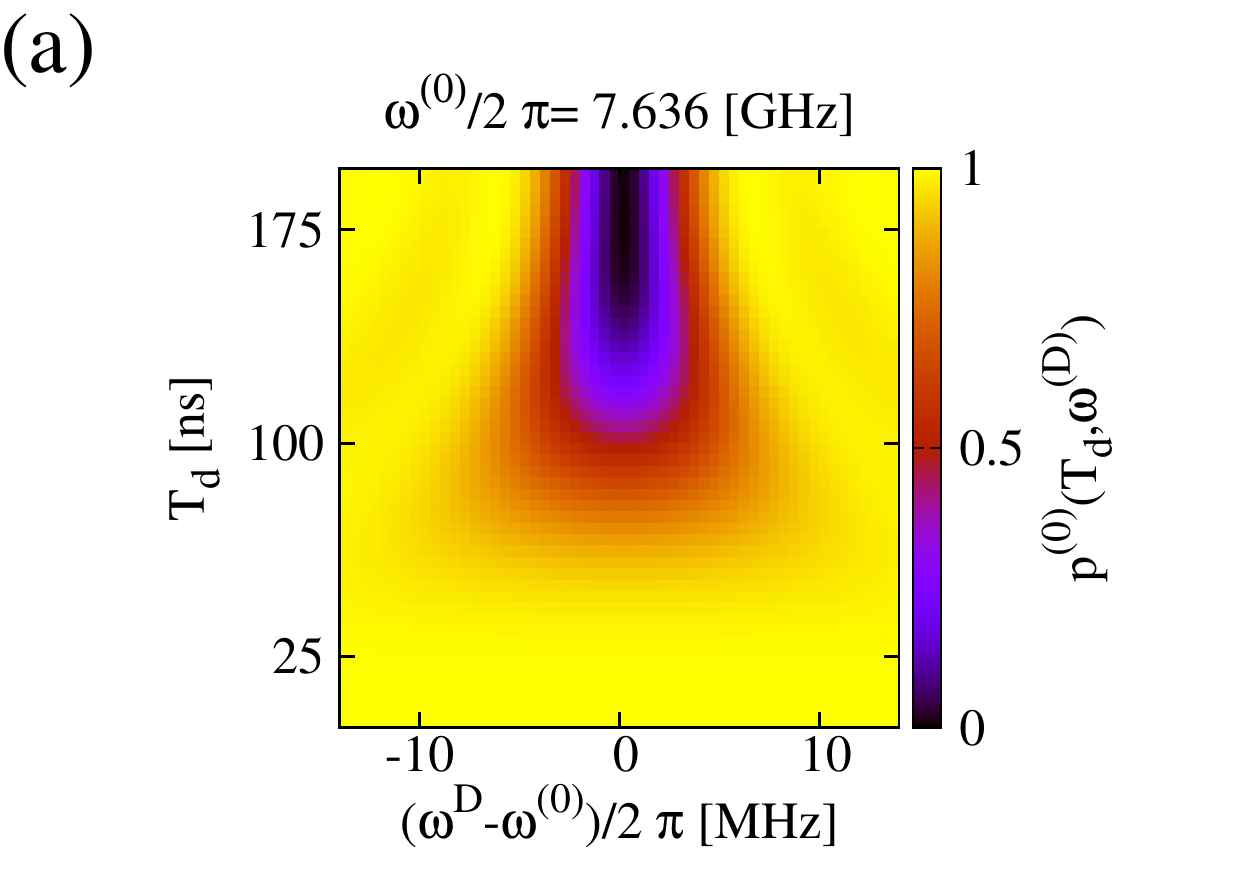}
    \end{minipage}\hfill
    \begin{minipage}{0.49\textwidth}
        \centering
        \includegraphics[width=\width\textwidth]{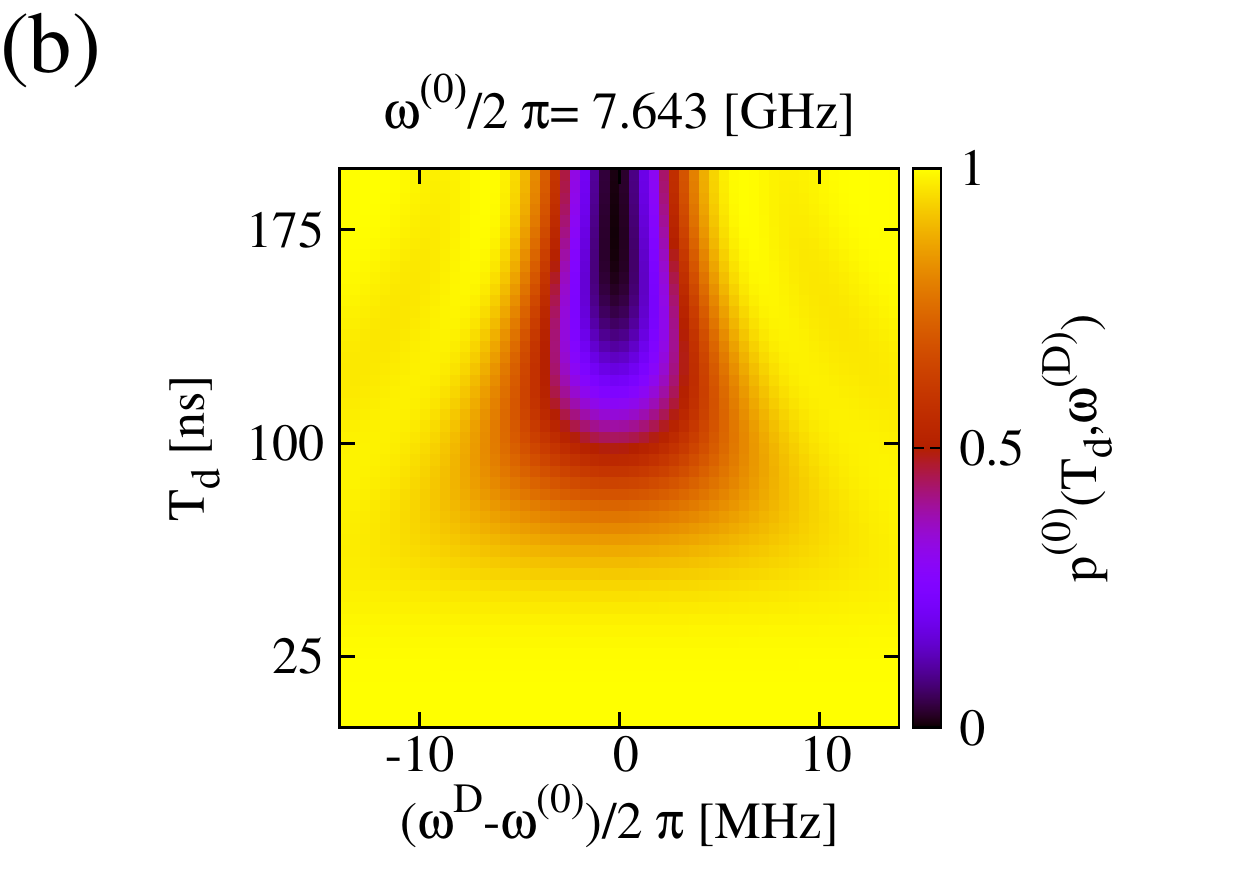}
    \end{minipage}
    \caption[Ground-state probabilities $p^{(0)}$ as functions of the pulse duration $\TD$ and the drive frequency $\DF$.]{Ground-state probabilities $p^{(0)}$ as functions of the pulse duration $\TD$ and the drive frequency $\DF$. We use the system parameters listed in \tabref{tab:device_parameter_flux_tunable_coupler_chip}, row $i=2$ and the pulse in \equref{eq:NA_control_pulse} with $\TRF=\TD/2$ and $\PATP=0.001$ to obtain the results. Here, we model a microwave pulse, see \figref{fig:NA_pulse_time_evo}(a). In \PANL{a} we solve the TDSE for the circuit Hamiltonian in \equref{eq:flux-tunable transmon recast}. Similarly, in \PANL{b} solve the TDSE for the effective Hamiltonian in \equref{eq:fft_eff_II}. At time $t=0$ we initialise the systems in the corresponding ground states. We center the two chevron patterns around the transition frequencies $\omega^{(0)}=7.636$ GHz(a) and $\omega^{(0)}=7.643$ GHz(b).}\label{fig:NA_RESMAPS}
\end{figure}

\graphicspath{{./FiguresAndData/NAPaper/CompareEffectiveCircuitHamiltonian/}}
\begin{figure}[!tbp]
    \centering
    \begin{minipage}{0.49\textwidth}
        \centering
        \includegraphics[width=\width\textwidth]{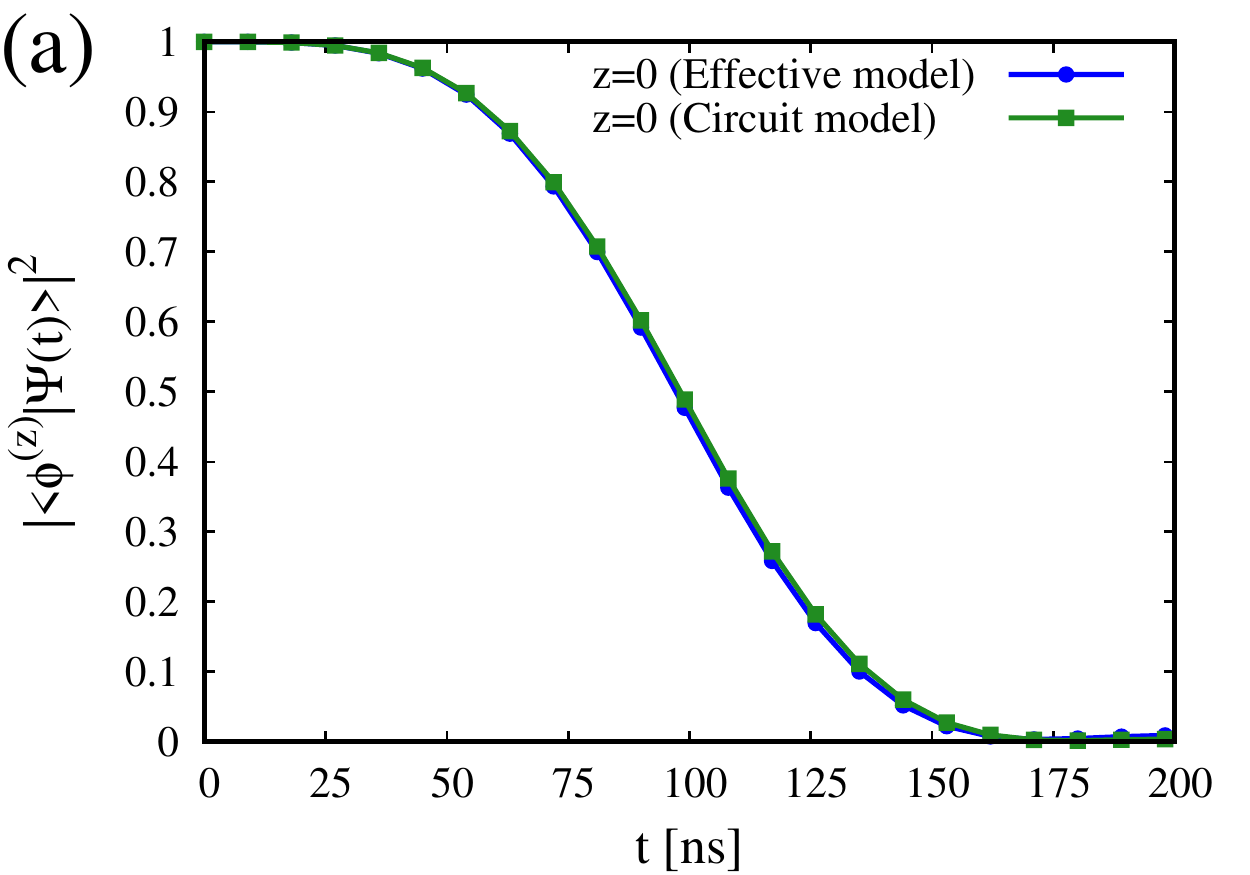}
    \end{minipage}\hfill
    \begin{minipage}{0.49\textwidth}
        \centering
        \includegraphics[width=\width\textwidth]{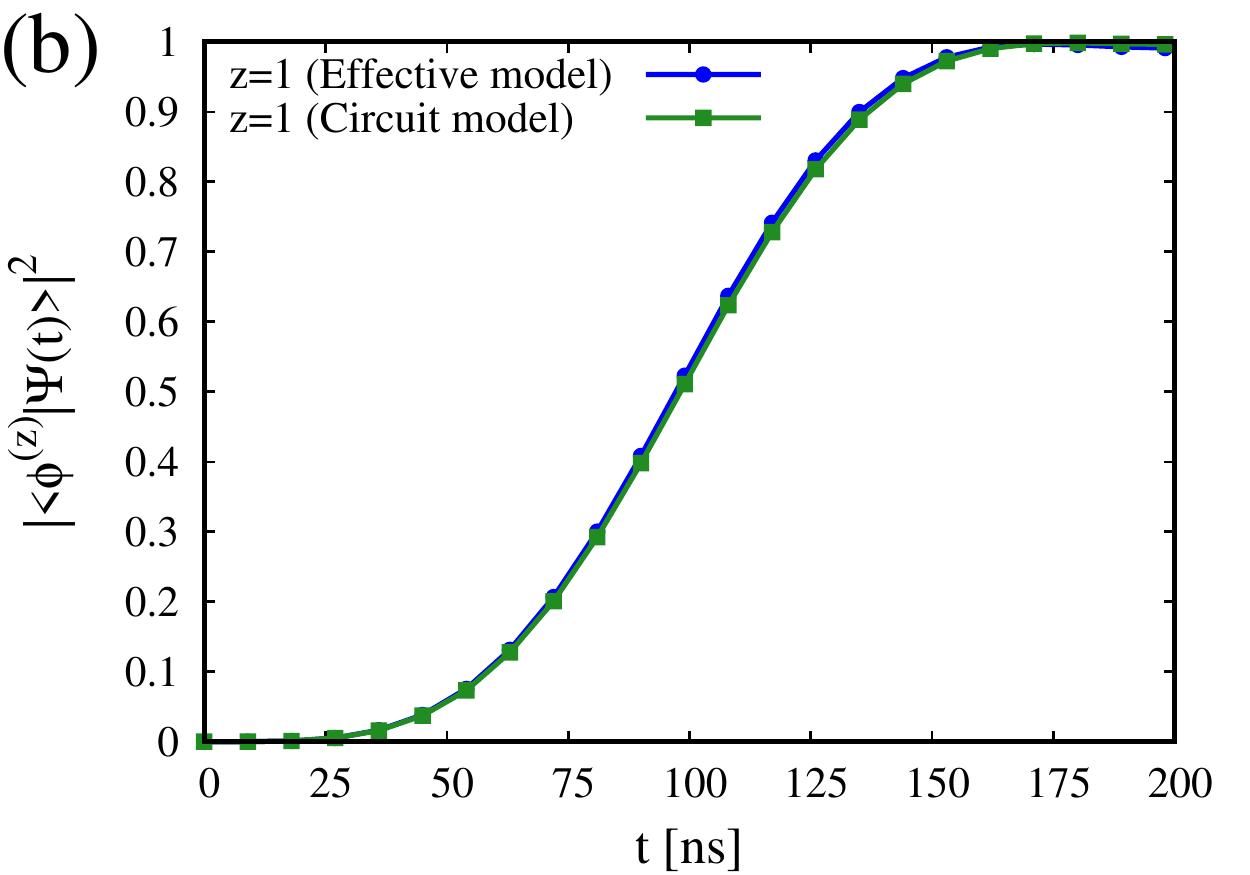}
    \end{minipage}
    \caption[Ground-state $p^{(0)}(t)$(a) and first-excited-state $p^{(1)}(t)$(b) probabilities as functions of time $t$ for the effective and the circuit model. We use the system parameters listed in \tabref{tab:device_parameter_flux_tunable_coupler_chip}, row $i=2$ and the pulse in \equref{eq:NA_control_pulse} with $\TRF=\TD/2$ and $\PATP=0.001$ to obtain the results.]{Ground-state $p^{(0)}(t)$(a) and first-excited-state $p^{(1)}(t)$(b) probabilities as functions of time $t$ for the effective and the circuit model. We use the system parameters listed in \tabref{tab:device_parameter_flux_tunable_coupler_chip}, row $i=2$ and the pulse in \equref{eq:NA_control_pulse} with $\TRF=\TD/2$ and $\PATP=0.001$ to obtain the results. Here, we model a microwave pulse, see \figref{fig:NA_pulse_time_evo}(a). \PANSC{a-b} show the results obtained with the circuit Hamiltonian \equref{eq:flux-tunable transmon recast} and the drive frequency $\DFTP=7.636$ GHz as well as the results obtained with the effective Hamiltonian \equref{eq:fft_eff_II} and the drive frequency $\DFTP=7.643$ GHz. This means we use the frequencies which cut through the chevron patterns in \figsref{fig:NA_RESMAPS}(a-b) as drive frequencies. At time $t=0$ we initialise the systems in the corresponding ground states.}\label{fig:NA_RESMAPS_CUTS}
\end{figure}
We now discuss how the Hamiltonians \equref{eq:fft_eff_II} and \equref{eq:flux-tunable transmon recast} react to a microwave pulse, see \figref{fig:NA_pulse_time_evo}(a). Figures \ref{fig:NA_RESMAPS}(a-b) show the ground-state probabilities $p^{(0)}$ as functions of the pulse duration $\TD$ and the drive frequency $\DF$. Here, we use a pulse with the amplitude $\PATP=0.001$ and the rise and fall time $\TRF=\TD/2$ to obtain the results. In \PANL{a} we solve the TDSE for the circuit Hamiltonian \equref{eq:flux-tunable transmon recast}. Similarly, in \PANL{b} we solve the TDSE for the effective Hamiltonian \equref{eq:fft_eff_II}. Furthermore, we center the two chevron patterns in \figsref{fig:NA_RESMAPS}(a-b) around the transition frequencies $\omega^{(0)}=7.636$ GHz(a) and $\omega^{(0)}=7.643$ GHz(b). As one can see, the results in \figsref{fig:NA_RESMAPS}(a-b) show a qualitative and quantitative similar behaviour if centered around the corresponding transition frequency.

Figures \ref{fig:NA_RESMAPS_CUTS}(a-b) show the time evolutions of the probabilities $p^{(0)}(t)$(a) and $p^{(1)}(t)$(b) as functions of the time $t$ for the effective and the circuit model. \PANSC{a-b} show the time evolution for the frequencies which cut through the chevron patterns in \figsref{fig:NA_RESMAPS}(a-b),\ie we use the frequencies $\omega^{(0)}=7.636$ GHz(a) and $\omega^{(0)}=7.643$ GHz(b) as drive frequencies. All the other simulation parameters are the same as for \figsref{fig:NA_RESMAPS}(a-b).

In \figsref{fig:NA_RESMAPS_CUTS}(a-b) we can observe a good qualitative and quantitative agreement between the time evolutions of the different probabilities for the circuit and the effective model.

Overall, for the microwave pulses we use to model transitions of the form $z=0 \rightarrow z=1$, the only distinct difference between the two models is the shift in the transition frequency. However, this shift is something we should expect. The fourth-order cosine expansion of the Hamiltonian \equref{eq:flux-tunable transmon recast} leads to an effective model with a spectrum which is not exact,\ie the spectra of both Hamiltonians do not match exactly.

We performed additional simulations (data not shown), where we include higher-order terms up to the 60th order in the cosine expansion of the Hamiltonian \equref{eq:flux-tunable transmon recast}. Here we find that the results for the chevron pattern in \figref{fig:NA_RESMAPS_CUTS}(b) stay the same as long as we centre the results for the ground-state probability around the corresponding transition frequency $\omega^{(0)}$. If we add enough terms to the cosine expansion, the results converge,\ie the transition frequency converges and no additional shifts at the order of MHz are noticeable anymore.

Also, we should emphasise that the adiabatic effective Hamiltonian \equref{eq:tunable-frequency eff} cannot, by definition, be used to model the $z=0 \rightarrow z=1$ transitions with a microwave pulse.

\graphicspath{{./FiguresAndData/NAPaper/CircuitHamiltonianGaugeSimulations/SingleFTT/20_04_2022_NonAdiaMapCirc/}{./FiguresAndData/NAPaper/EffectiveHamiltonianGaugeSimulations/SingleFTT/20_04_2022_NonAdiaMap/}}
\begin{figure}[!tbp]
    \centering
    \includegraphics[width=0.49\textwidth]{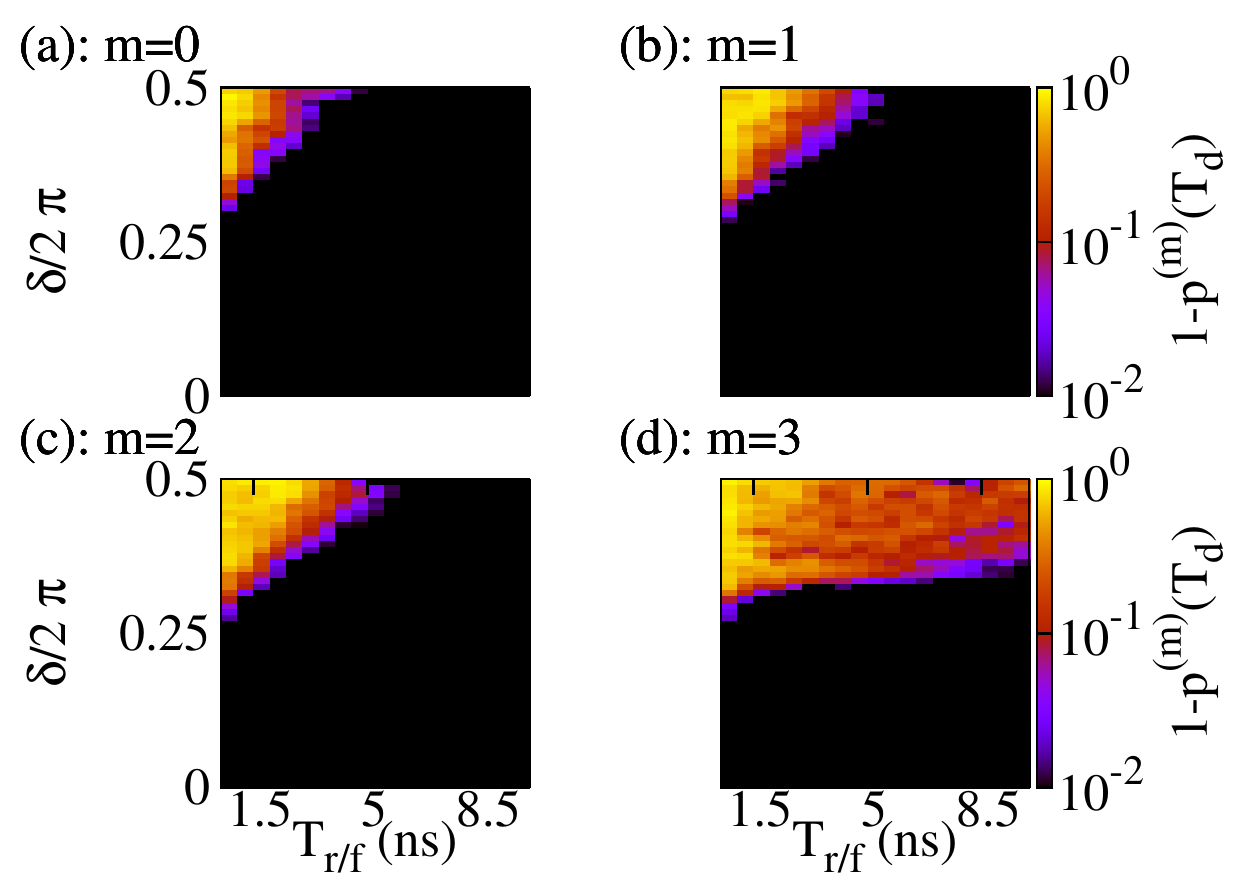}
    \includegraphics[width=0.49\textwidth]{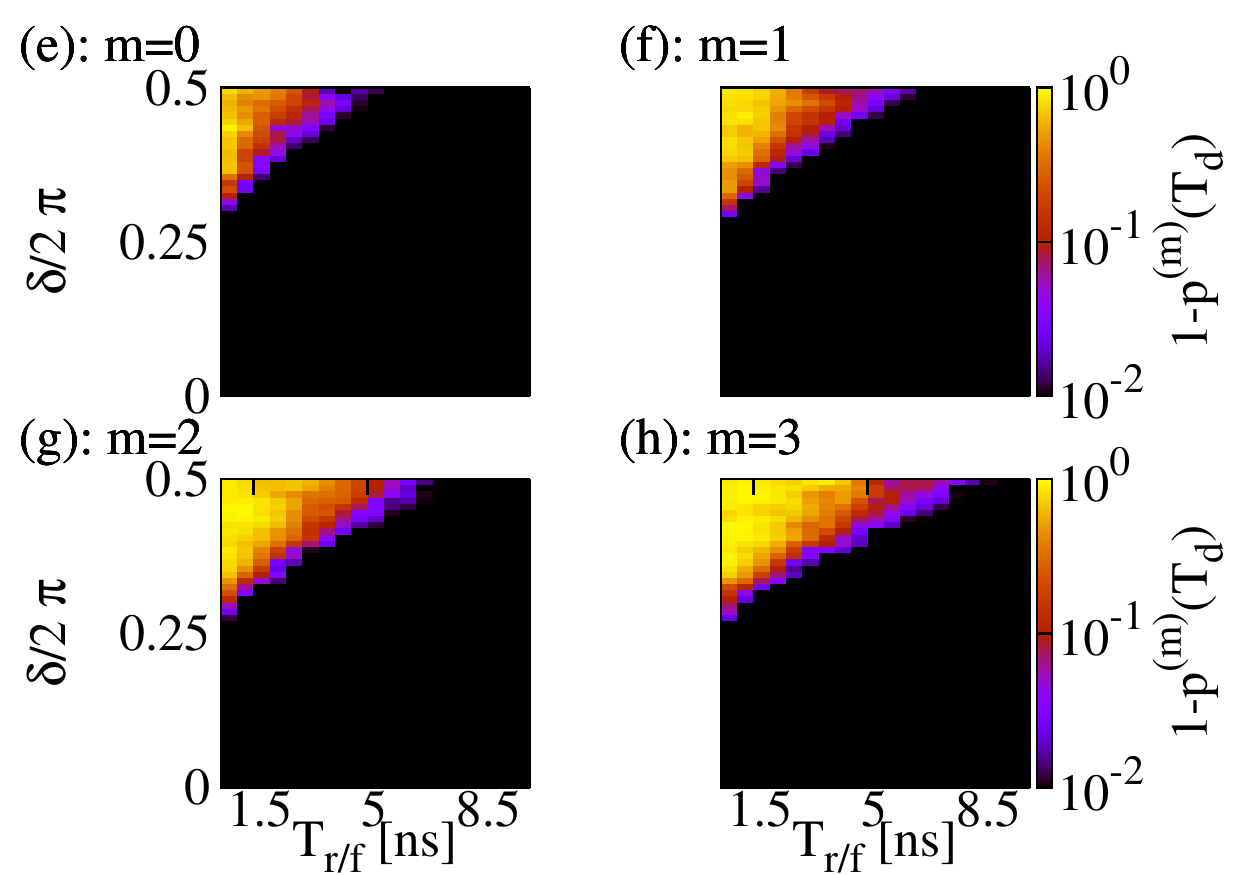}
    \caption[Probabilities $1-p^{(z)}$ for $z \in \{0,1,2,3\}$ as functions of the rise and fall time $\TRF$ and the pulse amplitude $\PA$.]{Probabilities $1-p^{(z)}$ for $z \in \{0,1,2,3\}$ as functions of the rise and fall time $\TRF$ and the pulse amplitude $\PA$. We use the system parameters listed in \tabref{tab:device_parameter_flux_tunable_coupler_chip}, row $i=2$ and the pulse in \equref{eq:NA_control_pulse} with $\TD=50$ ns and $\DF=0$ GHz to obtain the results. Here we model a unimodal pulse, see \figref{fig:NA_pulse_time_evo}(b). In \PANSL{a-d} we solve the TDSE for the circuit Hamiltonian \equref{eq:flux-tunable transmon recast}. Similarly, in \PANSL{e-h} we solve the TDSE for the effective Hamiltonian \equref{eq:fft_eff_II}. At time $t=0$ we initialise the systems in the eigenstates $\ket{z}$ for $z \in \{0,1,2,3\}$.}\label{fig:NA_non_adiabatic_transitions}
\end{figure}
We now discuss how the Hamiltonians \equaref{eq:flux-tunable transmon recast}{eq:fft_eff_II} react to a unimodal pulse, see \figref{fig:NA_pulse_time_evo}(b). Figures \ref{fig:NA_non_adiabatic_transitions}(a-h) show the probabilities $1-p^{(z)}$ at time $\TD$ for $z \in \{0,1,2,3\}$ as functions of the rise and fall time $\TRF$ and the pulse amplitude $\PA$. We use the duration $\TD=50$ ns and the drive frequency $\DF=0$ GHz to model the pulse $\FP$. In \PANSL{a-d} we solve the TDSE for the circuit Hamiltonian \equref{eq:flux-tunable transmon recast}. Similarly, in \PANSL{e-h} we solve the TDSE for the non-adiabatic effective Hamiltonian \equref{eq:fft_eff_II}. At time $t=0$ we initialise the systems in the corresponding eigenstates $\ket{z}$ for $z \in \{0,1,2,3\}$. Furthermore, we use twenty basis states to obtain the results presented in \figsref{fig:NA_non_adiabatic_transitions}(a-h).

The simulations test whether or not the systems have left the adiabatic regime for a given set of pulse parameters. The bright areas correspond to regions where the systems do not fully remain in their corresponding initial eigenstates and therefore have left the adiabatic regime. As expected, we can observe that this is the case for $\TRF \rightarrow 0$ and $\PATP \rightarrow 0.5$.

We also find that the circuit Hamiltonian results in \PANSL{a-c} and the effective Hamiltonian results in \PANSL{e-g} for the states $z \in \{0,1,2\}$ agree qualitatively. However, the results in \PANL{d} and \PANL{h} for $z=3$ show qualitative and quantitative differences.

Here too, we should emphasise that the adiabatic effective Hamiltonian \equref{eq:tunable-frequency eff} cannot, by definition, be used to model any of the transitions discussed above.

In \figsref{fig:NA_RESMAPS}(a-b) we can observe that the transition frequencies $\omega^{(0)}$ for the effective and the circuit model deviate. Therefore, we want to investigate how well the analytical expressions for the energies $E^{(z)}(\varphi(t))$ actually approximate the numerically exact values $E_{\text{exact.}}^{(z)}(\varphi(t))$. Since the circuit Hamiltonian \equref{eq:flux-tunable transmon recast} has two symmetry points, we only have to consider the flux variable on the interval $\varphi/2\pi \in [0,0.5]$.

If we follow the derivation of the Hamiltonian \equref{eq:fft_eff_II} presented in \secref{sec:Transmons}, we find that the energies of the flux-tunable transmon are approximated by the expression
\begin{equation}\label{eq:spec}
  \left(E^{(z)}(\varphi)-E^{(0)}(\varphi)\right) = \left(z \omega^{(q)}(\varphi)+\frac{\alpha^{(q_{0})}}{2}z(z-1)\right),
\end{equation}
where
\begin{equation}\label{eq: frequency der}
  \omega^{(q)}(\varphi)=\omega^{(q_{0})} \sqrt[4]{\cos\left(\frac{\varphi}{2}\right)^{2} + d^{2} \sin\left(\frac{\varphi}{2}\right)^{2}}+ \alpha^{(q_{0})}.
\end{equation}
Here $\omega^{(q_{0})}=\sqrt{E_{C}E_{\Sigma}}$ and $\alpha^{(q_{0})}=-E_{C}/4$ are expressed in terms of the capacitive and Josephson energies which characterise a flux-tunable transmon, see \secref{sec:Transmons}. We denote this type of parametrisation as approximation I.

In the literature one can find instances, see for example \REFS\cite{McKay16,Roth19,Ganzhorn20,Gu21}, where the tunable coupler frequency $\omega^{(q)}(\varphi)$ in \equref{eq:spec} is modelled differently. Here one uses
\begin{equation}\label{eq: frequency}
  \omega^{(q)}(\varphi)=\omega^{(q_{0})} \sqrt[4]{\cos\left(\frac{\varphi}{2}\right)^{2} + d^{2} \sin\left(\frac{\varphi}{2}\right)^{2}},
\end{equation}
with $\omega^{(q_{0})}=\const$, $\alpha^{(q_{0})}=\const$ and $d=\const$ to parameterise the energies of the effective flux-tunable transmon or an experimental flux-tunable transmon. This second parametrisation will be denoted as approximation II.

Additionally, the literature also provides us with more complex expressions for the tunable frequency and the anharmonicity in \equref{eq:spec}. One such example is given by the authors of \REF\cite{Didier}. The corresponding expression for the tunable frequency reads
\begin{equation}\label{eq: expansion frequency}
  \omega(\varphi)=\sqrt{2 E_{C} E_{J_{\text{eff}}}(\varphi)} - \frac{E_{C}}{4} \sum_{n=0}^{24} a_{n} \Xi(\varphi)^{n}.
\end{equation}
Similarly, the tunable anharmonicity is modelled with the expression
\begin{equation}\label{eq: expansion anharmonicity}
  \alpha(\varphi)= - \frac{E_{C}}{4} \sum_{n=0}^{24} b_{n} \Xi(\varphi)^{n}.
\end{equation}
Here $a_{n}$ and $b_{n}$ are real-valued coefficients and the function $\Xi(\varphi)$ can be expressed as
\begin{equation}
  \Xi(\varphi)=\sqrt{\frac{E_{C}}{2 E_{J_{\text{eff}}}(\varphi)}}.
\end{equation}
We emphasise that both expression are taken from \REF\cite{Didier}. This type of parametrisation is denoted as approximation III.

\renewcommand{\scale}{0.375}
\graphicspath{{.//FiguresAndData/NAPaper/Approx/}}
\begin{figure}[!tbp]
    \centering
    \begin{minipage}{0.30\textwidth}
        \centering
        \includegraphics[scale=\scale]{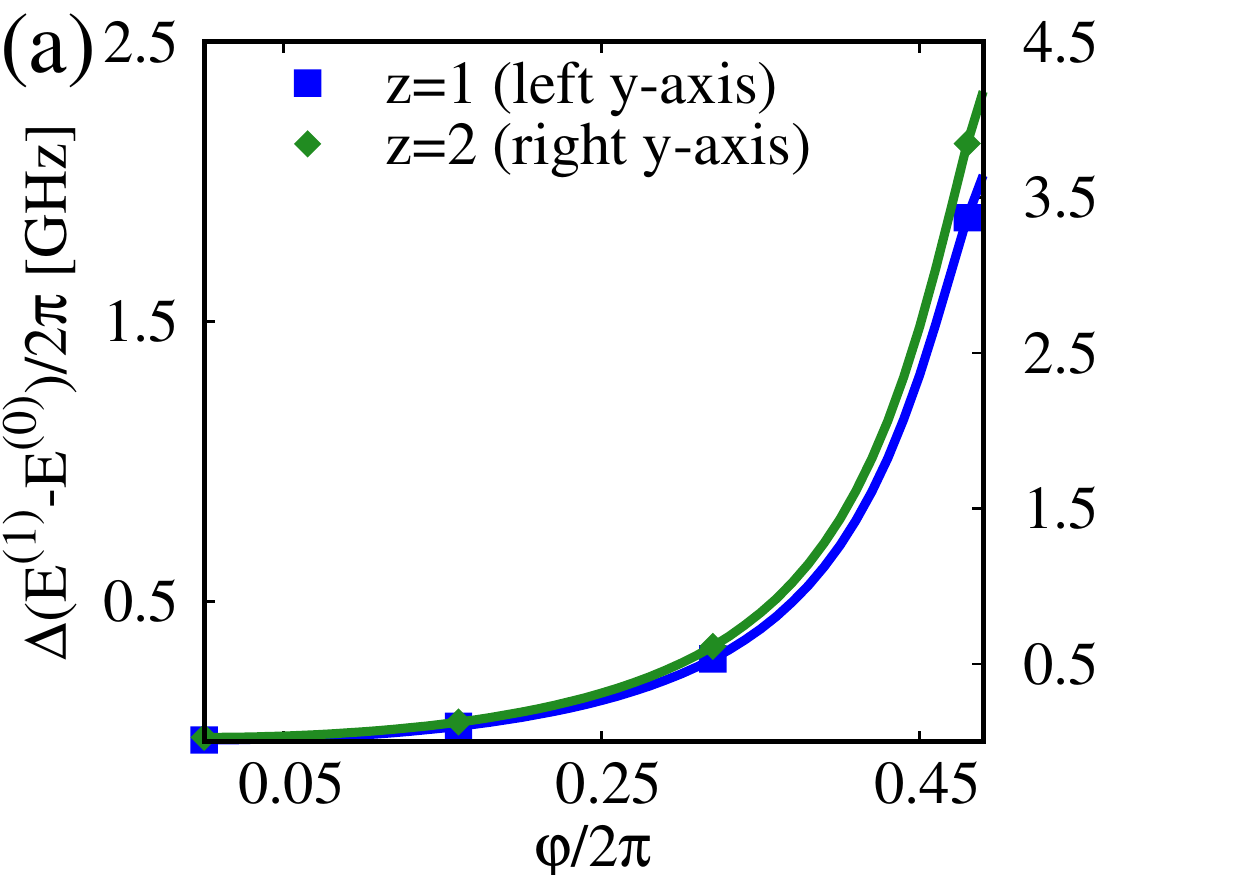}
    \end{minipage}
    \begin{minipage}{0.30\textwidth}
        \centering
        \includegraphics[scale=\scale]{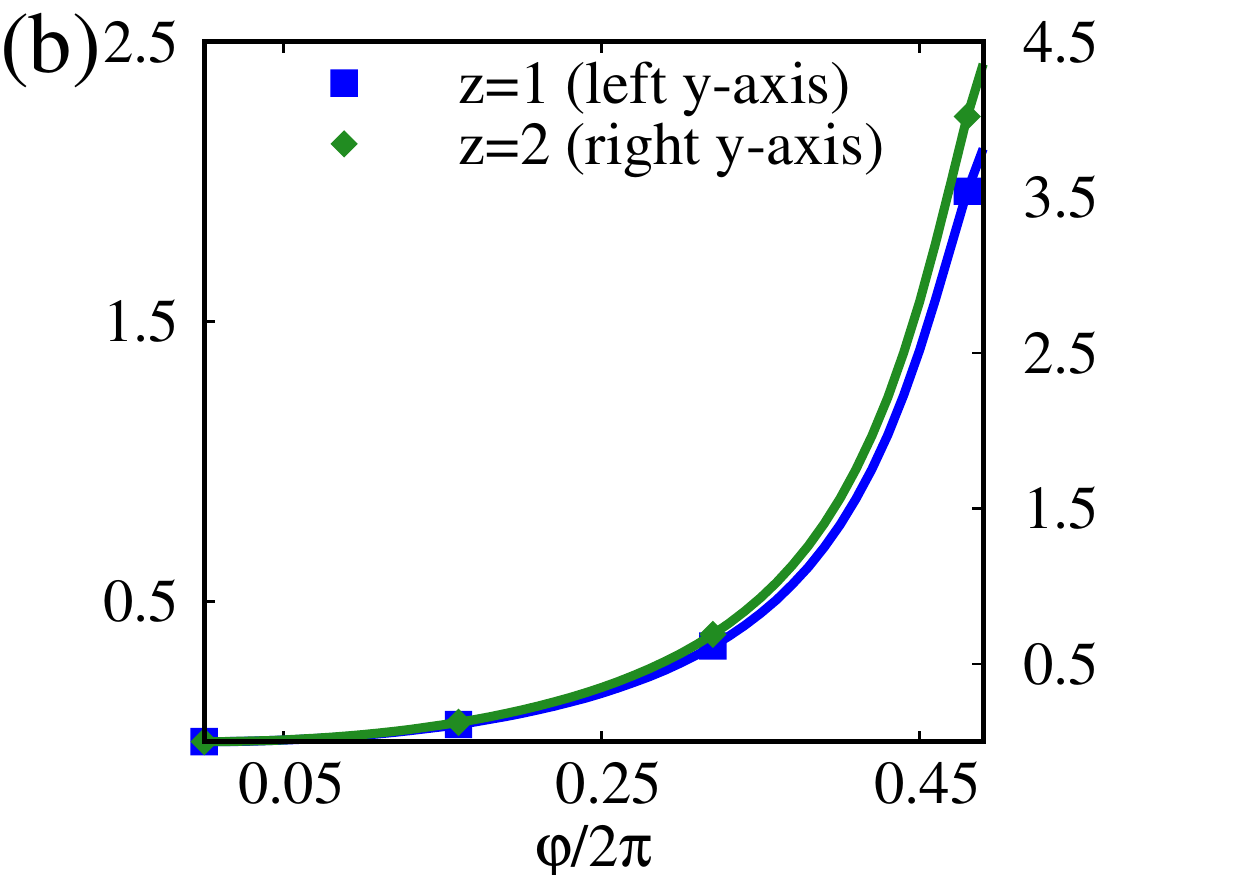}
    \end{minipage}
    \begin{minipage}{0.30\textwidth}
        \centering
        \includegraphics[scale=\scale]{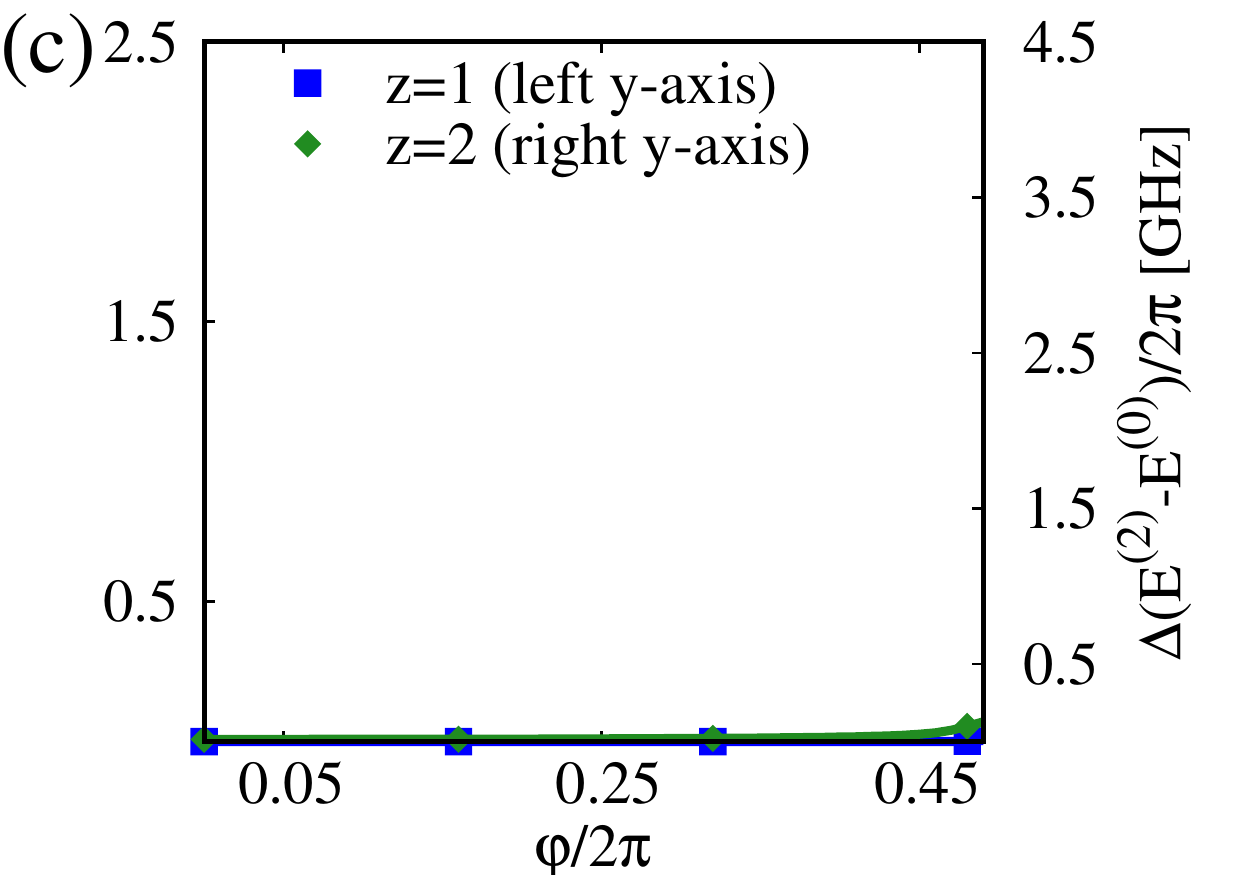}
    \end{minipage}
    \caption[Deviations $\Delta\BRR{E^{(z)}-E^{(0)}}$ given by \equref{eq:spec_deviation} between the numerically exact eigenvalues of the circuit Hamiltonian and three approximations I(a), II(b) and III(c) for the eigenvalues as functions of the flux $\varphi/2 \pi$.]{Deviations $\Delta\BRR{E^{(z)}-E^{(0)}}$ given by \equref{eq:spec_deviation} between the numerically exact eigenvalues of the circuit Hamiltonian \equref{eq:flux-tunable transmon recast} and three approximations I(a), II(b) and III(c) for the eigenvalues as functions of the flux $\varphi/2 \pi$. The left y-axis, square markers and blue lines, shows the deviations for $z=1$. Similarly, the right y-axis, diamond markers and green lines, shows the deviations for $z=2$. We use the energy parameters listed in \tabref{tab:device_parameter_flux_tunable_coupler_chip} row $i=2$ to obtain the results. The approximations I-III are discussed in the main text, see \equsref{eq:spec}{eq: expansion anharmonicity}.}\label{fig:NA_spec_approximation}
\end{figure}

We measure the deviations between the numerically exact values and the approximations I-III with the expression
\begin{align}\label{eq:spec_deviation}
\begin{split}
\Delta \left(E^{(z)}(\varphi)-E^{(0)}(\varphi)\right)&=\bigg|\left(E_{\text{exact.}}^{(z)}(\varphi)-E_\text{exact.}^{(0)}(\varphi)\right)\\
-&\left(E^{(z)}(\varphi)-E^{(0)}(\varphi)\right)\bigg|,
\end{split}
\end{align}
where $z \in \{1,2\}$. Figures \ref{fig:NA_spec_approximation}(a-c) show the deviations $\Delta \left(E^{(z)}(\varphi)-E^{(0)}(\varphi)\right)$ as functions of the flux $\varphi$ for the approximations I(a), II(b) and III(c). The left y-axis, square markers and blue lines, shows the deviations for $z=1$. Similarly, the right y-axis, diamond markers and green lines, shows the deviations for $z=2$. We use the circuit Hamiltonian \equref{eq:flux-tunable transmon recast} and $N_{c}=50$ charge basis states to obtain the numerically exact values.

Figure \ref{fig:NA_spec_approximation}(a) shows the deviations for approximation I. As one can see, the deviations start to grow slowly and then rapidly increase. In the end, at $\varphi/2\pi=0.5$ the first energy gap deviates by about $2$ GHz and the second energy gap by about $4$ GHz. Figure \ref{fig:NA_spec_approximation}(b), for approximation II, shows similar qualitative and quantitative features as \figref{fig:NA_spec_approximation}(a). Figure \ref{fig:NA_spec_approximation}(c) shows barely any deviations on the GHz energy scale. Only if $\varphi/2\pi \rightarrow 0.5$, we see some deviations for the second energy gap. The deviations for the first energy gap are at most $4$ MHz, for the case presented here.

We also computed the deviations for the flux-tunable transmon parameters listed in \tabref{tab:device_parameter_resonator_coupler_chip} row $i=0$ and $i=1$ (data not shown). Here we find similar qualitative features as in \figref{fig:NA_spec_approximation}(a-c). However, the numerical values for the deviations can vary by a couple of GHz for different parameters $E_{C}$, $E_{J,l}$ and $E_{J,r}$.

The deviations in terms of the Hamiltonian spectra we can see in \figref{fig:NA_spec_approximation}(a-c) can potentially affect the modelling of non-adiabatic quantum gates, see \REF\cite{DiCarlo2009,Foxen20}, in small NIGQC models. Here we use the pulse amplitude $\PA$ to tune the energies of two states into resonance. Furthermore, so far we only discussed the case for a single, isolated flux-tunable transmon. It might be the case that the corresponding errors, in terms of the spectra, for an interacting multi-transmon system are more complex and larger. In such systems, energy levels tend to repel each other such that the resulting behaviour with respect to an external parameter, like the external flux or the interaction strength, leads to complex energy level structures, see for example \REF\cite{Berke21}.

Additionally, the results in \figsref{fig:NA_RESMAPS}(a-b) and \figsref{fig:NA_non_adiabatic_transitions}(a-h) are obtained for the operating point $\varphi_{0}/2\pi=0.15$. This means, we should expect that the previously discussed deviations vary with the operating point we use for the simulations.

\section{Simulations of suppressed transitions in the adiabatic effective two-qubit model: architecture I}\label{sec:NA_E_suppressed}

In \secref{sec:NA_single_flux_tunable_transmon}, we discussed the case of a single flux-tunable transmon. Here, we use effective and circuit Hamiltonian models to study how such systems react to different external fluxes $\varphi(t)$, see also \equref{eq:NA_control_pulse} and \figref{fig:NA_pulse_time_evo}(a-b). We find that the adiabatic effective Hamiltonian \equref{eq:tunable-frequency eff} does not allow us, by definition, to model any transitions at all. However, the authors of \REFS\cite{McKay16,Roth19,Ganzhorn20} show that if one uses the adiabatic effective Hamiltonian \equref{eq:tunable-frequency eff} to model flux-tunable transmons in architecture I, see \figref{fig:arch_sketch}(a), one can at least model \ISWAP{} and \CZ{} two-qubit transitions. In this section, we investigate whether or not certain single-qubit type transitions are also absent.

As already suggested, we study a system of type architecture I, see \figref{fig:arch_sketch}(a), which contains two fixed-frequency transmons and these are both coupled to a single flux-tunable transmon. Here we use the positive discrete indices $z_{0}$, see \tabref{tab:device_parameter_flux_tunable_coupler_chip} row $i=0$ and $z_{1}$, see \tabref{tab:device_parameter_flux_tunable_coupler_chip} row $i=1$, to address the fixed-frequency transmons and the index $z_{2}$, see \tabref{tab:device_parameter_flux_tunable_coupler_chip} row $i=2$, to address the flux-tunable transmon. To this end, we define the basis state tuple $\mathbf{z}=(z_{2},z_{1},z_{0})$.

\graphicspath{{./FiguresAndData/NAPaper/CircuitHamiltonianGaugeSimulations/ArchitectureI/ResMapX01NEW/}{./FiguresAndData/NAPaper/CircuitHamiltonianGaugeSimulations/ArchitectureI/ResMapX10NEW/}}
\begin{figure}[!tbp]
    \centering
    \begin{minipage}{0.49\textwidth}
        \centering
        \includegraphics[width=\width\textwidth]{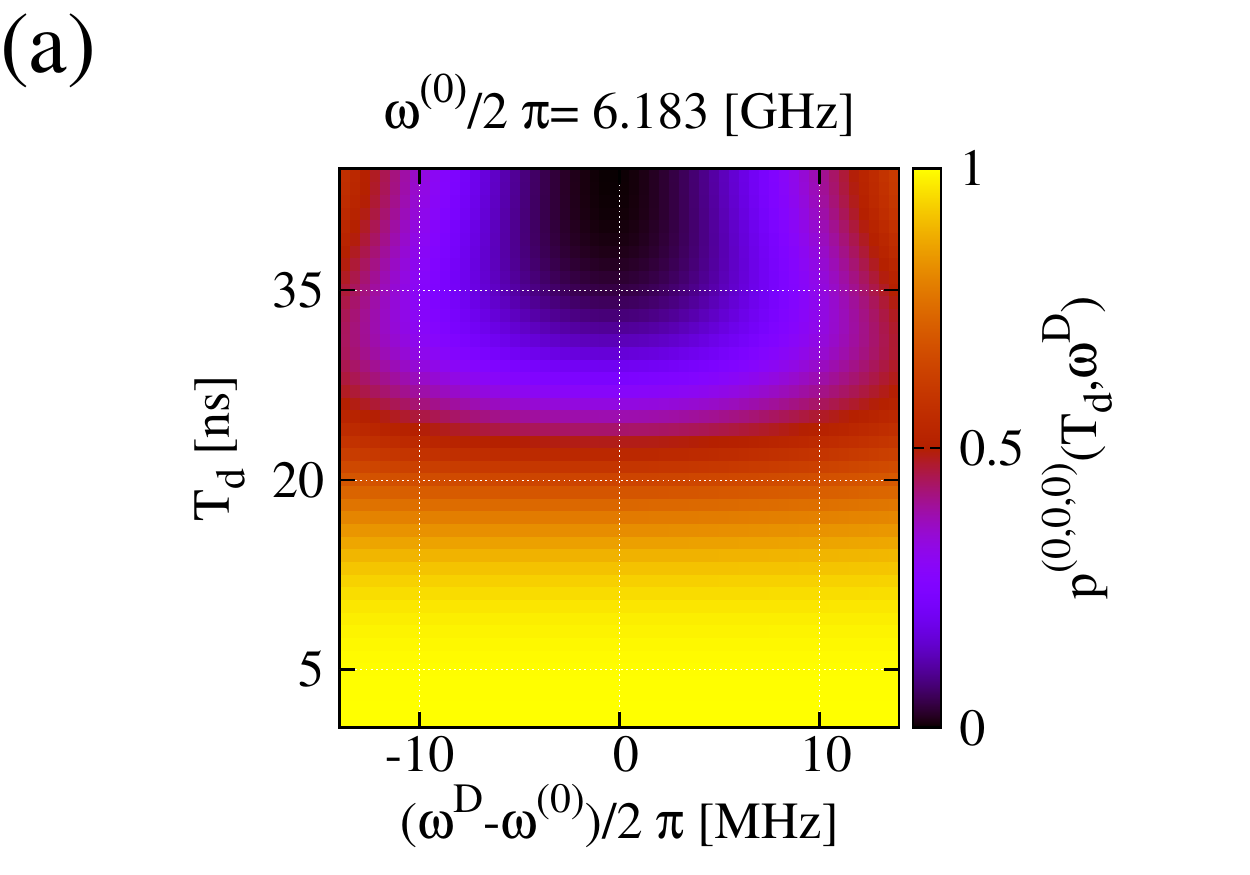}
    \end{minipage}\hfill
    \begin{minipage}{0.49\textwidth}
        \centering
        \includegraphics[width=\width\textwidth]{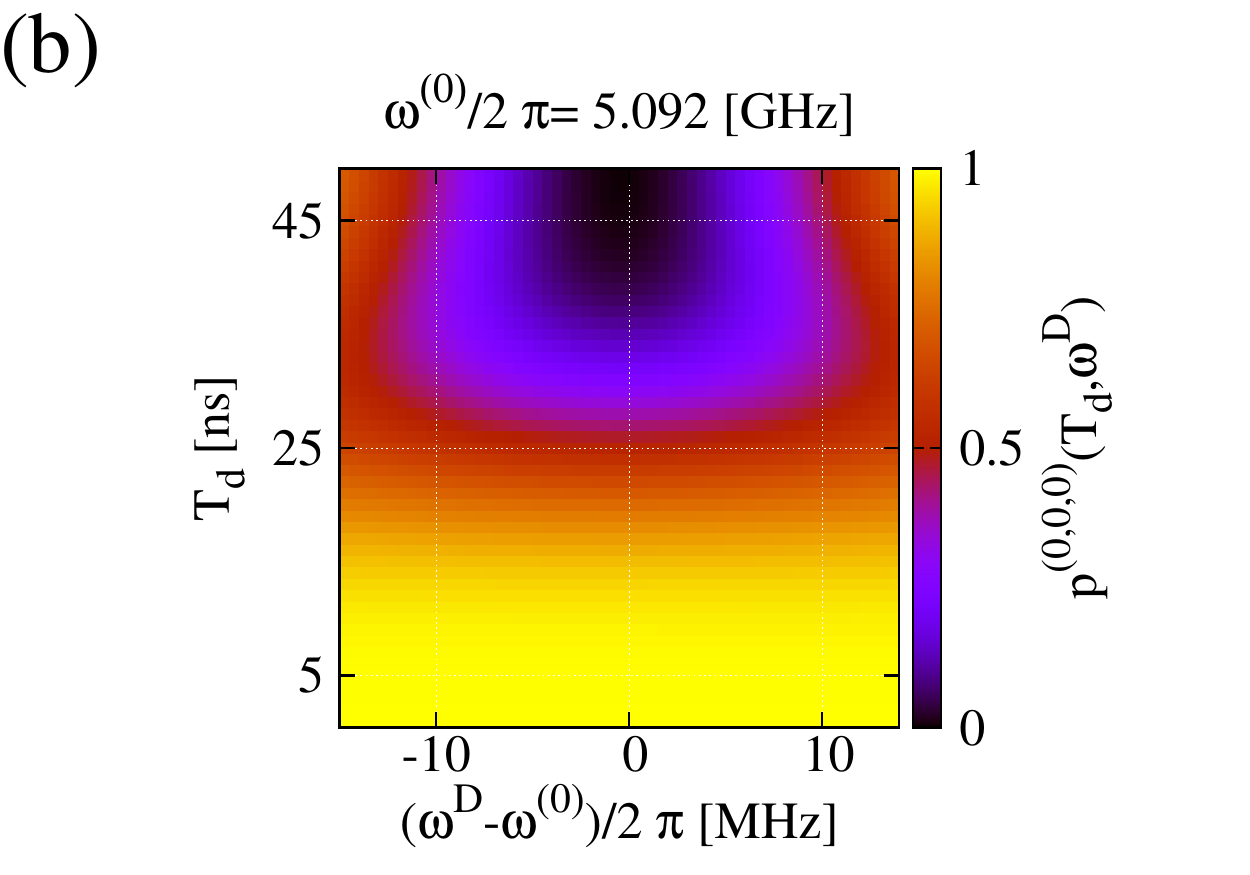}
    \end{minipage}
    \caption[Probabilities $p^{\mathbf{z}}$ for $\mathbf{z}=(0,0,0)$ as functions of the pulse duration $\TD$ and the drive frequency $\DF$ (circuit model architecture I).]{Probabilities $p^{\mathbf{z}}$ for $\mathbf{z}=(0,0,0)$ as functions of the pulse duration $\TD$ and the drive frequency $\DF$. We use the circuit Hamiltonian \equref{eq:CHM}, the initial state $\ket{0,0,0}$, the device parameters listed in \tabref{tab:device_parameter_flux_tunable_coupler_chip} and the control pulse given by \equref{eq:NA_control_pulse} with the rise and fall time $\TRF=\TD/2$ to obtain the results. In \PANL{a} we model the transition $\mathbf{z}=(0,0,0) \rightarrow \mathbf{z}=(0,1,0)$ and we use the pulse amplitude $\PATP=0.045$. In \PANL{b} we model the transition $\mathbf{z}=(0,0,0) \rightarrow \mathbf{z}=(0,0,1)$ and we use the pulse amplitude $\PATP=0.085$.}\label{fig:NA_supressed_chevron_pattern}
\end{figure}

Figures \ref{fig:NA_supressed_chevron_pattern}(a-b) show the ground-state probabilities $p^{(0,0,0)}$ as functions of the pulse duration $\TD$ and the drive frequency $\DF$. We use the circuit Hamiltonian \equref{eq:CHM}, the device parameters listed in \tabref{tab:device_parameter_flux_tunable_coupler_chip}, the pulse given by \equref{eq:NA_control_pulse} with $\TRF=\TD/2$ and three basis states for every transmon to obtain the results. In \figref{fig:NA_supressed_chevron_pattern}(a), we can observe how the system reacts to pulses with different drive frequencies $\DF$ and the constant pulse amplitude $\PATP=0.045$. Here, we model the transition $\mathbf{z}=(0,0,0) \rightarrow \mathbf{z}=(0,1,0)$. Similarly, in \figref{fig:NA_supressed_chevron_pattern}(b), we see how the system reacts to pulses with different drive frequencies $\DF$ and the constant pulse amplitude $\PATP=0.085$. Here, we model the transition $\mathbf{z}=(0,0,0) \rightarrow \mathbf{z}=(0,0,1)$. In \figsref{fig:NA_supressed_chevron_pattern}(a-b) we can observe how the system leaves the bare ground state $\mathbf{z}=(0,0,0)$ for various pulse parameters.

We find that the corresponding effective model, where the flux-tunable transmon is modelled adiabatically, does not allow us to qualitatively and/or quantitatively reproduce the results we show in \figsref{fig:NA_supressed_chevron_pattern}(a-b). For these simulations we use the adiabatic effective Hamiltonian \equref{eq:EHM}, the device parameters listed in \tabref{tab:device_parameter_flux_tunable_coupler_chip} and \tabref{tab:device_parameter_flux_tunable_coupler_chip_effective}, the pulse given by \equref{eq:NA_control_pulse} and four basis states for every transmon.

For this reason, we search for the missing transitions $\mathbf{z}=(0,0,0) \rightarrow \mathbf{z}=(0,0,1)$ and $\mathbf{z}=(0,0,0) \rightarrow \mathbf{z}=(0,1,0)$ in a more systematic manner. We define the expression
\begin{equation}\label{eq:cost_function}
  \epsilon=1-\min_{(\DF,\PA,t)\in \mathcal{G}}(p^{(0,0,0)}(\DF,\PA,t)),
\end{equation}
where $\mathcal{G}\subseteq\mathbb{R}^{3}$ and $p^{(0,0,0)}(\DF,\PA,t)=|\braket{0,0,0|\Psi(\DF,\PA,t)}|^{2}$ and evaluate it for a suitable parameter grid $\mathcal{G}$. If we initialise the system in the bare ground state $\ket{\Psi(0)}=\ket{0,0,0}$ and the resulting bare ground state probability $p^{(0,0,0)}(\DF,\PA,t)$ substantially deviates from one for at least one tuple $(\DF,\PA,t) \in \mathbb{R}^{3}$, we would find $\epsilon \rightarrow 1$. Therefore, we define search grids $\mathcal{G}$ which potentially contain the desired transitions.
\begin{table}[!tbp]
\caption[Table with grid search parameters and the results of the corresponding searches in terms of the values for $\epsilon$.]{\label{tab:grid_search}Table with grid search parameters and the results of the corresponding searches in terms of the values for $\epsilon=1-\min_{(\DF,\PA,t) \in \mathcal{G}}(p^{(0,0,0)}(\DF,\PA,t))$, where $\mathcal{G}\subseteq\mathbb{R}^{3}$. Here $p^{(0,0,0)}(\DF,\PA,t)=|\braket{0,0,0|\Psi(\DF,\PA,t)}|^{2}$ denotes the probability for finding the system in the bare ground state when a pulse of the form \equref{eq:NA_control_pulse} is applied to the system and $\ket{\Psi(0)}=\ket{0,0,0}$. We use the effective Hamiltonian \equref{eq:EHM} and the parameters listed in \tabref{tab:device_parameter_flux_tunable_coupler_chip_effective} to model a two-qubit system of type architecture I. The first column contains letters which allow us to enumerate the different searches we perform. The second column shows frequency intervals for the drive frequency $\DF$ in units of GHz. The third column shows the search interval for the pulse amplitude $\PA$. The fourth column shows the search interval for the pulse duration $\TD$ in ns. The fifth column shows the results for the different searches in terms of $\epsilon$. We use the step parameters $\Delta\DFTP=10^{-5}$ GHz, $\Delta\PATP=10^{-3}$ and $\Delta t=0.2$ ns to obtain the results.}
\begin{tabularx}{\textwidth}{ X X X X X }
\hline\hline
Case &$\DFTP$ & $\PATP$ & $T_{d}$ & $\epsilon$  \\
\hline
A & $[4.90,5.30]$ & $[0.000,0.110]$ & $[0,300]$ & $0.001$\\

B & $[6.00,6.40]$ & $[0.000,0.110]$ & $[0,300]$ & $0.001$\\

C & $[0.00,0.00]$ & $[0.000,0.000]$ & $[0,300]$ & $0.001$\\
\hline\hline
\end{tabularx}
\end{table}

Table \ref{tab:grid_search} shows the results for three different cases or search parameter grids $\mathcal{G}$. We define the search grids $\mathcal{G}$ by means of the step parameters $\Delta\DFTP=10^{-5}$ GHz, $\Delta\PATP=10^{-3}$ and $\Delta t=0.2$ ns and search intervals for the drive frequency $\DF$, the pulse amplitude $\PA$ and the pulse duration $\TD$. We use the effective Hamiltonian \equref{eq:EHM}, the device parameters listed in \tabref{tab:device_parameter_flux_tunable_coupler_chip_effective} and the pulse given by \equref{eq:NA_control_pulse} to obtain the results. Note that there exist several ways to modify the effective Hamiltonian model. We can choose between two different sets of functions to model the tunable transmon qubit frequency and anharmonicity and we can perform simulations with static and dynamic interaction strengths $g^{(c,c)}(t)$. Furthermore, we can mix the different cases. We performed simulations for all four scenarios and we find that the numerical results are identical, up to the third decimal in terms of $\epsilon$. Therefore, we only present the results for one case.

The first row shows the case where we search for the transition $\mathbf{z}=(0,0,0) \rightarrow \mathbf{z}=(0,0,1)$. Here we find the value $\epsilon=0.001$. The second row shows the case where we search for the transition $\mathbf{z}=(0,0,0) \rightarrow \mathbf{z}=(0,1,0)$. Here too, we find the value $0.001$ for $\epsilon$. The third row shows the case for the free time evolution and again we find $\epsilon=0.001$. Consequently, for the search grids we define in \tabref{tab:grid_search}, we find that the bare ground state probability $p^{(0,0,0)}(\omega^{D},\delta,t)$ reacts to various pulses, see row one and two, in the same way it reacts to the absence of a pulse, see third row. Note that the chevron patterns in \figref{fig:NA_supressed_chevron_pattern}(a-b) are several MHz wide. Therefore, we might conclude that these transitions are absent from the model specified above. Obviously, we cannot make this claim with absolute certainty. It is possible, that there exist other pulse parameters which might allow us to model these transitions. However, the results in \tabref{tab:grid_search} indicate the absence of the transitions we find in the circuit Hamiltonian model.

So far we only considered the transitions $\mathbf{z}=(0,0,0) \rightarrow \mathbf{z}=(0,0,1)$ and $\mathbf{z}=(0,0,0) \rightarrow \mathbf{z}=(0,1,0)$. However, we find that more such transitions are absent in the effective Hamiltonian model,\ie we can activate these transitions in the circuit Hamiltonian model but not in the effective Hamiltonian model. For example, similar to the results we discuss in \secref{sec:NA_single_flux_tunable_transmon}, we are able to excite the flux-tunable transmon itself $\mathbf{z}=(0,0,0) \rightarrow \mathbf{z}=(1,0,0)$ by applying a pulse with a drive frequency $\DF$ near $7.6$ GHz.

Furthermore, the results in \secref{sec:NA_single_flux_tunable_transmon} suggest that the non-adiabatic driving term in \equref{eq:drive_term_ftt_second_time} can possibly be used to model the suppressed transitions. Therefore, we included the driving term into the model and performed the corresponding simulations, see \figsref{fig:NA_supressed_chevron_pattern}(a-b), again. We find (data not shown) that the transitions $\mathbf{z}=(0,0,0) \rightarrow \mathbf{z}=(0,0,1)$ and $\mathbf{z}=(0,0,0) \rightarrow \mathbf{z}=(0,1,0)$ and others can be modelled by including the non-adiabatic driving term in \equref{eq:drive_term_ftt_second_time}.

\section{Simulations of unsuppressed transitions in the adiabatic effective two-qubit model: architecture I}\label{sec:NA_unsuppressed_arch_I}

In \secref{sec:NA_single_flux_tunable_transmon} we study a single flux-tunable transmon, see \tabref{tab:device_parameter_flux_tunable_coupler_chip} row $i=2$ and we find that various transitions between the states of this system can be modelled with the circuit Hamiltonian \equref{eq:flux-tunable transmon recast} but not with the adiabatic effective Hamiltonian \equref{eq:tunable-frequency eff}. Similarly, in \secref{sec:NA_E_suppressed}, we study a two-qubit system of type architecture I, see \figref{fig:arch_sketch}(a), consisting of two fixed-frequency transmons and a single flux-tunable transmon. Here, we identify transitions which can be modelled with the circuit Hamiltonian model. However, it seems to be the case that these transitions are absent in the associated adiabatic effective model, where the flux-tunable transmon is modelled by means of \equref{eq:tunable-frequency eff}. In this section, we study the same two-qubit system, but here we focus on two different transitions which can be described with both Hamiltonian models.

The system we model is of type architecture I, see \figref{fig:arch_sketch}(a). We use the positive discrete indices $z_{0}$ and $z_{1}$ to address the two fixed-frequency transmons, see \tabref{tab:device_parameter_flux_tunable_coupler_chip} row $i=0$ and $i=1$, respectively. These are the transmons which are usually used as qubits in this particular architecture, see for example \REFS\cite{McKay16,Roth19,Ganzhorn20}. Similarly, we use the index $z_{2}$ to address the flux-tunable transmon, see \tabref{tab:device_parameter_flux_tunable_coupler_chip} row $i=2$, which acts as a coupler element,\ie this subsystem is supposed to convey interactions between the two fixed-frequency transmons we use as qubits. Finally, we address the different bare basis states $\ket{\mathbf{z}}$ of the system, see also \secaref{sec:SOTEO_HB}{sec:SOTEO_MB}, by means of the tuples $\mathbf{z}=(z_{2},z_{1},z_{0})$.

In the following, we study two different transitions: the $\mathbf{z}=(0,0,1) \rightarrow \mathbf{z}=(0,1,0)$ transition is sometimes used to implement $\ISWAP{}$ gates and the $\mathbf{z}=(0,1,1) \rightarrow \mathbf{z}=(0,2,0)$ transition is often used to implement $\CZ{}$ gates. References \cite{McKay16,Roth19,Roth20,Ganzhorn20} discuss systems of type architecture I in this context.

\renewcommand{\hold}{}
\graphicspath{{./FiguresAndData/NAPaper/CircuitHamiltonianGaugeSimulations/ArchitectureI/}}
\begin{figure}[!tbp]
    \centering
    \begin{minipage}{0.49\textwidth}
        \centering
        \includegraphics[width=\width\textwidth]{TimeEvolISWAPStates3Beta_0_5}
        \includegraphics[width=\width\textwidth]{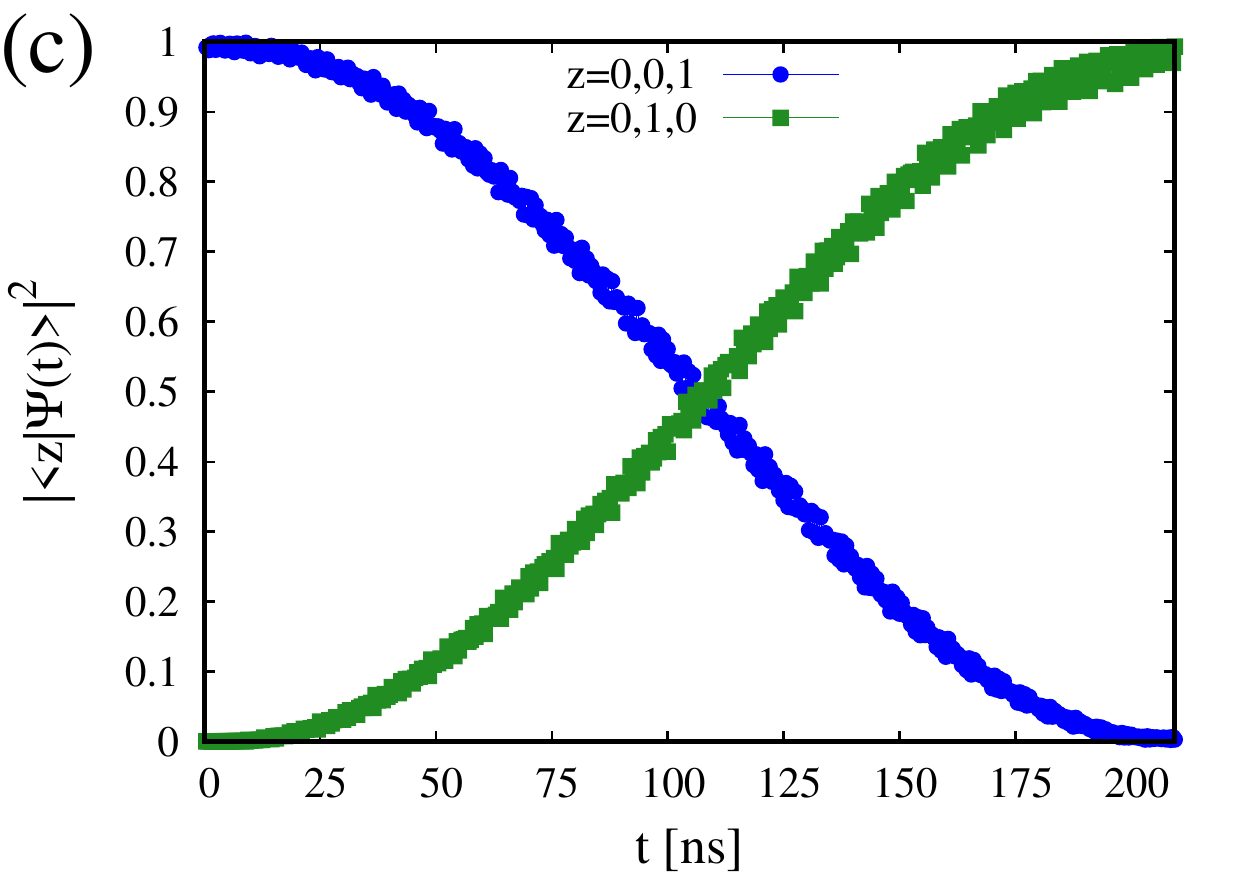}
    \end{minipage}\hfill
    \begin{minipage}{0.49\textwidth}
        \centering
        \includegraphics[width=\width\textwidth]{TimeEvolISWAPStates4Beta_0_5}
        \includegraphics[width=\width\textwidth]{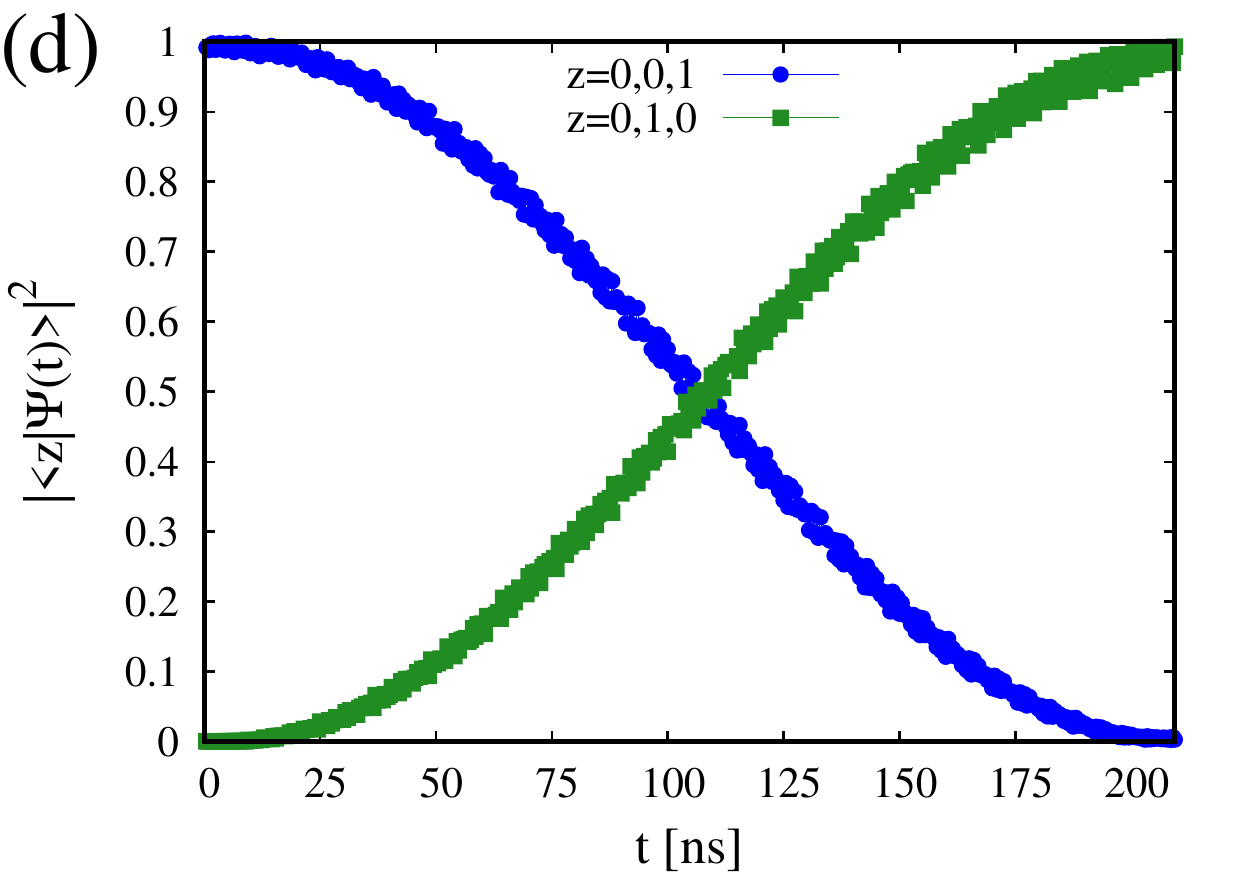}
    \end{minipage}
    \caption[Probabilities $p^{\mathbf{z}}(t)=|\braket{\mathbf{z}|\Psi^{\hold}(t)}|^{2}$ as functions of time $t$ for $\mathbf{z}=(0,0,1)$ and $\mathbf{z}=(0,1,0)$ (circuit model architecture I).]{Probabilities $p^{\mathbf{z}}(t)=|\braket{\mathbf{z}|\Psi^{\hold}(t)}|^{2}$ as a functions of time $t$ for $\mathbf{z}=(0,0,1)$ and $\mathbf{z}=(0,1,0)$. Here, $\ket{\Psi^{\hold}(0)}=\ket{0,0,1}$. We use the circuit Hamiltonian \equref{eq:CHM}, the device parameters listed in \tabref{tab:device_parameter_flux_tunable_coupler_chip} and the pulse given by \equref{eq:NA_control_pulse} to generate the results. The pulse parameters are the drive frequency $\DFTP=1.089$ GHz, the amplitude $\PATP=0.075$, the rise and fall $\TRF=13.00$ ns and the pulse duration $\TD=209.40$ ns. Furthermore, $n_{J}$ denotes the number of flux-tunable transmon basis states and we use $n_{J}=3$ in \PANL{a}, $n_{J}=4$ in \PANL{b}, $n_{J}=6$ in \PANL{c} and $n_{J}=15$ in \PANL{d} to model the dynamics of the system. All pulse parameters are listed in \tabref{tab:summary_circuit_hamiltonian_results}.}\label{fig:NA_cir_ISWAP_cases_chalmers}
\end{figure}

\graphicspath{{./FiguresAndData/NAPaper/CircuitHamiltonianGaugeSimulations/ArchitectureI/}}
\begin{figure}[!tbp]
    \centering
    \begin{minipage}{0.49\textwidth}
        \centering
        \includegraphics[width=\width\textwidth]{TimeEvolCZStates3Beta_0_5}
        \includegraphics[width=\width\textwidth]{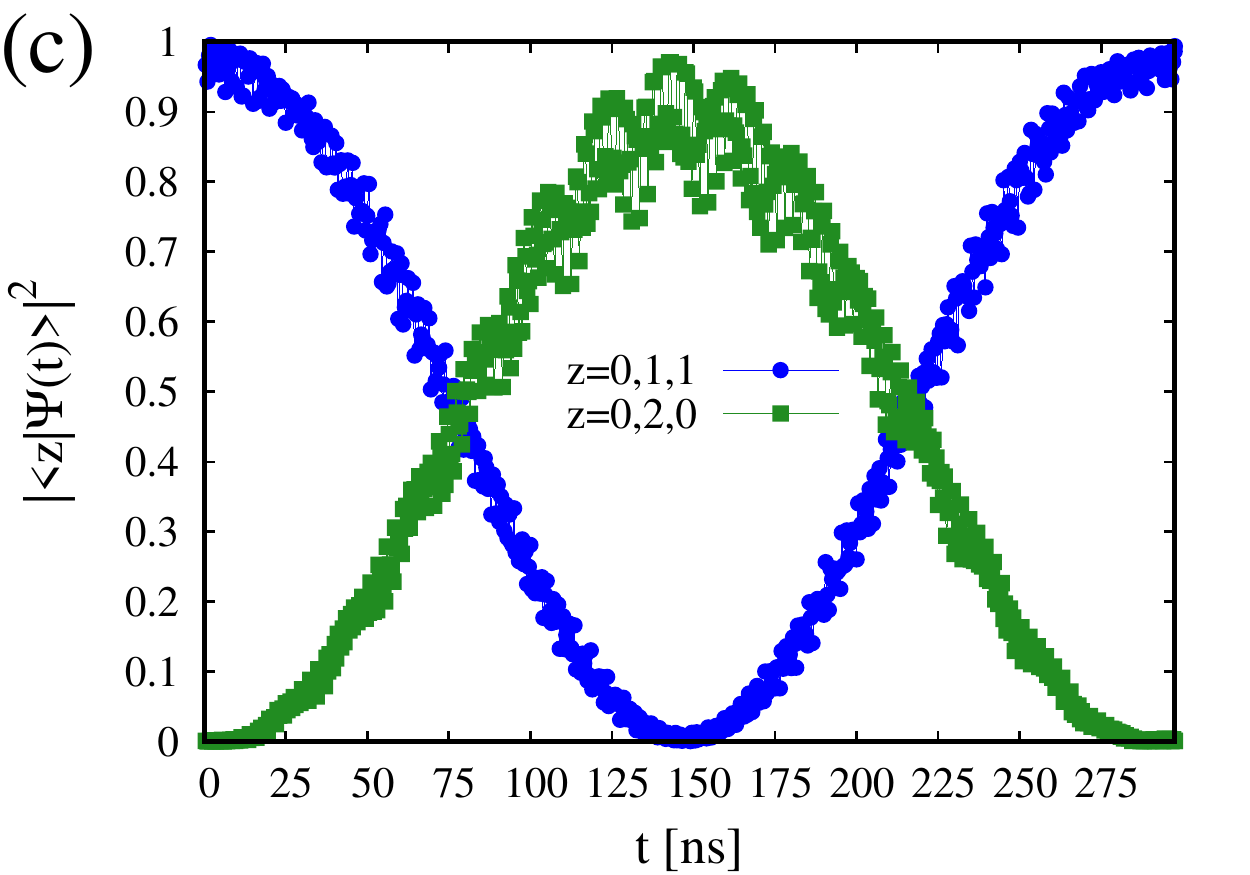}
    \end{minipage}\hfill
    \begin{minipage}{0.49\textwidth}
        \centering
        \includegraphics[width=\width\textwidth]{TimeEvolCZStates4Beta_0_5}
        \includegraphics[width=\width\textwidth]{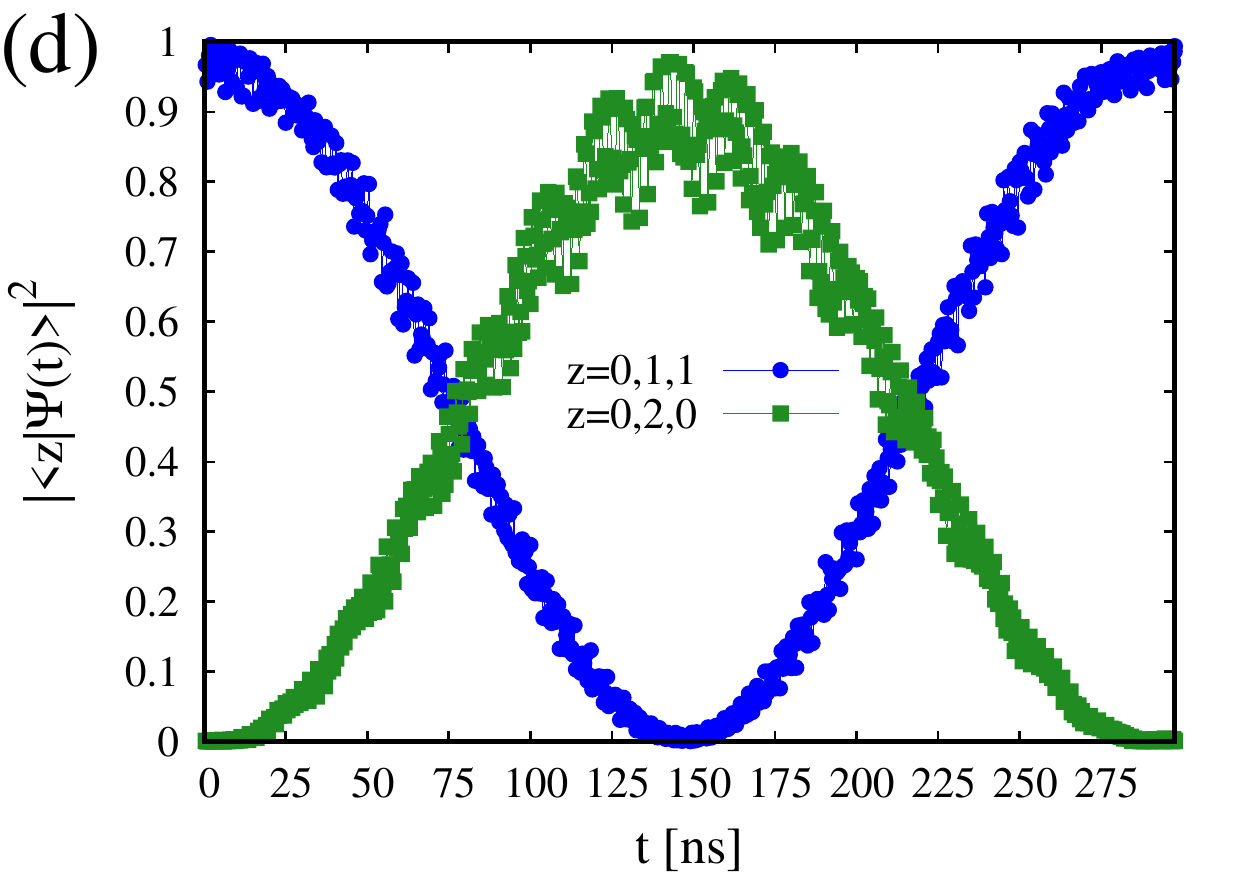}
    \end{minipage}
    \caption[Probabilities $p^{\mathbf{z}}(t)=|\braket{\mathbf{z}|\Psi^{\hold}(t)}|^{2}$ as functions of time $t$ for $\mathbf{z}=(0,1,1)$ and $\mathbf{z}=(0,2,0)$ (circuit model architecture I).]{Probabilities $p^{\mathbf{z}}(t)=|\braket{\mathbf{z}|\Psi^{\hold}(t)}|^{2}$ as a functions of time $t$ for $\mathbf{z}=(0,1,1)$ and $\mathbf{z}=(0,2,0)$. Here, $\ket{\Psi^{\hold}(0)}=\ket{0,1,1}$.  We use the circuit Hamiltonian \equref{eq:CHM}, the device parameters listed in \tabref{tab:device_parameter_flux_tunable_coupler_chip} and the pulse given by \equref{eq:NA_control_pulse} to obtain the results. The pulse parameters are the drive frequency $\DFTP=0.809$ GHz, the amplitude $\PATP=0.085$, the rise and fall time $\TRF=13.00$ ns and the pulse duration $\TD=297.55$ ns. Furthermore, $n_{J}$ denotes the number of flux-tunable transmon basis states and we use $n_{J}=3$ in \PANL{a}, $n_{J}=4$ in \PANL{b}, $n_{J}=8$ in \PANL{c} and $n_{J}=15$ in \PANL{d} to model the dynamics of the system. All pulse parameters are listed in \tabref{tab:summary_circuit_hamiltonian_results}.}\label{fig:NA_cir_cz_cases_chalmers}
\end{figure}

Figures \ref{fig:NA_cir_ISWAP_cases_chalmers}(a-d) show the probabilities $p^{\mathbf{z}}(t)$ for $\mathbf{z}=(0,0,1)$ and $\mathbf{z}=(0,1,0)$ as functions of time $t$. Here, $p^{\mathbf{z}}(t)=|\braket{\mathbf{z}|\Psi^{\hold}(t)}|^{2}$ and $\ket{\Psi^{\hold}(0)}=\ket{0,0,1}$. We use the circuit Hamiltonian \equref{eq:CHM}, the device parameters listed in \tabref{tab:device_parameter_flux_tunable_coupler_chip} and the pulse given by \equref{eq:NA_control_pulse} to describe the dynamics of the system. The pulse is modelled with the drive frequency $\DFTP=1.089$ GHz, the pulse amplitude $\PATP=0.075$, the rise and fall time $\TRF=13.00$ ns and the pulse duration $\TD=209.40$ ns. We model the transition with different numbers of basis states $n_{J}$. We use $n_{J}=3$ in \figref{fig:NA_cir_ISWAP_cases_chalmers}(a), $n_{J}=4$ in \figref{fig:NA_cir_ISWAP_cases_chalmers}(b), $n_{J}=6$ in \figref{fig:NA_cir_ISWAP_cases_chalmers}(c) and $n_{J}=15$ \figref{fig:NA_cir_ISWAP_cases_chalmers}(d).

As can be seen, \figsref{fig:NA_cir_ISWAP_cases_chalmers}(a-d) show an exchange of population between the states $\mathbf{z}=(0,0,1)$ and $\mathbf{z}=(0,1,0)$. However, we observe that the time evolution of the probabilities changes with the number of basis states $n_{J}$ we use to model the dynamics of the system. Figures \ref{fig:NA_cir_ISWAP_cases_chalmers}(a-b) show that at time $T_{d}$ there is no full swap between the states $\mathbf{z}=(0,0,1)$ and $\mathbf{z}=(0,1,0)$. In \figref{fig:NA_cir_ISWAP_cases_chalmers}(c) the solution of the TDSE has converged up to the third decimal. Figure \ref{fig:NA_cir_ISWAP_cases_chalmers}(d) functions as a graphical reference solution,\ie we cannot find any noticeable differences between  \figref{fig:NA_cir_ISWAP_cases_chalmers}(c) and \figref{fig:NA_cir_ISWAP_cases_chalmers}(d). In this case it suffices to describe the dynamics of the system with $n_{J}=6$ basis states. Furthermore, the functions $p^{\mathbf{z}}(t)$ in \figsref{fig:NA_cir_ISWAP_cases_chalmers}(a-d) exhibit small oscillations at the order of nanoseconds. The oscillations seem to increase with the amount of population which resides in the corresponding state $\mathbf{z}$.

Figures \ref{fig:NA_cir_cz_cases_chalmers}(a-d) show the probabilities $p^{\mathbf{z}}(t)=|\braket{\mathbf{z}|\Psi^{\hold}(t)}|^{2}$ for $\mathbf{z}=(0,1,1)$ and $\mathbf{z}=(0,2,0)$ as functions of time $t$. The initial state of the system is $\ket{\Psi^{\hold}(0)}=\ket{0,1,1}$. We use the same Hamiltonian, device parameters and pulse as in \figsref{fig:NA_cir_ISWAP_cases_chalmers}(a-d). The pulse is modelled with the drive frequency $\DFTP=0.809$ GHz, the pulse amplitude $\PATP=0.085$, the rise and fall time $\TRF=13.00$ ns and the pulse duration $\TD=297.55$ ns. Furthermore, we use $n_{J}=3$ in \figref{fig:NA_cir_cz_cases_chalmers}(a), $n_{J}=4$ in \figref{fig:NA_cir_cz_cases_chalmers}(b), $n_{J}=8$ in \figref{fig:NA_cir_cz_cases_chalmers}(c) and $n_{J}=15$ \figref{fig:NA_cir_cz_cases_chalmers}(d).

Figures \ref{fig:NA_cir_cz_cases_chalmers}(a-d) exhibit similar features as \figsref{fig:NA_cir_ISWAP_cases_chalmers}(a-d). First, we observe that we need more than three or four basis states to model the dynamics of the system. In this case we have to use at least $n_{J}=8$ basis states. Second, we see that the functions $p^{\mathbf{z}}(t)$ oscillate at the order of nanoseconds. We also observe that the oscillations in \figsref{fig:NA_cir_cz_cases_chalmers}(c-d) at around $150$ ns seem to be stronger than everything we observe in \figsref{fig:NA_cir_ISWAP_cases_chalmers}(c-d).

We now turn our attention to the effective Hamiltonian model. As previously discussed, see \secaref{sec:NA_single_flux_tunable_transmon}{sec:NA_E_suppressed}, we might use different functional dependencies to model the tunable energies and interaction strengths. For example, in \REFS\cite{McKay16,Roth19,Roth20,Ganzhorn20}, the authors use models where the interaction strengths are assumed to be constant. Furthermore, here the energies are modelled with $\omega^{(q)}(\varphi)$ given by \equref{eq: frequency} and $\alpha^{(q_{0})}(\varphi)=\const{}$ The simulation software allows us to switch between different model options or rather assumptions and in the following we intend to make use of this functionality. This means we perform simulations of all four modelling scenarios: simulations with and without an adjusted spectrum as well as simulations with and without time-dependent interaction strengths. Here we use the adiabatic effective Hamiltonian \equref{eq:EHM}, the pulse given by \equref{eq:NA_control_pulse} and four basis states for every transmon to model the two-qubit system with the effective model. Furthermore, we either use the device parameters listed in \tabref{tab:device_parameter_flux_tunable_coupler_chip} or the parameters listed in \tabref{tab:device_parameter_flux_tunable_coupler_chip_effective} to model the dynamics. The parameters in \tabref{tab:device_parameter_flux_tunable_coupler_chip} are necessary once we use \equref{eq: expansion frequency} and \equref{eq: expansion anharmonicity} to model the tunable energies of the system or once we use \equref{eq:eff_int_trans_trans_second_time} to model the time-dependent interaction strengths. Otherwise, we can use the parameters in \tabref{tab:device_parameter_flux_tunable_coupler_chip_effective}.

For the device parameters and the control pulse we use, we find (data not shown) that the adjusted spectrum barely affects the simulation results. Therefore, we only present the results for cases where $\omega^{(q)}(\varphi)$ given by \equref{eq: frequency} and $\alpha^{(q_{0})}(\varphi)=\const$ are used to model the tunable energies of the flux-tunable transmon $z_{2}$. Still, we present the results for both interaction strength cases.

\begin{figure}[!tbp]
    \centering
    \begin{minipage}{0.49\textwidth}
        \centering
        \graphicspath{{./FiguresAndData/NAPaper/InteractionStrength/}}
        \includegraphics[width=\width\textwidth]{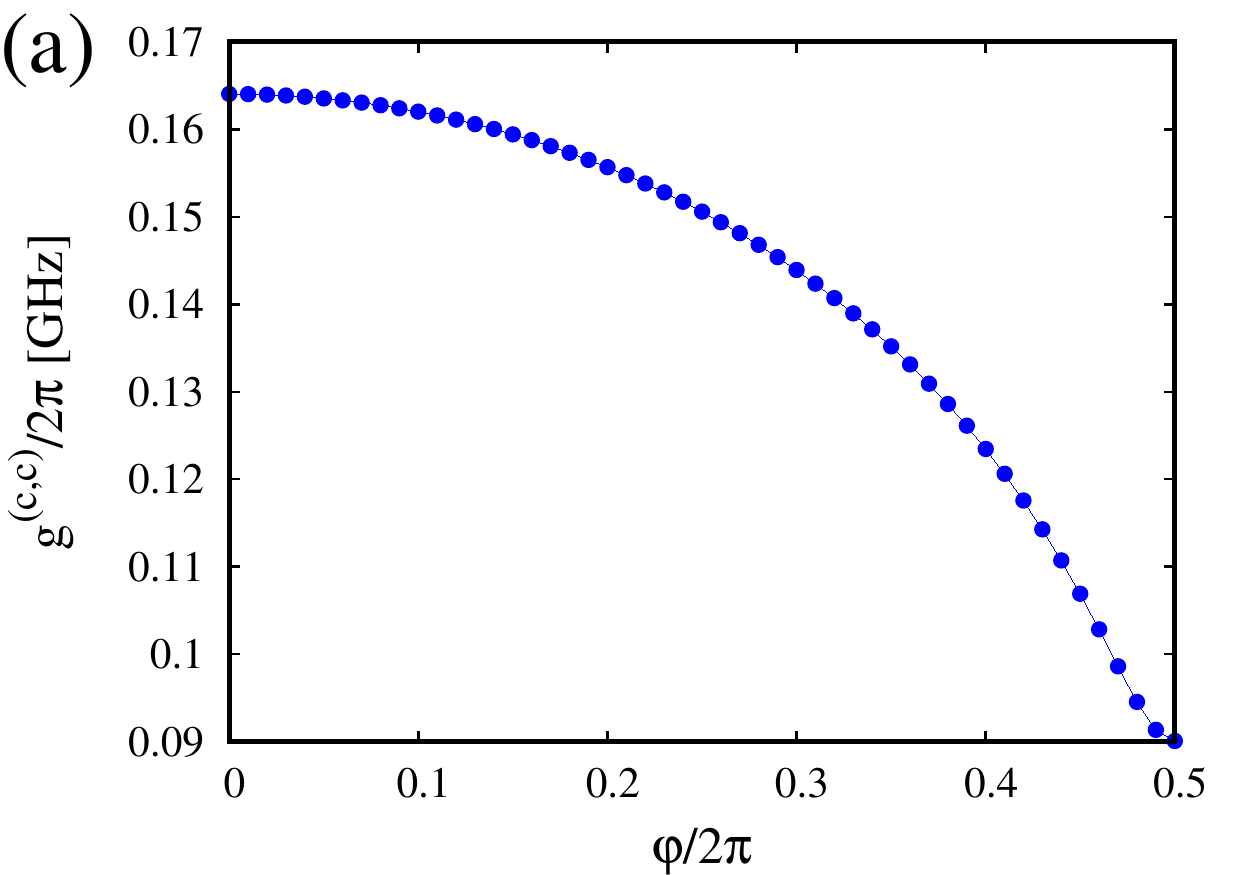}
    \end{minipage}\hfill
    \begin{minipage}{0.49\textwidth}
        \centering
        \graphicspath{{./FiguresAndData/NAPaper/EffIntAndPulse/}}
        \includegraphics[width=\width\textwidth]{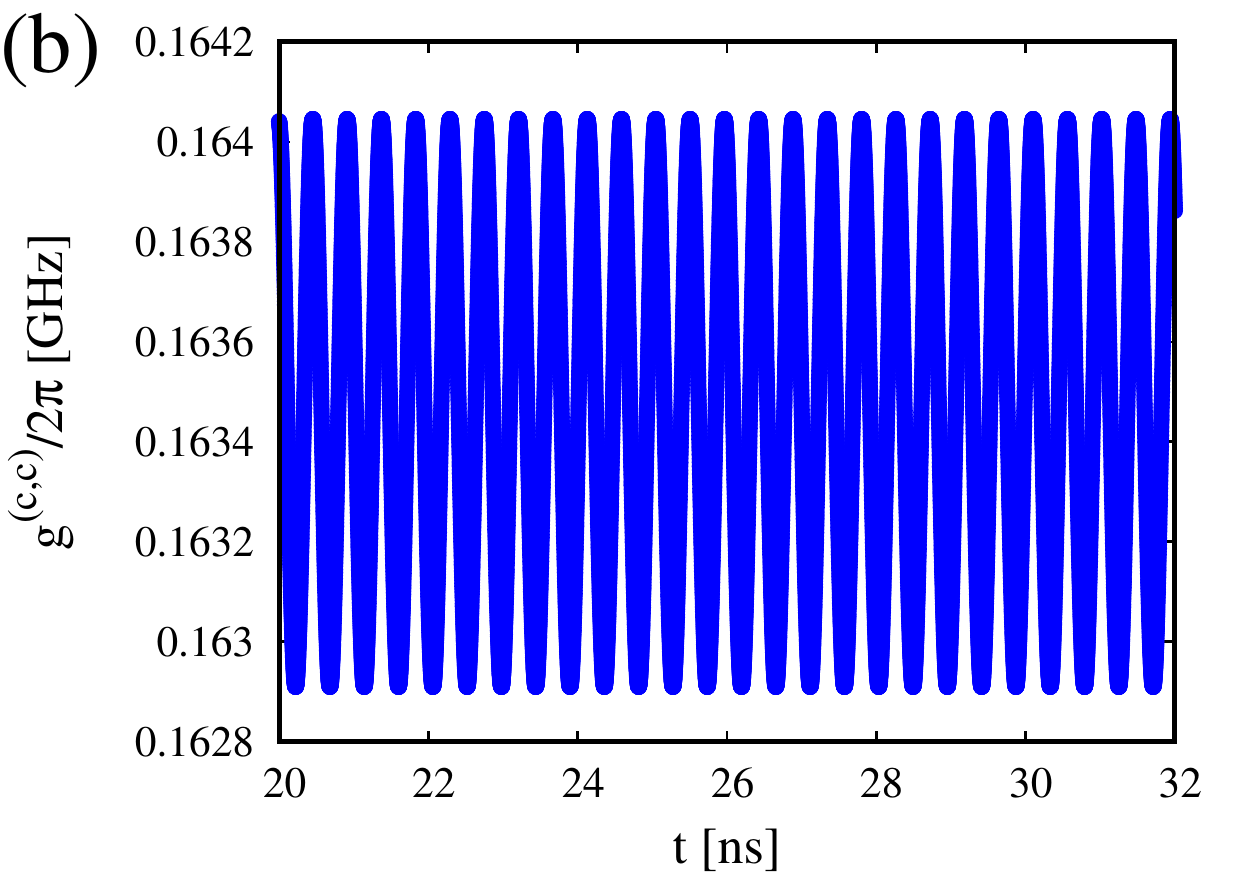}
    \end{minipage}
    \caption[Effective interaction strength $g^{(c,c)}$ between a fixed-frequency and a flux-tunable transmon as a function of the external flux $\varphi$ in \PANL{a} and as a function of time $t$ in \PANL{b}.]{Effective interaction strength $g^{(c,c)}$ between a fixed-frequency and a flux-tunable transmon as a function of the external flux $\varphi$ in \PANL{a} and as a function of time $t$ in \PANL{b}. We use \equref{eq:eff_int_trans_trans_second_time} and the energies listed in \tabref{tab:device_parameter_flux_tunable_coupler_chip}, row $i=1$ and $i=2$, to obtain the numerical values for $g^{(c,c)}$. In \PANL{a} we compute the effective interaction strength $g^{(c,c)}(\varphi)$ for the interval $\varphi/2\pi \in [0,0.5]$. In \PANL{b} we use the control pulse $\varphi(t)$ given by \equref{eq:NA_control_pulse} with the flux-offset value $\varphi_{0}/2\pi=0.15$, the drive frequency $\DFTP=1.089$ GHz, the pulse amplitude $\PATP=0.075$, the rise and fall time $\TRF=13.0$ ns and the pulse duration $\TD=205.4$ ns to obtain the time evolution of the effective interaction strength.}\label{fig:NA_eff_int_flux_evolution_archI}
\end{figure}

Figure \ref{fig:NA_eff_int_flux_evolution_archI}(a) shows the interaction strength $g^{(c,c)}$ as a function of the flux variable $\varphi$. Similarly, in \figref{fig:NA_eff_int_flux_evolution_archI}(b) we show the interaction strength $g^{(c,c)}$ as a function of time $t$. We use \equref{eq:eff_int_trans_trans_second_time} and the energy parameters listed in \tabref{tab:device_parameter_flux_tunable_coupler_chip}, row $i=1$ and $i=2$ to obtain $g^{(c,c)}$ for various flux values $\varphi$. The time evolution in \figref{fig:NA_eff_int_flux_evolution_archI}(b) is modelled with the pulse $\varphi(t)$ given by \equref{eq:NA_control_pulse}. Here we use the flux-offset value $\varphi_{0}/2\pi=0.15$, the drive frequency $\DFTP=1.089$ GHz, the pulse amplitude $\PATP=0.075$, the rise and fall time $\TRF=13.0$ ns and the pulse duration $\TD=205.4$ ns.

\renewcommand{\hold}{}
\graphicspath{{./FiguresAndData/NAPaper/EffectiveHamiltonian/TwoQubitsWithCoupler/}}
\begin{figure}[!tbp]
    \centering
    \begin{minipage}{0.49\textwidth}
        \centering
        \includegraphics[width=\width\textwidth]{NoTimeDep/te_ISWAP}
        \includegraphics[width=\width\textwidth]{NoTimeDep/te_cz}
    \end{minipage}\hfill
    \begin{minipage}{0.49\textwidth}
        \centering
        \includegraphics[width=\width\textwidth]{TimeDep/te_ISWAP}
        \includegraphics[width=\width\textwidth]{TimeDep/te_cz}
    \end{minipage}
    \caption[Probabilities $p^{\mathbf{z}}(t)=|\braket{\mathbf{z}|\Psi^{\hold}(t)}|^{2}$ as functions of time $t$ (effective model architecture I).]{Probabilities $p^{\mathbf{z}}(t)=|\braket{\mathbf{z}|\Psi^{\hold}(t)}|^{2}$ as functions of time $t$. In \PANSL{a,b} we model the transitions $\mathbf{z}=(0,0,1) \rightarrow \mathbf{z}=(0,1,0)$. In \PANSL{c,d} we model the transitions $\mathbf{z}=(0,1,1) \rightarrow \mathbf{z}=(0,2,0)$. We use the effective Hamiltonian \equref{eq:EHM}, the device parameters listed in \tabref{tab:device_parameter_flux_tunable_coupler_chip_effective} and the pulse given by \equref{eq:NA_control_pulse} with the rise and fall time $\TD=13$ ns to obtain the results. The remaining, non-specified pulse parameters are the drive frequency $\DF$, the amplitude $\PA$ and the pulse duration $\TD$. Most of these parameters change for every panel we show. We use $\DFTP=1.088$ GHz, $\PATP=0.075$ and $\TD=139.6$ ns in \PANL{a}, $\DFTP=1.089$ GHz, $\PATP=0.075$ and $\TD=205.4$ ns in \PANL{b}, $\DFTP=0.807$ GHz, $\PATP=0.085$ and $\TD=196.5$ ns in \PANL{c} as well as $\DFTP=0.807$ GHz, $\PATP=0.085$ and $\TD=272.0$ ns in \PANL{d}. The first-order series expansion is used to model the tunable qubit frequency for all panels. In \PANSL{a,c} we use $g^{(c,c)}(t)=\const$ to model a static effective interaction strength. Similarly, in \PANSL{b,d} we use $g^{(c,c)}(t)$ given by \equref{eq:eff_int_trans_trans_second_time} to model a dynamic effective interaction strength, see also \figref{fig:NA_eff_int_flux_evolution_archI}(b). All pulse parameters and cases are listed in \tabref{tab:summary_effective_hamiltonian_results}.}\label{fig:NA_eff_cz_ISWAP_cases_chalmers}
\end{figure}

As can be seen in \figref{fig:NA_eff_int_flux_evolution_archI}(a), the interaction strength varies by about $75$ MHz over the range $\varphi/2\pi \in [0,0.5]$. Furthermore, in \figref{fig:NA_eff_int_flux_evolution_archI}(b) we observe a variation of the interaction strength which is at the order of $1$ MHz. The control pulse $\varphi(t)$ we use to model the time evolution of $g^{(c,c)}$ is also used to model the transition $\mathbf{z}=(0,0,1) \rightarrow \mathbf{z}=(0,1,0)$ with the effective model, see \tabref{tab:summary_effective_hamiltonian_results} row six. The results in \figref{fig:NA_eff_int_flux_evolution_archI}(b) suggest that the assumption of a static interaction strength, see \REFS\cite{McKay16,Roth19,Roth20,Ganzhorn20}, might be justified. In the following, we show two counterexamples.

Figures \ref{fig:NA_eff_cz_ISWAP_cases_chalmers}(a-d) show the probabilities $p^{\mathbf{z}}(t)=|\braket{\mathbf{z}|\Psi^{\hold}(t)}|^{2}$ as functions of time $t$. In \figsref{fig:NA_eff_cz_ISWAP_cases_chalmers}(a-b) we model the $\mathbf{z}=(0,0,1) \rightarrow \mathbf{z}=(0,1,0)$ transition with, see \figref{fig:NA_eff_cz_ISWAP_cases_chalmers}(b) and without, see \figref{fig:NA_eff_cz_ISWAP_cases_chalmers}(a), a time-dependent interaction strength. Similarly, in \figsref{fig:NA_eff_cz_ISWAP_cases_chalmers}(c-d) we model the $\mathbf{z}=(0,1,1) \rightarrow \mathbf{z}=(0,2,0)$ transition with, see \figref{fig:NA_eff_cz_ISWAP_cases_chalmers}(d) and without, see \figref{fig:NA_eff_cz_ISWAP_cases_chalmers}(c), a time-dependent interaction strength.

We determined the pulse parameters for \figsref{fig:NA_eff_cz_ISWAP_cases_chalmers}(a-d) as follows. First, we fix the pulse amplitude $\PA$ and rise and fall time $\TRF$. Here, we use the parameters we found for the circuit Hamiltonian model, see \tabref{tab:summary_circuit_hamiltonian_results}. Second, we perform a drive frequency $\DF$ spectroscopy and determined the frequency at the center of the chevron pattern, see also \figref{fig:NA_supressed_chevron_pattern}(a-b). This means we use the frequency at the center of the chevron pattern as the drive frequency. At last, we determine the pulse duration $T_{d}$ for this set of parameters. The pulse parameters are the rise and fall time $\TRF=13.0$ ns in \figsref{fig:NA_eff_cz_ISWAP_cases_chalmers}(a-d), $\DFTP=1.088$ GHz, $\PATP=0.075$ and $\TD=139.6$ ns in \figref{fig:NA_eff_cz_ISWAP_cases_chalmers}(a), $\DFTP=1.089$ GHz, $\PATP=0.075$ and $\TD=205.4$ ns in \figref{fig:NA_eff_cz_ISWAP_cases_chalmers}(b), $\DFTP=0.807$ GHz, $\PATP=0.085$ and $\TD=196.5$ ns in \figref{fig:NA_eff_cz_ISWAP_cases_chalmers}(c) as well as $\DFTP=0.807$ GHz, $\PATP=0.085$ and $\TD=272.0$ ns in \figref{fig:NA_eff_cz_ISWAP_cases_chalmers}(d). A summary of these parameters is provided in \tabref{tab:summary_effective_hamiltonian_results}, row five to eight.

We first the discuss the results for the $\mathbf{z}=(0,0,1) \rightarrow \mathbf{z}=(0,1,0)$ transition. As one can see, the probabilities in \figsref{fig:NA_eff_cz_ISWAP_cases_chalmers}(a-b) show a qualitatively similar behaviour. However, we can observe a quantitative difference between both results, namely that the pulse duration $\TD$ in \figref{fig:NA_eff_cz_ISWAP_cases_chalmers}(b) is roughly $65.8$ ns longer. Note that in \figref{fig:NA_eff_cz_ISWAP_cases_chalmers}(b) we model the system with a time-dependent interaction strength. Furthermore, the circuit Hamiltonian model predicts a gate duration $\TD$ of $209.4$ ns. Consequently, there is a $4$ ns discrepancy between the circuit Hamiltonian model and the effective model with a time-dependent interaction strength and a $69.8$ ns difference between the circuit Hamiltonian model and the effective model without a time-dependent interaction strength.

Now we turn our attention to the $\mathbf{z}=(0,1,1) \rightarrow \mathbf{z}=(0,2,0)$ transition. Figures \ref{fig:NA_eff_cz_ISWAP_cases_chalmers}(c-d) exhibit similar features as \figsref{fig:NA_eff_cz_ISWAP_cases_chalmers}(a-b). First, we find that \figsref{fig:NA_eff_cz_ISWAP_cases_chalmers}(c-d) show a similar qualitative behaviour in terms of the time evolution of the probabilities $p^{\mathbf{z}}(t)$. Second, we see a shift in the pulse duration $\TD$ between both effective models. The pulse duration difference between both effective models is $75.5$ ns. For this case, the circuit Hamiltonian predicts a gate duration of $\TD=297.55$ ns. Therefore, we find a deviation of $25.55$ ns between the effective model with time-dependent interaction strength and the circuit model and a difference of $101.05$ ns between the effective model without a time-dependent interaction strength and the circuit model.

\renewcommand{\width}{1.0}
\begin{figure}[!tbp]
  \begin{minipage}{\textwidth}
    \centering
    \begin{tikzpicture}[thick,scale=0.45, every node/.style={transform shape}]
	       \begin{pgfonlayer}{nodelayer}
		\node [style=none] (8) at (6, 7) {};
		\node [style=none] (9) at (10, 7) {};
		\node [style=none] (10) at (8, 9) {};
		\node [style=none] (11) at (8, 5) {};
		\node [style=none] (16) at (-2, 0) {};
		\node [style=none] (17) at (2, 0) {};
		\node [style=none] (18) at (0, 2) {};
		\node [style=none] (19) at (0, -2) {};
		\node [style=none] (20) at (14, 0) {};
		\node [style=none] (21) at (18, 0) {};
		\node [style=none] (22) at (16, 2) {};
		\node [style=none] (23) at (16, -2) {};
		\node [style=none] (26) at (8, 7) {\Huge $\omega_{2}^{(q)}(t)$};
		\node [style=none] (29) at (0, 0) {\Huge $\omega_{0}^{(q_{0})}$};
		\node [style=none] (30) at (16, 0) {};
		\node [style=none] (32) at (16, 0) {\Huge $\omega_{1}^{(q_{0})}$};
		\node [style=none] (46) at (8, 10) {};
		\node [style=none] (47) at (8, 10) {\Huge \textbf{Flux-tunable transmon} $i=2$};
		\node [style=none] (49) at (0, -3) {\Huge \textbf{Fixed-frequency transmon} $i=0$};
		\node [style=none] (51) at (16, -3) {\Huge \textbf{Fixed-frequency transmon} $i=1$};
		\node [style=none] (60) at (4, 3) {};
		\node [style=none] (61) at (3, 3) {\Huge $g_{2,0}^{(c,c)}(t)$};
		\node [style=none] (62) at (12, 3) {};
		\node [style=none] (63) at (13, 3) {\Huge $g_{2,1}^{(c,c)}(t)$};
		\node [style=none] (65) at (0, 7) {};
		\node [style=none] (66) at (0, 7) {\Huge $\varphi^{*}(t)=\varphi_{0}+\delta^{*} e(t) \cos(\omega^{(D)} t)$};
		\node [style=none] (67) at (16, 7) {};
		\node [style=none] (68) at (16, 7) {\Huge $\varphi(t)=\varphi_{0}+\delta e(t) \cos(\omega^{(D)} t)$};
		\node [style=none] (71) at (10.75, 3.0) {};
		\node [style=none] (72) at (1.5, 3.0) {};
		\node [style=none] (73) at (-4.5, 6.5) {};
		\node [style=none] (74) at (9, 7) {};
		\node [style=none] (75) at (11.0, 7) {};
		\node [style=none] (76) at (16, 6) {};
		\node [style=none] (77) at (-2.25, 8.5) {\Huge $\frac{\delta^{*} }{2\pi} \in [0,0.125]$};
		\node [style=none] (78) at (13.5, 8.5) {\Huge $\frac{\delta}{2\pi} =$const.};
	\end{pgfonlayer}
	       \begin{pgfonlayer}{edgelayer}
      		\draw [bend left=45,line width=\lw] (8.center) to (10.center);
      		\draw [bend left=45,line width=\lw] (11.center) to (8.center);
      		\draw [bend left=45,line width=\lw] (10.center) to (9.center);
      		\draw [bend left=45,line width=\lw] (9.center) to (11.center);
      		\draw [bend left=45,line width=\lw] (16.center) to (18.center);
      		\draw [bend left=45,line width=\lw] (19.center) to (16.center);
      		\draw [bend left=45,line width=\lw] (18.center) to (17.center);
      		\draw [bend left=45,line width=\lw] (17.center) to (19.center);
      		\draw [bend left=45,line width=\lw] (20.center) to (22.center);
      		\draw [bend left=45,line width=\lw] (23.center) to (20.center);
      		\draw [bend left=45,line width=\lw] (22.center) to (21.center);
      		\draw [bend left=45,line width=\lw] (21.center) to (23.center);
      		\draw [line width=\lw]              (11.center) to (17.center);
      		\draw [line width=\lw]              (11.center) to (20.center);
      		\draw [->,line width=\lw]              (73.center) to (72.center);
      		\draw [->,line width=\lw]              (73.center) to (71.center);
      		\draw [<-,line width=\lw]             (74.center) to (75.center);
	\end{pgfonlayer}
    \end{tikzpicture}
  \end{minipage}
  \caption[Sketch of the simulation scenario we use to obtain the results shown in \figsref{fig:eff_int_scaling}(a-b).]{Sketch of the simulation scenario we use to obtain the results shown in \figsref{fig:eff_int_scaling}(a-b). The sketch shows architecture I, see \figref{fig:arch_sketch}(a), with different time dependencies  $\varphi^{*}(t)$ and $\varphi(t)$ given by \equref{eq:NA_control_pulse} for the control pulse or external flux. We model the time-dependent interaction strengths $g^{(c,c)}(t)$ given by \equref{eq:eff_int_trans_trans_second_time} with the pulse $\varphi^{*}(t)$ and the tunable frequency $\omega^{(q)}(t)$ with the pulse $\FP$. The only difference between these two pulses is that we model the pulse $\varphi^{*}(t)$ with the amplitudes $\delta^{*}/2\pi \in [0,0.125]$ and the pulse $\varphi(t)$ with the amplitude $\PATP=\text{const}$. This allows us to observe the transitioning behaviour from a model with a static interaction strength $\delta^{*}=0$ to a model with a dynamic interaction strength $\delta^{*}=\delta$. We show the results for this scenario because the combined effects of changing the coupler amplitude $\delta$ and the interaction strength amplitude $\delta^{*}$ together mixes up two different mechanisms which affect the \ISWAP{} and \CZ{} transitions. Furthermore, in \figref{fig:eff_int_scaling}(a) we use the same amplitude $\PATP=0.075$ we use to model the \ISWAP{} transitions in \figref{fig:NA_eff_cz_ISWAP_cases_chalmers}(a-b). Similarly, in \figref{fig:eff_int_scaling}(b) we use the same amplitude $\PATP=0.085$ we use to model the \CZ{} transitions in \figref{fig:NA_eff_cz_ISWAP_cases_chalmers}(c-d).}\label{fig:sketch_int_strength_sim}
\end{figure}

Next, we investigate how the time-dependent interaction strength $g^{(c,c)}(t)$ affects the gate duration $\TD$ for the \ISWAP{} and \CZ{} transitions in more detail. To this end, we implement the simulation scenario presented in \figref{fig:sketch_int_strength_sim}. The results for this scenario are displayed in \figsref{fig:eff_int_scaling}(a-b). The sketch in \figref{fig:sketch_int_strength_sim} shows architecture I with two different control pulses $\varphi(t)$ and $\varphi^{*}(t)$. Both pulses only deviate in terms of the amplitudes,\ie we model the pulse $\varphi^{*}(t)$ with the amplitudes $\delta^{*} \in [0,0.125]$ and the pulse $\varphi(t)$ with the amplitude $\delta=\const$. The pulse $\varphi^{*}(t)$ is used to control the time-dependent interaction strengths $g^{(c,c)}(t)$ given by \equref{eq:eff_int_trans_trans_second_time}. We use the other pulse $\varphi(t)$ to control the tunable frequency $\omega^{q}(t)$ of the coupler element. In order to obtain the results in \figsref{fig:eff_int_scaling}(a-b), we use the amplitudes $\PATP=0.075$(a) and $\PATP=0.085$(b). This means, if the pulse amplitude is set to zero $\delta^{*}=0$, we model the cases we present in \figsref{fig:NA_eff_cz_ISWAP_cases_chalmers}(a,c). Furthermore, if we set both pulse amplitudes equal $\delta^{*}=\delta$, we model the cases we present in \figsref{fig:NA_eff_cz_ISWAP_cases_chalmers}(b,d). All pulses are modelled with the rise and fall time $\TRF=13$ ns and the pulse duration $\TD=300$ ns. This simulation scenario allows us to separate the two mechanisms which affect the \ISWAP{} and \CZ{} transitions. On the one hand, we have the time-dependent coupler which can be used to activate the \ISWAP{} and \CZ{} transitions in the adiabatic effective model without a time-dependent interaction strength. On the other hand, we have the time-dependent interaction strength which seems to affect the time evolution of the system.

\renewcommand{\width}{1.0}
\graphicspath{{./FiguresAndData/NAPaper/TdInterStrength/CZ/}{./FiguresAndData/NAPaper/TdInterStrength/ISWAP/}}
\begin{figure}[!tbp]
  \begin{minipage}{0.5\textwidth}
    \centering
    \includegraphics[width=\width\textwidth]{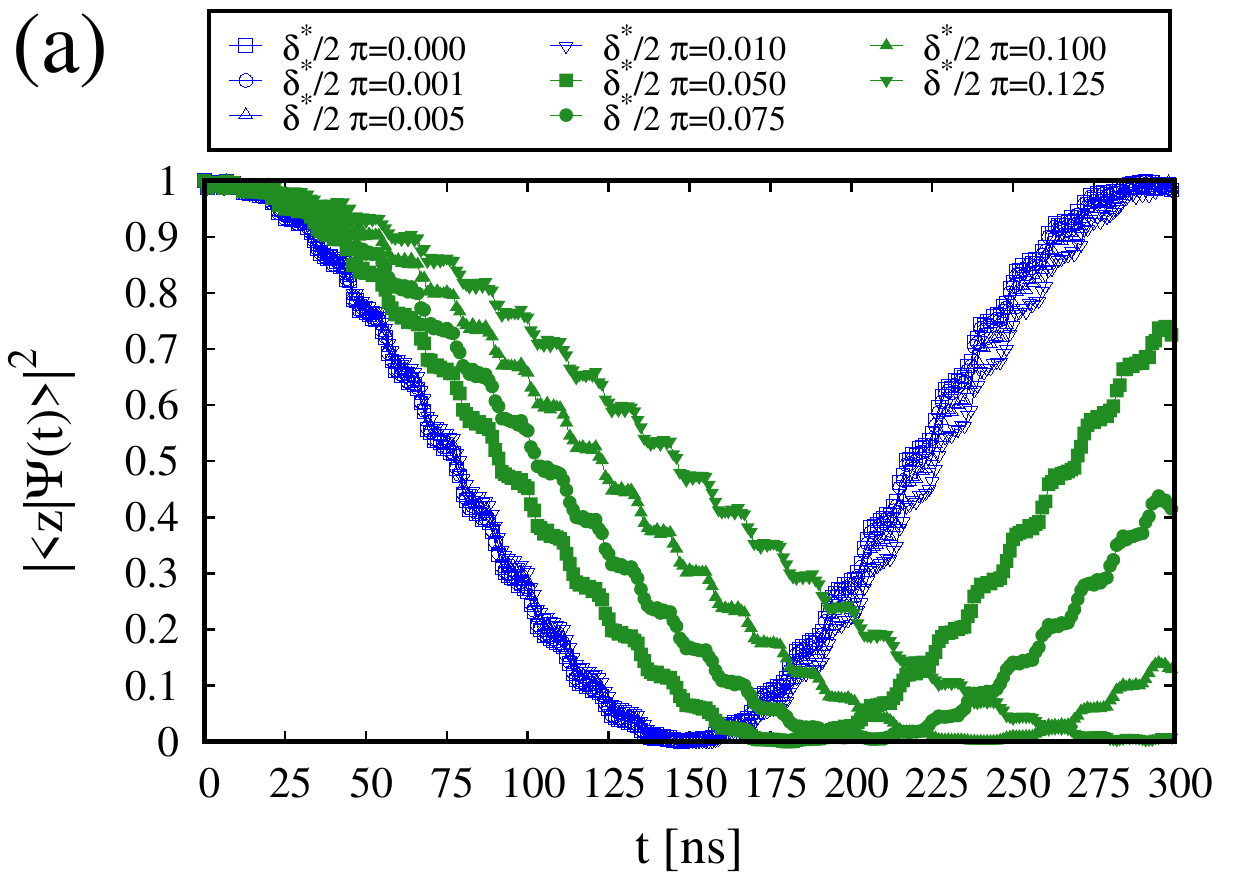}
  \end{minipage}
  \begin{minipage}{0.5\textwidth}
    \centering
    \includegraphics[width=\width\textwidth]{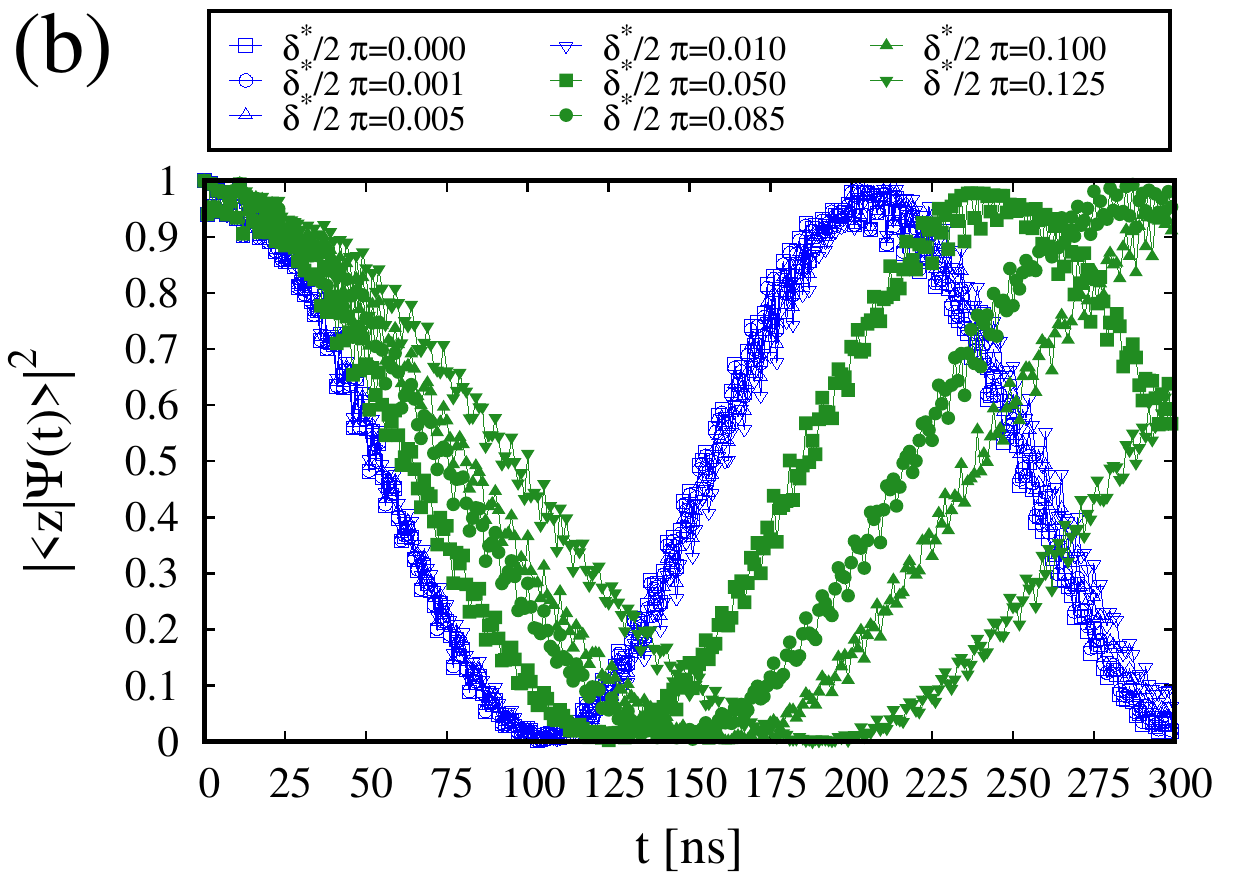}
  \end{minipage}
  \caption[Probabilities $p^{\mathbf{z}}(t)=|\braket{\mathbf{z}|\Psi^{\hold}(t)}|^{2}$ for $\mathbf{z}=(0,0,1)$(a) and $\mathbf{z}=(0,1,1)$(b) as functions of time $t$ (effective model architecture I).]{Probabilities $p^{\mathbf{z}}(t)=|\braket{\mathbf{z}|\Psi^{\hold}(t)}|^{2}$ for $\mathbf{z}=(0,0,1)$(a) and $\mathbf{z}=(0,1,1)$(b) as functions of time $t$. Here, we model the transition from a model with a static interaction strength to a model with a dynamic interaction strength as illustrated in \figref{fig:sketch_int_strength_sim}. This means we model the time-dependent interaction strength given by \equref{eq:eff_int_trans_trans_second_time} with the pulse $\varphi^{*}(t)$ in \equref{eq:NA_control_pulse} and the pulse amplitudes $\delta^{*}/2\pi \in [0,0.125]$. The tunable frequency of the coupler is modelled with the pulse $\varphi(t)$ in \equref{eq:NA_control_pulse}. However, here we use different constant pulse amplitudes $\PATP=0.075$(a) and $\PATP=0.085$(b). The remaining pulse parameters are the same for both pulses. The rise and fall time $\TRF=13$ ns and the pulse duration $\TD=300$ ns are the same for all pulses. In \PANL{a} we use the drive frequencies $\DFTP=1.088$, blue lines and unfilled markers and $\DFTP=1.089$, green lines and filled markers. Similarly, in \PANL{b} we use the drive frequencies $\DFTP=0.807$, blue lines and unfilled markers and $\DFTP=0.808$, green lines and filled markers. If $\delta^{*}=0$, we model the system with a static interaction strength. Therefore, we model the scenarios in \figsref{fig:NA_eff_cz_ISWAP_cases_chalmers}(a,c). If $\delta^{*}=\delta$, we model the scenarios in \figsref{fig:NA_eff_cz_ISWAP_cases_chalmers}(b,d).}\label{fig:eff_int_scaling}
\end{figure}

Figures \ref{fig:eff_int_scaling}(a-b) show the probabilities $p^{\mathbf{z}}(t)=|\braket{\mathbf{z}|\Psi(t)}|^{2}$ for $\mathbf{z}=(0,0,1)$(a) and $\mathbf{z}=(0,1,1)$(b) as functions of time $t$. In \PANL{a} we model $\mathbf{z}=(0,0,1) \rightarrow \mathbf{z}=(0,1,0)$ \ISWAP{} transitions with the drive frequencies $\DFTP=1.088$ GHz, blue lines and unfilled markers and $\DFTP=1.089$ GHz, green lines and filled markers. In \PANL{b} we model $\mathbf{z}=(0,1,1) \rightarrow \mathbf{z}=(0,2,0)$ \CZ{} transitions with the drive frequencies $\DFTP=0.807$ GHz, blue lines and unfilled markers and $\DFTP=0.808$ GHz, green lines and filled markers.

In \figref{fig:eff_int_scaling}(a), we see that for the pulse amplitudes $\delta^{*}/2\pi \in [0,0.01]$ there is barely a change in the first minima of function $p^{(0,0,1)}(t)$. However, for the pulse amplitudes $\delta^{*}/2\pi \in (0.01,0.125]$ we can observe increases of roughly $25$ ns for the different amplitudes. In \figref{fig:eff_int_scaling}(b), we find similar features for the function $p^{(0,1,1)}(t)$.

The results in \figsref{fig:NA_eff_cz_ISWAP_cases_chalmers}(a-d) and \figsref{fig:eff_int_scaling}(a-b) raise the question, why do the oscillations of the interaction strength $g^{(c,c)}(t)$ affect the gate duration $\TD$ so strongly? Despite performing additional simulations, we were not able to find a conclusive answer to this question. Consequently, we have to leave this question for future research.

There remains one other obvious open question in this section, namely does the non-adiabatic driving term in \equref{eq:drive_term_ftt_second_time} affect the \ISWAP{} and \CZ{} transitions we model in this section. Here, we talk about the driving term that we implicitly investigated in \secref{sec:NA_single_flux_tunable_transmon}. Broadly speaking, we can answer this question as follows: we find (data not shown) that the non-adiabatic driving term in \equref{eq:drive_term_ftt_second_time} barely affects the \ISWAP{} and \CZ{} transitions we model in this section. If we want to be precise, we have to say that the probabilities at time $\TD$ are at most affected by the third decimal and the overall qualitative transitioning behaviour does not seem to be affected at all. Note that we only verified this for the relevant pulse parameters in \tabref{tab:summary_effective_hamiltonian_results}. It is possible that larger drive frequencies $\DF$ and pulse amplitudes $\PA$ can cause larger deviations between the adiabatic and non-adiabatic effective models. Furthermore, if the drive frequency $\DF$ comes energetically near various other transitions, see \secref{sec:NA_E_suppressed}, we can expect to see additional deviations between the adiabatic and non-adiabatic effective Hamiltonian.

In summary, we find that for short time scales around $250$ ns, the adiabatic effective model with time-dependent interaction strength and the circuit Hamiltonian model predict similar outcomes in terms of the probabilities $p^{\mathbf{z}}(t)=|\braket{\mathbf{z}|\Psi^{\hold}(t)}|^{2}$ which result from the state vector $\ket{\Psi^{\hold}(t)}$ for the corresponding model. Additionally, the deviations between the effective model with static interaction strength and the circuit Hamiltonian model for the same scenarios seem to be to strong to be neglected.

\section{Simulations of suppressed transitions in the adiabatic effective two-qubit model: architecture II}\label{sec:NA_E_suppressed_AII}

In \secaref{sec:NA_E_suppressed}{sec:NA_unsuppressed_arch_I}, we studied the relation between the circuit and effective Hamiltonian models for a two-qubit system of type architecture I. In this section, we do the same, but we consider another two-qubit system, namely a system of type architecture II, see \figref{fig:arch_sketch}(b).

The system consists of two flux-tunable transmons which are coupled to a single resonator element. We use the positive discrete indices $z_{0}$ and $z_{1}$ to index the basis states of the first flux-tunable transmon, see \tabref{tab:device_parameter_resonator_coupler_chip} row $i=0$ and the second flux-tunable transmon, see \tabref{tab:device_parameter_resonator_coupler_chip} row $i=1$, respectively. The coupler basis states are indexed by means of the positive discrete index $z_{2}$, see \tabref{tab:device_parameter_resonator_coupler_chip} row $i=2$. The individual transmon and harmonic oscillator basis states are discussed in \secaref{sec:ResAndTLS}{sec:Transmons}, respectively. We address the bare basis states $\ket{\mathbf{z}}=\ket{z_{2},z_{1},z_{0}}$ of the two-qubit system by means of the tuples $\mathbf{z}=(z_{2},z_{1},z_{0})$. The ground $z_{0/1}=0$ and first-excited $z_{0/1}=1$ transmon states are used as qubit states. This choice is motivated by the work in \REFS\cite{Rol19,Blais2020circuit,Krinner2020}.

The simulations in this section are motivated by three observations. First, while performing simulations with the circuit Hamiltonian \equref{eq:CHM} to obtain the results described in \secref{sec:NA_unsuppressed_arch_II}, we noticed (data now shown) that the coupler states can be excited considerably. Note that the results in \secref{sec:NA_unsuppressed_arch_II} are obtained with a unimodal pulse, see \figref{fig:NA_pulse_time_evo}(b). Excitations of the coupling resonator usually occurred for pulses with fast rising (falling) pulse flanks. Second, the authors of \REF\cite{Rol19} neglect the coupler states by assumption. Note that the authors of \REF\cite{Rol19} investigate a system that consists of two flux-tunable transmons coupled via a coupling resonator. Here, a so-called bimodal pulse is used to drive the system. Third, in \secref{sec:NA_unsuppressed_arch_I} we found that certain transitions between the states of the system illustrated in \figref{fig:arch_sketch}(a) are seemingly suppressed in the adiabatic effective model for architecture I. In this section we study transitions which are seemingly suppressed in the adiabatic effective model for architecture II.
\renewcommand{\hold}{0.40}
\graphicspath{{./FiguresAndData/NAPaper/CircuitHamiltonianGaugeSimulations/ArchitectureII/SidebandTransitionsArchII/}}
\begin{figure}[!tbp]
    \centering
    \begin{minipage}{0.31\textwidth}
        \centering
        \includegraphics[scale=\hold]{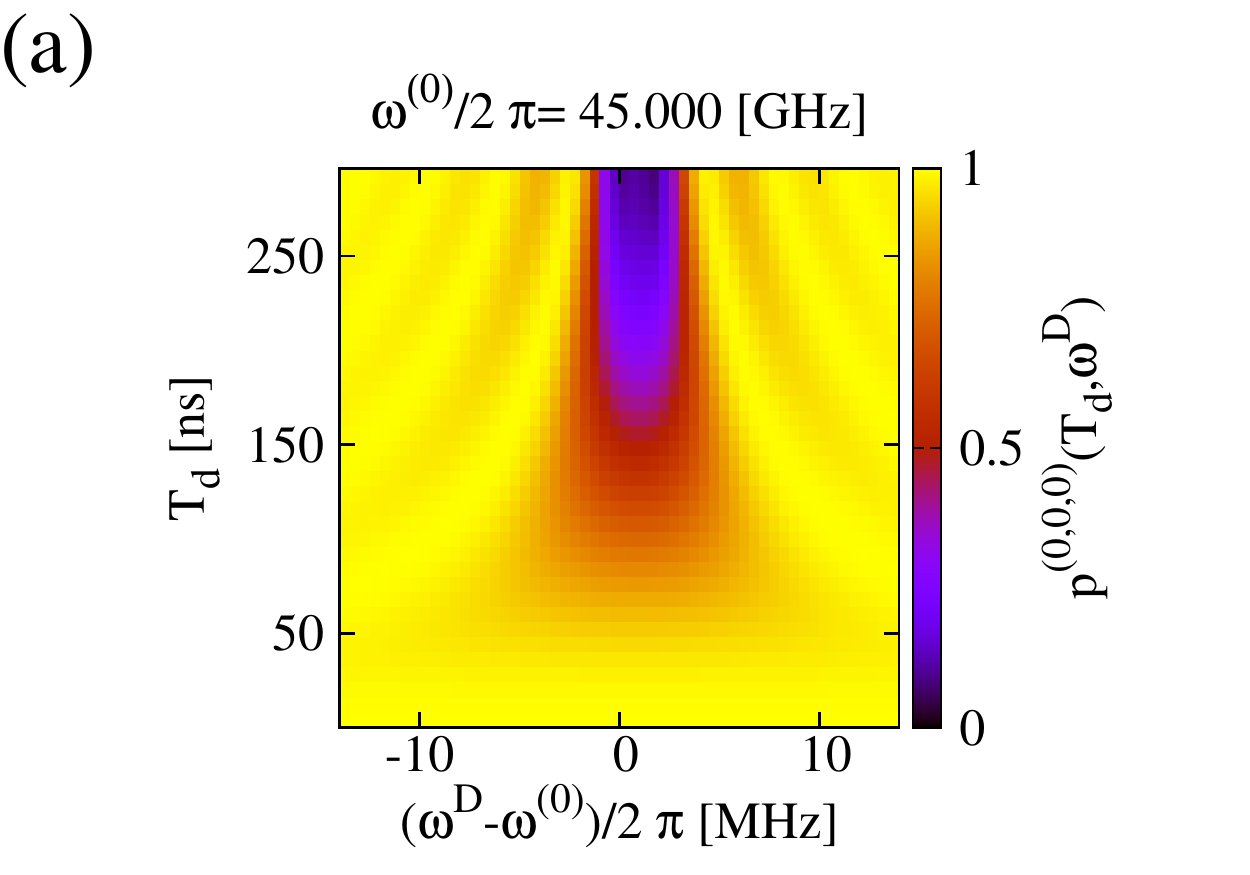}
    \end{minipage}
    \begin{minipage}{0.31\textwidth}
        \centering
        \includegraphics[scale=\hold]{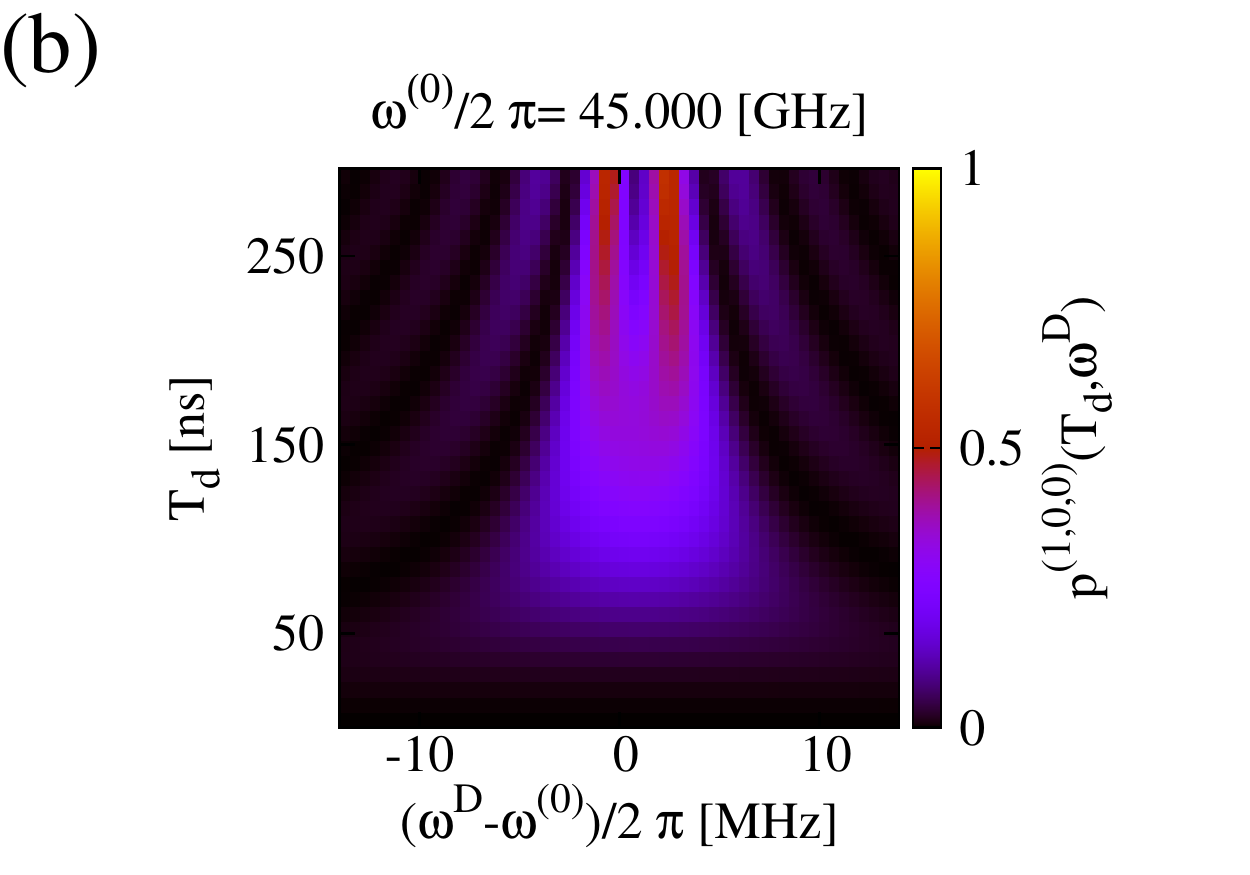}
    \end{minipage}
    \begin{minipage}{0.31\textwidth}
        \centering
        \includegraphics[scale=\hold]{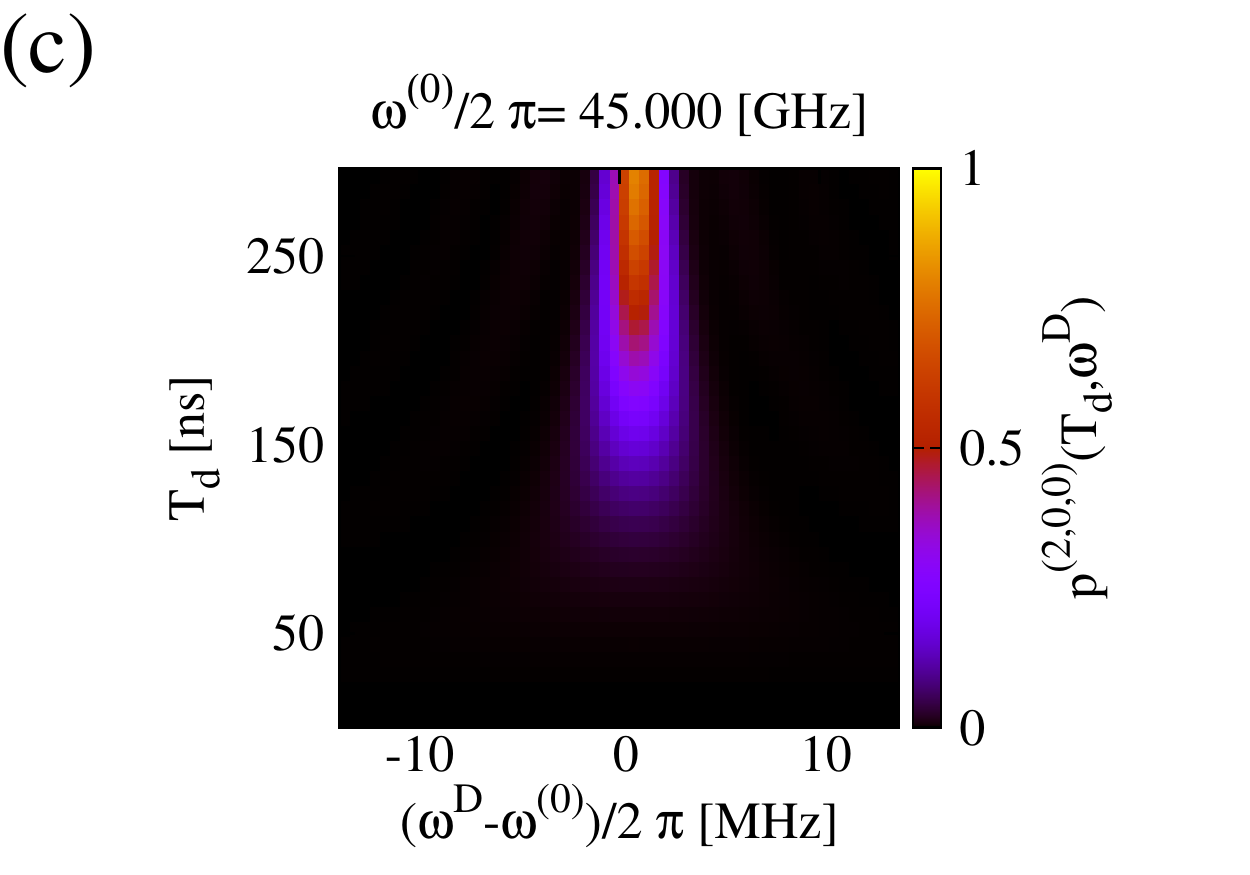}
    \end{minipage}
    \caption[Probabilities $p^{\mathbf{z}}$ for $\mathbf{z}=(0,0,0)$(a), $\mathbf{z}=(1,0,0)$(b) and $\mathbf{z}=(2,0,0)$(c) as functions of the pulse duration $\TD$ and the drive frequency $\DF$ (circuit model architecture II).]{Probabilities $p^{\mathbf{z}}$ for $\mathbf{z}=(0,0,0)$(a), $\mathbf{z}=(1,0,0)$(b) and $\mathbf{z}=(2,0,0)$(c) as functions of the pulse duration $\TD$ and the drive frequency $\DF$. We use the circuit Hamiltonian \equref{eq:CHM}, the initial state $\ket{0,0,0}$, the device parameters listed in \tabref{tab:device_parameter_resonator_coupler_chip} and the control pulse given by \equref{eq:NA_control_pulse} with the rise and fall time $\TRF=\TD/2$ and the pulse amplitude $\PA/2\pi=0.020$ to obtain the results. The results are centered around the eigenfrequency of the coupling resonator in \figref{fig:arch_sketch}(b).}\label{fig:NA_supressed_AII}
\end{figure}

Figures~\ref{fig:NA_supressed_AII}(a-c) show the probabilities $p^{\mathbf{z}}$ for $\mathbf{z}=(0,0,0)$(a), $\mathbf{z}=(1,0,0)$(b) and $\mathbf{z}=(2,0,0)$(c) as functions of the pulse duration $\TD$ and the drive frequency $\DF$. Here, we use the circuit Hamiltonian \equref{eq:CHM}, the device parameters listed in \tabref{tab:device_parameter_resonator_coupler_chip} and the control pulse given by \equref{eq:NA_control_pulse} to obtain the results. The control pulse parameters are $\PATP=0.020$ and $\TRF=\TD/2$, see \figref{fig:NA_pulse_time_evo}(a). The system is initialised in the ground state $\ket{0,0,0}$. Furthermore, we use $n_{J}=4$ and $n_{K}=4$ basis states to model the dynamics of the system.

We can clearly observe transitions from the ground state $\mathbf{z}=(0,0,0)$ to the first-excited $\mathbf{z}=(1,0,0)$ and second-excited $\mathbf{z}=(2,0,0)$ states of the coupling resonator. The results for the third-excited state $\mathbf{z}=(3,0,0)$ are omitted for simplicity. Note that the energy gap between the ground state and the first-excited coupler state are roughly the same as the energy gap between the first-excited coupler state and the second-excited coupler state.

Additionally, we performed analogous simulations with the adiabatic and non-adiabatic effective Hamiltonian \equref{eq:EHM}. We find (data not shown) that the non-adiabatic effective Hamiltonian \equref{eq:EHM} allows us to model the transitions displayed in \figsref{fig:NA_supressed_AII}(a-c). Here, we modelled the non-adiabatic effective Hamiltonian \equref{eq:EHM} with and without a time-dependent interaction strength. Furthermore, we adjusted the spectrum as discussed in \secref{sec:NA_single_flux_tunable_transmon}. The transitions in \figref{fig:NA_supressed_AII}(a-c) can be activated with all four effective models. However, the adiabatic effective Hamiltonian \equref{eq:EHM} does not react to the control pulse in the same manner,\ie we cannot find any excitations of the ground state. Here again, we simulated all four non-adiabatic effective models. These results are in agreement with the findings in \secaref{sec:NA_single_flux_tunable_transmon}{sec:NA_E_suppressed}. Note that in this case we did not look more closely for the missing transitions as in \secref{sec:NA_E_suppressed}.

The results presented in this section suggest that modelling flux-tunable transmons as adiabatic anharmonic oscillators can suppress potential excitations of the coupling resonator. Obviously, this is only the case if we consider scenarios where an external flux drive is present.

\section{Simulations of unsuppressed transitions in the adiabatic effective two-qubit model: architecture II}\label{sec:NA_unsuppressed_arch_II}

In \secref{sec:NA_E_suppressed_AII} we studied transitions which are suppressed in the adiabatic effective model for architecture II, see \figref{fig:arch_sketch}(b). In this section, we consider transitions which are unsuppressed in the adiabatic effective model for architecture II. Note that we introduced the notation we used to address the states of the system at the beginning of \secref{sec:NA_E_suppressed_AII}.

In the following, we investigate two different transitions, namely the $\mathbf{z}=(0,0,1) \rightarrow \mathbf{z}=(0,1,0)$ and $\mathbf{z}=(0,1,1) \rightarrow \mathbf{z}=(0,0,2)$ transitions. The former (latter) transitions can potentially be used to implement $\ISWAP{}$ ($\CZ{}$) gates. As in \secref{sec:NA_unsuppressed_arch_I}, we first discuss the results of the circuit Hamiltonian model and then study the results of the effective models. We use the Hamiltonian \equref{eq:CHM}, the device parameters listed in \tabref{tab:device_parameter_resonator_coupler_chip} and the pulse $\varphi(t)$ given by \equref{eq:NA_control_pulse} to model the dynamics of the two-qubit system with the circuit model.

\renewcommand{\hold}{}
\graphicspath{{./FiguresAndData/NAPaper/CircuitHamiltonianGaugeSimulations/ArchitectureII/}}
\begin{figure}[!tbp]
    \centering
    \begin{minipage}{0.49\textwidth}
        \centering
        \includegraphics[width=\width\textwidth]{TimeEvolISWAPStates3Beta_0_5}
        \includegraphics[width=\width\textwidth]{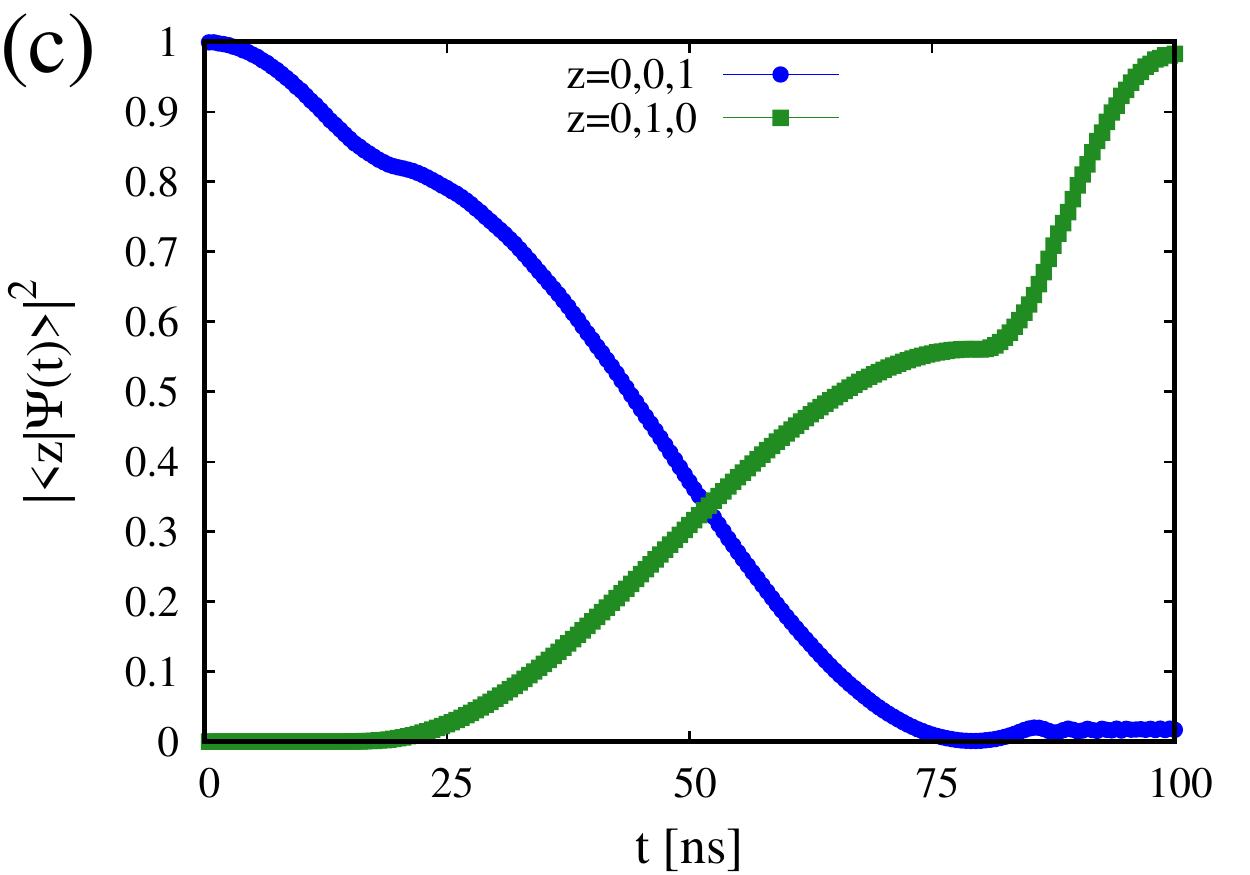}
    \end{minipage}\hfill
    \begin{minipage}{0.49\textwidth}
        \centering
        \includegraphics[width=\width\textwidth]{TimeEvolISWAPStates4Beta_0_5}
        \includegraphics[width=\width\textwidth]{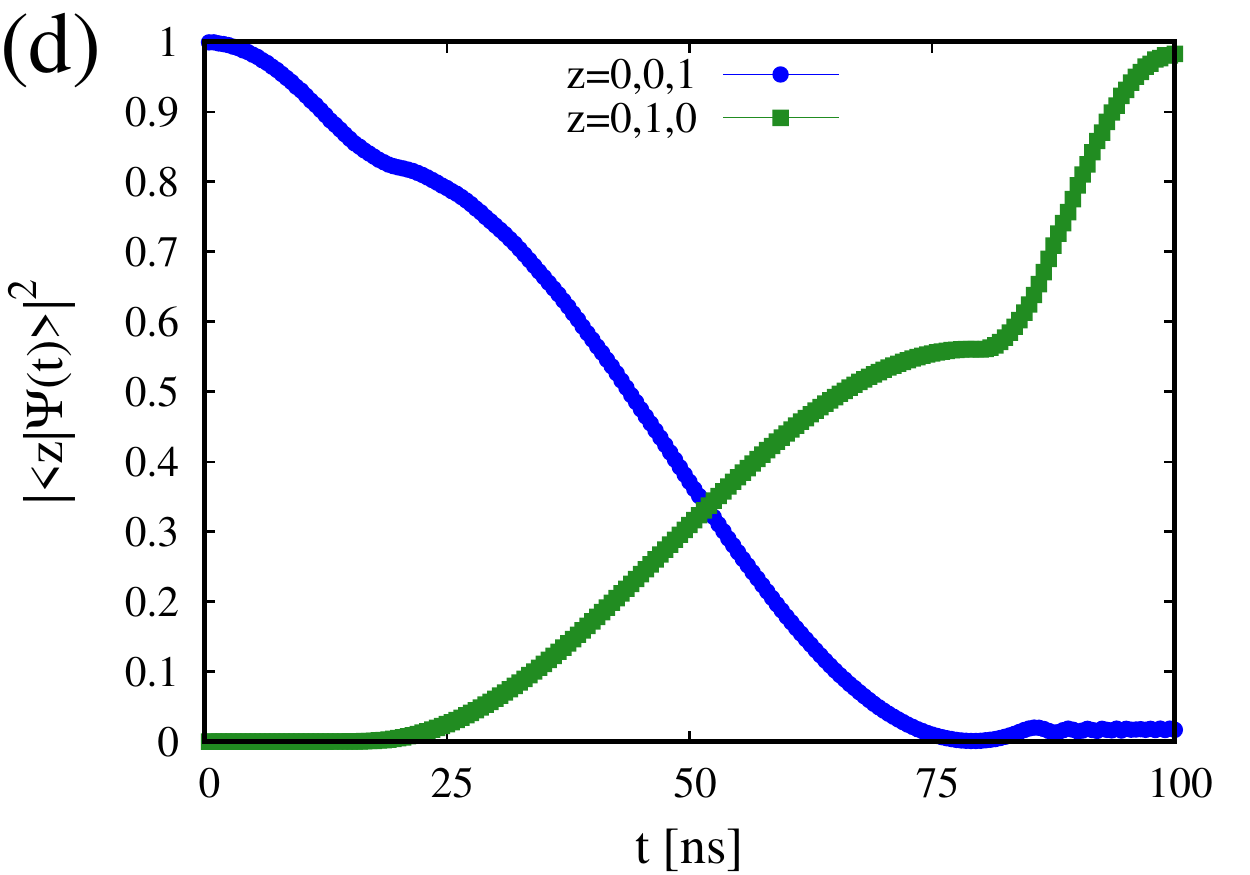}
    \end{minipage}
    \caption[Probabilities $p^{\mathbf{z}}(t)=|\braket{\mathbf{z}|\Psi^{\hold}(t)}|^{2}$  as functions of time $t$ for $\mathbf{z}=(0,0,1)$ and $\mathbf{z}=(0,1,0)$ (circuit model architecture II).]{Probabilities $p^{\mathbf{z}}(t)=|\braket{\mathbf{z}|\Psi^{\hold}(t)}|^{2}$  as functions of time $t$ for $\mathbf{z}=(0,0,1)$ and $\mathbf{z}=(0,1,0)$. We use the circuit Hamiltonian \equref{eq:CHM}, the device parameters listed in \tabref{tab:device_parameter_resonator_coupler_chip} and the pulse given by \equref{eq:NA_control_pulse} to obtain the results. The pulse parameters are the drive frequency $\DFTP=0$ GHz, the amplitude $\PATP=0.289$, the rise and fall time $\TRF=20.0$ ns and the pulse duration $\TD=100.0$ ns. Furthermore, $n_{J}$ denotes the number of flux-tunable transmon basis states and we use $n_{J}=3$ in \PANL{a}, $n_{J}=4$ in \PANL{b}, $n_{J}=14$ in \PANL{c} and $n_{J}=25$ in \PANL{d}. The resonator is modelled with $n_{K}=4$ harmonic basis states. All pulse parameters are listed in \tabref{tab:summary_circuit_hamiltonian_results}.}\label{fig:NA_cir_ISWAP_cases_eth}
\end{figure}

Figures \ref{fig:NA_cir_ISWAP_cases_eth}(a-d) show the probabilities $p^{\mathbf{z}}(t)$ for $\mathbf{z}=(0,0,1)$ and $\mathbf{z}=(0,1,0)$ as functions of time $t$. Here $p^{\mathbf{z}}(t)=|\braket{\mathbf{z}|\Psi^{\hold}(t)}|^{2}$ and the initial state of the system is $\ket{\Psi^{\hold}(0)}=\ket{0,0,1}$. We use the drive frequency $\DFTP=0$ GHz, the amplitude $\PATP=0.289$, the rise and fall time $\TRF=20.0$ ns and the pulse duration $\TD=100.0$ ns to model the control pulse $\varphi(t)$. Furthermore, we simulate the system with different numbers of basis states $n_{J}$ to model the flux-tunable transmons in the system. We use $n_{J}=3$ in \figref{fig:NA_cir_ISWAP_cases_eth}(a), $n_{J}=4$ in \figref{fig:NA_cir_ISWAP_cases_eth}(b), $n_{J}=14$ in \figref{fig:NA_cir_ISWAP_cases_eth}(c) and $n_{J}=25$ in \figref{fig:NA_cir_ISWAP_cases_eth}(d). The resonator is modelled with $n_{K}=4$ basis states only.

On the one hand, we observe that the time evolutions of the probabilities $p^{\mathbf{z}}(t)$ in \figsref{fig:NA_cir_ISWAP_cases_eth}(a-b) and \figsref{fig:NA_cir_ISWAP_cases_eth}(c-d) exhibit qualitative and quantitative deviations. On the other hand, we see that the time evolutions of the probabilities $p^{\mathbf{z}}(t)$ in \figsref{fig:NA_cir_ISWAP_cases_eth}(a) and \figsref{fig:NA_cir_ISWAP_cases_eth}(b) as well as \figsref{fig:NA_cir_ISWAP_cases_eth}(c) and \figsref{fig:NA_cir_ISWAP_cases_eth}(d) are similar. Additionally, in \figsref{fig:NA_cir_ISWAP_cases_eth}(a-b) we can observe that the states $\mathbf{z}=(0,0,1)$ and $\mathbf{z}=(0,1,0)$ do not exchange any population. Instead, we see that after the application of the pulse, the system mainly returns to its initial state. Only if we increase the number of basis states to $n_{J}=14$, see \figref{fig:NA_cir_ISWAP_cases_eth}(c), we find that the probabilities $p^{\mathbf{z}}(t)$ have converged to the third decimal. Here we use the solution for $n_{J}=25$ as a graphical reference solution,\ie there are no noticeable differences between the results in \figsref{fig:NA_cir_ISWAP_cases_eth}(c-d).

\graphicspath{{./FiguresAndData/NAPaper/CircuitHamiltonianGaugeSimulations/ArchitectureII/}}
\begin{figure}[!tbp]
    \centering
    \begin{minipage}{0.49\textwidth}
        \centering
        \includegraphics[width=\width\textwidth]{TimeEvolCZStates3Beta_0_5}
        \includegraphics[width=\width\textwidth]{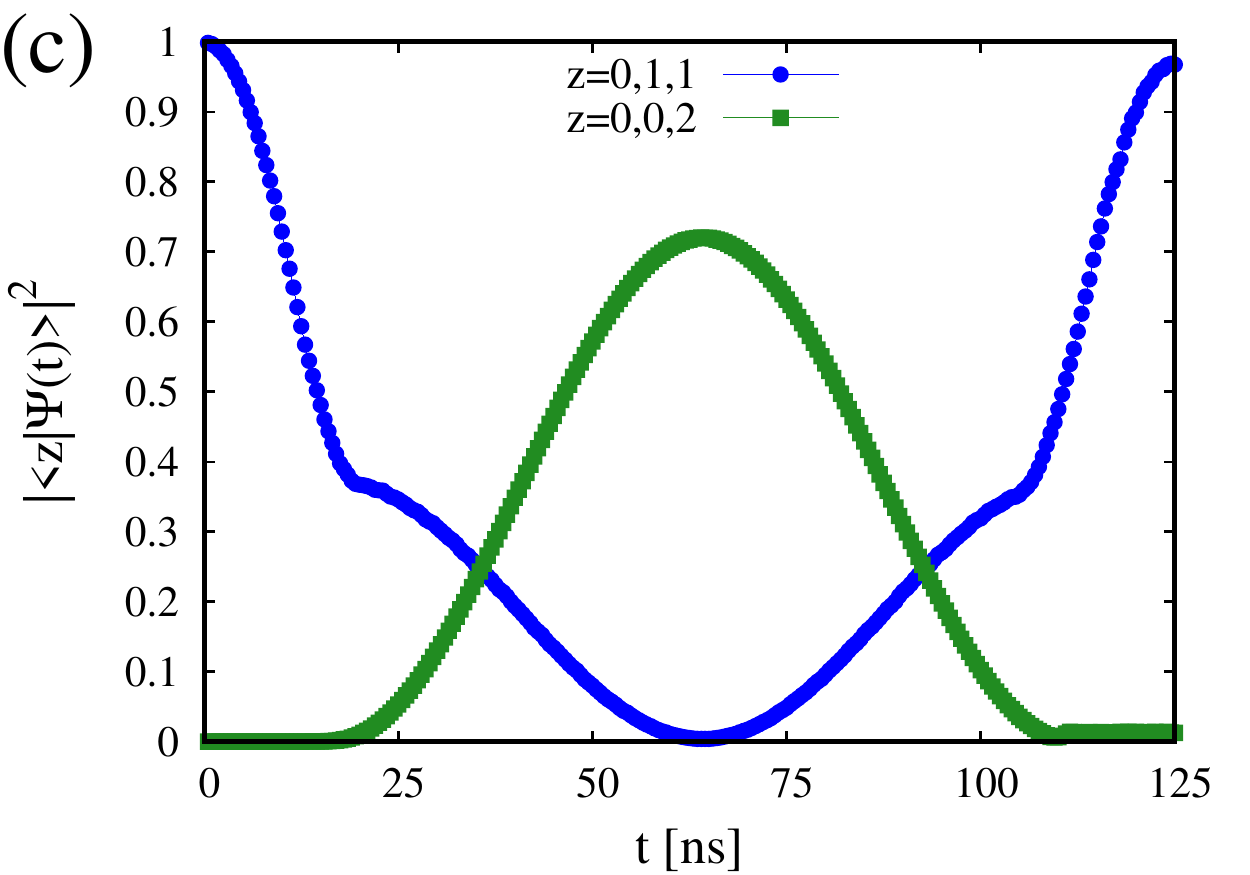}
    \end{minipage}\hfill
    \begin{minipage}{0.49\textwidth}
        \centering
        \includegraphics[width=\width\textwidth]{TimeEvolCZStates4Beta_0_5}
        \includegraphics[width=\width\textwidth]{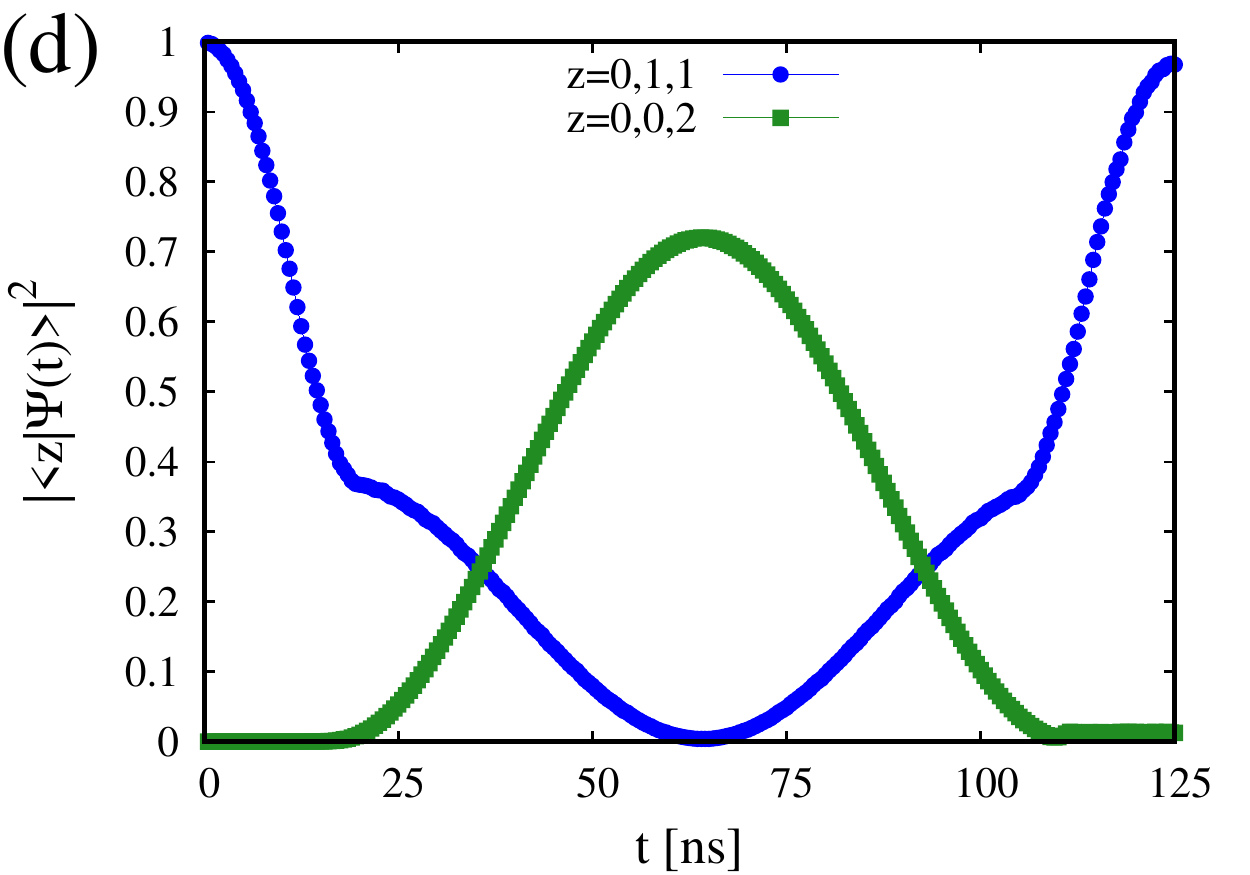}
    \end{minipage}
    \caption[Probabilities $p^{\mathbf{z}}(t)=|\braket{\mathbf{z}|\Psi^{\hold}(t)}|^{2}$ as functions of time $t$ for $\mathbf{z}=(0,1,1)$ and $\mathbf{z}=(0,0,2)$ (circuit model architecture II).]{Probabilities $p^{\mathbf{z}}(t)=|\braket{\mathbf{z}|\Psi^{\hold}(t)}|^{2}$ as functions of time $t$ for $\mathbf{z}=(0,1,1)$ and $\mathbf{z}=(0,0,2)$. We use the circuit Hamiltonian \equref{eq:CHM}, the device parameters listed in \tabref{tab:device_parameter_resonator_coupler_chip} and the pulse given by \equref{eq:NA_control_pulse} to obtain the results. The pulse parameters are the drive frequency $\DFTP=0$ GHz, the amplitude $\PATP=0.3335$, the rise and fall time $\TRF=20.0$ ns and the pulse duration $\TD=125.00$ ns. Furthermore, $n_{J}$ denotes the number of flux-tunable transmon basis states and we use $n_{J}=3$ in \PANL{a}, $n_{J}=4$ in \PANL{b}, $n_{J}=16$ in \PANL{c} and $n_{J}=25$ in \PANL{d}. The resonator is modelled with $n_{K}=4$ harmonic basis states. All pulse parameters are listed in \tabref{tab:summary_circuit_hamiltonian_results}.}\label{fig:NA_cir_cz_cases_eth}
\end{figure}

Figures~\ref{fig:NA_cir_cz_cases_eth}(a-d) show the probabilities $p^{\mathbf{z}}(t)=|\braket{\mathbf{z}|\Psi^{\hold}(t)}|^{2}$ for $\mathbf{z}=(0,1,1)$ and $\mathbf{z}=(0,0,2)$ as functions of time $t$. The initial state of the system is $\ket{\Psi^{\hold}(0)}=\ket{0,1,1}$. Here we use the drive frequency $\DF=0$ GHz, the amplitude $\PATP=0.3335$, the rise and fall time $\TRF=20.0$ ns and the pulse duration $\TD=125.0$ ns to model the pulse $\varphi(t)$. Furthermore, the pulse duration $\TD$ is chosen such that the population exchange between the states $\mathbf{z}=(0,1,1)$ and $\mathbf{z}=(0,0,2)$ is not optimal, see \figsref{fig:NA_cir_cz_cases_eth}(c-d) at time $\TD$. This is done for the following reason. The system has the tendency to return back to its initial state if we model the system with not enough basis states $n_{J}$, see \figsref{fig:NA_cir_cz_cases_eth}(a-b). Consequently, it would be difficult to detect changes in the probabilities $p^{\mathbf{z}}(t)$ at time $\TD$ if we calibrate the system such that $p^{\mathbf{z}}(\TD) \rightarrow 1$ for $\mathbf{z}=(0,1,1)$.

Figures~\ref{fig:NA_cir_cz_cases_eth}(a-d) show similar features as \figref{fig:NA_cir_ISWAP_cases_eth}(a-d). The results for $n_{J}=3$ and $n_{J}=4$, see \figsref{fig:NA_cir_cz_cases_eth}(a-b), exhibit qualitative and quantitative differences and we have to increase the number of basis states to $n_{J}=16$ until the probabilities $p^{\mathbf{z}}(T_{d})$ converge to the desired decimal.

We now focus on the corresponding effective Hamiltonian models. For these simulations, we use the adiabatic effective Hamiltonian \equref{eq:EHM}, the pulse $\varphi(t)$ given by \equref{eq:NA_control_pulse} and four basis states for every transmon and resonator in the system. Here, as in \secref{sec:NA_unsuppressed_arch_I}, we have several options in specifying the effective Hamiltonian model. First, we can either use a time-dependent or a time-independent interaction strength to model the coupling between the flux-tunable transmons and the resonator. Second, we can choose between two sets of functions, see \secref{sec:NA_single_flux_tunable_transmon}, to model the tunable energies of the transmons. The effective model we define, determines the device parameters we need to perform the simulations. If we model the system with a time-dependent interaction strength and/or the tunable energies with the functions \equaref{eq: expansion frequency}{eq: expansion anharmonicity}, we have to use the device parameters listed in \tabref{tab:device_parameter_resonator_coupler_chip}. The last remaining case can be modelled with the parameters listed in \tabref{tab:device_parameter_resonator_coupler_chip_effective}.

\begin{figure}[!tbp]
    \centering
    \begin{minipage}{0.49\textwidth}
        \centering
        \graphicspath{{./FiguresAndData/NAPaper/InteractionStrength/}}
        \includegraphics[width=\width\textwidth]{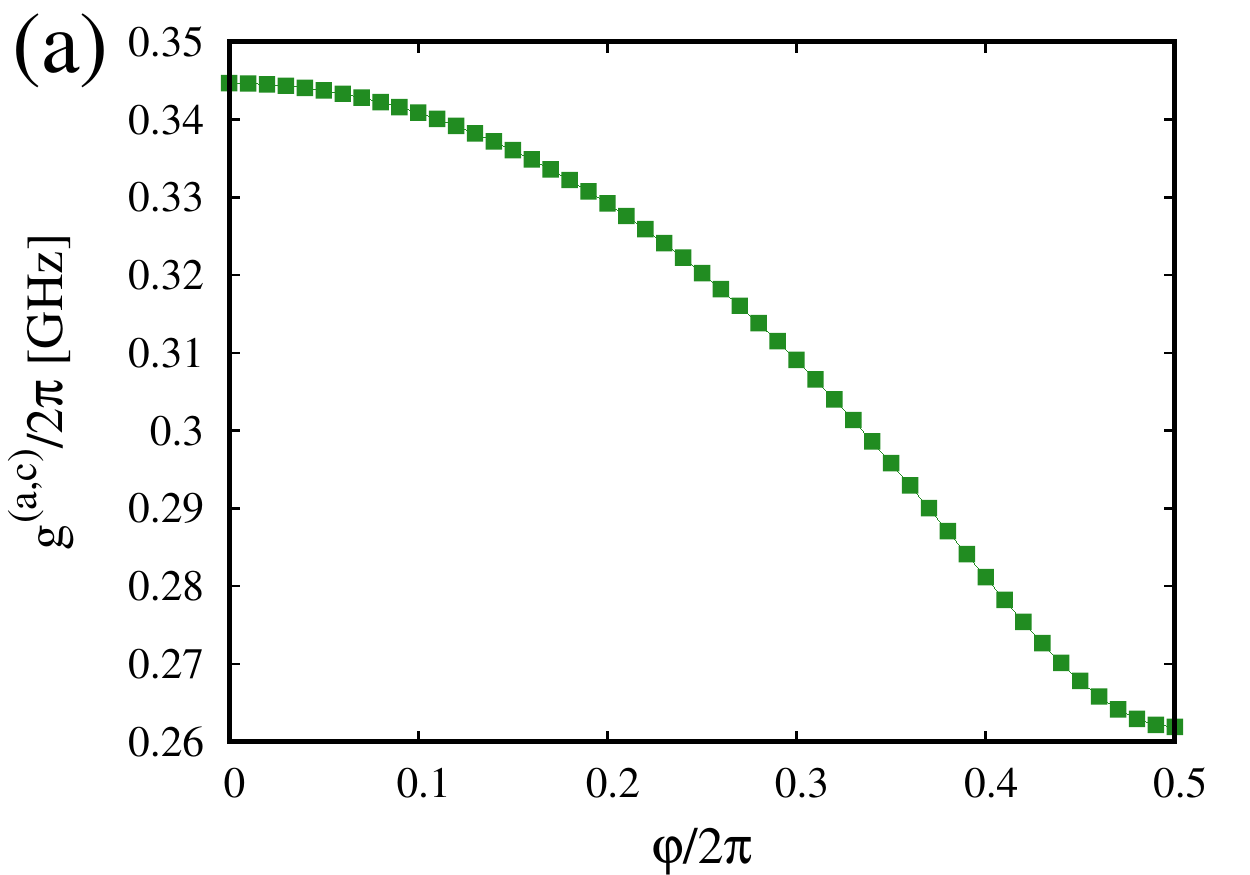}
    \end{minipage}\hfill
    \begin{minipage}{0.49\textwidth}
        \centering
        \graphicspath{{./FiguresAndData/NAPaper/EffIntAndPulse/}}
        \includegraphics[width=\width\textwidth]{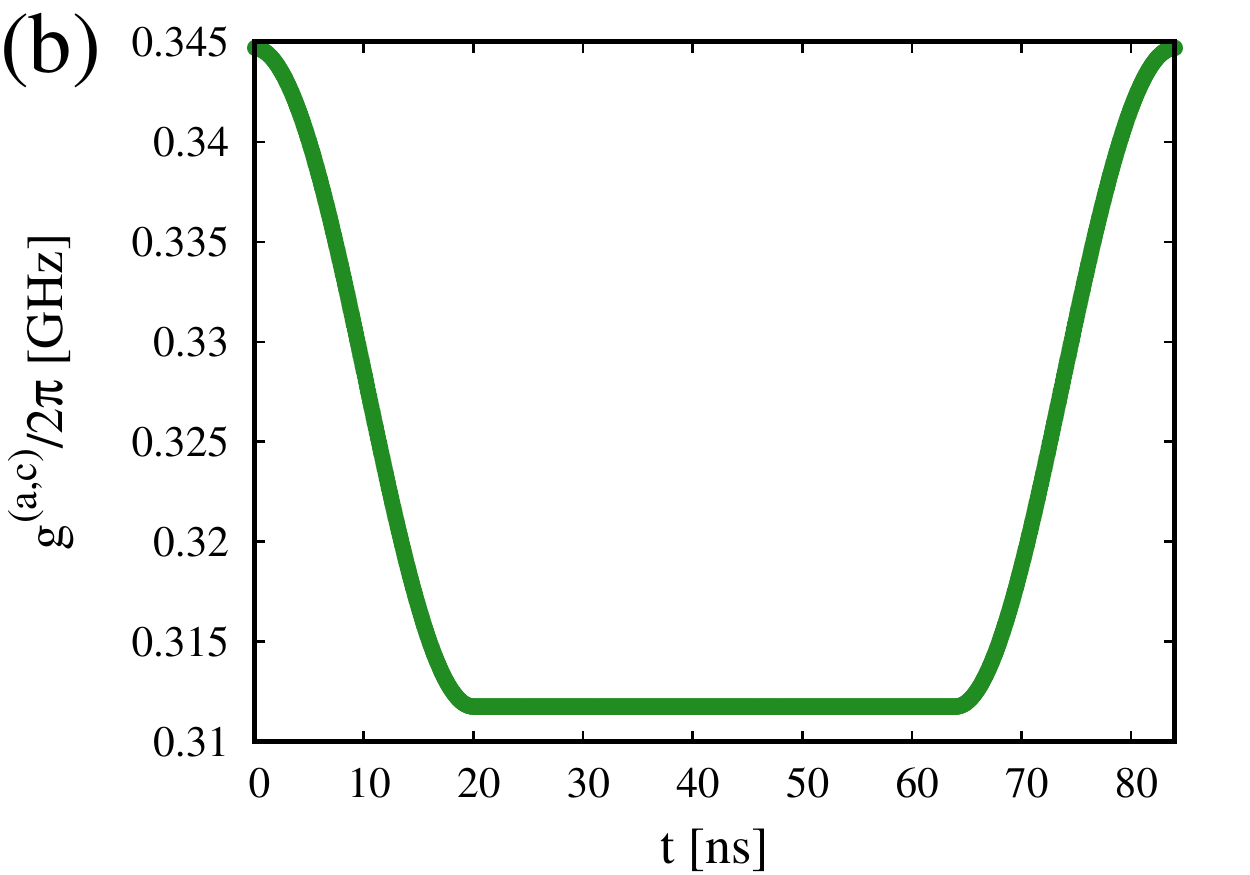}
    \end{minipage}
    \caption[Effective interaction strength $g^{(a,c)}$ between a resonator and a flux-tunable transmon as a function of the external flux $\varphi$ in \PANL{a} and as a function of time $t$ in \PANL{b}.]{Effective interaction strength $g^{(a,c)}$ between a resonator and a flux-tunable transmon as a function of the external flux $\varphi$ in \PANL{a} and as a function of time $t$ in \PANL{b}. We use \equref{eq:eff_int_trans_res_second_time} and the energies listed in \tabref{tab:device_parameter_resonator_coupler_chip} row $i=1$ to obtain the numerical values for $g^{(a,c)}$. In \PANL{a} we compute the effective interaction strength $g^{(a,c)}(\varphi)$ for the interval $\varphi/2\pi \in [0,0.5]$. In \PANL{b} we use the control pulse $\varphi(t)$ given by \equref{eq:NA_control_pulse} with the flux-offset value $\varphi_{0}/2\pi=0$, the drive frequency $\DFTP=0$ GHz, the pulse amplitude $\PATP=0.297$, the rise and fall time $\TRF=20.0$ ns and the pulse duration $\TD=84.0$ ns to obtain the effective interaction strength $g^{(a,c)}(\varphi(t))$ as a function of time $t$.}\label{fig:NA_eff_int_flux_evolution_archII}
\end{figure}

Figure~\ref{fig:NA_eff_int_flux_evolution_archII}(a) shows the effective interaction strength $g^{(a,c)}$ as a function of the flux $\varphi$. Figure~\ref{fig:NA_eff_int_flux_evolution_archII}(b) shows the effective interaction strength $g^{(a,c)}$ for a flux pulse $\varphi(t)$ given by \equref{eq:NA_control_pulse} as a function of time $t$. We use the energies listed in \tabref{tab:device_parameter_resonator_coupler_chip}, row $i=1$, the drive frequency $\DF=0$ GHz, the amplitude $\PATP=0.297$, the rise and fall time $\TRF=20.0$ ns and the pulse duration $\TD=84.0$ ns to generate the data for \figsref{fig:NA_eff_int_flux_evolution_archII}(a-b). Note that the pulse parameters are only necessary to generate the data for \figref{fig:NA_eff_int_flux_evolution_archII}(b).

In \figref{fig:NA_eff_int_flux_evolution_archII}(a) we can observe that the interaction strength $g^{(a,c)}$ varies by about $85$ MHz over the interval $\varphi/2\pi \in [0,0.5]$. If we use the pulse $\varphi(t)$ to compute the interaction strength $g^{(a,c)}$, specified above, we find that $g^{(a,c)}$ drops by about $35$ MHz for a period of roughly $45$ ns, afterwards $g^{(a,c)}$ returns back to its initial value. Note that in \secref{sec:NA_unsuppressed_arch_II} we found that oscillations of about $1$ MHz can shift the pulse duration $\TD$ up to about $65$ ns.

\renewcommand{\hold}{}
\graphicspath{{./FiguresAndData/NAPaper/EffectiveHamiltonian/TwoQubitsWithResonator/}}
\begin{figure}[!tbp]
    \centering
    \begin{minipage}{0.49\textwidth}
        \centering
        \includegraphics[width=\width\textwidth]{NoTimeDep/te_ISWAP}
        \includegraphics[width=\width\textwidth]{NoTimeDep/te_cz}
    \end{minipage}\hfill
    \begin{minipage}{0.49\textwidth}
        \centering
        \includegraphics[width=\width\textwidth]{TimeDep/te_ISWAP}
        \includegraphics[width=\width\textwidth]{TimeDep/te_cz}
    \end{minipage}
    \caption[Probabilities $p^{\mathbf{z}}(t)=|\braket{\mathbf{z}|\Psi^{\hold}(t)}|^{2}$ as functions of time $t$. (effective model architecture II)]{Probabilities $p^{\mathbf{z}}(t)=|\braket{\mathbf{z}|\Psi^{\hold}(t)}|^{2}$ as functions of time $t$. In \PANSL{a,b} we model the transitions $\mathbf{z}=(0,0,1) \rightarrow \mathbf{z}=(0,1,0)$. In \PANSL{c,d} we model the transitions $\mathbf{z}=(0,1,1) \rightarrow \mathbf{z}=(0,0,2)$. We use the effective Hamiltonian \equref{eq:EHM}, the device parameters listed in \tabref{tab:device_parameter_resonator_coupler_chip_effective} and the pulse given by \equref{eq:NA_control_pulse} with the drive frequency $\DFTP=0$ GHz and the rise and fall time $\TD=20$ ns to obtain the results. The remaining, non-specified pulse parameters are the amplitude $\PA$ and the pulse duration $\TD$. Both parameters are different for every panel. We use $\PATP=0.297$ and $\TD=84.0$ ns in \PANL{a}, $\PATP=0.289$ and $\TD=96.0$ ns in \PANL{b}, $\PATP=0.343$ and $\TD=105.0$ ns in \PANL{c} as well as $\PATP=0.334$ and $\TD=121.0$ ns in \PANL{d}. In \PANSL{a,c} we use the first-order series expansion to model the tunable qubit frequency. Here, we also use $g^{(a,c)}(t)=\const$ to model a static effective interaction strength. Similarly, in \PANSL{b,d} we use the higher-order series expansion to model the energies of the flux-tunable transmons and $g^{(a,c)}(t)$ given by \equref{eq:eff_int_trans_res_second_time} to model a dynamic effective interaction strength, see \figref{fig:NA_eff_int_flux_evolution_archII}(b). All pulse parameters and cases are listed in \tabref{tab:summary_effective_hamiltonian_results}.}\label{fig:NA_eff_cz_ISWAP_cases_eth}
\end{figure}

Figures~\ref{fig:NA_eff_cz_ISWAP_cases_eth}(a-d) show the probabilities $p^{\mathbf{z}}(t)=|\braket{\mathbf{z}|\Psi^{\hold}(t)}|^{2}$ as functions of time $t$. In \figsref{fig:NA_eff_cz_ISWAP_cases_eth}(a-b) we model $\mathbf{z}=(0,0,1) \rightarrow \mathbf{z}=(0,1,0)$ \ISWAP{} transitions with, see \figref{fig:NA_eff_cz_ISWAP_cases_eth}(b) and without, see \figref{fig:NA_eff_cz_ISWAP_cases_eth}(a), a time-dependent interaction strength. Similarly, in \figsref{fig:NA_eff_cz_ISWAP_cases_eth}(c-d) we model $\mathbf{z}=(0,1,1) \rightarrow \mathbf{z}=(0,0,2)$ \CZ{} transitions with, see \figref{fig:NA_eff_cz_ISWAP_cases_eth}(d) and without, see \figref{fig:NA_eff_cz_ISWAP_cases_eth}(c), a time-dependent interaction strength.

We determined the pulse parameters as follows. First, we fix the rise and fall time $\TRF$. Here, we use the rise and fall time $\TRF=20$ ns we found for the corresponding circuit Hamiltonian transitions. Next, we perform a spectroscopy simulation for the pulse amplitude $\PA$. Here we use the amplitude in the center of the corresponding chevron pattern. At last, we adjust the pulse duration such that the corresponding operation can be realised with the two previously specified pulse parameters. Furthermore, we use the drive frequency $\DF=0$ GHz for all pulses.

The results in \figref{fig:NA_eff_cz_ISWAP_cases_eth}(a) are obtained with the pulse amplitude $\PATP=0.297$ GHz and the pulse duration $\TD=84.0$ ns. Here we use the tunable frequency given by \equref{eq: frequency} and a constant anharmonicity to model the spectrum of the flux-tunable transmon. If we compare these results with the one of the corresponding circuit Hamiltonian simulations, see \tabref{tab:summary_circuit_hamiltonian_results}, we find that the pulse amplitudes deviate by $\Delta \PATP=0.008$. Furthermore, if we compare the pulse durations $\TD$ for both cases, we find a difference of $16$ ns.

The results in \figref{fig:NA_eff_cz_ISWAP_cases_eth}(b) are obtained with the pulse amplitude $\PATP=0.289$ GHz and the pulse duration $\TD=96.0$ ns. Here we use the tunable frequency given by \equref{eq: expansion frequency} and the tunable anharmonicity in \equref{eq: expansion anharmonicity} to model the spectrum of the flux-tunable transmon. Consequently, if we use the adjusted spectrum to model the flux-tunable transmon, the deviations between the effective and the circuit model practically disappear. Furthermore, the difference in terms of the pulse duration has reduced to $4$ ns. Interestingly, even though the nominal changes in interaction strength $g^{(a,c)}$ in \figref{fig:NA_eff_cz_ISWAP_cases_eth}(b) are substantially larger than for the case $g^{(c,c)}$ in \figref{fig:NA_eff_int_flux_evolution_archI}(b) in \secref{sec:NA_unsuppressed_arch_I}, we find much smaller deviations between the two effective models,\ie the one with and the one without a time-dependent interaction strength.

The results in \figref{fig:NA_eff_cz_ISWAP_cases_eth}(c-d) are obtained with the same simulation settings as the results in \figref{fig:NA_eff_cz_ISWAP_cases_eth}(a-b). Here we find similar features as before. The large differences in terms of the pulse amplitude disappear if we adjust the spectrum of the flux-tunable transmon in the effective model. Furthermore, the deviations in terms of the pulse duration $\TD$ are reduced to $4$ ns if we model the effective model with a time-dependent interaction strength. It seems peculiar that we often find a difference of $4$ ns for the pulse duration $\TD$ if we compare the effective and the circuit model. However, to the best knowledge of the author, this finding is just a coincidence.

In \secref{sec:NA_unsuppressed_arch_I}, we investigate whether or not the non-adiabatic driving term in \equref{eq:drive_term_ftt_second_time} affects the \ISWAP{} and \CZ{} transitions if we simulate architecture I with the effective model. Here we find that the transitions are barely affected. Similarly, we added the driving term in \equref{eq:drive_term_ftt_second_time} to the adiabatic effective model for the flux-tunable transmons in architecture II and repeated the simulations again. We find (data not shown) that the \ISWAP{} and \CZ{} transitions for the relevant pulse parameters listed in \tabref{tab:summary_effective_hamiltonian_results} are barely affected. This means that the probabilities $p^{\mathbf{z}}$ at time $\TD$ are at most affected by the third decimal. Furthermore, the overall transitioning behaviour does not seem to be affected at all. As before, we emphasise that different pulse parameters can potentially cause larger deviations.

\section{Summary, conclusions and outlook}\label{sec:NA_SummaryConclusionsAndOutlook}

In this chapter we investigated how different approximations, which result in different effective Hamiltonians, affect the transitions from one state to another. To this end, we studied the probability amplitudes which are associated with the different states. Furthermore, we compared the results of the effective Hamiltonian models to the associated circuit Hamiltonian models. Note that here we only considered transitions which can be activated by means of an external flux.

In \secref{sec:NA_single_flux_tunable_transmon} we investigated a single flux-tunable transmon. Here we compared how the non-adiabatic effective Hamiltonian \equref{eq:fft_eff_II}, the adiabatic effective Hamiltonian \equref{eq:tunable-frequency eff} and the circuit Hamiltonian \equref{eq:flux-tunable transmon recast} react to various microwave pulses, see \figref{fig:NA_pulse_time_evo}(a) and unimodal pulses, see \figref{fig:NA_pulse_time_evo}(b).

Obviously, the adiabatic effective Hamiltonian \equref{eq:tunable-frequency eff} cannot, by definition, be used to model any transitions. Note that the adiabatic model for the flux-tunable transmon is often used to describe flux-tunable transmons in multi-qubit systems. Conversely, we found that the non-adiabatic effective Hamiltonian \equref{eq:fft_eff_II} and the circuit Hamiltonian \equref{eq:flux-tunable transmon recast} can be used to model various resonant and non-adiabatic transitions. Additionally, for the case of resonant transitions, driven by a microwave pulse, we showed an example where the predictions for both models match qualitatively and quantitatively with minor deviations in one of the pulse parameters. For the case of non-adiabatic transitions, driven by a unimodal pulse, we showed an example where the predictions made by both models match qualitatively for most instances. Furthermore, we showed that the spectrum of the flux-tunable transmon in question, modelled with the tunable frequencies given by \equaref{eq: frequency der}{eq: frequency} can deviate up to several GHz from the numerically exact spectrum. In conclusion, the non-adiabatic effective Hamiltonian \equref{eq:tunable-frequency eff} can cover at least some of the aspects of the associated circuit Hamiltonian \equref{eq:flux-tunable transmon recast}.

In \secaref{sec:NA_E_suppressed}{sec:NA_unsuppressed_arch_I} we considered a two-qubit NIGQC model, see \secref{sec:FromStaticsToDynamics}, which we denote as architecture I, see \figref{fig:arch_sketch}(a). This device consists of two fixed-frequency transmons whose ground and first-excited states are supposed to function as qubit states and a flux-tunable transmon which works as a time-dependent coupler. The transitions in this device architecture are usually activated by means of microwave pulses, see \figref{fig:NA_pulse_time_evo}(a).

We showed evidence that the adiabatic effective Hamiltonian \equref{eq:EHM} is not able to describe certain single-qubit (X) type transitions with a flux drive, see for example \tabref{tab:summary_effective_hamiltonian_results} rows three and four. However, we also found that the non-adiabatic effective Hamiltonian \equref{eq:EHM} can be used to model exactly these transitions. Note that the definitions of the adiabatic and non-adiabatic effective Hamiltonians only differ by the non-adiabatic driving term in \equref{eq:drive_term_ftt_second_time}. These results are discussed in \secref{sec:NA_E_suppressed}. Consequently, when studying the single-qubit (X) type transitions with an effective model for architecture I, one should consider using the driving term in \equref{eq:drive_term_ftt_second_time} to do so.

Furthermore, for architecture I we found that modelling the interaction strength of the adiabatic effective Hamiltonian \equref{eq:EHM} with the time-dependent function given by \equref{eq:eff_int_trans_trans_second_time} leads to large shifts in the pulse duration of up to $75$ ns for \ISWAP{} and \CZ{} gate transitions, see \figsref{fig:NA_eff_cz_ISWAP_cases_chalmers}(a-d), \figsref{fig:eff_int_scaling}(a-b) and \tabref{tab:summary_effective_hamiltonian_results}. Note that the time dependence in \equref{eq:eff_int_trans_trans_second_time} induces oscillations of about $1$ MHz in the interaction strength, see \figref{fig:NA_eff_int_flux_evolution_archI}(b). Therefore, a time-dependent interaction strength of the form \equref{eq:eff_int_trans_trans_second_time} can have a non-negligible impact on the pulse duration for \ISWAP{} and \CZ{} transitions if we model these transitions with the effective model.

In \secref{sec:NA_unsuppressed_arch_I}, we compared the results of the effective models with and without time-dependent interaction strength to the results of the associated circuit model. If we model architecture I with a time-independent interaction strength and the adiabatic effective Hamiltonian \equref{eq:EHM}, we find deviations of up to $100$ ns for the pulse duration with regard to the associated circuit Hamiltonian \equref{eq:CHM}. If we model architecture I with a time-dependent interaction strength and the adiabatic effective Hamiltonian \equref{eq:EHM}, we find deviations of up to $25$ ns for the pulse duration with regard to the associated circuit Hamiltonian \equref{eq:CHM}.

Additionally, we found that the non-adiabatic driving term in \equref{eq:drive_term_ftt_second_time} barely affects the \ISWAP{} and \CZ{} transitions for the relevant pulse parameters in \tabref{tab:summary_effective_hamiltonian_results}. Note that different pulse parameters can potentially lead to more severe deviations.

In \secaref{sec:NA_E_suppressed_AII}{sec:NA_unsuppressed_arch_II} we studied a two-qubit NIGQC model which we denote as architecture II, see \figref{fig:arch_sketch}(b). This device consists of two flux-tunable transmons whose ground and first-excited states are supposed to function as qubit states and a resonator element which works as a time-independent coupler. The transitions in this device architecture are usually activated by means of a unimodal pulses, see \figref{fig:NA_pulse_time_evo}(b).

We found that we can excite the coupling resonator with a microwave control pulse if we use the circuit Hamiltonian \equref{eq:CHM} and the non-adiabatic effective Hamiltonian \equref{eq:EHM}. However, as in \secref{sec:NA_E_suppressed} we found that the adiabatic effective Hamiltonian \equref{eq:EHM} does not allow us to model the targeted transitions. These results are discussed in \secref{sec:NA_E_suppressed_AII}. Consequently, when studying excitations of the coupler states with an effective model for architecture II, one should consider using the driving term in \equref{eq:drive_term_ftt_second_time} to do so.

We also found that modelling the interaction strength of the adiabatic effective Hamiltonian \equref{eq:EHM} with the time-dependent function given by \equref{eq:eff_int_trans_res_second_time} leads to modest shifts in the pulse durations of about $15$ ns for \ISWAP{} and \CZ{} gate transitions, see \figsref{fig:NA_eff_cz_ISWAP_cases_eth}(a-d) and \tabref{tab:summary_effective_hamiltonian_results}. Note that the time dependence in \equref{eq:eff_int_trans_res_second_time} induces a square pulse like reduction of roughly $35$ MHz of the interaction strength, see \figref{fig:NA_eff_int_flux_evolution_archII}(b).

For architecture II, we also compared the results of the effective models with and without time-dependent interaction strength to the results of the associated circuit model. If we use the time-dependent model given by \equref{eq:eff_int_trans_trans_second_time} for the interaction strength, we find deviations of up to $4$ ns for the pulse duration with regard to the circuit model. If we use the time-independent model for the interaction strength, we find deviations of up to $20$ ns for the pulse duration with regard to the circuit model.

Additionally, we found that if we use the adiabatic effective Hamiltonian \equref{eq:EHM} to model \ISWAP{} and \CZ{} type transitions, we have to use the tunable frequency in \equref{eq: expansion frequency} and the tunable anharmonicity in \equref{eq: expansion anharmonicity} to model the spectra of flux-tunable transmons, assuming we want the pulse amplitudes for the effective and circuit model to match. The findings with regard to the spectra in \secref{sec:NA_single_flux_tunable_transmon} make these results seem plausible.

For architecture II, we also found that the non-adiabatic driving term in \equref{eq:drive_term_ftt_second_time} barely affects the \ISWAP{} and \CZ{} transitions for the relevant pulse parameters in \tabref{tab:summary_effective_hamiltonian_results}. Here too, note that different pulse parameters can potentially lead to more severe deviations. These results are discussed in \secref{sec:NA_unsuppressed_arch_II}.

In conclusion, as long as we cannot proof that two time-evolution operators, see \equref{eq:TDSE_OP}, for two different models are the same or close for the moments in time which are of interest to the physical scenario we investigate, we should not assume that the two different models make similar predictions for the same physical scenario.

So far, we only considered how different approximations,\ie simplifications, affect the time evolution of selected probabilities or probability amplitudes. In \chapref{chap:GET}, we investigate how different approximations change the gate-error signatures for complete quantum circuits, see \REF\cite{Wi17}.


\chapter{On the fragility of gate-error metrics in simulations of flux-tunable transmon quantum computers}\label{chap:GET}
\newcommand{\ROT}{R^{(x)}}
\newcommand{\ROTZ}{R^{(z)}}
\newcommand{\CNOT}{\text{CNOT}}
\newcommand{\TP}{2 \pi}
\newcommand{\HA}{\text{H}}

Building a gate-based quantum computer in the real world requires a tremendous amount of precision in terms of control over a quantum system. Therefore, the need to somehow quantify deviations between the states of the ideal gate-based quantum computer (IGQC), see \secsref{sec:MathematicalFramework}{sec:Algorithms} and prototype gate-based quantum computers (PGQCs), see \secref{sec:Prototype gate-based quantum computers}, arises. Additionally, many theoretical studies, see for example \REFS\cite{Gu21,Roth19,Yan18,McKay16,Wittler21}, focus on deviations between the states of the IGQC and non-ideal gate-based quantum computers (NIGQCs), see \secref{sec:FromStaticsToDynamics}.

In this chapter we study how susceptible gate-error quantifiers, that aim to quantify these deviations, are to the assumptions (approximations) which make up the underlying model. Here, we consider NIGQC models which describe certain aspects of the time evolution of PGQCs. A review of the literature suggests that PGQCs are susceptible to many time-dependent factors, see \secref{sec:Prototype gate-based quantum computers}. Consequently, it seems to be worth knowing to what extent the gate-error quantifiers are affected by changes in the model once we describe the time evolution of a system as a real-time process. In \chapref{chap:NA} we discussed two device architectures, see \figsref{fig:arch_sketch}(a-b). A detailed discussion of only one device architecture already takes up a large part of this thesis. Therefore, in this chapter we focus on one device architecture only. The author of this thesis decided to present the results for architecture II, which seemed to be more understandable.

This chapter is structured as follows. In \secsref{sec:GET_SystemSpecificationAndSimulationParameters}{sec:GET_errormeasures} we specify the model. First, in \secref{sec:GET_SystemSpecificationAndSimulationParameters}, we introduce the device architecture and the device parameters. Next, in \secref{sec:GET_ControlPulsesAndGateImplementation}, we introduce the control pulses we use to realise the elementary gate sets. Then, in \secref{sec:GET_errormeasures}, we introduce the gate-error quantifiers we use to assess to what extend the state of a NIGQC and the state of the IGQC deviate from one another for a given sequence of gates. The main results of this chapter are discussed in \secsref{sec:GET_SpectrumOfAFourQubitNIGQC}{sec:GET_Aprox}. First, in \secref{sec:GET_SpectrumOfAFourQubitNIGQC}, we study the spectrum of a four-qubit NIGQC and its relevance for the implementation of two-qubit gates. Next, in \secref{sec:GET_GateMetricsForTheElementaryGateSet}, we discuss the gate-error metrics for the elementary gate sets we use to implement all gate sequences with our NIGQCs. Then, in \secref{sec:GET_HigherStates}, we investigate how modelling the dynamics of a NIGQC with more and less basis states affects the gate-error quantifiers we compute. Afterwards, in \secref{sec:GET_Para}, we study how changes in the control pulse parameters affect the gate-error quantifiers we compute. Next, in \secref{sec:GET_Aprox}, we look at how modelling flux-tunable transmons as adiabatic and non-adiabatic anharmonic oscillators affects the gate-error quantifiers. Finally, in \secref{sec:GET_SummaryAndConclusions}, we summarise and discuss the findings. Here, we also present a series of conclusions drawn from our findings. Note that we use $\hbar=1$ throughout this chapter.

\section{System specification and simulation parameters}\label{sec:GET_SystemSpecificationAndSimulationParameters}

In this section we introduce the device architecture and the model parameters we use to obtain the results in this chapter. The device parameters, control pulses and some technical details relevant to the implementation of different gates were provided by the Quantum Device Lab which is affiliated with the Eidgenoessische Technische Hochschule Zurich. Note that \REFS\cite{Lacroix2020,Andersen2020,Krinner2020} discuss a series of experiments with PGQCs characterised by similar device parameters.

\begin{figure}[tbp!]
\renewcommand{\hold}{0.65}
\centering
\begin{minipage}{0.30\textwidth}
    \centering
    \scalebox{\hold}{
    \begin{tikzpicture}[node distance={20.0mm}, thick, main/.style = {draw, circle}]
    \node[main]    (0)            {$\omega_{0}^{(Q)}$};
    \node[main]    (1)            [above right of=0]{$\omega_{2}^{(R)}$};
    \node[main]    (2)            [above right of=1]{$\omega_{1}^{(Q)}$};

    \path[-]
    (0)            edge[bend right=0] node[midway,below,rotate=45] {$G_{0,2}^{(4)}$}    (1)
    (1)            edge[bend right=0] node[midway,below,rotate=45] {$G_{1,2}^{(4)}$}    (2);
    \end{tikzpicture}
    }\\
    \vspace{2.3cm}{\LARGE (a)}
  \end{minipage}
\begin{minipage}{0.30\textwidth}
      \centering
      \scalebox{\hold}{
      \begin{tikzpicture}[node distance={20.0mm}, thick, main/.style = {draw, circle}]
      \node[main]    (0)            {$\omega_{0}^{(Q)}$};
      \node[main]    (1)            [above right of=0]{$\omega_{3}^{(R)}$};
      \node[main]    (2)            [above right of=1]{$\omega_{1}^{(Q)}$};
      \node[main]    (3)            [above left of=2]{$\omega_{4}^{(R)}$};
      \node[main]    (4)            [above left of=3]{$\omega_{2}^{(Q)}$};

      \path[-]
      (0)            edge[bend right=0] node[midway,below,rotate=45] {$G_{0,3}^{(4)}$}    (1)
      (1)            edge[bend right=0] node[midway,below,rotate=45] {$G_{1,3}^{(4)}$}    (2)
      (2)            edge[bend left=0]  node[midway,above,rotate=-45] {$G_{1,4}^{(4)}$}    (3)
      (3)            edge[bend right=0] node[midway,above,rotate=-45] {$G_{2,4}^{(4)}$}    (4);
      \end{tikzpicture}
      }\\
      \vspace{0.45cm}{\LARGE (b)}
  \end{minipage}
\begin{minipage}{0.30\textwidth}
    \centering
    \scalebox{\hold}{
    \begin{tikzpicture}[node distance={20.0mm}, thick, main/.style = {draw, circle}]
    \node[main]    (0)            {$\omega_{0}^{(Q)}$};
    \node[main]    (1)            [above right of=0]{$\omega_{4}^{(R)}$};
    \node[main]    (2)            [above right of=1]{$\omega_{1}^{(Q)}$};
    \node[main]    (3)            [above left of=2]{$\omega_{5}^{(R)}$};
    \node[main]    (4)            [above left of=3]{$\omega_{2}^{(Q)}$};
    \node[main]    (5)            [below left of=4]{$\omega_{6}^{(R)}$};
    \node[main]    (6)            [below left of=5]{$\omega_{3}^{(Q)}$};
    \node[main]    (7)            [below right of=6]{$\omega_{7}^{(R)}$};
    \path[-]
    (0)            edge[bend right=0] node[midway,below,rotate=45] {$G_{0,4}^{(4)}$}    (1)
    (0)            edge[bend left=0]  node[midway,below,rotate=-45] {$G_{0,7}^{(4)}$}    (7)
    (1)            edge[bend right=0] node[midway,below,rotate=45] {$G_{1,4}^{(4)}$}    (2)
    (2)            edge[bend left=0]  node[midway,above,rotate=-45] {$G_{1,5}^{(4)}$}    (3)
    (3)            edge[bend right=0] node[midway,above,rotate=-45] {$G_{2,5}^{(4)}$}    (4)
    (4)            edge[bend left=0]  node[midway,above,rotate=45] {$G_{6,2}^{(4)}$}    (5)
    (5)            edge[bend right=0] node[midway,above,rotate=45] {$G_{3,6}^{(4)}$}    (6)
    (6)            edge[bend left=0]  node[midway,below,rotate=-45] {$G_{3,7}^{(4)}$}    (7);
    \end{tikzpicture}
    }\\
    \vspace{0.48cm}{\LARGE (c)}
  \end{minipage}
\caption[Illustrations of a two-qubit(a), three-qubit(b) and four-qubit(c) NIGQC.]{Illustrations of a two-qubit(a), three-qubit(b) and four-qubit(c) NIGQC. The different NIGQCs consist of LC resonators, or resonators for short, with resonance frequencies $\omega_{i}^{(R)}$ and flux-tunable transmon qubits, or transmon qubits for short, with park frequencies $\omega_{i}^{(Q)}$, see \secaref{sec:ResAndTLS}{sec:Transmons}, respectively. Here, $i$ denotes the discrete index we use to address the different subsystems. The parameters $G_{i,j}^{(4)}$ denote interaction strength constants, see \equref{eq:CHMDEF}(e). The interactions between the resonators and transmon qubits are modelled as dipole-dipole interactions. We use the model Hamiltonians \equaref{eq:CHM}{eq:EHM} to describe the dynamics of the different systems. The devices parameters are listed in \tabref{tab:device_para}.\label{fig:device_sketch}}
\end{figure}
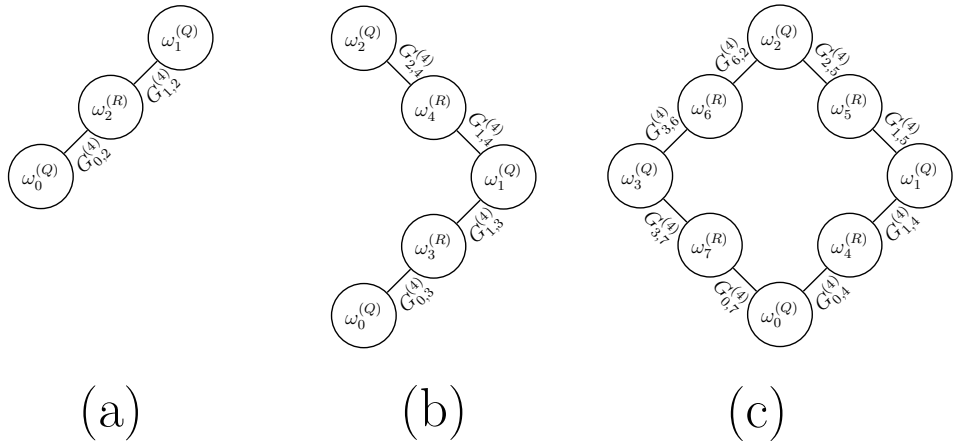
\begin{table}[!tbp]
\caption[Device parameters for the two-qubit, three-qubit and four-qubit NIGQCs illustrated in \figsref{fig:device_sketch}(a-c).]{Device parameters for the two-qubit, three-qubit and four-qubit NIGQCs illustrated in \figsref{fig:device_sketch}(a-c). The first column contains the discrete indices $i$ we use to index the different subsystems. The second column contains the resonator frequencies $\omega_{i}^{(R)}$. The third and fourth column contain the transmon qubit frequencies $\omega_{i}^{(Q)}$ and anharmonicities $\alpha_{i}^{(Q)}$. The fifth column contains the capacitive energies $E_{C_{i}}$ for the transmon qubits. The sixth and seventh column contain the left $E_{J_{i},l}$ and right $E_{J_{i},r}$ Josephson energies. The eighth column contains the flux offset values $\varphi_{0,i}=\varphi_{j}(t)$ at time $t=0$. These values define operating points for the different transmon qubits. All units, except for the flux offset, are in GHz. We do not list the asymmetry factors $d_{i}=0.5$ and the interaction strength constants $G_{i,j}^{(4)}=0.300$ because they are constant for all $i,j \in \mathbb{N}^{0}$. Note that we use $\hbar=1$ throughout this chapter.}\label{tab:device_para}
\centering
{\small
\setlength{\tabcolsep}{4pt}
\begin{tabularx}{\textwidth}{X X X X X X X X X}

\hline\hline

$i$&              $\omega_{i}^{(R)}/2\pi$&              $\omega_{i}^{(Q)}/2\pi$&              $\alpha_{i}^{(Q)}/2\pi$&              $E_{C_{i}}/2\pi$&              $E_{J_{i},l}/2\pi$&              $E_{J_{i},r}/2\pi$&              $\varphi_{0,i}/2\pi$&                             \\
\hline
$0$              &              $\text{n/a}$   &              $4.200$        &              $-0.320$       &              $1.068$        &              $2.355$        &              $7.064$        &              $0$            &                             \\

$1$              &              $\text{n/a}$   &              $5.200$        &              $-0.295$       &              $1.037$        &              $3.612$        &              $10.837$       &              $0$            &                             \\

$2$              &              $\text{n/a}$   &              $5.700$        &              $-0.285$       &              $1.017$        &              $4.374$        &              $13.122$       &              $0$            &                             \\

$3$              &              $\text{n/a}$   &              $4.960$        &              $-0.300$       &              $1.045$        &              $3.281$        &              $9.843$        &              $0$            &                             \\

$4-7$              &              $45.000$       &              $\text{n/a}$   &              $\text{n/a}$   &              $\text{n/a}$   &              $\text{n/a}$   &              $\text{n/a}$   &              $\text{n/a}$   &                             \\

\hline\hline
\end{tabularx}
}
\end{table}

Figures~\ref{fig:device_sketch}(a-c) display illustrations of the three different NIGQCs we model in this chapter. The $\omega_{i}^{(R)}$ parameters denote resonator frequencies. Similarly, the $\omega_{i}^{(Q)}$ parameters denote transmon qubit park frequencies. These are the frequencies at which the transmon qubits reside if no external fluxes are applied to the system. Furthermore, the $G_{i,j}^{(4)}$ parameters denote the interaction strength constants, see \equref{eq:CHMDEF}(e). We assume that interactions are of the dipole-dipole type, see \secaref{sec:TheQuantumComputerCircuitHamiltonianModel}{sec:TheQuantumComputerEffectiveHamiltonianModel}. In this chapter we model two-qubit (N=2), three-qubit (N=3) and four-qubit (N=4) NIGQCs, see \figsref{fig:device_sketch}(a-c), respectively. Here $N \in \mathbb{N}$ denotes the number of qubits.

The device parameters for the Hamiltonians in \equaref{eq:CHM}{eq:EHM} are listed in \tabref{tab:device_para}. We were provided with the measured qubit frequencies, anharmonicities, flux offset values and asymmetry factors, see \secref{sec:Prototype gate-based quantum computers}, \secref{sec:Transmons} and \REF\cite{Naghiloo19}. Furthermore, the resonator frequencies are given. We use the relations
\begin{subequations}
  \begin{align}
    \omega_{i}^{(Q)}&=(E_{i}^{(1)}-E_{i}^{(0)}),\\
    \alpha_{i}^{(Q)}&=(E_{i}^{(2)}-E_{i}^{(0)})-2\omega_{i}^{(Q)},\\
    d_{i}&=\frac{(E_{J_{r,i}}-E_{J_{l,i}})}{(E_{J_{r,i}}+E_{J_{l,i}})},
  \end{align}
\end{subequations}
to fit the given experimental parameters to the lowest three energy levels of the circuit Hamiltonian \equref{eq:flux-tunable transmon_beta}. This procedure yields the capacitive and Josephson energies $E_{C_{i}}$, $E_{J_{r,i}}$ and $E_{J_{l,i}}$ as a result. The model parameters used deviate at most tens of MHz from the ones which were provided. Note that we use the constant asymmetry factor $d_{i}=0.5$ for all transmon qubits as well as the constant frequency $\omega_{i}^{(R)}=45.0$ GHz for all coupling resonators. Furthermore, we use a fixed value of $300$ MHz for the interaction strength constants $G_{i,j}^{(4)}$. This value for the interaction strength constants roughly reproduces the gate durations for the two-qubit $\CZ$ gates which were observed in experiments.

For almost all simulations of the circuit Hamiltonian \equref{eq:CHM}, we use sixteen bare basis states for all transmon qubits which experience a flux drive. Furthermore, the transmon qubits which do not experience a flux drive and the coupling resonators are modelled with four basis states only.

For the simulations of the effective Hamiltonian \equref{eq:EHM}, we use four bare basis states for all transmon qubits and coupling resonators. Also, we use a time-dependent interaction strength to model the dipole-dipole interactions. Additionally, we use the higher-order series expansions \equaref{eq: expansion frequency}{eq: expansion anharmonicity} to model the flux-tunable frequencies and anharmonicities, see \chapref{chap:NA} for more details.

\section{Control pulses and gate implementation}\label{sec:GET_ControlPulsesAndGateImplementation}
\newcommand{\Cctl}{n(t)}
\newcommand{\CctlAp}{a}
\newcommand{\CctlDr}{b}
\newcommand{\CctlDf}{\omega^{(D)}}
\newcommand{\CctlSg}{\sigma}
\newcommand{\CctlTd}{T_{d}}
\newcommand{\Fctl}{\varphi(t)}
\newcommand{\FctlAp}{\delta}
\newcommand{\FctlSg}{\sigma}
\newcommand{\FctlTp}{T_{p}}
\newcommand{\FctlTd}{T_{d}}

In this section we introduce the control pulses we use to implement different gates with our NIGQCs.

\graphicspath{{.//FiguresAndData/GETPaper/PulseData/}}
\renewcommand{\scale}{0.35}
\begin{figure}[!tbp]
    \centering
    \begin{minipage}{0.32\textwidth}
        \centering
        \includegraphics[scale=\scale]{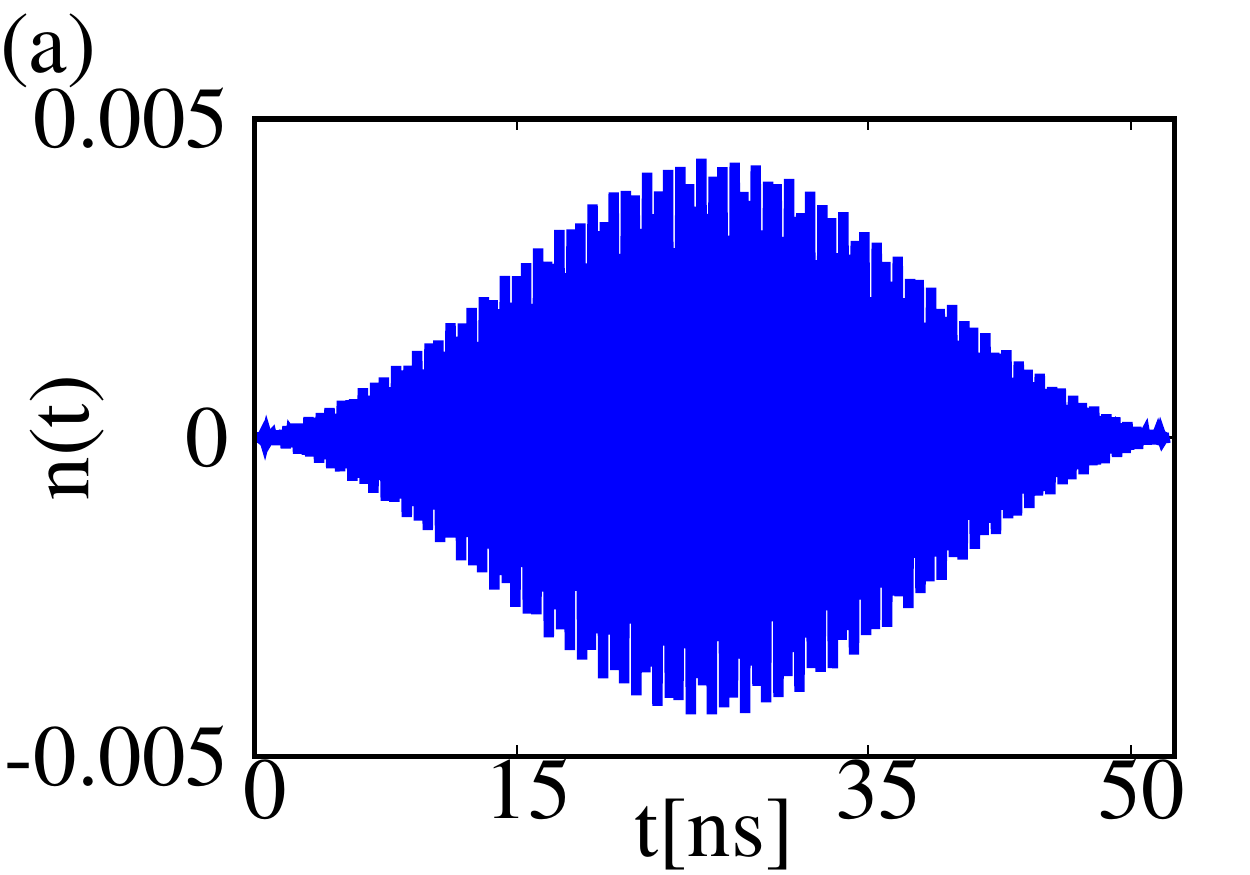}
    \end{minipage}
    \begin{minipage}{0.32\textwidth}
        \centering
        \includegraphics[scale=\scale]{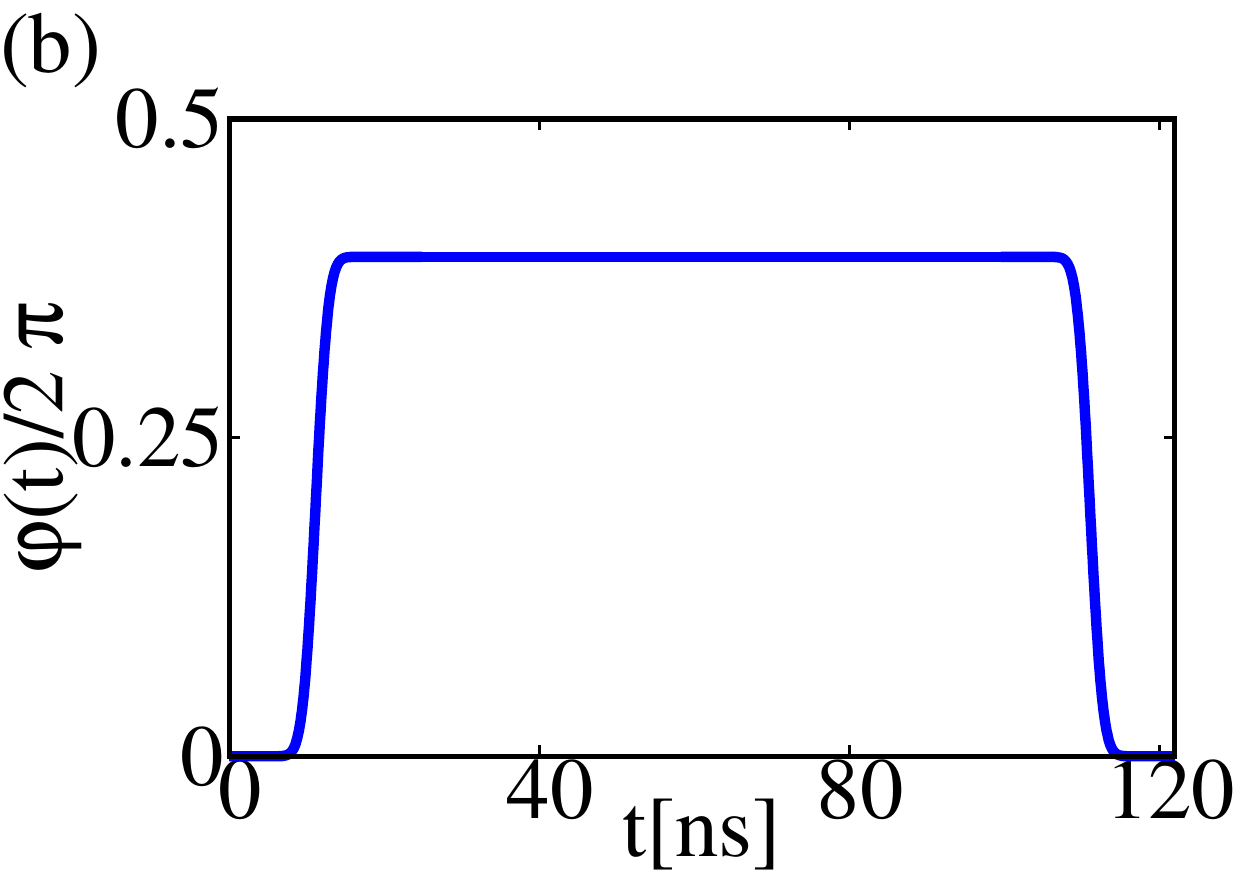}
    \end{minipage}
    \begin{minipage}{0.32\textwidth}
        \centering
        \includegraphics[scale=\scale]{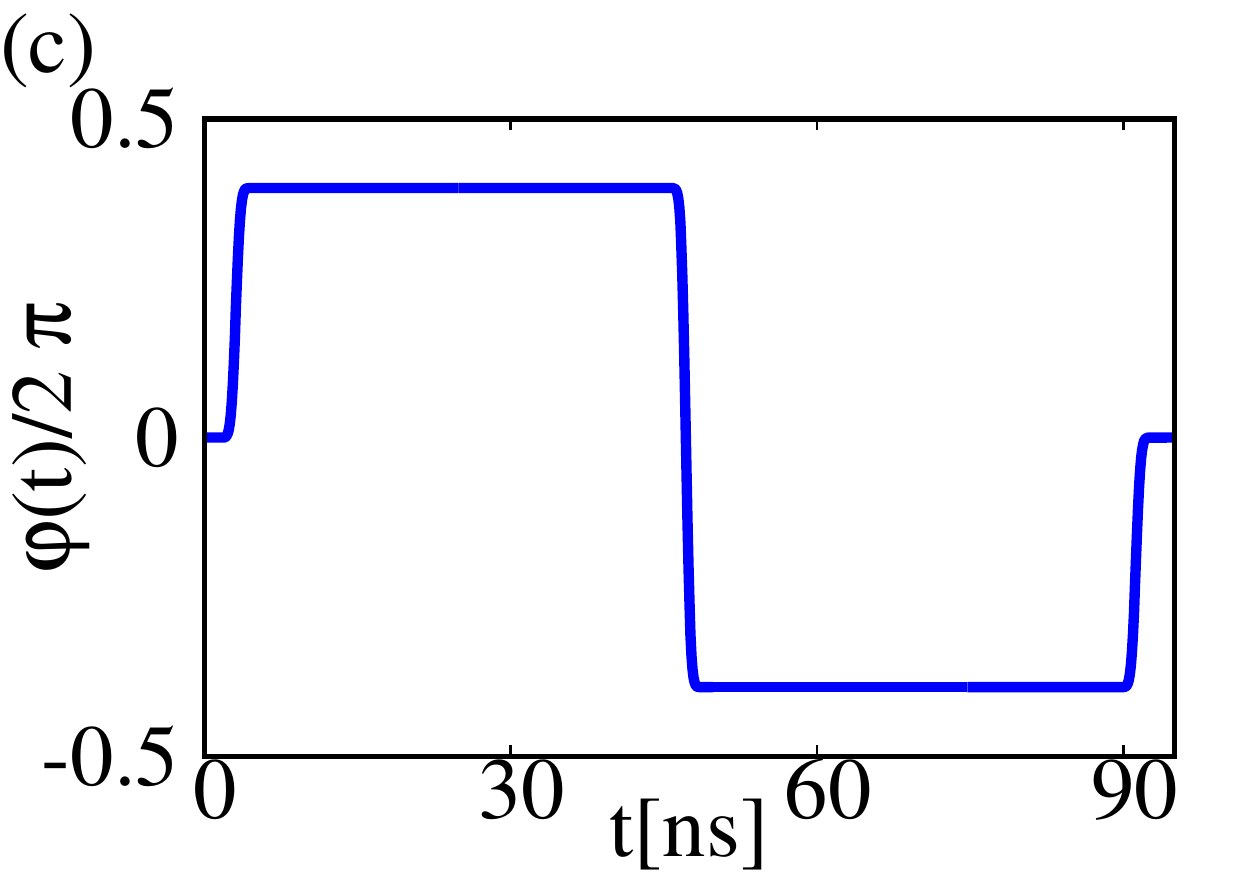}
    \end{minipage}
    \caption[External charge $\Cctl$(a) and external flux $\Fctl$(b-c) as functions of time $t$.]{External charge $\Cctl$(a) and external flux $\Fctl$(b-c) as functions of time $t$. In \PANL{a} we show a microwave pulse (MP). In \PANSL{b-c} we display a unimodal pulse (UMP) and a bimodal pulse (BMP), respectively.}\label{fig:GET_pulse_time_evo}
\end{figure}

Figures~\ref{fig:GET_pulse_time_evo}(a-c) show the external charge $n(t)$(a) and external flux $\varphi(t)$(b-c) as functions of time $t$. The pulse in \figref{fig:GET_pulse_time_evo}(a) is a microwave pulse (MP). The pulse in \figref{fig:GET_pulse_time_evo}(c) is usually referred to as a unimodal pulse (UMP). Similarly, we denote the pulse in \figref{fig:GET_pulse_time_evo}(c) as a bimodal pulse (BMP). Note that this pulse is also sometimes referred to as net-zero flux pulse, see \REF\cite{Rol19}.

We use the MP
\begin{equation}\label{eq:charge_ctl}
    \Cctl= \CctlAp G(t,\CctlSg,\CctlTd) \cos\BRR{\CctlDf t-\phi} + \CctlDr \dot{G}(t,\CctlSg,\CctlTd) \sin\BRR{\CctlDf t-\phi},
\end{equation}
to implement single-qubit $\ROT(\pi/2)$ rotations, see \equref{eq:single_qubit_rot}(a), by making use of the charge-dependent linear driving terms in \equref{eq:CHMDEF}(b) and \equref{eq:drive_term_charge}. Here $\CctlAp$ denotes the pulse amplitude, $\CctlDf$ is the drive frequency and $\CctlDr$ is the amplitude of the DRAG component of the pulse, see \REF\cite{Motzoi09}. Furthermore, the real-valued function $G(t,\CctlSg,\CctlTd)$ denotes a Gaussian envelope that is centered around half of the pulse duration $\CctlTd$. The parameter $\CctlSg$ determines the shape of the envelope.

The variable $\phi$ denotes the phase of the MP. We use this phase to implement so-called virtual $\ROTZ(\phi)$ rotations, see \equref{eq:single_qubit_rot}(c) or virtual $Z$ gates for short, see \REF\cite{McKay17}. The simulation code is used to compute the phases $\phi$ for the MPs in \equref{eq:charge_ctl} has to be specifically designed for the two-qubit gates we implement on our NIGQCs. This algorithm preprocesses every program,\ie every sequence of gates and computes the correct pulse phases $\phi$ for the different MPs in the pulse sequence. Since we only implement two-qubit $\CZ$ gates on our NIGQCs, we find that the preprocessing is rather simple. Note that the $\CZ$ gate commutes with the single-qubit z-axis rotation. We omit a detailed discussion of the preprocessing algorithm which is part of the compiler module in \figref{fig:SoftwareDiagram} to shorten this chapter.

We use the UMP
\begin{equation}\label{eq:flux_ctl_ump}
  \Fctl= \frac{\FctlAp}{2}\BRR{\text{erf}\BRR{\frac{t}{\sqrt{2}\FctlSg}} - \text{erf}\BRR{\frac{(t-\FctlTp)}{\sqrt{2}\FctlSg}}},
\end{equation}
and BMP
\begin{equation}\label{eq:flux_ctl_bmp}
  \begin{split}
    \Fctl= \frac{\FctlAp}{2}  \BRR{ \text{erf}\BRR{\frac{t}{\sqrt{2}\FctlSg}} - 2\text{erf}\BRR{\frac{(t-\FctlTp/2)}{\sqrt{2}\FctlSg}} +\text{erf}\BRR{\frac{(t-\FctlTp)}{\sqrt{2}\FctlSg}}}.
  \end{split}
\end{equation}
to implement two-qubit $\CZ$ gates by making use of the flux-dependent driving terms in \equref{eq:CHMDEF}(b) as well as \equaref{eq:transeff}{eq:drive_term_flux}. Here, the parameter $\FctlAp$ denotes the pulse amplitude, $\FctlSg$ refers to a parameter that allows us to control how fast the pulse flanks rise (fall) and $\FctlTp$ denotes the pulse time. We use another parameter $\FctlTd$, the pulse duration, to add some buffer time to the control pulses. The pulse duration $\FctlTd$ is not part of \equaref{eq:flux_ctl_ump}{eq:flux_ctl_bmp} but only part of the simulation software. Note that the Hamiltonian \equref{eq:flux-tunable transmon recast} is $2\pi$ periodic. Therefore, in this chapter we give the external flux $\varphi(t)$ and the pulse amplitude $\PA$ always in units of $2\pi$.

Every flux pulse is followed by single-qubit z-axis rotations $\ROTZ_{i}(\phi_{i})$ applied to every transmon qubit in the system, see \REF\cite[Section VII B 2]{Blais2020circuit}. In the end, all z-axis rotations are executed as virtual $Z$ gates by means of the MPs as discussed above. Finally, at the end of the program, \ie the pulse sequence, we only once transform the frame of reference. This transformation can be expressed as $c_{\mathbf{z}} \mapsto e^{i \phi_{\mathbf{z}} } c_{\mathbf{z}}$ for $\mathbf{z} \in \{0,1\}^{N}$, where the different phase factors $\phi_{\mathbf{z}}$ are computed by the preprocessing algorithm, see also \REF\cite[Section 3.3.2]{Willsch2020}.

\section{Computation of gate-error quantifiers}\label{sec:GET_errormeasures}
\newcommand{\DNV}{\mu_{\diamond}}
\newcommand{\FV}{\mu_{\text{F}_{\text{avg}}}}
\newcommand{\IFV}{\mu_{\text{IF}_{\text{avg}}}}
\newcommand{\LNV}{\mu_{\text{Leak}}}
\newcommand{\SDV}{\mu_{\text{SD}}}
\newcommand{\idlabel}{\text{U}}
\newcommand{\aclabel}{\text{M}}
\newcommand{\idxEFF}{\text{Eff.}}
\newcommand{\idxCIR}{\text{Cir.}}

In this section we introduce the gate-error quantifiers used to assess the deviations between the states of the IGQC and various NIGQC models for given sequences of gates.

In \secref{sec:MathematicalFramework} we described the computations performed with the IGQC as unitary mappings of the form
\begin{equation}
  \ket{\psi} \mapsto \OP{U}\ket{\psi},
\end{equation}
where $\OP{U} \in \mathbb{U}(\mathcal{H}^{2^{N}})$ is a unitary operator defined on the $2^{N}$-dimensional Hilbert space $\mathcal{H}^{2^{N}}$. We can use a simple gate-error quantifier like the statistical distance
\begin{equation}\label{eq:stat_dis}
  \SDV(\mathbf{p},\mathbf{\tilde{p}})=\frac{1}{2} \sum_{\mathbf{z} \in \{0,1\}^{N}} \norm{p_{\mathbf{z}}-\tilde{p}_{\mathbf{z}}}_{1},
\end{equation}
where $\mathbf{p}=(p_{\mathbf{z}})_{\mathbf{z} \in \{0,1\}^{N}}$ and $\mathbf{\tilde{p}}=(\tilde{p}_{\mathbf{z}})_{\mathbf{z} \in \{0,1\}^{N}}$ denote discrete probability distributions, to quantify how well a computation $\OP{U}$ was realised with a PGQC or NIGQC. Here, $\norm{\mdot}_{1}$ denotes the absolute value, $\mathbf{p}$ is defined in terms of the probability amplitudes $p_{\mathbf{z}}=\norm{c_{\mathbf{z}}}_{1}^{2}$ of the IGQC for a given initial state $\ket{\psi}$ and $\mathbf{\tilde{p}}$ denotes the reference distribution we compare against. We can define $\mathbf{\tilde{p}}$ to be the relative frequencies obtained in a PGQC experiment or the probability amplitudes obtained from the truncated state vector of an NIGQC. The statistical distance is also known as the total variation distance, see \REFS\cite{Sanders2015,Sanders16,Nielsen:2011:QCQ:1972505}. Note that if we use the statistical distances to quantify gate errors, we completely ignore the phase of the system. Furthermore, the statistical distance only takes into account a single initial state. Hence, often theoretically more sophisticated gate-error quantifiers are considered. However, in order to provide a proper definition of these quantifiers, we have to extend the mathematical framework we use.

So far we have considered computations as mappings between two state vectors. However, there exists an alternative formalism to describe the state of a quantum system, the density operator formalism. Note that this formalism is used widely in physics,\eg in statistical physics see \REF\cite{Balian91,Balian07}. The density operator formalism also relies on four postulates and/or assumptions which set the stage for the complete mathematical framework.

If we consider an $N$-qubit IGQC, the first postulate states that the state of a quantum system is completely described by a density operator $\OP{\rho} \in \mathbb{P}(\mathcal{H}^{2^{N}})$. The set of density operators $\mathbb{P}(\mathcal{H}^{2^{N}})\subseteq \mathbb{L}(\mathcal{H}^{2^{N}})$ contains operators $\OP{\rho}$ which satisfy the three conditions: $\trace{}(\OP{\rho})=1$, $\OP{\rho}=\OP{\rho}^{\dagger}$ and for all $\ket{\psi} \in \mathcal{H}^{2^{N}}$ the inequality $\braket{\psi|\OP{\rho}|\psi}\geq 0$ holds true. Operators which satisfy the last condition are also known as positive (semi-definite) operators and one often writes $\OP{\rho}\geq 0$ as a shorthand for $\forall \ket{\psi} \in \mathcal{H}^{2^{N}} \colon \braket{\psi|\OP{\rho}|\psi}\geq 0$. The second postulate states that the evolution of a closed system is described by a unitary transformation of the form
\begin{equation}
  \OP{\rho} \mapsto \OP{U} \OP{\rho} \OP{U}^{\dagger}.
\end{equation}
As for the state vector formalism, the third postulate provides us with relations which allow us to determine the event probabilities
\begin{equation}
  p(i)=\trace{}\BRR{\OP{E}_{i}^{\dagger}\OP{E}_{i}\OP{\rho}},
\end{equation}
where the set of operators $\{\OP{E}_{i}\}\subseteq \mathbb{L}(\mathcal{H}^{2^{N}})$ allows us to distinguish between different discrete events $i \in I \subseteq \mathbb{N}^{0}$,\ie measurement outcomes. This set has to satisfy the relation
\begin{equation}
    \sum_{i  \in I}\OP{E}_{i}^{\dagger}\OP{E}_{i} = \OP{I}.
\end{equation}
Furthermore, the density operator
\begin{equation}
  \OP{\rho}^{\prime}=\frac{\OP{E}_{i}\OP{\rho}\OP{E}_{i}^{\dagger}}{p(i)},
\end{equation}
describes the state after the measurement if we find that event $i$ has taken place. The fourth postulate states that composed systems are described by a tensor product structure. Here we assume that if the individual systems $n \in \mathbb{N}$ are prepared in the states $\OP{\rho}_{n}$, then the state of the composite system is given by
\begin{equation}
  \OP{\rho}=\tensupper{n=0}{N-1}\OP{\rho}_{n}.
\end{equation}
The density operator formalism is a convenient theoretical tool which allows us to describe certain physical scenarios with much more ease, compared to the state vector formalism. For example, we might describe situations where the initial state of the system $\ket{\psi} \in \mathcal{H}^{2^{N}}$ is not known definitively but only with a certain probability. Also, density operators often simplify the analysis of individual subsystems which are part of a larger system, in the sense of postulate four. We do not make much use of the formalism in this thesis. Therefore, we cut the discussion of this subject short. A much more detailed introduction can be found in \REFS\cite{Nielsen:2011:QCQ:1972505,Watrous2018}. In fact, the only reason why we have to consider density operators and the corresponding mathematical framework is that some gate-error quantifiers are defined in terms of so-called quantum operations.

In \secref{sec:MathematicalFramework} we introduce operators,\ie linear functions which map one vector to another vector. Since we find that the set $\mathbb{L}(\mathcal{H}^{2^{N}})$ together with the addition of operators, scalar multiplication and the field of complex numbers $\mathbb{C}$ forms a vector space of its own, we can simply define linear functions which map operators to other operators and retain some of the old terminology. We denote these functions as superoperators $\SUPOP{\mathcal{E}}$. Superoperators, like every other function, can possess features like symmetries. Additionally, we can restrict the domain and range of superoperators such that we map only certain types of operators to others. In this thesis we only consider superoperators which map density operators to other density operators. The superoperators we are interested in are usually denoted as quantum operations and we impose the following constraints on them. First, we require them to be Hermiticity preserving, which means $\SUPOP{\mathcal{E}}(\OP{\rho})^{\dagger}=\SUPOP{\mathcal{E}}(\OP{\rho}^{\dagger})$ for all $ \OP{\rho} \in \mathbb{P}(\mathcal{H}^{2^{N}})$. Second, we would like to work with so-called completely positive superoperators. A completely positive superoperator $\SUPOP{\mathcal{E}}$ maps all positive $\OP{\rho}  \in \mathcal{P}(\mathcal{H}^{D} \otimes \mathcal{H}^{2^{N}})$, where $D\in \mathbb{N}$, to positive operators by means of the map $\SUPOP{I}\otimes\SUPOP{\mathcal{E}}$, where $\SUPOP{I}$ is the identity operator associated with the Hilbert space $\mathcal{H}^{D}$. The statements made in the last sentence are not as accurate as required. However, since we do not make much use of quantum operations in this thesis, we abstain from extending the mathematical framework any further. A much more extensive discussion of the subject is provided by \REFS\cite{Watrous2018,Nielsen:2011:QCQ:1972505}. Additionally, if we find that a quantum operation $\SUPOP{\mathcal{E}}$ is also trace preserving $\trace{}(\SUPOP{\mathcal{E}}(\OP{\rho}))=\trace{}(\OP{\rho})$, we denote this quantum operation as quantum or error channel.

With the mathematical framework discussed, we can define the quantum operation
\begin{equation}\label{eq:quantumoperation_id}
  \SUPOP{\mathcal{E}}_{\idlabel}(\OP{\rho})=\OP{U} \OP{\rho} \OP{U}^{\dagger},
\end{equation}
which describes the ideal target operation. Furthermore, we can define the quantum operation
\begin{equation}\label{eq:quantumoperation_ac}
  \SUPOP{\mathcal{E}}_{\aclabel}(\OP{\rho})=\OP{M} \OP{\rho} \OP{M}^{\dagger},
\end{equation}
which describes the actual operation performed with the NIGQCs we model in this chapter. Here, the operator $\OP{M}$ is defined as
\begin{equation}
  \OP{M}= \OP{P}\OP{\mathcal{U}}(t,t_{0})\OP{P},
\end{equation}
where $\OP{\mathcal{U}}(t,t_{0})$ denotes the time-evolution operator
\begin{equation}\label{eq:time_evo_op}
  \hat{\mathcal{U}}(t,t_{0}) = \mathcal{T} \exp\left( -i \int_{t_{0}}^{t} \hat{H}(t^{\prime}) dt^{\prime} \right),
\end{equation}
at time $t$ and $\OP{P}$ denotes a projection onto the computational subspace. Note that the definitions in \equaref{eq:quantumoperation_id}{eq:quantumoperation_ac} are taken from the author of \REF\cite[Section 6.1]{Willsch2020}.

We can use the quantum operations in \equaref{eq:quantumoperation_id}{eq:quantumoperation_ac} to define the average fidelity as
\begin{equation}
  \FV=\int \braket{\psi| \SUPOP{\mathcal{E}}_{\aclabel} \SUPOP{\mathcal{E}}_{\idlabel}^{-1} \BRR{\ketbra{\psi}{\psi}}|\psi} \dx{\ket{\psi}},
\end{equation}
where the integral is taken over all normalised states $\ket{\psi}\in\mathcal{H}^{2^{N}} \colon \braket{\psi|\psi}=1$. If we define the auxiliary operator
\begin{equation}
  \OP{V}=\OP{U}\OP{M}^{\dagger},
\end{equation}
we can express the average fidelity as
\begin{equation}\label{eq:fid_avg}
  \FV=\frac{\norm{  \trace{}(\OP{V})}_{1}^{2} +\trace{}(\OP{M} \OP{M}^{\dagger})}{D\BRR{D+1}},
\end{equation}
where $D=2^{N}$. A derivation of \equref{eq:fid_avg} can be found in \REF\cite[Section 7]{Jin21}. Furthermore, we define the average infidelity
\begin{equation}\label{eq:in_fid_avg}
  \IFV=1-\mu_{\text{F}_{\text{Avg}}}.
\end{equation}
and a leakage measure
\newcommand{\labelleak}{\text{Leak}}
\begin{equation}\label{eq:leak}
  \LNV=1-\BRR{\frac{\trace{}(\OP{M} \OP{M}^{\dagger})}{D}},
\end{equation}
whose definition is motivated by the average fidelity in \equref{eq:fid_avg} and the fact that $\trace{}(\OP{M} \OP{M}^{\dagger})$ can be expressed solely in terms of the probability amplitudes such that the relation $0 \leq  \trace{}(\OP{M} \OP{M}^{\dagger}) \leq D$ holds true. Consequently, we find that $\LNV \in [0,1]$. The last gate-error quantifier we define is the diamond distance
\begin{equation}
        \DNV= \frac{1}{2} \norm{\SUPOP{\mathcal{E}}_{\aclabel} \SUPOP{\mathcal{E}}_{\idlabel}^{-1} - \OP{I}}_{\diamond},
\end{equation}
where the diamond norm
\begin{equation}\label{eq:diamond_norm_def}
  \norm{\SUPOP{\mathcal{A}}}_{\diamond}=\sup_{\mathcal{H}^{\tilde{D}}} \left\{ \sup_{\OP{\rho} \in \mathcal{P}(\mathcal{H}^{\tilde{D}} \otimes \mathcal{H}^{D})}\left\{ \norm{\SUPOP{I}\otimes\SUPOP{\mathcal{A}}\BRR{\OP{\rho}}}_{\trace} \right\} \right\},
\end{equation}
for an arbitrary superoperator $\SUPOP{\mathcal{A}}$ associated with the Hilbert space $\mathcal{H}^{D}$ can be defined, see \REF\cite[Section 2.1.1]{Sanders16}, in terms of a supremum over the finite-dimensional Hilbert spaces $\mathcal{H}^{\tilde{D}}$ and the trace norm
\begin{equation}
  \norm{X}_{\trace}=\trace{}\BRR{\sqrt{\OP{X}^{\dagger}\OP{X}}}.
\end{equation}
We can express the diamond distance as an infimum
\renewcommand{\hold}{Q}
\begin{equation}\label{eq:diamond_norm_inf}
  \begin{split}
    \DNV=\frac{1}{2}\inf_{\hold \in \text{GL}_{4}\BRR{\mathbb{C}}} \left\{\right. &\norm{\BRR{\OP{V},\OP{I}} \OP{\hold}^{-\dagger}\OP{\hold}^{-1} \BRR{\OP{V},\OP{I}}^{T}}_{2}^{\frac{1}{2}}\\                                                              &\norm{\BRR{\OP{V},-\OP{I}} \OP{\hold}^{\dagger}\OP{\hold}^{1} \BRR{\OP{V},-\OP{I}}^{T}}_{2}^{\frac{1}{2}} \left.\right\},
  \end{split}
\end{equation}
over the set of complex, invertible two by two matrices, see \REF\cite{Johnston09} or as a supremum
\renewcommand{\hold}{\psi}
\begin{equation}\label{eq:diamond_norm_sup}
  \begin{split}
    \DNV=\frac{1}{2}\sup_{\ket{\hold} \in \mathcal{H}^{2^{N}}} \left\{\right. \norm{\BRR{\OP{V}^{\dagger}\otimes\OP{I}} \ketbra{\hold}{\hold} \BRR{\OP{V}^{\dagger}\otimes\OP{I}}^{\dagger} - \ketbra{\hold}{\hold}}_{\trace} \left.\right\} ,
  \end{split}
\end{equation}
over the set of all state vectors $\ket{\hold} \in \mathcal{H}^{2^{N}}$ of the IGQC, see \REF\cite{Watrous2018}. To the best knowledge of the author, a closed form expression for the diamond distance $\DNV$ is not known to the research community. Therefore, we use the relation
\begin{equation}
    \DNV^{(\text{inf})} \leq \DNV \leq \DNV^{(\text{sup})},
\end{equation}
and the algorithms discussed in \REF\cite[Section 6.1.2]{Willsch2020} to obtain the value of $\DNV$ up to the fourth decimal.

\begin{figure}[tbp!]
\begin{minipage}{1.0\textwidth}
  \centering
  \adjustbox{scale=0.75,center}{\begin{tikzcd}[column sep=normal,arrows=rightarrow]
                                          & \mathcal{U}_{\idxCIR}(t) \arrow[bend left=30,dr] &                                                                                   & &\\
    \ket{\Psi^{(z)}(t_{\text{start}})}=\ket{\mathbf{z}} \arrow[ bend left=30,ur] \arrow[ bend right=30,dr] &                                  &  \ket{\Psi^{(z)}(t_{\text{final}})} \arrow["\text{Proj.} \OP{P}",r]  & \ket{\Psi^{(z)}_{\text{comp.}}} \arrow[r] & M(t_{\text{final}})  \arrow[r] & \mu_{x}\\
                                          & \mathcal{U}_{\idxEFF}(t) \arrow[bend right=30,ur] &                                          &                                                                      &                                  & &\\
    & & & & &\\
  \end{tikzcd}}
\end{minipage}
\caption[Illustration of the computations we perform to obtain a gate-error quantifier $\mu_{x}$, where $x$ denotes an arbitrary gate-error quantifier label.]{Illustration of the computations we perform to obtain a gate-error quantifier $\mu_{x}$. Here, $x$ denotes an arbitrary gate-error quantifier label. First, we simulate the time evolution $\ket{\Psi^{(z)}(t_{\text{final}})}$ for all $2^{N}$ initial states $\ket{\Psi^{(z)}(t_{\text{start}})}=\ket{\mathbf{z}}$ of a $N$-qubit NIGQC with the algorithms discussed in \chapref{chap:IV}. Next, we apply the projection matrix $P$ to obtain the computational states $\ket{\Psi^{(z)}_{\text{comp.}}}=P \ket{\Psi^{(z)}(t_{\text{final}})}$. Then, we construct the matrix $M$ given by \equref{eq:prop_matrix},\ie we store the different computational states as the columns of the matrix M. Note that we use the programming paradigm MPI to parallelise the simulations of the $2^{N}$ time evolutions. Furthermore, all simulation are performed on the supercomputer JUWELS, see \REF\cite{JUWELS}. Finally, we can use the matrix $U$ of some target operation and the matrix $M$ to compute $V=U M^{\dagger}$ and consequently $\mu_{x}$ by means of the expressions \equref{eq:in_fid_avg}, \equref{eq:leak} and \equaref{eq:diamond_norm_inf}{eq:diamond_norm_sup}.}\label{fig:comp_sketch}
\end{figure}
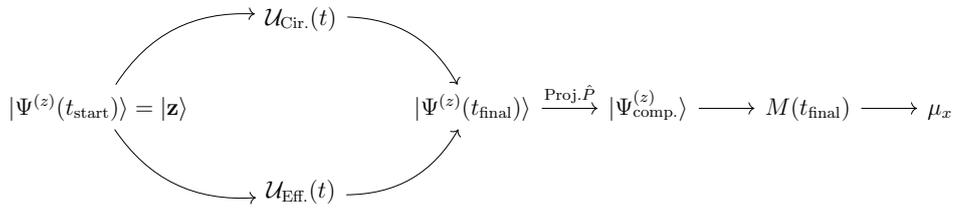
Figure~\ref{fig:comp_sketch} shows an illustration of the compuations we perform to obtain the gate-error quantifiers average infidelity $\IFV$ given by \equref{eq:in_fid_avg}, the leakage measure $\LNV$ in \equref{eq:leak} and the diamond distance $\DNV$ given by \equaref{eq:diamond_norm_inf}{eq:diamond_norm_sup}. Here the gate-error quantifier $\mu_{x}$ denotes an arbitrary gate-error quantifier with the label $x$. First, we simulate the time evolution of the system $\ket{\Psi^{(z)}(t_{\text{final}})}$ for all $2^{N}$ computational basis states $\ket{\Psi^{(z)}(t_{\text{start}})}=\ket{\mathbf{z}}$ of an $N$-qubit NIGQC with the algorithms discussed in \chapref{chap:IV}. Then, we compute the computational states $\ket{\Psi^{(z)}_{\text{comp.}}}=P \ket{\Psi^{(z)}(t_{\text{final}})}$ by applying the projection matrix $P$. Afterwards, we compute the matrix
\begin{equation}\label{eq:prop_matrix}
   M=\sum_{\mathbf{z} \in \{0,1\}^{N}} \ketbra{\Psi_{\text{comp.}}^{(z)}}{\mathbf{z}},
\end{equation}
where $\ket{\mathbf{z}}$ denotes one of the $2^{N}$ Cartesian unit vectors in the computer program. This means we store the computational states $\ket{\Psi^{(z)}_{\text{comp.}}}$ as the columns of the matrix $M$. In the end, we compute the matrix $V=U M^{\dagger}$ for the target operation $U$. Note that we parallelise the computations of the $2^{N}$ time evolutions with the programming paradigm MPI. Furthermore, the computations are performed on the supercomputer JUWELS, see \REF\cite{JUWELS}.

\section{Spectrum of the four-qubit NIGQC obtained with the circuit Hamiltonian}\label{sec:GET_SpectrumOfAFourQubitNIGQC}
\newcommand{\ListAdiabticModels}{\cite{Rol19,Foxen20,Yan18} }

The main subject of this section is the spectrum of the four-qubit NIGQC, illustrated in \figref{fig:GET_device_spectrum}(c), and its importance for the realisation of two-qubit $\CZ$ gates. Hence, we begin this section with a general discussion of the gate implementation mechanism and then turn our attention to the spectrum itself.

In order to realise a $\CZ$ gate, we tune a target computational state like $\ket{0,0,1,1}$ into resonance with a non-computational state like $\ket{0,0,0,2}$. Then, we wait some time until the population between the two states has swapped back and forth. Hopefully, in the end, we find that the target computational state has gained an additional phase of $e^{i \pi}$ with respect to the remaining computational states of the NIGQC. Furthermore, since the remaining computational states have presumably acquired only a so-called dynamic phase, see \REF[Section 6.6]\cite{Weinberg2015}, we apply z-axis rotations $\ROTZ_{i}(\phi_{i})$ to the different transmon qubits, see \REF\cite[Section VII B]{Blais2020circuit}. In this context, the word \Quote{tune} means that we tune the energies of the instantaneous basis states. Therefore, in our discussion of the gate implementation mechanism we refer to the instantaneous basis states of the system.

The flux control pulses we use to implement the $\CZ$ gates have pulse durations $\FctlTd$ between $80$ and $125$ ns and the time it takes for the population to swap back and forth lies roughly between $75$ and $100$ ns. The swap time corresponds to the plateau time of a control pulse, see for example \figsref{fig:GET_pulse_time_evo}(b-c). We can control the plateau or swap time by changing the interaction strength constant $G$ accordingly. For the simulations in this chapter, we use a constant interaction strength of $300$ MHz for all dipole-dipole interactions. This constant roughly reproduces the gate times of the experiments. The flux control pulse amplitude $\FctlAp$, corresponds to the plateau height of a pulse. The pulse amplitude is determined by the condition that the states involved in the population exchange should have approximately the same energies. Furthermore, both pulse flanks play a crucial role in the gate implementation. The role of the pulse flanks is discussed together with the spectrum of the four-qubit NIGQC later in this section.

The authors of \REF\cite[Section VII B 2]{Blais2020circuit} distinguish between so-called adiabatic, see \REF\cite{DiCarlo2009}, and non-adiabatic gates, see \REF\cite{DiCarlo2010}. However, in quantum theory the words adiabatic and non-adiabatic are usually only used in the context of the adiabatic approximation, see \REF\cite[Section 6.6]{Weinberg2015}. Clearly, since two instantaneous basis states exchange population, we find that the gate implementation mechanism described above is not an adiabatic process. Therefore, in the literature, the words adiabatic and non-adiabatic are used differently in this context. To the best knowledge of the author, in this particular context, gates implemented with shorter pulses are referred to as non-adiabatic and gates implemented with longer pulses are referred to as adiabatic gates. Additionally and also to the best knowledge of the author, in most theoretical studies, transmon qubits are modelled as adiabatic anharmonic oscillators or as adiabatic two-level systems, see for example \REFS\ListAdiabticModels. Furthermore, we have to take into account an additional complication, namely that assigning labels to the energies $E^{(z)}(\varphi)$ and eigenstates $\ket{\phi^{(z)}(\varphi)}$ of a continuous set of Hermitian matrices $\{H(\varphi)\}$ is a non-trivial problem in itself, see \REFS\cite{Hund1927,vonNeumann1993,Uhlig2020,Srinivasan2020}. Here, $z \in \mathbb{N}^{0}$ and $\varphi/\TP \in [0,1]$. Also, once we apply flux control pulses like the once displayed in \figsref{fig:GET_pulse_time_evo}(b-c), we cannot exclude the possibility that additional transitions between the states of the system occur,\ie more transitions than the ones needed to implement the $\CZ$ gates.

Since we would like to provide clarity on the issues mentioned above, we avoid using the nomenclature used in \REF\cite[Section VII B 2]{Blais2020circuit}, while discussing the spectrum of the four-qubit NIGQC.

\graphicspath{{FiguresAndData/GETPaper/Spectrum/}}
\renewcommand{\hold}{0.55}
\begin{figure}[!tbp]
  \centering
  \begin{minipage}{1.0\textwidth}
      \centering
      \includegraphics[scale=\hold]{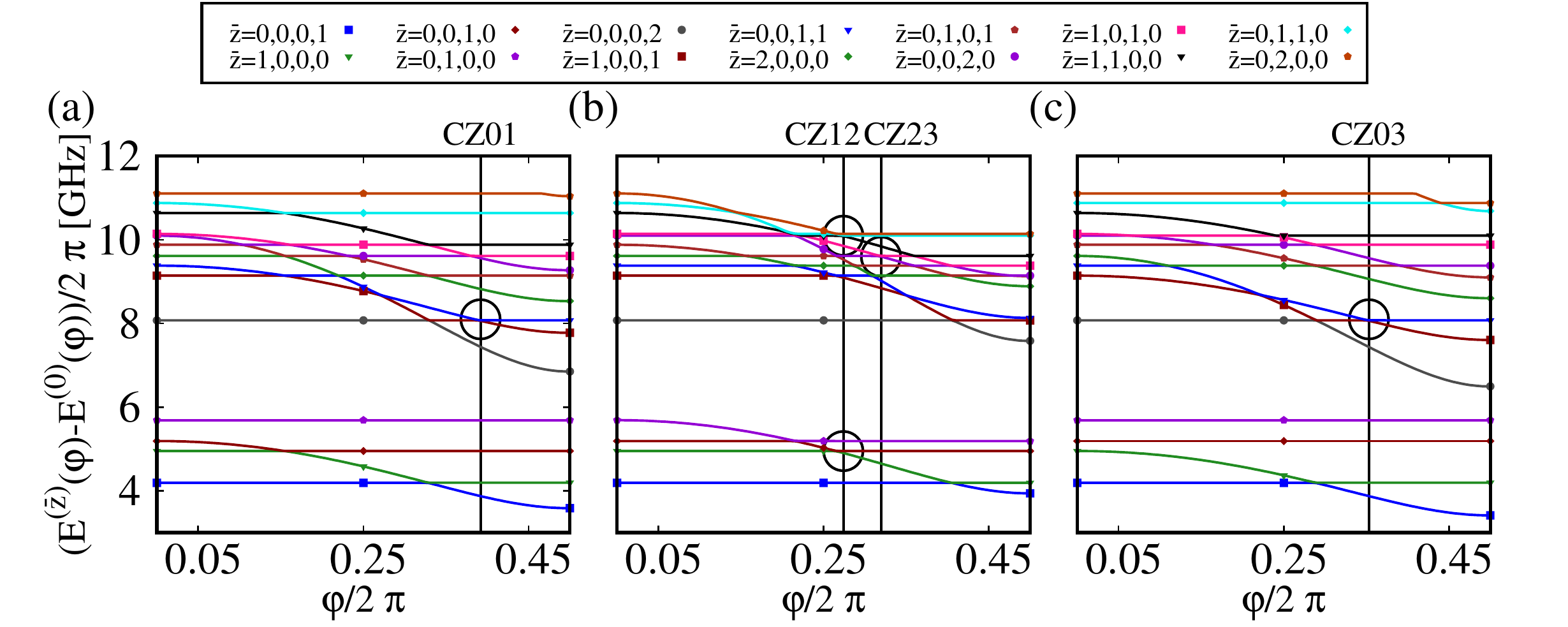}
  \end{minipage}
  \caption[Energy levels as functions of the external flux $\varphi$.]{Energy levels as functions of the external flux $\varphi$. The lowest fourteen energy levels of the four-qubit NIGQC illustrated in \figref{fig:device_sketch}(c) are computed. In \PANSL{a,b,c} we tune the external flux $\varphi$ of the second ($i=1$), third ($i=2$) and fourth ($i=3$) transmon qubit, respectively. We use the device parameters listed in \tabref{tab:device_para}, the circuit Hamiltonian \equref{eq:CHM} and a standard diagonalisation algorithm, see \REFS\cite{MKL09,PACK99}, to obtain the results. The energies $E^{(\mathbf{\bar{z}})}(\varphi)$ of the interacting Hamiltonian \equref{eq:CHM} are labelled according to the sorted energies $E^{(\mathbf{z})}(\varphi)$ of the non-interacting Hamiltonian \equref{eq:CHM} at the operating points $\varphi_{0,i}=0$.\label{fig:GET_device_spectrum}}
\end{figure}

Figures~\ref{fig:GET_device_spectrum}(a-c) show the lowest fourteen energy levels of the four-qubit NIGQC illustrated in \figref{fig:device_sketch}(c) as functions of the external flux offset $\varphi$. In \PANSL{a-c} we drive the second ($i=1$), third ($i=2$) and fourth ($i=3$) transmon qubit, respectively. The results are obtained with the device parameters listed in \tabref{tab:device_para}, the circuit Hamiltonian \equref{eq:CHM} and a standard diagonalisation algorithm, see \REFS\cite{MKL09,PACK99}. All transmon qubits are modelled with three basis states. Similarly, all coupling resonators are modelled with two basis states only. Note that the circuit Hamiltonian in \equref{eq:flux-tunable transmon recast} has two symmetry points $\varphi/\TP=0.5$ and $\varphi/\TP=1$.

The energy levels $E^{(\mathbf{\bar{z}})}(\varphi)$ of the interacting Hamiltonian with $G\neq0$ are labeled according to the sorted energy levels $E^{(\mathbf{z})}(\varphi)$ of the non-interacting Hamiltonian with $G=0$ at the flux offset value $\varphi=0$. For clarity and better visibility, we add three markers to each energy level. Furthermore, we mark points of particular importance by means of black vertical lines and circles. The vertical lines indicate the pulse amplitudes $\varphi=\FctlAp$ used to implement the $\CZ$ gates, see \tabref{tab:CtlTqgIVMB}. The circles indicate the energy level repulsions (ELRs) used to implement the $\CZ$ gates as well as ELRs that can lead to problems for the implementation of two-qubit $\CZ$ gates, see $\varphi/\TP=0.275$ in \figref{fig:GET_device_spectrum}(b) for the $\CZ_{1,2}$ gate.

The operating point for all transmon qubits is $\varphi=0$. This means, we always start from this point and tune the external flux to the point $\varphi=\delta$, see \figsref{fig:GET_device_spectrum}(a-c). On our way to the target ELR,\ie the ELR that we use to activate population exchange between a target computational state and a non-computational state as described above, we have to pass through many unused ELRs with sufficient speed. Unused ELRs are the ones we do not use to activate population exchange between the target computational state and the non-computational state. If we do not pass through the unused ELRs sufficiently fast, we can observe (data not shown) population exchange between the two states whose energy levels repel each other. If we pass through the unused ELRs too fast, we can observe (data not shown) a variety of other transitions between the states of the system. For example, we can potentially observe that the coupler leaves its ground state. We can make these observations either by studying the matrix in \equref{eq:prop_matrix} during the optimisation of the control pulse parameters or by explicitly studying the probabilities during the time evolution of the system. The algorithm we use to optimise the control pulse parameters seems to search for a balance between these two mechanisms. In order to find this balance, the optimisation algorithm mainly optimises the parameters $\FctlSg$, $\FctlTp$ and $\FctlTd$. The pulse amplitude $\FctlAp$ is usually only fine tuned to achieve nearly perfect population exchange and alignment of the phases. Note that we have to take into account all $2^{N}$ computational states and not only the two states which exchange population.

With this in mind, in \figref{fig:GET_device_spectrum}(b), we can point out a problem that cannot be solved by optimising the control pulse parameters. If we consider the flux offset value $\varphi/\TP=0.275$ which is used to implement the $\CZ_{1,2}$ gate by means of the ELR marked in the energy band between $8$ and $12$ GHz, we can see a second ELR in the energy band between $4$ and $6$ GHz that is close to the flux offset value $\varphi/\TP=0.275$. This second ELR leads to unwanted transitions between different computational states of the NIGQC. Some authors, see \REF\cite[Supplementary Material Section 1]{Andersen2020}, suggest to remedy this issue by additionally driving non-interacting transmon qubits. Other authors, see \REF\cite[Supplementary Material Section VI C 1]{Arute19}, suggest to solve this problem by redefining the target two-qubit gate such that the time evolution of the system fits more naturally to the target gate.

Furthermore, if we want to grasp the whole difficulty of the problem, we also have to take into account that driving different transmon qubits results in different energy levels, see \figsref{fig:GET_device_spectrum}(a-c). Consequently, if we begin to design a device or system, we are already confronted with an optimisation problem of exponential size. We have to avoid ELRs between all $2^{N}$ computational basis states of the NIGQC and at the same time take into account that driving different transmon qubits leads to different energy levels. To the best knowledge of the author, the largest PGQC in existence, based on the device architecture discussed in this chapter, has seventeen transmon qubits, see \REF\cite{Krinner21}.

So far in this section we only considered two-qubit gates. However, single-qubit gates suffer from similar problems as two-qubit gates. The MP control pulses we use to implement the $\ROT(\pi/2)$ rotations are not monochromatic, see \figref{fig:GET_pulse_time_evo}(a). Therefore, adding a single transmon qubit to the system reduces the number of available frequencies that NIGQC computational basis states can be separated by. Since an IGQC with $N$ qubits can execute single-qubit gates on every qubit, we potentially have to take into account the frequency bandwidths of all $N$ MPs and the energy levels of all $2^{N}$ computational basis states. Consequently, we are confronted with another optimisation problem of exponential size.

\section{Gate metrics for the elementary gate set}\label{sec:GET_GateMetricsForTheElementaryGateSet}

In this section we discuss the results of the control pulse optimisation for the single-qubit $\ROT(\pi/2)$ rotations and the $\CZ$ gates. These gates form the elementary gate sets of our NIGQCs. This means we express all gates in a sequence of gates,\ie a program, in terms of the elementary gate set and the virtual $\ROTZ(\phi)$ rotations as discussed in \secref{sec:TheMultiQubitSpace}. A useful table with gate decompositions can be found in \REF\cite[Appendix B]{Willsch2020}. Note that the results for the effective model presented in this section are obtained with the adiabatic effective Hamiltonian \equref{eq:EHM}. Furthermore, the dipole-dipole interactions are time dependent and we use the series expansions in \equaref{eq: expansion frequency}{eq: expansion anharmonicity} to model the flux-tunable frequencies and anharmonicities in \equref{eq:transeff}, see \chapref{chap:NA} for more details.

\graphicspath{{FiguresAndData/GETPaper/BoxPlot/}}
\renewcommand{\hold}{0.65}
\begin{figure}[!tbp]
  \begin{minipage}{1.0\textwidth}
  \centering
  \includegraphics[scale=\hold]{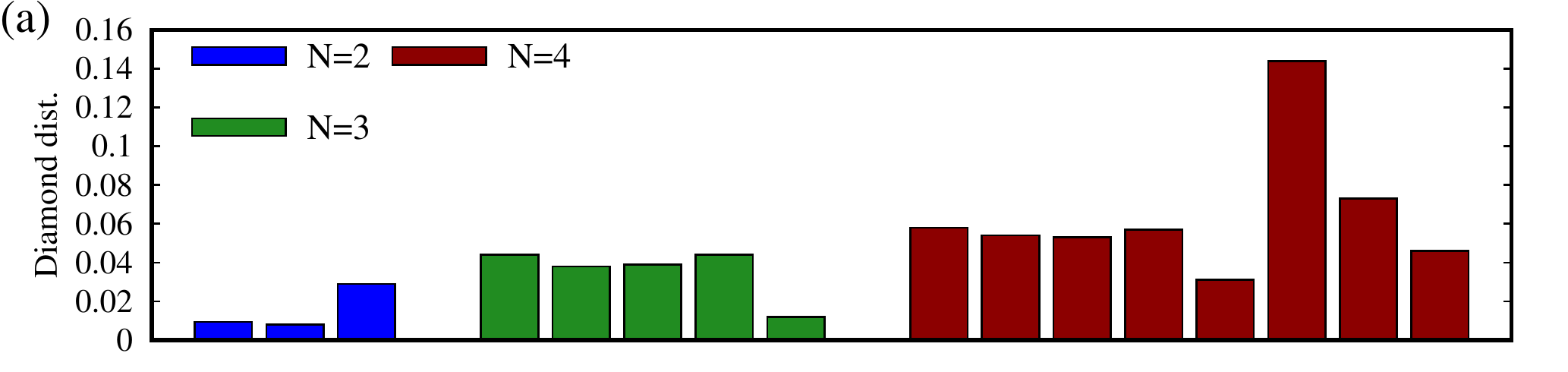}
  \includegraphics[scale=\hold]{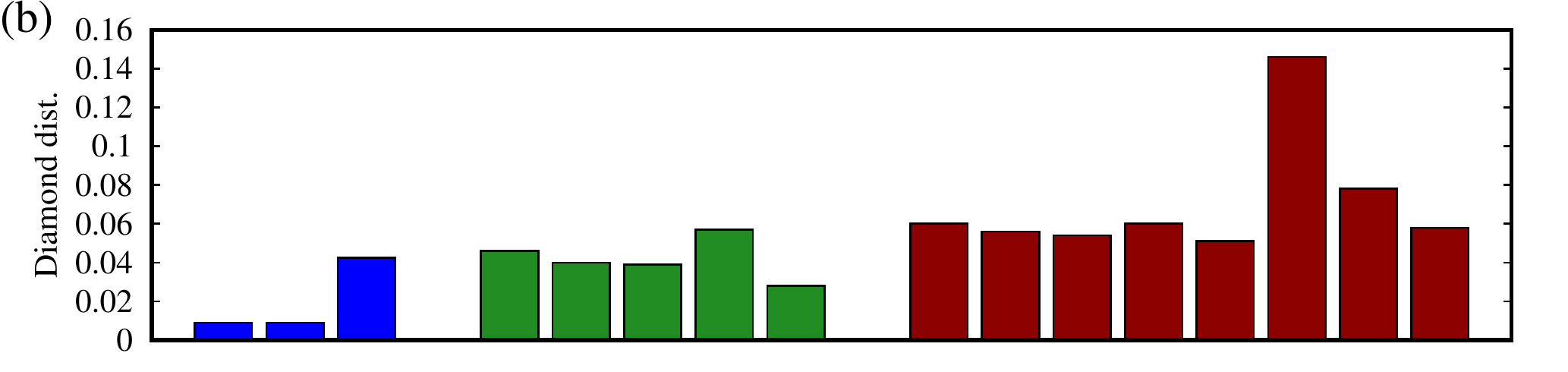}
  \includegraphics[scale=\hold]{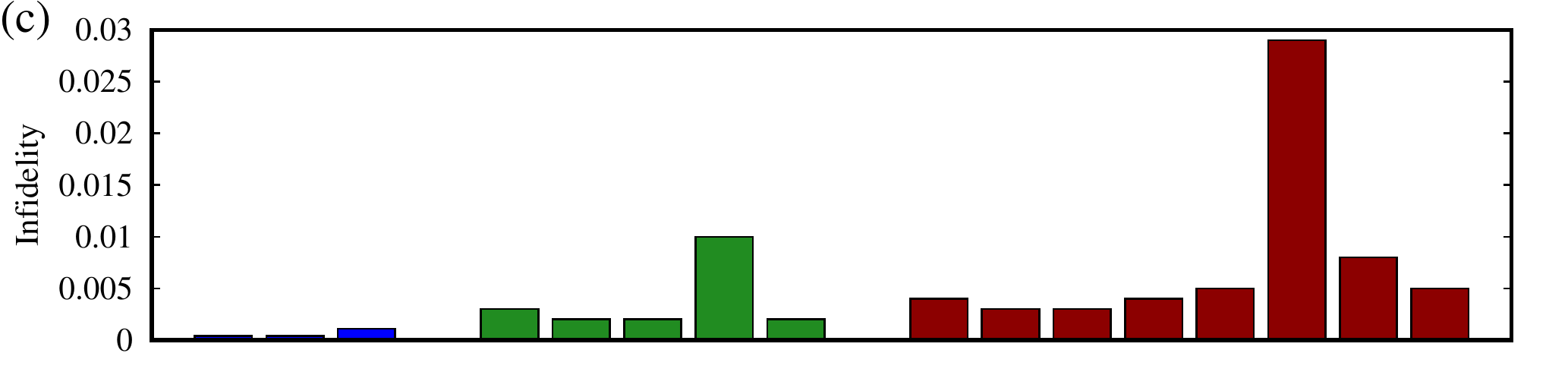}
  \includegraphics[scale=\hold]{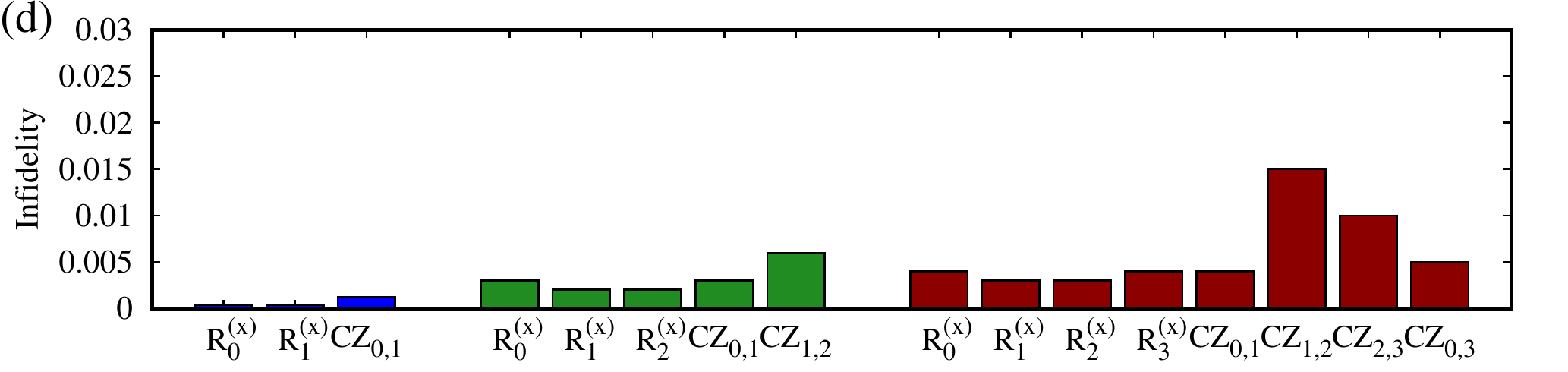}
  \end{minipage}
  \caption[Gate-error metrics for single-qubit $\ROT(\pi/2)$ and two-qubit $\CZ$ gates for the two-qubit (in blue), three-qubit (in green) and four-qubit (in blue) NIGQCs illustrated in \figsref{fig:device_sketch}(a-c), respectively.]{Gate-error metrics for single-qubit $\ROT(\pi/2)$ and two-qubit $\CZ$ gates for the two-qubit (in blue), three-qubit (in green) and four-qubit (in red) NIGQCs illustrated in \figsref{fig:device_sketch}(a-c), respectively. In \PANSL{a-b} we show the diamond distance $\DNV$ given by \equaref{eq:diamond_norm_inf}{eq:diamond_norm_sup} for the different NIGQCs obtained with the circuit Hamiltonian \equref{eq:CHM}(a) and the effective Hamiltonian \equref{eq:EHM}(b). Similarly, in \PANSL{c-d} we show the average infidelity $\IFV$ given by \equref{eq:in_fid_avg} for the different NIGQCs obtained with the circuit Hamiltonian \equref{eq:CHM}(c) and the effective Hamiltonian \equref{eq:EHM}(d). We use the device parameters listed in \tabref{tab:device_para} and the control pulse parameters listed in \tabsref{tab:CtlSqgIIMB}{tab:CtlTqgIVHB} to obtain the results. Note that the gate-error metrics are also listed in \tabsref{tab:EMIIMB}{tab:EMIVHB}.\label{fig:GET_metrics}}
\end{figure}

Figures~\ref{fig:GET_metrics}(a-d) show the diamond distance $\DNV$(a,b) given by \equaref{eq:diamond_norm_inf}{eq:diamond_norm_sup} and the average infidelity $\IFV$(c,d) given by \equref{eq:in_fid_avg} for the two-qubit (in blue), three-qubit (in green) and four-qubit (in red) NIGQCs illustrated in \figsref{fig:device_sketch}(a-c), respectively. We use the optimisation algorithms provided by the NLopt library, see \REF\cite{NLopt}, the device parameters listed in \tabref{tab:device_para}, the control pulse parameters listed in \tabsref{tab:CtlSqgIIMB}{tab:CtlTqgIVMB} (\tabsref{tab:CtlSqgIIHB}{tab:CtlTqgIVHB}) and the circuit Hamiltonian \equref{eq:CHM} (effective Hamiltonian \equref{eq:EHM}) to obtain the results displayed in \PANSL{a,c} (\PANSL{b,d}). Note that the results are also listed in \tabsref{tab:EMIIMB}{tab:EMIVHB}.

The lowest value for the diamond distance $\DNV$ (the average infidelity $\IFV$) obtained with the circuit Hamiltonian \equref{eq:CHM} is $0.0080$ ($0.0004$). These gate-error metrics were obtained with the control pulse parameters for the $\ROT_{1}(\pi/2)$ gate. However, when we optimise the control pulse parameters for the $\ROT_{1}(\pi/2)$ gate by simulating the circuit Hamiltonian \equref{eq:CHM} with the device parameters of the four-qubit NIGQC, the best value we can obtain is $0.058$ ($0.004$) for the diamond distance $\DNV$ (the average infidelity $\IFV$).

Similarly, the lowest value for the diamond distance $\DNV$ (the average infidelity $\IFV$) obtained with the effective Hamiltonian \equref{eq:EHM} is $0.0089$ ($0.0004$). These gate-error metrics were obtained with the effective Hamiltonian \equref{eq:CHM} and the control pulse parameters for the $\ROT_{0}(\pi/2)$ gate. However, when we optimise the control pulse parameters for the $\ROT_{0}(\pi/2)$ gate by simulating the effective Hamiltonian \equref{eq:EHM} with the device parameters of the four-qubit NIGQC, the best value we can obtain is $0.060$ ($0.004$) for the diamond distance $\DNV$ (the average infidelity $\IFV$).

Therefore, adding additional circuit elements to the system does not allow us to reproduce the good results for the case of the two-qubit NIGQCs.

The largest value for the diamond distance $\DNV$ (the average infidelity $\IFV$) obtained with the circuit Hamiltonian \equref{eq:CHM} is $0.144$ ($0.029$). Similarly, the largest value for the diamond distance $\DNV$ (the average infidelity $\IFV$) obtained with the effective Hamiltonian \equref{eq:EHM} is $0.146$ ($0.015$). All these gate-error metrics were obtained with the control pulse parameters for the $\CZ_{1,2}$ gates by simulating the circuit Hamiltonian \equref{eq:CHM} (effective Hamiltonian \equref{eq:EHM}) for the four-qubit NIGQC illustrated in \figref{fig:device_sketch}(c).

We may explain these results by studying the energy levels in \figref{fig:GET_device_spectrum}(b). The pulse amplitude we use to activate the $\CZ_{1,2}$ gate transitions is $\FctlAp/\TP=0.275$. This value for the pulse amplitude activates not only the desired transitions which can be associated with the ELR, see black circle, in the energy band between $8$ and $12$ GHz but also transitions between computational states in the energy band between $4$ and $6$ GHz, see black circle. Ultimately, the optimisation algorithm cannot find a value for the pulse amplitude that only activates the transitions of the upper ELR. Note that in order to reproduce this problem with the effective Hamiltonian \equref{eq:EHM}, we have to use the higher-order series expansions in \equaref{eq: expansion frequency}{eq: expansion anharmonicity} to model the flux-tunable frequency and anharmonicity of the transmon qubits,\ie a first-order expansion does not lead to similar results (data not shown).

Furthermore, \figsref{fig:GET_metrics}(a-d) also show a tendency to larger gate-error metrics for NIGQCs with more transmon qubits. Note that both models show this tendency. The largest system consists of eight circuit elements and the smallest system is made up of three circuit elements.

While obtaining the control pulse parameters for the two-qubit gates for NIGQCs of increasing size, we noticed increasing difficulty in navigating through the various ELRs in \figref{fig:GET_device_spectrum}(a-c). The task is to pass through most ELRs sufficiently fast (slow) such that undesirable transitions do not occur. However, the increasing complexity of the energy levels makes this a difficult challenge. On the one hand, we can observe (data not shown) that using control pulses with fast rising (falling) pulse flanks can potentially cause all sorts of undesirable transitions. For example, for such pulses we could observe that the coupler leaves its ground state, see also the discussion in \secref{sec:NA_E_suppressed_AII}. On the other hand, we can observe (data not shown) that using control pulses with slow rising (falling) pulse flanks leads to transitions between states caused by the ELRs we have to pass through before we reach the one ELR we use to  implement the actually desired transitions. The single-qubit $\ROT(\pi/2)$ rotations suffer from a similar problem in the frequency space. The MPs we use to implement the $\ROT(\pi/2)$ rotations are not monochromatic, this can potentially cause transitions between different states. Both these issues may explain the trend to larger gate-error metrics for NIGQCs of increasing size. Note that we discuss these issues also in \secref{sec:GET_SpectrumOfAFourQubitNIGQC}.

Another explanation for the tendency to larger gate-error metrics might be the increasing difficulty of the problem passed to the optimisation algorithm.
Optimising the control pulse parameters of an $N$-qubit NIGQC amounts to influencing the time evolution of the system such that the matrix $M$ in \equref{eq:prop_matrix} nearly perfectly aligns with the matrix $U$ of the target gate. This means perfectly aligning $2^{2 N+1}$ double-precision numbers without knowing how the system responds to a change in one or more of the control pulse parameters. Note that this task has to be carried out for all gates in the elementary gate set. To the best knowledge of the author, there exists no optimisation algorithm that can solve this task with guarantee of
success.

Finally, we find (data not shown) that using the control pulse parameters optimised for the circuit (effective) model does not allow us to obtain comparable results with the effective (circuit) model. If we compute the gate-error metrics for one model with the control pulse parameters optimised for the other model, we obtain values close to one for the diamond distance and the average infidelity. There are two reasons which can potentially explain this finding. First, in \chapref{chap:NA}, we found that the circuit Hamiltonian \equref{eq:CHM} and the effective Hamiltonian \equref{eq:EHM} predict similar outcomes if we define the effective model accordingly. However, there still remain deviations sufficiently large to cause the mismatch between both models. Second, if we intend to reproduce the results presented in \figsref{fig:GET_metrics}(a-d), we have to fine-tune control pulse parameters like the drive frequency $\CctlDf$ ($\hbar\CctlDf$) for the MP and the pulse amplitude $\FctlAp$ ($\hbar \omega^{(q)}(\FctlAp)$) for the UMP up to the sixth decimal (a couple of KHz) within one model. Furthermore, even within one model these parameters are highly susceptible to changes in the device parameters. For example, if we simulate the circuit and the effective model, we find that changing the interaction strength constant $G$ from $300$ to $301$ MHz forces us to repeat the control pulse parameter optimisation for all gates.

\section{Influence of higher states on gate-error quantifiers obtained with the circuit Hamiltonian}\label{sec:GET_HigherStates}
\newcommand{\ListStates}{\cite{Gu21,Wittler21,McKay16,Roth19,Rol19,Yan18}}
In this section we discuss how higher states affect the computation of gate-error quantifiers obtained with the circuit Hamiltonian \equref{eq:CHM} and the device parameters listed in \tabref{tab:device_para}. Here we only consider the four-qubit NIGQC illustrated in \figref{fig:device_sketch}(c) and the execution of $\ROT_{0}(\pi/2)$ gates.

\tabref{tab:GET_States} shows the diamond distance $\DNV$ given by \equaref{eq:diamond_norm_inf}{eq:diamond_norm_sup}, the average infidelity $\IFV$ in \equref{eq:in_fid_avg}, the leakage measure $\LNV$ given by \equref{eq:leak} and the statistical distance $\SDV$ in \equref{eq:stat_dis} for the $\ROT_{0}(\pi/2)$ gate. The statistical distance is obtained for the initial state $\ket{0,0,0,0}$. Furthermore, the results in the first (second) row are obtained with the device parameters listed in \tabref{tab:device_para}, the control pulse parameters listed in \tabref{tab:CtlSqgIVMB} row one, the circuit Hamiltonian \equref{eq:CHM} and four (sixteen) basis states for every transmon qubit. Note that the coupling resonators are modelled with four basis states only.

\begin{table}[H]
\caption[Gate-error quantifiers for the $\ROT_{0}(\pi/2)$ gate executed on the four-qubit NIGQC illustrated in \figref{fig:device_sketch}(c).]{Gate-error quantifiers for the $\ROT_{0}(\pi/2)$ gate executed on the four-qubit NIGQC illustrated in \figref{fig:device_sketch}(c). The first column shows the target gate. The second column shows the number of basis states $n_{J}$ we use to model the dynamics of the transmon qubits. Note that all coupling resonators are modelled with four basis states only. The third column shows the diamond distance $\DNV$ given by \equaref{eq:diamond_norm_inf}{eq:diamond_norm_sup}. The fourth column shows the average infidelity $\IFV$ given by \equref{eq:in_fid_avg}. The fifth column shows the leakage measure $\LNV$ given by \equref{eq:leak}. The sixth column shows the statistical distance $\SDV$ given by \equref{eq:stat_dis} for the initial state $\ket{0,0,0,0}$. We use the device parameters listed in \tabref{tab:device_para}, the control pulse parameters listed in \tabref{tab:CtlSqgIVMB} first row and the circuit Hamiltonian \equref{eq:CHM} to obtain the results.\label{tab:GET_States}}
\centering
{\small
\setlength{\tabcolsep}{4pt}
\begin{tabularx}{\textwidth}{X X X X X X}
\hline\hline

$\text{Gate}$& $\text{States}$ $n_{J}$&     $\DNV$&              $\IFV$&              $\LNV$&              $\SDV$\\

\hline

$\ROT_{0}(\pi/2)$ & $4$   &              $0.0505$       &              $0.0037$       &              $0.0024$       &              $0.0014$       \\

$\ROT_{0}(\pi/2)$ & $16$  &          $0.0584$       &              $0.0040$       &              $0.0024$       &              $0.0014$       \\

\hline\hline
\end{tabularx}
}
\end{table}

We can see that if we use sixteen basis states, the diamond distance and average infidelity show an increase in the third decimal. However, the leakage measure and the statistical distance are unaffected up to the fourth decimal. This finding can potentially be explained by the fact that the leakage measure and the statistical distance can be expressed solely in terms of the squares of the state vector amplitudes. Therefore, both these quantifiers do not take into account the phase of the system.

\graphicspath{{FiguresAndData/GETPaper/STATES/STATES_GPU_ETH_4_MB/}}
\renewcommand{\hold}{1.0}
\begin{figure}[!tbp]
  \centering
  \begin{minipage}{1.0\textwidth}
      \centering
      \includegraphics[scale=\hold]{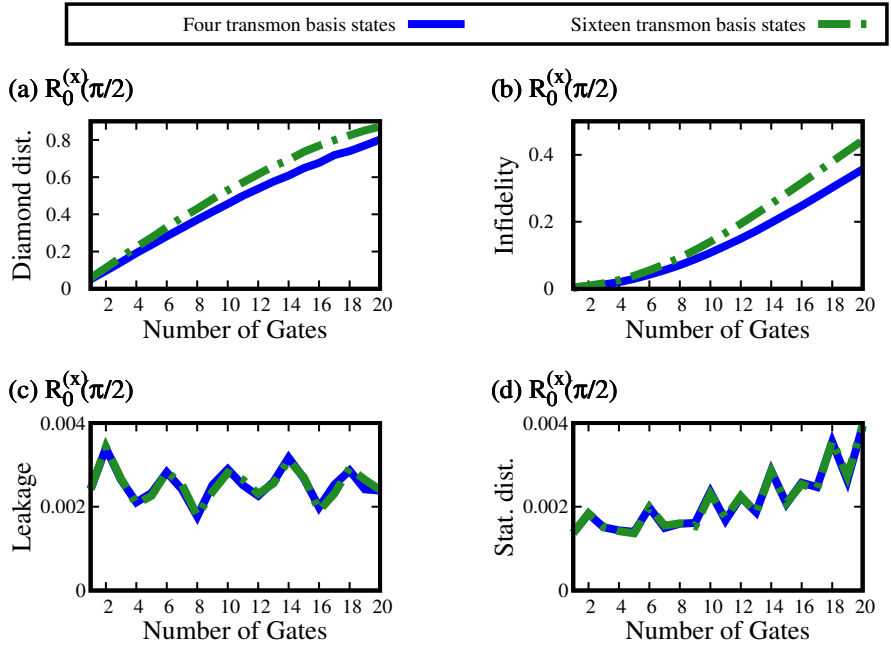}
  \end{minipage}
  \caption[Gate-error quantifiers as functions of the number of $\ROT_{0}(\pi/2)$ gates executed on a four-qubit NIGQC.]{Gate-error quantifiers as functions of the number of $\ROT_{0}(\pi/2)$ gates executed on the four-qubit NIGQC illustrated in \figref{fig:device_sketch}(c). All resonators are modelled with four basis states and the transmon qubits are modelled with four (blue solid line) and sixteen (green dashed line) basis states $n_{J}$. Here we use the device parameters listed in \tabref{tab:device_para}, the control pulse parameters listed in \tabref{tab:CtlSqgIVMB} first row and the circuit Hamiltonian \equref{eq:CHM} to obtain the results. In \PANSL{a-d} we show the diamond distance $\DNV$(a) given by \equaref{eq:diamond_norm_inf}{eq:diamond_norm_sup}, the average infidelity $\IFV$(b) in \equref{eq:in_fid_avg}, the leakage measure $\LNV$(c) given by \equref{eq:leak} and the statistical distance $\SDV$(d) in \equref{eq:stat_dis} for the initial state $\ket{0,0,0,0}$.\label{fig:GET_States}}
\end{figure}

Figures~\ref{fig:GET_States}(a-d) show the diamond distance $\DNV$(a) in \equaref{eq:diamond_norm_inf}{eq:diamond_norm_sup}, the average infidelity $\IFV$(b) given by \equref{eq:in_fid_avg}, the leakage measure $\LNV$(c) in \equref{eq:leak} and the statistical distance $\SDV$(d) given \equref{eq:stat_dis} for the computational state $\ket{0,0,0,0}$ as functions of the number of $\ROT_{0}(\pi/2)$ gates executed on the four-qubit NIGQC illustrated in \figref{fig:device_sketch}(c). Here we use the same simulation model and the same simulation parameters as before, see \tabref{tab:GET_States}. The blue solid (green dashed) line represents the results for the case where all transmon qubits are modelled with four (sixteen) basis states. As before, all coupling resonators are modelled with four basis states only.

We can see that, after the execution of a few $\ROT_{0}(\pi/2)$ gates, the diamond distance and the average infidelity start to exhibit deviations if we model the time evolution of the system with four and sixteen basis states. The deviations grow up to about 10\% after the execution of twenty $\ROT_{0}(\pi/2)$ gates. As before, we can observe that the leakage measure and the statistical distance are less affected. Here, we find that both quantifiers are roughly the same up to the fourth decimal. To the best knowledge of the author, in the literature, it seems to be common practice to use two or three basis states only to model the dynamics of NIGQCs and to compute gate errors, see for example \REFS\ListStates.

The results presented in this section lead the author of this thesis to advocate the view that gate-error metrics like the diamond distance and the average infidelity are only valid for the number of basis states they are obtained for. Consequently, changing the number of basis states leads to a new NIGQC model in the sense that the gate-error metrics obtained for the old and the new model are not guaranteed to be the same. Here we exclude the unlikely case that the truncated time-evolution operators, see \equaref{eq:prop_matrix}{eq:time_evo_op}, are known to be the same for both models.

\section{Influence of parameter changes on gate-error quantifiers obtained with the circuit Hamiltonian}\label{sec:GET_Para}
In this section we discuss how changes in the control pulse parameters affect the computation of gate-error quantifiers obtained with the circuit Hamiltonian \equref{eq:CHM} and the device parameters listed in \tabref{tab:device_para}. To this end, we execute simple gate repetition programs that consist of twenty $\CNOT_{0,1}=\HA_{0} \CZ_{0,1} \HA_{0}$ gates on the two-qubit, three-qubit and four-qubit NIGQCs illustrated in \figref{fig:device_sketch}(a-c) and vary the control pulse amplitude $\delta=\delta_{0}+\Delta\FctlAp$ of the UMP that realises the $\CZ_{0,1}$ gate. Here $\delta_{0}$ denotes the optimised control pulses amplitude of the UMP and $\Delta\FctlAp/\TP \in \{0, 10^{-6},10^{-5},10^{-4}\}$ refers to the offset value we use to model the parameter change.

\tabref{tab:GET_Para} shows the diamond distance $\DNV$ given by \equaref{eq:diamond_norm_inf}{eq:diamond_norm_sup}, the average infidelity $\IFV$ in \equref{eq:in_fid_avg}, the leakage measure $\LNV$ given by \equref{eq:leak} and the statistical distance $\SDV$ in \equref{eq:stat_dis} for $\CNOT_{0,1}$  gates executed on the two-qubit NIGQC illustrated in \figref{fig:device_sketch}(a). The statistical distance is obtained for the initial state $\ket{0,0,0,0}$. The results are obtained with the device parameters listed in \tabref{tab:device_para}, the control pulse parameters listed in \tabaref{tab:CtlSqgIIMB}{tab:CtlTqgIIMB} first rows and the circuit Hamiltonian \equref{eq:CHM}. The different rows show the gate-error quantifiers obtained with slightly different $\Delta\FctlAp$ control pulse amplitudes.

\begin{table}[H]
\caption[Gate-error quantifiers for the $\CNOT_{0,1}$ gate executed on the two-qubit NIGQC illustrated in \figref{fig:device_sketch}(c).]{Gate-error quantifiers for the $\CNOT_{0,1}$ gate executed on the two-qubit NIGQC illustrated in \figref{fig:device_sketch}(c). Here we use UMPs to implement the $\CZ$ gates. The first column shows the target gate. The second column shows pulse amplitude offset value $\Delta\FctlAp$ we use to model parameter changes. The third column shows the diamond distance $\DNV$ given by \equaref{eq:diamond_norm_inf}{eq:diamond_norm_sup}. The fourth column shows the average infidelity $\IFV$ in \equref{eq:in_fid_avg}. The fifth column shows the leakage measure $\LNV$ given by \equref{eq:leak}. The sixth column shows the statistical distance $\SDV$ in \equref{eq:stat_dis} for the initial state $\ket{0,0,0,0}$. We use the device parameters listed in \tabref{tab:device_para}, the control pulse parameters listed in \tabaref{tab:CtlSqgIIMB}{tab:CtlTqgIIMB} first rows and the circuit Hamiltonian \equref{eq:CHM} to obtain the results.\label{tab:GET_Para}}
\centering
{\small
\setlength{\tabcolsep}{4pt}
\begin{tabularx}{\textwidth}{X X X X X X}
\hline\hline

$\text{Gate}$& $\Delta\FctlAp/\TP$&              $\DNV$&              $\IFV$&              $\LNV$&              $\SDV$\\

\hline

$\CNOT_{0,1}$& $0$      &       $0.0386$       &              $0.0018$       &              $0.0012$       &              $0.0013$       \\

$\CNOT_{0,1}$& $10^{-6}$          &   $0.0390$       &              $0.0018$       &              $0.0012$       &              $0.0013$       \\

$\CNOT_{0,1}$& $10^{-5}$          &   $0.0456$       &              $0.0022$       &              $0.0012$       &              $0.0013$       \\

$\CNOT_{0,1}$& $10^{-4}$           &   $0.1594$       &              $0.0156$       &              $0.0018$       &              $0.0038$       \\

\hline\hline
\end{tabularx}
}
\end{table}
As one can see, the diamond distance is affected by the third decimal if we add an offset value of $\Delta\FctlAp/\TP=10^{-6}$ to the optimised pulse amplitude $\delta_{0}$. Furthermore, the remaining gate-error quantifiers stay the same up to the fourth decimal for this offset value. If we increase the offset value by a factor of ten to $\Delta\FctlAp/\TP=10^{-5}$, we can observe that the diamond distance is affected by the second decimal and the average infidelity is affected by the third decimal. In this case too, the leakage measure and the statistical distance stay the same up to the fourth decimal. If we increase the offset value again such that $\Delta\FctlAp/\TP=10^{-4}$, we find that all gate-error quantifiers are affected. The diamond distance, average infidelity, leakage measure and the statistical distance are affected by the first, second, fourth and third decimal, respectively.

\renewcommand{\hold}{1.0}
\begin{figure}[!tbp]
  \centering
  \begin{minipage}{1.0\textwidth}
      \centering
      \graphicspath{{FiguresAndData/GETPaper/PARA/PARA_CPU_ETH_2_MB/}}
      \includegraphics[scale=\hold]{error}
      \graphicspath{{FiguresAndData/GETPaper/PARA/PARA_GPU_ETH_3_MB/}}
      \includegraphics[scale=\hold]{error}
      \graphicspath{{FiguresAndData/GETPaper/PARA/PARA_GPU_ETH_4_MB/}}
      \includegraphics[scale=\hold]{error}
  \end{minipage}
  \caption[Gate-error metrics diamond distance $\DNV$(a,c,e) in \equref{eq:diamond_norm_inf} and \equref{eq:diamond_norm_sup} and average infidelity $\IFV$(b,d,f) in \equref{eq:in_fid_avg} as functions of the number of gates executed on the two-qubit(a-b), three-qubit(c-d) and four-qubit(e-f) NIGQCs illustrated in \figsref{fig:device_sketch}(a-c), respectively (parameter change study).]{Gate-error metrics diamond distance $\DNV$(a,c,e) given by \equaref{eq:diamond_norm_inf}{eq:diamond_norm_sup} and average infidelity $\IFV$(b,d,f) given by \equref{eq:in_fid_avg} as functions of the number of gates executed on the two-qubit(a-b), three-qubit(c-d) and four-qubit(e-f) NIGQCs illustrated in \figsref{fig:device_sketch}(a-c), respectively. The different NIGQCs execute the same program, namely a program with twenty $\CNOT_{0,1}$ instructions. Note that all $\CNOT$ gates are implemented by means of two MPs and one UMP. The program is executed four times on each NIGQC. Here, we vary the pulse amplitude $\delta=\delta_{0}+\Delta\FctlAp$, where $\delta_{0}$ denotes the optimised control pulse amplitude and $\Delta\delta$ denotes the offset value we use to model the parameter change. We use $\Delta\FctlAp/\TP \in \{0,10^{-6},10^{-5},10^{-4}\}$ (blue solid line, green dashed line, red dashed line, violet solid line), the device parameter listed in \tabref{tab:device_para}, the control pulse parameters listed in \tabsref{tab:CtlSqgIIMB}{tab:CtlTqgIVMB} first rows and the circuit Hamiltonian \equref{eq:CHM} to obtain the results.\label{fig:GET_Para}}
\end{figure}

Figures~\ref{fig:GET_Para}(a-f) show the diamond distance $\DNV$(a,c,e) in \equaref{eq:diamond_norm_inf}{eq:diamond_norm_sup} and the average infidelity $\IFV$(b,d,f) given by \equref{eq:in_fid_avg} as functions of the number of executed gates on the two-qubit(a-b), the three-qubit(c-d) and the four-qubit(e-f) NIGQCs illustrated in \figsref{fig:device_sketch}(a-c), respectively. The program we execute consists of twenty $\CNOT_{0,1}$ gates. Since we are interested in how the gate errors behave as a function of the number of executed gates, we do not optimise the circuit by removing identities. The results are obtained with the device parameters listed in \tabref{tab:device_para}, the control pulse parameters listed in \tabsref{tab:CtlSqgIIMB}{tab:CtlTqgIVMB} first rows and the circuit Hamiltonian \equref{eq:CHM}. The blue solid (green dashed, red dashed, violet solid) line shows the results for the offset value $\Delta\FctlAp=0$ ($\Delta\FctlAp=10^{-6}$, $\Delta\FctlAp=10^{-5}$, $\Delta\FctlAp=10^{-4}$).

We can see that the offset value $\Delta\FctlAp/\TP=10^{-6}$ only slightly affects the qualitative and quantitative behaviour of the gate-error trajectories displayed in \figsref{fig:GET_Para}(a-f). Furthermore, deviations usually only become visible at the end of the gate repetition programs. If we increase the offset value to $\Delta\FctlAp/\TP=10^{-5}$, we find that the qualitative and quantitative behaviour of the gate-error trajectories becomes increasingly affected by the control pulse parameter change. The deviations seem to build up over time,\ie the deviations seem to grow with the number of executed gates. If we use the offset value $\Delta\FctlAp/\TP=10^{-4}$ to model the parameter change, we can observe some type of tipping behaviour in \figsref{fig:GET_Para}(a-f). The qualitative and quantitative behaviour of all gate-error trajectories show substantial non-linear change. Note that we can determine similar results (data not shown) if we model the parameter change by adding offset values $\Delta\CctlDf \in \{0, 10^{-6}, 10^{-5}, 10^{-4}\}$ to the optimised drive frequency $\CctlDf_{0}$ of the $\ROT_{0}(\pi/2)$ pulse which allows us to realise the Hadamard gates $\HA_{0}$.

The results discussed in this section show the fragile nature of the diamond distance and the average infidelity obtained with the circuit Hamiltonian \equref{eq:CHM} and the optimised control pulse parameters. The results rely on the assumption that we are able to control the flux-tunable qubit frequencies (the pulse amplitudes) at least up to a few KHz ($\Delta\FctlAp/\TP=10^{-6}$). Furthermore, we should emphasise that we only discussed the case where one of the many control pulse parameters of a NIGQC was changed, namely the pulse amplitude of the UMP which realises the $\CZ_{0,1}$ gate. This means we implicitly assumed that the remaining control pulse parameters are constant. Therefore, we should not be surprised by an increased susceptibility in cases where this assumption is not justified.

In order to emphasise the fragile nature of the gate-error metrics, we convert the offset value $\Delta\FctlAp/\TP=10^{-6}$ for the external flux $\Delta\Phi_{e}= (\FQ \Delta\delta)/\TP$ from Weber to Tesla and compare the associated magnetic field strength $\Delta|B|$ with the earth's magnetic field. Here, we consider an area of ten by ten micrometer and assume that the external flux is given by $\Phi_{e}=|B| A$, where $|B|$ is the magnetic field strength and $A$ is the area of the surface we consider. The assumption about the area size is motivated by \REF\cite{Roth22}. The result $\Delta|B|$ for the offset value $\Delta\FctlAp/\TP=10^{-6}$ is proportional to $10^{-11}$ Tesla. The earth's magnetic field is about $10^{-5}$ Tesla strong.

In our discussion so far, we ignored another feature of the gate-error trajectories displayed in \figsref{fig:GET_Para}(a-f). We can observe that the initial values for the gate-error metrics are poor predictors for the future behaviour of the gate error sequence which develops over time. Additionally, most gate-error trajectories presented in this section certainly do not behave linearly. The results presented in this section confirm previous findings obtained for a different device architecture and therefore add additional evidence to the conclusion stated in \REF\cite{Wi17}.

In this thesis, we can provide an explanation for this finding. Even for very small gate-error metrics $\mu_{x}$, where $x$ denotes a label for an arbitrary gate-error metric, the time evolution $\ket{\Psi(t)}$ of the system is still governed by the TDSE and not only by the gate-error metric $\mu_{x}$ obtained for one particular moment in time $t^{\prime}$, see \figref{fig:comp_sketch}. For the opposite to be true, we would need to show that a sufficiently small value for the gate-error metric $\mu_{x}$ imposes constraints on the time evolution of the system described by the state vector $\ket{\Psi(t)}$. Furthermore, since the time evolution of an arbitrary system, modelled with the Hamiltonian $\OP{H}(t)$, is by assumption generated by the time-evolution operator in \equref{eq:time_evo_op} (the TDSE), we would need to prove that this statement is true for all Hamiltonians $\OP{H}(t)$. However, to the best knowledge of the author, there exists no such proof in the literature.

The results presented in this section lead the author of this thesis to advocate the view that gate-error metrics like the diamond distance and the average infidelity are at most measures of closeness at one particular point in time $t^{\prime}$. How the gate errors in a NIGQC model develop over time is primarily determined by the TDSE and the model Hamiltonian.

\section{Influence of the adiabatic approximation on gate-error quantifiers obtained with the effective Hamiltonian}\label{sec:GET_Aprox}
\newcommand{\ListFastFluxPulses}{\cite{Foxen20,Gu21,Rol19,Roth19,McKay16,Baker22,Yan18,Blais2020circuit} }
In this section we discuss how a seemingly justified adiabatic approximation affects the computation of gate-error quantifiers obtained with the effective Hamiltonian \equref{eq:EHM} and the device parameters listed in \tabref{tab:device_para}. Since we are interested in the dynamic features of gate errors that emerge in NIGQC models, we execute simple gate repetition programs that consist of forty $\CZ_{0,1}$ and forty $\CNOT_{0,1}=\HA_{0} \CZ_{0,1} \HA_{0}$ gates on the two-qubit, three-qubit and four-qubit NIGQCs illustrated in \figref{fig:device_sketch}(a-c), respectively. We perform all simulations twice. During the first run (the adiabatic case), we model transmon qubits as adiabatic anharmonic oscillators, see \secaref{sec:TheQuantumComputerEffectiveHamiltonianModel}{sec:NA_single_flux_tunable_transmon}. During the second run (the non-adiabatic case), we model transmon qubits as non-adiabatic anharmonic oscillators, see \secaref{sec:TheQuantumComputerEffectiveHamiltonianModel}{sec:NA_single_flux_tunable_transmon}. The approximation we discuss in this section is often used in the literature, see for example \REFS\ListFastFluxPulses.

In our simulations, we realise the $\CZ_{0,1}$ gates with two types of control pulses, namely the UMP given by \equref{eq:flux_ctl_ump} and BMP given by \equref{eq:flux_ctl_bmp}, see \figref{fig:GET_pulse_time_evo}(b-c), respectively. Since the BMP in \figref{fig:GET_pulse_time_evo}(c) exhibits a fast falling pulse flank at around half of the pulse time $\FctlTp$, we expect the deviations between the adiabatic and non-adiabatic case to be larger for the BMP. Furthermore, for both control pulses we find (data not shown) that the time derivative $\dot{\varphi}(t)$ of the external flux is practically zero for most of the pulse duration. Therefore, the deviations between both models should originate from the pulse flanks of the control pulses, see \figref{fig:GET_pulse_time_evo}(b-c).

We first discuss the simulation results obtained with the UMP and the control pulse parameters listed in \tabsref{tab:CtlSqgIIHB}{tab:CtlTqgIIHB} first rows and then the simulation results acquired with the BMP and the control pulse parameters listed in \tabsref{tab:CtlSqgIIHB}{tab:CtlTqgIIHB} first and second rows.

\tabref{tab:GET_EMAPPROXUMPHB} shows the diamond distance $\DNV$ given by \equaref{eq:diamond_norm_inf}{eq:diamond_norm_sup}, the average infidelity $\IFV$ in \equref{eq:in_fid_avg}, the leakage measure $\LNV$ given by \equref{eq:leak} and the statistical distance $\SDV$ in \equref{eq:stat_dis} for $\CZ_{0,1}$ gates executed on the two-qubit (first and second row), three-qubit (third and fourth row) and four-qubit (fifth and second sixth) NIGQCs illustrated in \figref{fig:device_sketch}(a-c), respectively. The statistical distance is obtained for the computational state $\ket{0,0,0,0}$. The odd (even) row numbers show the results obtained with the adiabatic (non-adiabatic) effective Hamiltonian \equref{eq:EHM} and the UMP. The simulation parameters we use are discussed at the beginning of this section.

\begin{table}[H]
\caption[Gate-error quantifiers for the $\CZ_{0,1}$ gate executed on the two-qubit, three-qubit and four-qubit NIGQCs illustrated in \figref{fig:device_sketch}.]{Gate-error quantifiers for the $\CZ_{0,1}$ gate executed on the two-qubit, three-qubit and four-qubit NIGQCs illustrated in \figsref{fig:device_sketch}(a-c), respectively. Here we use the UMP to implement the $\CZ$ gates. The first column shows the target gate. The second column shows the type of effective Hamiltonian we use to describe the transmon qubits, see \equaref{eq:fft_eff_II}{eq:tunable-frequency eff}. The third column shows figure references for the different NIGQCs we model. The fourth column shows the diamond distance $\DNV$ given by \equaref{eq:diamond_norm_inf}{eq:diamond_norm_sup}. The fifth column shows the average infidelity $\IFV$ in \equref{eq:in_fid_avg}. The sixth column shows the leakage measure $\LNV$ given by \equref{eq:leak}. The seventh column shows the statistical distance $\SDV$ in \equref{eq:stat_dis} for the initial state $\ket{0,0,0,0}$. We use the device parameters listed in \tabref{tab:device_para}, the control pulse parameters listed in \tabref{tab:CtlTqgIIHB}, \tabref{tab:CtlTqgIIIHB} and \tabref{tab:CtlTqgIVHB} first rows and the effective Hamiltonian \equref{eq:EHM} to obtain the results.\label{tab:GET_EMAPPROXUMPHB}}
\centering
{\small
\setlength{\tabcolsep}{4pt}
\begin{tabularx}{\textwidth}{X X X X X X X}
\hline\hline

$\text{Gate}$& $\text{Adiabatic}$& $\text{System}$ &      $\DNV$&              $\IFV$&              $\LNV$&              $\SDV$\\

\hline

$\CZ_{0,1}$& $\text{Yes}$   & \figsref{fig:device_sketch}(a) &              $0.0424$       &              $0.0012$       &              $0.0005$       &              $0.0012$       \\

$\CZ_{0,1}$& $\text{No}$    & \figsref{fig:device_sketch}(a) &             $0.0425$       &              $0.0012$       &              $0.0005$       &              $0.0012$       \\

$\CZ_{0,1}$& $\text{Yes}$   & \figsref{fig:device_sketch}(b) &             $0.0569$       &              $0.0033$       &              $0.0017$       &              $0.0026$       \\

$\CZ_{0,1}$& $\text{No}$    & \figsref{fig:device_sketch}(b) &             $0.0574$       &              $0.0033$       &              $0.0017$       &              $0.0026$       \\

$\CZ_{0,1}$& $\text{Yes}$   & \figsref{fig:device_sketch}(c) &              $0.0514$       &              $0.0040$       &              $0.0028$       &              $0.0025$       \\

$\CZ_{0,1}$& $\text{No}$    & \figsref{fig:device_sketch}(c) &              $0.0509$       &              $0.0040$       &              $0.0028$       &              $0.0025$       \\

\hline\hline
\end{tabularx}
}
\end{table}
We can see (first and second rows) that the diamond distance for the two-qubit NIGQC modelled with the non-adiabatic effective Hamiltonian \equref{eq:EHM} is affected by the fourth decimal. Furthermore, the diamond distances for the three-qubit and four-qubit NIGQCs modelled with the non-adiabatic effective Hamiltonian \equref{eq:EHM} are affected by the third decimal. All other gate-error quantifiers are not affected up to the fourth decimal.

\renewcommand{\hold}{0.66}
\begin{figure}[!tbp]
  \begin{minipage}{1.0\textwidth}
      \centering
      \graphicspath{{FiguresAndData/GETPaper/APPROX/UMP/APPROX_UMP_GPU_ETH_2_HB/}}
      \includegraphics[scale=\hold]{error}
      \graphicspath{{FiguresAndData/GETPaper/APPROX/UMP/APPROX_UMP_GPU_ETH_3_HB/}}
      \includegraphics[scale=\hold]{error}
      \graphicspath{{FiguresAndData/GETPaper/APPROX/UMP/APPROX_UMP_GPU_ETH_4_HB/}}
      \includegraphics[scale=\hold]{error}
  \end{minipage}
  \caption[Gate-error metrics diamond distance $\DNV$(a,e,i,c,g,k) given by \equaref{eq:diamond_norm_inf}{eq:diamond_norm_sup} and average infidelity $\IFV$(b,f,j,d,h,l) given by \equref{eq:in_fid_avg} as functions of the number of gates executed on the two-qubit(a-d), three-qubit(e-h) and four-qubit(i-l) NIGQCs illustrated in \figsref{fig:device_sketch}(a-c), respectively (approximation study with UMP pulse).]{Gate-error metrics diamond distance $\DNV$(a,e,i,c,g,k) given by \equaref{eq:diamond_norm_inf}{eq:diamond_norm_sup} and average infidelity $\IFV$(b,f,j,d,h,l) given by \equref{eq:in_fid_avg} as functions of the number of gates executed on the two-qubit(a-d), three-qubit(e-h) and four-qubit(i-l) NIGQCs illustrated in \figsref{fig:device_sketch}(a-c), respectively. The different NIGQCs execute the same programs, namely a program with forty $\CZ_{0,1}$(a,b,e,f,i,j) instructions and forty $\CNOT_{0,1}$(c,d,g,h,k,l) instructions. Note that all $\CNOT$ gates are implemented by means of two MPs and one UMP. The programs are executed two times on each NIGQC. During the first run (blue solid line) we model the transmon qubits as adiabatic anharmonic oscillators, see \equref{eq:tunable-frequency eff}. During the second run (green dashed line) we model the transmon qubits as non-adiabatic anharmonic oscillators, see \equref{eq:fft_eff_II}. We use the device parameter listed in \tabref{tab:device_para}, the control pulse parameters listed in \tabsref{tab:CtlSqgIIHB}{tab:CtlTqgIVHB} first rows and the effective Hamiltonian \equref{eq:EHM} to obtain the results.\label{fig:GET_APPROX_UMP}}
\end{figure}

Figures~\ref{fig:GET_APPROX_UMP}(a-l) show the diamond distance $\DNV$(a,c,e,g,i,k) given by \equaref{eq:diamond_norm_inf}{eq:diamond_norm_sup} and the average infidelity $\IFV$(b,d,f,h,j,l) in \equref{eq:in_fid_avg} as functions of the number of gates executed on the two-qubit(a-d), three-qubit(e-h) and four-qubit(i-l) NIGQCs illustrated in \figsref{fig:device_sketch}(a-c), respectively. The results for the $\CZ$ and $\CNOT$ gate repetition programs are displayed in \PANSL{a,b,e,f,i,j} and \PANSL{c,d,g,h,k,l}, respectively. Here we use UMPs to realise the $\CZ_{0,1}$ gates. The results obtained with the adiabatic (non-adiabatic) effective Hamiltonian \equref{eq:EHM} are presented as a solid (dashed) blue (green) line. The simulation parameters we use are discussed at the beginning of this section.

Overall, we can observe that the results for the adiabatic and non-adiabatic case show a good qualitative agreement. In \PANSL{a,c,d,e,g,i} deviations between both models seem to grow with the number of gates executed on the various NIGQCs. In \PANSL{b,f,h,j,k,l} the deviations are barely noticeable.

We should probably emphasise that the UMPs we use in this section are rather long if we compare them with other instances which can be found in the literture, see for example \REF\cite{Foxen20}. Obviously, one cannot generalise the results presented in this section and conclude that modelling transmon qubits without the non-adiabatic drive term in \equref{eq:drive_term_flux} is justified. We now turn our attention to the results for BMPs.

\tabref{tab:GET_EMAPPROXUMPHB} shows the diamond distance $\DNV$ in \equaref{eq:diamond_norm_inf}{eq:diamond_norm_sup}, the average infidelity $\IFV$ given by \equref{eq:in_fid_avg}, the leakage measure $\LNV$ in \equref{eq:leak} and the statistical distance $\SDV$ given by \equref{eq:stat_dis} for $\CZ_{0,1}$ gates executed on the two-qubit (first and second row), three-qubit (third and fourth row) and four-qubit (fifth and second sixth) NIGQCs illustrated in \figref{fig:device_sketch}(a-c), respectively. The statistical distance is obtained for the computational state $\ket{0,0,0,0}$. The odd (even) row numbers show the results obtained with the adiabatic (non-adiabatic) effective Hamiltonian \equref{eq:EHM} and the BMP. The simulation parameters we use are discussed at the beginning of this section.

\begin{table}[H]
\caption[Gate-error quantifiers for the $\CZ_{0,1}$ gate executed on the two-qubit, three-qubit and four-qubit NIGQCs illustrated in \figref{fig:device_sketch}.]{Gate-error quantifiers for the $\CZ_{0,1}$ gate executed on the two-qubit, three-qubit and four-qubit NIGQCs illustrated in \figsref{fig:device_sketch}(a-c), respectively. Here we use the BMP to implement the $\CZ$ gates.  The rows and columns show the same unit-less quantities as \tabref{tab:GET_EMAPPROXUMPHB}. We use the device parameters listed in \tabref{tab:device_para}, the control pulse parameters listed in \tabref{tab:CtlTqgIIHB}, \tabref{tab:CtlTqgIIIHB} and \tabref{tab:CtlTqgIVHB} second rows and the effective Hamiltonian \equref{eq:EHM} to obtain the results.\label{tab:GET_EMAPPROXBMPHB}}
\centering
{\small
\setlength{\tabcolsep}{4pt}
\begin{tabularx}{\textwidth}{X X X X X X X}

\hline\hline

$\text{Gate}$& $\text{Adiabatic}$& $\text{System}$ & $\DNV$&              $\IFV$&              $\LNV$&              $\SDV$\\

\hline

$\CZ_{0,1}$& $\text{Yes}$  & \figsref{fig:device_sketch}(a) &              $0.0167$       &              $0.0006$       &              $0.0005$       &              $0.0004$       \\

$\CZ_{0,1}$& $\text{No}$   & \figsref{fig:device_sketch}(a) &              $0.0195$       &              $0.0007$       &              $0.0005$       &              $0.0004$       \\

$\CZ_{0,1}$& $\text{Yes}$  & \figsref{fig:device_sketch}(b) &              $0.0306$       &              $0.0042$       &              $0.0036$       &              $0.0020$       \\

$\CZ_{0,1}$& $\text{No}$   & \figsref{fig:device_sketch}(b) &              $0.0336$       &              $0.0042$       &              $0.0035$       &              $0.0019$       \\

$\CZ_{0,1}$& $\text{Yes}$  & \figsref{fig:device_sketch}(c) &              $0.0415$       &              $0.0043$       &              $0.0035$       &              $0.0024$       \\

$\CZ_{0,1}$& $\text{No}$   & \figsref{fig:device_sketch}(c) &              $0.0435$       &              $0.0044$       &              $0.0035$       &              $0.0024$       \\

\hline\hline
\end{tabularx}
}
\end{table}
If we consider the results for the two-qubit, three-qubit and four-qubit cases, we can see that the diamond distances are all affected by the third decimal. Furthermore, most average infidelities are affected by the fourth decimal. The leakage measure and the statistical distance are only affected in one case, \ie for the three-qubit case.

\renewcommand{\hold}{0.66}
\begin{figure}[!tbp]
  \begin{minipage}{1.0\textwidth}
    \centering
    \graphicspath{{FiguresAndData/GETPaper/APPROX/BMP/APPROX_BMP_GPU_ETH_2_HB/}}
    \includegraphics[scale=\hold]{error}
    \graphicspath{{FiguresAndData/GETPaper/APPROX/BMP/APPROX_BMP_GPU_ETH_3_HB/}}
    \includegraphics[scale=\hold]{error}
    \graphicspath{{FiguresAndData/GETPaper/APPROX/BMP/APPROX_BMP_GPU_ETH_4_HB/}}
    \includegraphics[scale=\hold]{error}
  \end{minipage}
  \caption[Gate-error metrics diamond distance $\DNV$(a,e,i,c,g,k) given by \equaref{eq:diamond_norm_inf}{eq:diamond_norm_sup} and average infidelity $\IFV$(b,f,j,d,h,l) given by \equref{eq:in_fid_avg} as functions of the number of gates executed on the two-qubit(a-d), three-qubit(e-h) and four-qubit(i-l) NIGQCs illustrated in \figsref{fig:device_sketch}(a-c), respectively (approximation study with BMP pulse).]{Gate-error metrics diamond distance $\DNV$(a,e,i,c,g,k) given by \equaref{eq:diamond_norm_inf}{eq:diamond_norm_sup} and average infidelity $\IFV$(b,f,j,d,h,l) given by \equref{eq:in_fid_avg} as functions of the number of gates executed on the two-qubit(a-d), three-qubit(e-h) and four-qubit(i-l) NIGQCs illustrated in \figsref{fig:device_sketch}(a-c), respectively. The different NIGQCs execute the same programs, namely a program with forty $\CZ_{0,1}$(a,b,e,f,i,j) instructions and forty $\CNOT_{0,1}$(c,d,g,h,k,l) instructions. Note that all $\CNOT$ gates are implemented by means of two MPs and one BMP. The programs are executed two times on each NIGQC. During the first run (blue solid line) we model the transmon qubits as adiabatic anharmonic oscillators, see \equref{eq:tunable-frequency eff}. During the second run (green dashed line) we model the transmon qubits as non-adiabatic anharmonic oscillators, see \equref{eq:fft_eff_II}. We use the device parameter listed in \tabref{tab:device_para}, the control pulse parameters listed in \tabsref{tab:CtlSqgIIHB}{tab:CtlTqgIVHB} and the effective Hamiltonian \equref{eq:EHM} to obtain the results.\label{fig:GET_APPROX_BMP}}
\end{figure}
Figures~\ref{fig:GET_APPROX_BMP}(a-l) show the diamond distance $\DNV$(a,c,e,g,i,k) given by \equaref{eq:diamond_norm_inf}{eq:diamond_norm_sup} and the average infidelity $\IFV$(b,d,f,h,j,l) in \equref{eq:in_fid_avg} as functions of the number of gates executed on the two-qubit(a-d), three-qubit(e-h) and four-qubit(i-l) NIGQCs illustrated in \figsref{fig:device_sketch}(a-c), respectively. The results for the $\CZ$ and $\CNOT$ gate repetition programs are displayed in \PANSL{a,b,e,f,i,j} and \PANSL{c,d,g,h,k,l}, respectively. Here we use BMPs to realise the $\CZ_{0,1}$ gates. The results obtained with the adiabatic (non-adiabatic) effective Hamiltonian \equref{eq:EHM} are presented as a solid (dashed) blue (green) line. The simulation parameters we use are discussed at the beginning of this section.

As can be seen, the small numerical deviations we discussed in the context of \tabref{tab:GET_EMAPPROXBMPHB} can over time substantially affect the quantitative and qualitative behaviour of the gate-error trajectories acquired with the adiabatic effective Hamiltonian \equref{eq:EHM}. Additionally, the deviations between the adiabatic and non-adiabatic case are larger for the simulation scenario where we realise the $\CZ_{0,1}$ gates with BMPs instead of the UMPs. Here again, we should emphasise that the UMPs and BMPs are very similar pulses, except for the fast falling pulse flank at around half of the pulse time $\FctlTp$, see \figref{fig:GET_pulse_time_evo}(b-c).

If we compare \figsref{fig:GET_APPROX_UMP}(a-l) and \figsref{fig:GET_APPROX_BMP}(a-l), we find that the qualitative and quantitative behaviour of the gate-error trajectories obtained with UMPs and BMPs and the corresponding independently optimised control pulse parameters is often rather different. Note that the remaining simulation parameters are identical for both cases. We can potentially explain this observation by considering the results discussed in \secref{sec:GET_Para}. Here we found that the qualitative and quantitative behaviour of gate-error trajectories is often affected by small changes in the control pulse parameters. Consequently, since we optimised the control pulse parameters for the UMPs and BMPs independently, the deviations between \figsref{fig:GET_APPROX_UMP}(a-l) and \figsref{fig:GET_APPROX_BMP}(a-l) can also be caused by the different control pulse parameters we use to obtain the results. Obviously, this explanation is not conclusive.

In this section we discussed results located at the small border between the adiabatic and non-adiabatic effective models. We emphasise that the BMPs we use in this section have rather long pulse durations. For example, the authors of \REF\cite{Rol19} investigate a device that consists of two transmon qubits coupled via a resonator. Here, the transmon qubits are modelled as adiabatic anharmonic oscillators and the BMPs which are used to realise $\CZ$ gates are about $40$ ns long. The numerical results presented in \REF\cite{Rol19} are likely to change once the transmon qubits are no longer modelled as adiabatic anharmonic oscillators.  Furthermore, we emphasise that we only discussed the case where one of the transmon qubits was modelled non-adiabatically. Consequently, we may observe larger deviations for programs where more than one non-adiabatically modelled transmon qubit experiences a flux drive.

The results discussed in this section are only concerned with a single seemingly justified and often applied approximation. However, it is usually the case that an effective Hamiltonian is the result of numerous approximations, see \chapref{chap:NA}. Therefore, it seems impossible to decide whether or not an approximation (assumption) that can substantially affect the time-evolution of a system is justified or not. Obviously, here we consider scenarios where gate-error metrics like the diamond distance or the average infidelity are obtained by simulating the time-evolution of a system. As in one of the previous sections, we exclude the unlikely case that the truncated time-evolution operators, see \equaref{eq:prop_matrix}{eq:time_evo_op}, are known to be the same for both models.

The data presented in this section lead the author of this thesis to advocate the view that every additional approximation or assumption leads to a new independent NIGQC model. Furthermore, the results presented in this section lead the author of this thesis to conjecture that the qualitative and quantitative behaviour of gate-error trajectories emerges due to a complex interplay of small deviations with respect to the target gates which occur over time. Note that the adiabatic and non-adiabatic effective Hamiltonian \equref{eq:EHM} only deviate for small time intervals, where $\dot{\varphi}(t)\neq 0$ and still we can observe how the gate-error trajectories for the different cases often diverge as time progresses. Certainly, gate errors in NIGQC models usually do not add up linearly.

\section{Summary, Discussion and Conclusions}\label{sec:GET_SummaryAndConclusions}

In this chapter we studied sequences of gate errors, in the form of gate-error quantifiers, that emerge if one repeatedly executes a gate on one of the two-qubit, three-qubit and four-qubit NIGQCs illustrated in \figsref{fig:device_sketch}(a-c), respectively. We call these sequences gate-error trajectories. The state of a NIGQC is by assumption completely determined by a time-dependent state vector $\ket{\Psi(t)}$ and the time-evolution of this state vector is, also by assumption, governed by the TDSE and a time-dependent model Hamiltonian $\OP{H}(t)$. We used the circuit Hamiltonian \equref{eq:CHM} and the effective Hamiltonian \equref{eq:EHM} to model the time-evolution of various NIGQCs. In \secsref{sec:GET_SystemSpecificationAndSimulationParameters}{sec:GET_errormeasures} we specified the model. First, in \secref{sec:GET_SystemSpecificationAndSimulationParameters}, we introduced the device architecture and the device parameters, see \figsref{fig:device_sketch}(a-c) and \tabref{tab:device_para}. Next, in \secref{sec:GET_ControlPulsesAndGateImplementation}, we introduced the control pulses used to realise the different gates. The single-qubit $\ROT(\pi/2)$ rotations are implemented with microwave pulses (MPs), see \figref{fig:GET_pulse_time_evo}(a). The two-qubit $\CZ$ gates are realised with unimodal pulses (UMPs), see \figref{fig:GET_pulse_time_evo}(b) and bimodal pulses (BMPs), see \figref{fig:GET_pulse_time_evo}(c). Then, in \secref{sec:GET_errormeasures}, we introduced the gate-error quantifiers used to assess to what extent the state of a NIGQC and the state of the IGQC deviate from one another for a given sequence of gates. Here, we discussed the diamond distance given by \equaref{eq:diamond_norm_inf}{eq:diamond_norm_sup}, the average infidelity in \equref{eq:in_fid_avg}, the leakage measure given by \equref{eq:leak} and the statistical distance in \equref{eq:stat_dis}. The main results in this chapter are presented in \secsref{sec:GET_SpectrumOfAFourQubitNIGQC}{sec:GET_Aprox}.

In \secref{sec:GET_SpectrumOfAFourQubitNIGQC} we discussed the energy levels of the four-qubit NIGQC illustrated in \figsref{fig:device_sketch}(c) as functions of the external flux $\varphi$. Here we used the circuit Hamiltonian \equref{eq:CHM} to obtain the results presented in \figsref{fig:GET_device_spectrum}(a-c). We also discussed the relevance of the spectrum for the two-qubit gates we implement and identified a problem that can potentially limit the scaling capabilities of the device architecture we studied in this chapter.

In \secref{sec:GET_GateMetricsForTheElementaryGateSet} we discussed the gate-error metrics for the elementary gate sets, see \figref{fig:GET_metrics}, obtained with the optimised control pulse parameters, see \tabsref{tab:CtlSqgIIMB}{tab:CtlTqgIVHB}. Here we used the circuit Hamiltonian \equref{eq:CHM} and the effective Hamiltonian \equref{eq:EHM} to obtain the results. We found that the gate-error metrics increase with the system size and discussed two potential, not mutually, exclusive causes for this trend.

In \secref{sec:GET_HigherStates} we discussed how modelling the four-qubit NIGQC illustrated in \figsref{fig:device_sketch}(c) with four and sixteen basis states affects the gate-error quantifiers. Here we used the circuit Hamiltonian \equref{eq:CHM} and MP for the simulations of $\ROT(\pi/2)$ single-qubit gates. The results are displayed in \tabref{tab:GET_States} and \figsref{fig:GET_States}(a-d). We found that after one (twenty) $\ROT(\pi/2)$ gates the diamond distance and the average infidelity for the two different numbers of basis states exhibit deviations at the order of 0.1\% (10\%). Furthermore, we found that the deviations for the leakage measure and the statistical distance are smaller by about a factor of one hundred. These results can potentially be explained by the fact that the leakage measure and the statistical distance can be expressed solely in terms of the squares of the state vector amplitudes. Note that gate-error quantifiers are often modelled with two or three basis states only, see for example \REFS\ListStates.

In \secref{sec:GET_Para} we discussed how modelling the two-qubit, three-qubit and four-qubit NIGQCs illustrated in \figsref{fig:device_sketch}(a-c) with slightly different control pulse parameters affects the gate-error quantifiers. Here we used the circuit Hamiltonian \equref{eq:CHM} for the simulations of $\CNOT_{0,1}=\HA_{0}\CZ_{0,1}\HA_{0}$ gates. The results are displayed in \tabref{tab:GET_Para} and \figsref{fig:GET_Para}(a-f). We used the pulse amplitude $\FctlAp=\FctlAp_{0}+\Delta\FctlAp$, where $\delta_{0}$ refers to the optimised control pulse amplitude and $\Delta\delta/\TP \in \{0,10^{-6},10^{-5},10^{-4}\}$ denotes the offset value used to model the parameter change, to realise the $\CZ_{0,1}$ gates with UMPs. We found that all non-zero offset values affect the diamond distance to some extent and most non-zero offset values affect the average infidelity. The leakage measure and the statistical distance were less affected. Furthermore, we noticed some type of tipping behaviour for the offset value $\Delta\delta = 10^{-4}$,\ie, for this offset value, the qualitative and quantitative behaviour of the gate-error trajectories changed substantially. Note that we found this type of tipping behaviour for all three NIGQCs, see \figref{fig:device_sketch}(a-c). We also converted the offset value $\Delta\delta/\TP = 10^{-6}$ for the external flux $\Delta\Phi_{e}= (\FQ \Delta\delta)/\TP$ from Weber to Tesla and found that gate-error metrics are sensitive to field strengths of about $10^{-11}$ Tesla. Here, we used an area size of ten by ten micrometer, see \REF\cite{Roth22}. In comparison, the earth's magnetic field is about $10^{-5}$ Tesla strong. Additionally, we found that the gate-error trajectories in \figsref{fig:GET_Para}(a-f) exhibit interesting non-linear behaviour.

In \secref{sec:GET_Aprox} we discussed how modelling the transmon qubits which constitute the two-qubit, three-qubit and four-qubit NIGQCs illustrated in \figsref{fig:device_sketch} (a-c) as adiabatic and non-adiabatic anharmonic oscillators, see \equaref{eq:tunable-frequency eff}{eq:fft_eff_II}, affects the gate-error quantifiers. Here we used the effective Hamiltonian \equref{eq:EHM} for the simulations of $\CZ_{0,1}$ and $\CNOT_{0,1}=\HA_{0}\CZ_{0,1}\HA_{0}$ gates with UMPs and BMPs. The results are displayed in \tabaref{tab:GET_EMAPPROXUMPHB}{tab:GET_EMAPPROXBMPHB} as well as \figsref{fig:GET_APPROX_UMP}(a-l) and \figsref{fig:GET_APPROX_BMP}(a-l). For the cases where we realised the $\CZ_{0,1}$ gates with UMPs we found deviations between 0.001\% (for the execution of a single $\CZ_{0,1}$ gate, compare \tabref{tab:GET_EMAPPROXUMPHB} first and second rows) and 20\% (for the execution of forty $\CZ_{0,1}$ gates, see \figref{fig:GET_APPROX_UMP}(a)) for the diamond distance. For the remaining gate-error quantifiers we found smaller deviations. Additionally, for most cases the qualitative behaviour of the gate-error trajectories is barely affected by the approximation. The BMP seems to justify the approximation to a lesser extent, see \figsref{fig:GET_pulse_time_evo}(b-c) and in fact we found larger qualitative and quantitative deviations once we used BMPs to realise $\CZ_{0,1}$ gates. Furthermore, we found that the gate-error trajectories for the $\CZ$ and $\CNOT$ gate repetition programs show interesting non-linear behaviour. Additionally, the qualitative behaviour of the gate-error trajectories that emerges by executing both programs can be substantially different. Therefore, the author of this thesis stated the conjecture that the qualitative and quantitative behaviour of gate-error trajectories emerges due to a complex interplay of small deviations with respect to the target gates which occur over time. Certainly, gate errors in the form of gate-error metrics, in NIGQC models usually do not add up linearly.

The results presented in this chapter, see \secsref{sec:GET_GateMetricsForTheElementaryGateSet}{sec:GET_Aprox}, show that making even minor changes to a NIGQC model can affect the computation of gate-error metrics like the diamond distance and the average infidelity. On the one hand, the mathematical relations used to obtain the gate-error metrics make the data we saw seem trivial, see \figref{fig:comp_sketch}. On the other hand, it is worth knowing to what extent the gate-error metrics are affected by changes in the model. Additionally, in \secsref{sec:GET_HigherStates}{sec:GET_Aprox} we focused on individual aspects of the model that can affect the computation of gate-error metrics. However, it is not too difficult to imagine what occurs once we start to combine the individual aspects by changing several aspects of the model at the same time, in such a case it seems highly unlikely that we are still able to establish a root cause for the varying gate-error metrics,\ie the gate errors.

Therefore, the data discussed in this chapter leads the author of this thesis to conclude that almost all assumptions (approximations) we make with regard to the model can substantially affect the gate-error metrics. For this reason, the author of this thesis advocates the view that every additional approximation or assumption leads to a new independent NIGQC model. Here we exclude the unlikely case that the truncated time-evolution operators, see \equaref{eq:prop_matrix}{eq:time_evo_op}, are known to be the same for the different models. The results presented in this chapter were selected to emphasise the small border between different NIGQC models. The author of this thesis is aware of various simulation scenarios where the deviations between different NIGQC models are of greater extent. However, the data presented was also selected to add evidence to the conjecture that the qualitative and quantitative behaviour of gate-error trajectories emerges due to a complex interplay of small deviations with respect to the target gates which occur over time.

In \secsref{sec:GET_HigherStates}{sec:GET_Aprox} we found that close values for the diamond distance and average infidelity for a target gate can result in substantially different qualitative behaviour of the gate-error metrics that emerge over time if we repeat the target gate several times. This leads the author of this thesis to conclude that the gate-error metrics alone cannot be used to predict the future behaviour of the sequence of gate-error metrics that emerges over time, let alone the performance of actual programs (algorithms). The data we discussed was obtained with the circuit Hamiltonian \equref{eq:CHM} and the effective Hamiltonian \equref{eq:EHM}. Consequently, this is not a feature of one particular NIGQC model. Furthermore, the authors of \REFS\cite{Wi17,Willsch2020} found similar evidence for a different device architecture modelled with a circuit Hamiltonian. Therefore, this is not a feature of one particular device architecture either. Hence, the author of this thesis advocates the view that gate-error metrics are at most a measure of closeness for a selected point in time. Certainly, gate-error metrics are not good predictors for future behaviour of a system. This finding is something we could have anticipated since the gate-error metrics have been derived in the context of the IGQC model. Changes of the state vector in this model are by assumption modelled as if they occur instantaneously. However, the future state of the system is always, by assumption, governed by the TDSE. Consequently, from a theoretical point of view there seems to be no reason to believe that gate-error metrics alone can be used to predict the time evolution of a system.

The discussion so far leads to the following line of reasoning. If we cannot even conclusively compare two NIGQC models and determine the root cause for the deviations between two models, how can we actually investigate gate errors in the form of gate-error metrics like the diamond distance and the average infidelity in PGQCs? Additionally, in \secref{sec:Prototype gate-based quantum computers} and the introduction to this chapter we already discussed several issues which affect the time evolution of PGQCs. Note that even the seemingly more complicated circuit Hamiltonian \equref{eq:CHM} which is based on the lumped-element approximation, see \secref{sec:TheLumpedElementApproximation}, does not take into account all of the issues discussed in the literature. Also, so far we did not take into account that measuring the state of the system is a non-trivial problem in itself, see \secref{sec:Prototype gate-based quantum computers}. Adding all these factors to the uncertainties that the complex experimental setup, again see \secref{sec:Prototype gate-based quantum computers}, already provides makes it seem impossible that a robust study of gate errors in the form of gate-error metrics like the diamond distance and the average infidelity is possible. All these issues lead the author of this thesis to advocate the view that the assessment of PGQCs should not be based on the seemingly fragile gate-error metrics considered in this chapter but on benchmark protocols like the ones discussed in \REFS\cite{Wi17,Michielsen17}.


\chapter{Summary, concluding remarks and outlook}\label{chap:End}
\newcommand{\TLSList}{\cite{McKay16,Roth19,Yan18,Foxen20,Gu21,Ganzhorn20,McKay17,Didier,Motzoi09,Rol19}}

In \chapref{chap:I} we provided an introduction to this thesis. Next, in \chapref{chap:II}, we reviewed the model of the ideal gate-based quantum computer (IGQC), see \secsref{sec:MathematicalFramework}{sec:Algorithms} and provided a distinction between the IGQC model, prototype gate-based quantum computers (PGQCs) and non-ideal gate-based quantum computers (NIGQCs),\ie the time-dependent Hamiltonian models used to describe the time evolution of transmon systems in this thesis.

Then, in \chapref{chap:III}, we discussed the lumped-element approximation which provides the mathematical relations used to obtain circuit Hamiltonians, see \secref{sec:TheLumpedElementApproximation}. Furthermore, we reviewed the circuit quantisation formalism, see \secref{sec:CircuitQuantisationFormalism} and derived circuit Hamiltonians for LC resonators, fixed-frequency and flux-tunable transmons, see \secaref{sec:ResAndTLS}{sec:Transmons}. Additionally, we provided clear and concise derivations of effective Hamiltonians for fixed-frequency and flux-tunable transmons, see \secref{sec:Transmons}. At the end of this chapter, we used these subsystems to define a many-particle circuit Hamiltonian \eqref{eq:CHM} and an associated effective Hamiltonian \equref{eq:EHM}, see \secaref{sec:TheQuantumComputerCircuitHamiltonianModel}{sec:TheQuantumComputerEffectiveHamiltonianModel} respectively. The interactions between the different subsystems are modelled as dipole-dipole interactions.

Afterwards, in \chapref{chap:IV}, we reviewed the fundamental problems that we face if we numerically solve the TDSE for most time-dependent Hamiltonians, see \secref{sec:TDSEIntro}. Furthermore, we provided a general discussion of the product-formula approach which can potentially be used to mitigate these problems, see \secref{sec:TheProductFormulaAlgorithm} and discussed the product-formula algorithms used to solve the TDSE for the effective Hamiltonian \equref{eq:EHM} and the circuit Hamiltonian \equref{eq:CHM}, see \secaref{sec:SOTEO_HB}{sec:SOTEO_MB} respectively. Additionally, we reported on attempts to numerically solve the TDSE for the circuit Hamiltonian \equref{eq:CHM} in two alternative bases. At the end of this chapter, we discussed the update rules used to implement the product-formula algorithms in the simulation software and provided an overview of the simulation software itself, see \secaref{sec:ImplementationOfTheTime-evolutionOperator}{sec:StructureOfTheSimulationSoftware} respectively.

Next, in \chapref{chap:NA}, we studied how various approximations affect the time evolution of transmon systems modelled with the TDSE and the effective Hamiltonian \equref{eq:EHM}. Additionally, we compared the time evolution of the circuit Hamiltonian \equref{eq:CHM} with the one of the effective Hamiltonian \equref{eq:EHM} for selected scenarios. Note that in this chapter we only considered the probability amplitudes relevant to the transitions which we study and transitions which can be activated by means of an external flux. We focused on three systems: a single flux-tunable transmon and two two-qubit device architectures, see \figsref{fig:arch_sketch}(a-b). For the case of the single flux-tunable transmon, we found that the non-adiabatic effective Hamiltonian \equref{eq:fft_eff_II} covers at least some of the dynamic aspects of the associated circuit Hamiltonian \equref{eq:flux-tunable transmon recast}, see \secref{sec:NA_single_flux_tunable_transmon}. For device architecture I, see \figref{fig:arch_sketch}(a), we identified resonant transitions which are seemingly suppressed in the adiabatic effective model, see \secref{sec:NA_E_suppressed}. Note that these transitions can be modelled with the non-adiabatic effective Hamiltonian \equref{eq:EHM} and the circuit Hamiltonian \equref{eq:CHM}. Furthermore, we found that nominal small time-dependent oscillations, see \figref{fig:NA_eff_int_flux_evolution_archI}(b), of the dipole-dipole interaction strength can substantially affect the $\ISWAP{}$ and $\CZ{}$ transitions modelled in \secref{sec:NA_unsuppressed_arch_I}. Additionally, we also found that the $\ISWAP{}$ and $\CZ{}$ transitions,\ie the corresponding probability amplitudes, modelled in \secref{sec:NA_unsuppressed_arch_I} are barely affected by the non-adiabatic driving term in \equref{eq:drive_term_flux}. For device architecture II, see \figref{fig:arch_sketch}(b), we also identified resonant transitions which are seemingly suppressed in the adiabatic effective model, see \secref{sec:NA_E_suppressed_AII}. Again, these transitions can be modelled with the non-adiabatic effective Hamiltonian \equref{eq:EHM} and the circuit Hamiltonian \equref{eq:CHM}. Additionally, we found that nominal large time-dependent reductions, see \figref{fig:NA_eff_int_flux_evolution_archII}(b), of the dipole-dipole interaction strength affect the $\ISWAP{}$ and $\CZ{}$ transitions modelled in \secref{sec:NA_unsuppressed_arch_II} to a lesser extent, cf.~the results discussed in \secaref{sec:NA_unsuppressed_arch_I}{sec:NA_unsuppressed_arch_II}. Furthermore, we found that the $\ISWAP{}$ and $\CZ{}$ transitions,\ie the corresponding probability amplitudes, modelled in \secref{sec:NA_unsuppressed_arch_II} are barely influenced by the non-adiabatic driving term in \equref{eq:drive_term_flux}.

Then, in \chapref{chap:GET}, we investigated how susceptible certain gate-error quantifiers are to the approximations which make up the simulation model. Here, we considered the diamond distance, the average infidelity, a leakage measure and the statistical distance, see \secref{sec:GET_errormeasures}. Note that while the statistical distance and the leakage measure can be expressed solely in terms of probability amplitudes, the gate-error metrics diamond distance and average infidelity take into account the phases of the system. In this chapter we executed $\ROT(\pi/2)$, $\CZ$ and $\CNOT$ gate repetition programs on the two-qubit, three-qubit and four-qubit NIGQCs illustrated in \figsref{fig:device_sketch}(a-c). Note that in this chapter we only considered device architecture II, cf.~\chapref{chap:NA}. We found that several aspects of the simulation model clearly affect the gate-error metrics diamond distance and average infidelity, see \secsref{sec:GET_HigherStates}{sec:GET_Aprox}. First, in \secref{sec:GET_HigherStates} we considered how the number of basis states used to model the dynamics of the system can influence the gate-error quantifiers we compute, see \tabref{tab:GET_States} and \figsref{fig:GET_States}(a-d). Then, in \secref{sec:GET_Para}, we studied how changing the control pulse parameters can affect the gate-error quantifiers we compute, see \tabref{tab:GET_Para} and \figsref{fig:GET_Para}(a-f). Afterwards, in \secref{sec:GET_Aprox} we considered how neglecting the non-adiabatic driving term in \equref{eq:drive_term_flux} can influence the gate-error quantifiers we compute, see \tabaref{tab:GET_EMAPPROXUMPHB}{tab:GET_EMAPPROXBMPHB}, \figsref{fig:GET_APPROX_UMP}(a-l) and \figsref{fig:GET_APPROX_BMP}(a-l). The statistical distance and the leakage measure are affected to a lesser extent. Additionally, we found that gate errors in the form of gate-error metrics like the diamond distance and the average infidelity often do not behave linearly as a function of the number of gates executed on the different NIGQCs, see for example  \figsref{fig:GET_APPROX_UMP}(a-l). In this regard, we should mention that, like the authors of \REFS\cite{Wi17,Willsch2020}, we found evidence that gate-error metrics like the diamond distance and the average infidelity are poor predictors of future gate errors which emerge over time. We also provided a clear and concise explanation for this finding, see \secref{sec:GET_SummaryAndConclusions}. Furthermore, we identified potential problems for the scaling capabilities of the device architecture studied in \chapref{chap:GET}, see \secaref{sec:GET_SpectrumOfAFourQubitNIGQC}{sec:GET_GateMetricsForTheElementaryGateSet}.

The results presented in this work lead the author of this thesis to make the following concluding remarks. From the author's perspective the findings in \chapref{chap:NA} seem more than just plausible for the following reason. As long as we cannot proof that two time-evolution operators, see \equref{eq:TDSE_OP}, for two different models are the same or close for the moments in time which are of interest to the physical scenario we investigate, we should not assume that the two different models make similar predictions for the same physical scenario. We already discussed this argument at the end of \secaaref{sec:Transmons}{sec:NA_SummaryConclusionsAndOutlook}{sec:GET_SummaryAndConclusions}. The results regarding the time dependence of the interaction strength in the effective model presented in \secaref{sec:NA_unsuppressed_arch_I}{sec:NA_unsuppressed_arch_II} provide a striking example for a case where the intuitive argument that the nominal small oscillations in the dipole-dipole interaction strength, see \figref{fig:NA_eff_int_flux_evolution_archI}(b), are negligible fails. Furthermore, these results become even more intriguing once we highlight the fact that the nominal much larger reduction in the time-dependent interaction strength shown in \figref{fig:NA_eff_int_flux_evolution_archII}(b) affects the time evolution of a different system to a lesser extent. Also, dropping the driving term in \equref{eq:drive_term_flux} seems to result in the suppression of all sorts of transitions, see \secaref{sec:NA_E_suppressed}{sec:NA_E_suppressed_AII}. Overall, the results in \chapref{chap:NA} show that modelling the time evolution of quantum systems is a difficult task since we cannot decide beforehand which dependencies are relevant to the problem at hand. Obviously, making intuitive decisions regarding the model can lead to unintended consequences. Furthermore, in \chapref{chap:GET} we saw that theoretically sophisticated gate-error metrics like the diamond distance and the average infidelity are poor predictors for the future behaviour of the system. This finding seems trivial once we acknowledge the fact that the TDSE in \equref{eq:TDSE} still governs the time evolution of the system and not only the gate-error metrics derived in the theoretical context of the static IGQC model. Additionally, since most approximations (assumptions) seem to clearly affect gate-error metrics, the author of this thesis advocates the view that every additional approximation (assumption) leads to a new independent NIGQC model. In other words, in almost all cases, we simply cannot easily estimate how modifying a model Hamiltonian affects the time evolution which is generated by the time-evolution operator in \equref{eq:TDSE_OP}. Here, we exclude the unlikely case that the time-evolution operators, see \equref{eq:TDSE_OP}, are known to be the same for the different models. The author of this thesis would like to make one final remark regarding the results presented in this thesis. Some authors emphasise the relevance of their research regarding PGQCs or NIGQC models by highlighting the aspects which seem relevant to the field of gate-based quantum computing,\ie to the realisation of a fully functioning real-world gate-based quantum computer. The author of this thesis would like to highlight a different aspect of this type of work. The study of PGQCs and the associated NIGQC models allows us to learn more about the subtle issues that affect the time evolution of driven quantum systems.

The results presented in this work lead the author of this thesis to highlight the following future research opportunities. A review of the literature suggests, see \REFS\TLSList{} and more, that it is more common to simplify NIGQC models which aim to describe superconducting PGQCs based on transmon qubits than to add more complexity to them. If we use real-time simulations to describe the dynamics of PGQCs, it seems more plausible to find a more balanced approach to the problem at hand. On the one hand, we would like to work with models which take into account all the relevant factors, see \secref{sec:Prototype gate-based quantum computers}. On the other hand, we find that these models are usually not solvable, numerically and/or analytically. However, we are also aware of the fact that NIGQC models are rather fragile. Consequently, it seems plausible to strategically remove certain aspects of the model which prevent a solution but at the same time to add additional complexity in terms of variables like system size. However, we should have at least a rudimentary understanding of the approximations (assumptions) we make, see \chapref{chap:NA}. For example, we could use the effective Hamiltonian \equref{eq:EHM} as the generator of the dynamics and add the read-out resonators and the measurement process to the simulation model, see \secref{sec:Prototype gate-based quantum computers}. Since we use a computer model to do so, we can turn on and off this aspect of the model and study how the measurement process itself affects the data acquisition. Note that this is only one aspect which could be integrated into the model. We should also take into account how susceptible gate-error metrics like the diamond distance and the average infidelity are. Consequently, it seems plausible to focus on modelling more robust quantities first,\ie quantities which can be expressed in terms of probability amplitudes, see \chapref{chap:GET}. Additionally, the author of this thesis was not able to find a satisfactory theoretical explanation for the time-dependent interaction strength effect discussed in \secaref{sec:NA_unsuppressed_arch_I}{sec:NA_unsuppressed_arch_II}. Hopefully other researchers are more successful in solving this intriguing problem. Note that parts of the simulation software used to obtain the results in this thesis are available as open source software, see \REF\cite{JugitJUSQUACE}.

\backmatter
{
	\begin{appendices}

\begin{subappendices}
\renewcommand\thesection{\Alph{section}}
\renewcommand{\thetable}{\Alph{section}.\arabic{table}}
\chapter{Appendices}
\section{Tables with control pulse parameters for different NIGQC models}\label{app:ControlPulseParameters}


\begin{table}[H]
\caption[Control pulse parameters for single-qubit $\ROT(\pi/2)$ gates implemented with the microwave pulse (MP) in \equref{eq:charge_ctl} (two-qubit circuit Hamiltonian NIGQC model)]{Control pulse parameters for single-qubit $\ROT(\pi/2)$ gates implemented with the microwave pulse (MP) in \equref{eq:charge_ctl}. The first column shows the gate we model. The second column shows the pulse duration $\CctlTd$ in ns. The third column shows the drive frequency $\CctlDf$ in GHz. The fourth column shows the unit-less pulse amplitude parameter $\CctlAp$. The fifth column shows the nameless parameter $\CctlSg$ in ns. The sixth column displays the DRAG amplitude $\CctlDr$ in ns. We use these parameters to model the two-qubit NIGQC illustrated in \figref{fig:device_sketch}(a) with the circuit Hamiltonian \equref{eq:CHM}.\label{tab:CtlSqgIIMB}}
\centering
{\small
\setlength{\tabcolsep}{4pt}
\begin{tabularx}{\textwidth}{X X X X X X}
\hline\hline
$\text{Gate}$&              $\CctlTd$        &              $\CctlDf/2\pi$&              $\CctlAp$            &              $\CctlSg$       &              $\CctlDr$            \\

\hline

$\ROT_{0}(\pi/2)$      &              $52.250$       &              $4.196$        &              $0.004$        &              $12.082$       &              $0.072$        \\

$\ROT_{1}(\pi/2)$       &              $52.950$       &              $5.195$        &              $0.005$        &              $10.000$       &              $0.070$        \\

\hline\hline
\end{tabularx}
}
\end{table}
\begin{table}[H]
\caption[Control pulse parameters for two-qubit $\CZ$ gates implemented with the unimodal pulse (UMP) in \equref{eq:flux_ctl_ump} (two-qubit circuit Hamiltonian NIGQC model).]{Control pulse parameters for two-qubit $\CZ$ gates implemented with the unimodal pulse (UMP) in \equref{eq:flux_ctl_ump}. The first column shows the gate we model. The second column shows the pulse type we use. The third column shows the pulse time $\FctlTp$ in ns. The fourth column shows the pulse duration $\FctlTd$ in ns. The fifth column shows the unit-less pulse amplitude $\FctlAp$. The sixth column displays the nameless parameter $\FctlSg$ in ns. The parameter in column six allows us to control the pulse flanks. We use these parameters to model the two-qubit NIGQC illustrated in \figref{fig:device_sketch}(a) with the circuit Hamiltonian \equref{eq:CHM}.\label{tab:CtlTqgIIMB}}
\centering
{\small
\setlength{\tabcolsep}{4pt}
\begin{tabularx}{\textwidth}{X X X X X X}

\hline\hline

$\text{Gate}$ &              $\text{Pulse}$              &              $\FctlTp$        &              $\FctlTd$        &              $\FctlAp/2\pi$  &              $\FctlSg$       \\

\hline

$\CZ_{0,1}$&     $\text{UMP}$         &              $99.835$       &              $125.000$      &              $0.392$        &              $1.313$        \\

\hline\hline
\end{tabularx}
}
\end{table}

\begin{table}[H]
\caption[Control pulse parameters for single-qubit $\ROT(\pi/2)$ gates implemented with the microwave pulse (MP) in \equref{eq:charge_ctl} (three-qubit circuit Hamiltonian NIGQC model).]{Control pulse parameters for single-qubit $\ROT(\pi/2)$ gates implemented with the microwave pulse (MP) in \equref{eq:charge_ctl}. The units in this table are the same as in \tabref{tab:CtlSqgIIMB}. We use these parameters to model the three-qubit NIGQC illustrated in \figref{fig:device_sketch}(b) with the circuit Hamiltonian \equref{eq:CHM}.\label{tab:CtlSqgIIIMB}}
\centering
{\small
\setlength{\tabcolsep}{4pt}
\begin{tabularx}{\textwidth}{X X X X X X}

\hline\hline

$\text{Gate}$&              $\CctlTd$        &              $\CctlDf/2\pi$&              $\CctlAp$            &              $\CctlSg$       &              $\CctlDr$            \\

\hline

$\ROT_{0}(\pi/2)$      &              $52.250$       &              $4.196$        &              $0.004$        &              $12.093$       &              $0.168$        \\

$\ROT_{1}(\pi/2)$      &              $52.950$       &              $5.190$        &              $0.004$        &              $9.997$        &              $0.067$        \\

$\ROT_{2}(\pi/2)$      &              $52.950$       &              $5.695$        &              $0.004$        &              $10.011$       &              $0.066$        \\

\hline\hline
\end{tabularx}
}
\end{table}
\begin{table}[H]
\caption[Control pulse parameters for two-qubit $\CZ$ gates implemented with the unimodal pulse (UMP) in \equref{eq:flux_ctl_ump} (three-qubit circuit Hamiltonian NIGQC model).]{Control pulse parameters for two-qubit $\CZ$ gates implemented with the unimodal pulse (UMP) in \equref{eq:flux_ctl_ump}. The units in this table are the same as in \tabref{tab:CtlTqgIIMB}. We use these parameters to model the three-qubit NIGQC illustrated in \figref{fig:device_sketch}(b) with the circuit Hamiltonian \equref{eq:CHM}.\label{tab:CtlTqgIIIMB}}
\centering
{\small
\setlength{\tabcolsep}{4pt}
\begin{tabularx}{\textwidth}{X X X X X X}
\hline\hline

$\text{Gate}$ &              $\text{Pulse}$              &              $\FctlTp$        &              $\FctlTd$        &              $\FctlAp/2\pi$  &              $\FctlSg$       \\

\hline

$\CZ_{0,1}$&     $\text{UMP}$         &              $96.026$       &              $125.000$      &              $0.391$        &              $1.823$        \\

$\CZ_{1,2}$&     $\text{UMP}$         &              $75.367$       &              $110.000$      &              $0.276$        &              $0.513$        \\

\hline\hline
\end{tabularx}
}
\end{table}

\begin{table}[H]
\caption[Control pulse parameters for single-qubit $\ROT(\pi/2)$ gates implemented with the microwave pulse (MP) in \equref{eq:charge_ctl} (four-qubit circuit Hamiltonian NIGQC model).]{Control pulse parameters for single-qubit $\ROT(\pi/2)$ gates implemented with the microwave pulse (MP) in \equref{eq:charge_ctl}. The units in this table are the same as in \tabref{tab:CtlSqgIIMB}. We use these parameters to model the four-qubit NIGQC illustrated in \figref{fig:device_sketch}(c) with the circuit Hamiltonian \equref{eq:CHM}.\label{tab:CtlSqgIVMB}}
\centering
{\small
\setlength{\tabcolsep}{4pt}
\begin{tabularx}{\textwidth}{X X X X X X}

\hline\hline

$\text{Gate}$&              $\CctlTd$        &              $\CctlDf/2\pi$&              $\CctlAp$            &              $\CctlSg$       &              $\CctlDr$            \\

\hline

$\ROT_{0}(\pi/2)$      &              $52.250$       &              $4.193$        &              $0.004$        &              $12.378$       &              $0.047$        \\

$\ROT_{1}(\pi/2)$      &              $52.950$       &              $5.190$        &              $0.004$        &              $10.255$       &              $0.063$        \\

$\ROT_{2}(\pi/2)$      &              $52.950$       &              $5.689$        &              $0.004$        &              $10.312$       &              $0.065$        \\

$\ROT_{3}(\pi/2)$      &              $52.950$       &              $4.951$        &              $0.005$        &              $10.191$       &              $0.012$        \\

\hline\hline
\end{tabularx}
}
\end{table}
\begin{table}[H]
\caption[Control pulse parameters for two-qubit $\CZ$ gates implemented with the unimodal pulse (UMP) in \equref{eq:flux_ctl_ump} (four-qubit circuit Hamiltonian NIGQC model).]{Control pulse parameters for two-qubit $\CZ$ gates implemented with the unimodal pulse (UMP) in \equref{eq:flux_ctl_ump}. The units in this table are the same as in \tabref{tab:CtlTqgIIMB}. We use these parameters to model the four-qubit NIGQC illustrated in \figref{fig:device_sketch}(c) with the circuit Hamiltonian \equref{eq:CHM}.\label{tab:CtlTqgIVMB}}
\centering
{\small
\setlength{\tabcolsep}{4pt}
\begin{tabularx}{\textwidth}{X X X X X X}

\hline\hline

$\text{Gate}$ &              $\text{Pulse}$              &              $\FctlTp$        &              $\FctlTd$        &              $\FctlAp/2\pi$  &              $\FctlSg$       \\

\hline

$\CZ_{0,1}$&     $\text{UMP}$         &              $100.241$      &              $125.000$      &              $0.392$        &              $1.283$        \\

$\CZ_{1,2}$&     $\text{UMP}$         &              $68.046$       &              $90.000$       &              $0.275$        &              $0.182$        \\

$\CZ_{2,3}$&     $\text{UMP}$         &              $80.500$       &              $94.000$       &              $0.320$        &              $0.500$        \\

$\CZ_{0,3}$&     $\text{UMP}$         &              $97.708$       &              $116.000$      &              $0.353$        &              $1.458$        \\

\hline\hline
\end{tabularx}
}
\end{table}



\begin{table}[H]
\caption[Control pulse parameters for single-qubit $\ROT(\pi/2)$ gates implemented with the microwave pulse (MP) in \equref{eq:charge_ctl} (two-qubit effective Hamiltonian NIGQC model).]{Control pulse parameters for single-qubit $\ROT(\pi/2)$ gates implemented with the microwave pulse (MP) in \equref{eq:charge_ctl}. The units in this table are the same as in \tabref{tab:CtlSqgIIMB}. We use these parameters to model the two-qubit NIGQC illustrated in \figref{fig:device_sketch}(a) with the effective Hamiltonian \equref{eq:EHM}.\label{tab:CtlSqgIIHB}}
\centering
{\small
\setlength{\tabcolsep}{4pt}
\begin{tabularx}{\textwidth}{X X X X X X}

\hline\hline

$\text{Gate}$&              $\CctlTd$        &              $\CctlDf/2\pi$&              $\CctlAp$            &              $\CctlSg$       &              $\CctlDr$            \\

\hline

$\ROT_{0}(\pi/2)$     &              $52.250$       &              $4.196$        &              $0.058$        &              $12.082$       &              $0.072$        \\

$\ROT_{1}(\pi/2)$     &              $52.950$       &              $5.195$        &              $0.065$        &              $10.000$       &              $0.070$        \\

\hline\hline
\end{tabularx}
}
\end{table}
\begin{table}[H]
\caption[Control pulse parameters for two-qubit $\CZ$ gates implemented with the unimodal pulse (UMP) and bimodal pulse (BMP) in \equaref{eq:flux_ctl_ump}{eq:flux_ctl_bmp}, respectively (two-qubit effective Hamiltonian NIGQC model).]{Control pulse parameters for two-qubit $\CZ$ gates implemented with the unimodal pulse (UMP) and bimodal pulse (BMP) in \equaref{eq:flux_ctl_ump}{eq:flux_ctl_bmp}, respectively. The units in this table are the same as in \tabref{tab:CtlTqgIIMB}. We use these parameters to model the two-qubit NIGQC illustrated in \figref{fig:device_sketch}(a) with the effective Hamiltonian \equref{eq:EHM}.\label{tab:CtlTqgIIHB}}
{\small
\setlength{\tabcolsep}{4pt}
\begin{tabularx}{\textwidth}{X X X X X X}

\hline\hline

$\text{Gate}$ &              $\text{Pulse}$              &              $\FctlTp$        &              $\FctlTd$        &              $\FctlAp/2\pi$  &              $\FctlSg$       \\

\hline

$\CZ_{0,1}$&     $\text{UMP}$         &$87.258$       &              $95.000$       &              $0.391$        &              $0.459$        \\

$\CZ_{0,1}$&     $\text{BMP}$         &$88.570$       &              $95.000$       &              $0.392$        &              $0.394$        \\

\hline\hline
\end{tabularx}
}
\end{table}

\begin{table}[H]
\caption[Control pulse parameters for single-qubit $\ROT(\pi/2)$ gates implemented with the microwave pulse (MP) in \equref{eq:charge_ctl} (three-qubit effective Hamiltonian NIGQC model).]{Control pulse parameters for single-qubit $\ROT(\pi/2)$ gates implemented with the microwave pulse (MP) in \equref{eq:charge_ctl}. The units in this table are the same as in \tabref{tab:CtlSqgIIMB}. We use these parameters to model the three-qubit NIGQC illustrated in \figref{fig:device_sketch}(b) with the effective Hamiltonian \equref{eq:EHM}.\label{tab:CtlSqgIIIHB}}
{\small
\setlength{\tabcolsep}{4pt}
\begin{tabularx}{\textwidth}{X X X X X X}

\hline\hline

$\text{Gate}$&              $\CctlTd$        &              $\CctlDf/2\pi$&              $\CctlAp$            &              $\CctlSg$       &              $\CctlDr$            \\

\hline

$\ROT_{0}(\pi/2)$ &              $52.250$       &              $4.196$        &              $0.058$        &              $12.082$       &              $0.072$        \\

$\ROT_{1}(\pi/2)$     &              $52.950$       &              $5.189$        &              $0.065$        &              $10.000$       &              $0.070$        \\

$\ROT_{2}(\pi/2)$      &              $52.950$       &              $5.694$        &              $0.066$        &              $9.990$        &              $0.032$        \\

\hline\hline
\end{tabularx}
}
\end{table}
\begin{table}[H]
\caption[Control pulse parameters for two-qubit $\CZ$ gates implemented with the unimodal pulse (UMP) and bimodal pulse (BMP) in \equaref{eq:flux_ctl_ump}{eq:flux_ctl_bmp}, respectively (three-qubit effective Hamiltonian NIGQC model).]{Control pulse parameters for two-qubit $\CZ$ gates implemented with the unimodal pulse (UMP) and bimodal pulse (BMP) in \equaref{eq:flux_ctl_ump}{eq:flux_ctl_bmp}, respectively. The units in this table are the same as in \tabref{tab:CtlTqgIIMB}. We use these parameters to model the three-qubit NIGQC illustrated in \figref{fig:device_sketch}(b) with the effective Hamiltonian \equref{eq:EHM}.\label{tab:CtlTqgIIIHB}}
{\small
\setlength{\tabcolsep}{4pt}
\begin{tabularx}{\textwidth}{X X X X X X}

\hline\hline

$\text{Gate}$ &              $\text{Pulse}$              &              $\FctlTp$        &              $\FctlTd$        &              $\FctlAp/2\pi$  &              $\FctlSg$       \\

\hline

$\CZ_{0,1}$&     $\text{UMP}$         &              $87.252$       &              $95.006$       &              $0.391$        &              $0.494$        \\

$\CZ_{0,1}$&     $\text{BMP}$         &             $90.057$       &              $92.188$       &              $0.391$        &              $0.420$        \\

$\CZ_{1,2}$&     $\text{UMP}$         &              $68.831$       &              $80.000$       &              $0.276$        &              $0.554$        \\

\hline\hline
\end{tabularx}
}
\end{table}

\begin{table}[H]
\caption[Control pulse parameters for single-qubit $\ROT(\pi/2)$ gates implemented with the microwave pulse (MP) in \equref{eq:charge_ctl} (four-qubit effective Hamiltonian NIGQC model).]{Control pulse parameters for single-qubit $\ROT(\pi/2)$ gates implemented with the microwave pulse (MP) in \equref{eq:charge_ctl}. The units in this table are the same as in \tabref{tab:CtlSqgIIMB}. We use these parameters to model the four-qubit NIGQC illustrated in \figref{fig:device_sketch}(c) with the effective Hamiltonian \equref{eq:EHM}.\label{tab:CtlSqgIVHB}}
{\small
\setlength{\tabcolsep}{4pt}
\begin{tabularx}{\textwidth}{X X X X X X}

\hline\hline

$\text{Gate}$&              $\CctlTd$        &              $\CctlDf/2\pi$&              $\CctlAp$            &              $\CctlSg$       &              $\CctlDr$            \\

\hline

$\ROT_{0}(\pi/2)$       &              $52.250$       &              $4.191$        &              $0.058$        &              $12.082$       &              $0.072$        \\

$\ROT_{1}(\pi/2)$      &              $52.950$       &              $5.189$        &              $0.065$        &              $10.000$       &              $0.070$        \\

$\ROT_{2}(\pi/2)$      &              $52.950$       &              $5.688$        &              $0.066$        &              $9.990$        &              $0.032$        \\

$\ROT_{3}(\pi/2)$      &              $52.950$       &              $4.950$        &              $0.066$        &              $9.990$        &              $0.032$        \\

\hline\hline
\end{tabularx}
}
\end{table}
\begin{table}[H]
\caption[Control pulse parameters for two-qubit $\CZ$ gates implemented with the unimodal pulse (UMP) and bimodal pulse (BMP) in \equaref{eq:flux_ctl_ump}{eq:flux_ctl_bmp}, respectively (four-qubit effective Hamiltonian NIGQC model).]{Control pulse parameters for two-qubit $\CZ$ gates implemented with the unimodal pulse (UMP) and bimodal pulse (BMP) in \equaref{eq:flux_ctl_ump}{eq:flux_ctl_bmp}, respectively. The units in this table are the same as in \tabref{tab:CtlTqgIIMB}. We use these parameters to model the four-qubit NIGQC illustrated in \figref{fig:device_sketch}(c) with the effective Hamiltonian \equref{eq:EHM}.\label{tab:CtlTqgIVHB}}
{\small
\setlength{\tabcolsep}{4pt}
\begin{tabularx}{\textwidth}{X X X X X X}

\hline\hline

$\text{Gate}$ &              $\text{Pulse}$              &              $\FctlTp$        &              $\FctlTd$        &              $\FctlAp/2\pi$  &              $\FctlSg$       \\

\hline

$\CZ_{0,1}$&     $\text{UMP}$         &              $87.254$       &              $95.013$       &              $0.391$        &              $0.453$        \\

$\CZ_{0,1}$&     $\text{BMP}$         &              $89.925$       &              $98.114$       &              $0.392$        &              $0.400$        \\

$\CZ_{1,2}$&     $\text{UMP}$         &              $67.802$       &              $115.238$      &              $0.275$        &              $0.338$        \\

$\CZ_{2,3}$&     $\text{UMP}$         &              $71.620$       &              $98.197$       &              $0.320$        &              $0.543$        \\

$\CZ_{0,3}$&     $\text{UMP}$         &              $92.616$       &              $124.768$      &              $0.353$        &              $1.877$        \\

\hline\hline
\end{tabularx}
}
\end{table}


\section{Tables with gate-error quantifiers for different NIGQC models}\label{app:TablesWithGateErrorMetrics}
\setcounter{table}{0}

\begin{table}[H]
\caption[Gate-error quantifiers for $\ROT(\pi/2)$ and $\CZ$ gates (two-qubit circuit Hamiltonian NIGQC model).]{Gate-error quantifiers for $\ROT(\pi/2)$ and $\CZ$ gates. The first column shows the target gate. The second column shows the control pulse we use to obtain the actual gate. The third column shows the unit-less diamond distance $\DNV$ given by \equaref{eq:diamond_norm_inf}{eq:diamond_norm_sup}. The fourth column shows the unit-less average infidelity $\IFV$ given by \equref{eq:in_fid_avg}. The fifth column shows the leakage measure $\LNV$ given by \equref{eq:leak}. We use the circuit Hamiltonian \equref{eq:CHM}, the device parameters listed in \tabref{tab:device_para} and the control pulse parameters listed in \tabaref{tab:CtlSqgIIMB}{tab:CtlTqgIIMB} to obtain the gate-error quantifiers in this table and to model the two-qubit NIGQC illustrated in \figref{fig:device_sketch}(a).\label{tab:EMIIMB}}
{\small
\setlength{\tabcolsep}{4pt}
\begin{tabularx}{\textwidth}{X X X X X X}

\hline\hline

$\text{Gate}$ & $\text{Pulse}$ &               $\DNV$&              $\IFV$&              $\LNV$ \\

\hline

$\ROT_{0}(\pi/2)$ & $\text{MP}$ &              $0.0093$       &              $0.0004$       &              $0.0004$       \\

$\ROT_{1}(\pi/2)$ & $\text{MP}$ &              $0.0080$       &              $0.0004$       &              $0.0004$       \\

$\CZ_{0,1}$ & $\text{UMP}$ &              $0.0290$       &              $0.0011$       &              $0.0008$      \\

\hline\hline
\end{tabularx}
}
\end{table}
\begin{table}[H]
\caption[Gate-error quantifiers for $\ROT(\pi/2)$ and $\CZ$ gates (three-qubit circuit Hamiltonian NIGQC model).]{Gate-error quantifiers for $\ROT(\pi/2)$ and $\CZ$ gates. The rows and columns show the same unit-less quantities as \tabref{tab:EMIIMB}. We use the circuit Hamiltonian \equref{eq:CHM}, the device parameters listed in \tabref{tab:device_para} and the control pulse parameters listed in \tabaref{tab:CtlSqgIIIMB}{tab:CtlTqgIIIMB} to obtain the gate-error quantifiers in this table and to model the three-qubit NIGQC illustrated in \figref{fig:device_sketch}(b).\label{tab:EMIIIMB}}
{\small
\setlength{\tabcolsep}{4pt}
\begin{tabularx}{\textwidth}{X X X X X X}

\hline\hline

$\text{Gate}$ & $\text{Pulse}$ &               $\DNV$&              $\IFV$&              $\LNV$\\

\hline

$\ROT_{0}(\pi/2)$ & $\text{MP}$ &              $0.044$        &              $0.003$        &              $0.002$        \\

$\ROT_{1}(\pi/2)$ & $\text{MP}$ &              $0.038$        &              $0.002$        &              $0.001$        \\

$\ROT_{2}(\pi/2)$ & $\text{MP}$ &              $0.039$        &              $0.002$        &              $0.001$        \\

$\CZ_{0,1}$ & $\text{UMP}$ &              $0.044$        &              $0.010$        &              $0.010$        \\

$\CZ_{1,2}$ & $\text{UMP}$ &              $0.012$        &              $0.002$        &              $0.002$       \\

\hline\hline
\end{tabularx}
}
\end{table}
\begin{table}[H]
\caption[Gate-error quantifiers for $\ROT(\pi/2)$ and $\CZ$ gates (four-qubit circuit Hamiltonian NIGQC model).]{Gate-error quantifiers for $\ROT(\pi/2)$ and $\CZ$ gates. The rows and columns show the same unit-less quantities as \tabref{tab:EMIIMB}. We use the circuit Hamiltonian \equref{eq:CHM}, the device parameters listed in \tabref{tab:device_para} and the control pulse parameters listed in \tabaref{tab:CtlSqgIVMB}{tab:CtlTqgIVMB} to obtain the gate-error quantifiers in this table and to model the four-qubit NIGQC illustrated in \figref{fig:device_sketch}(c).\label{tab:EMIVMB}}
{\small
\setlength{\tabcolsep}{4pt}
\begin{tabularx}{\textwidth}{X X X X X X}

\hline\hline

$\text{Gate}$ & $\text{Pulse}$ &               $\DNV$&              $\IFV$&              $\LNV$\\

\hline

$\ROT_{0}(\pi/2)$ & $\text{MP}$ &              $0.058$        &              $0.004$        &              $0.002$        \\

$\ROT_{1}(\pi/2)$ & $\text{MP}$ &              $0.054$        &              $0.003$        &              $0.002$       \\

$\ROT_{2}(\pi/2)$ & $\text{MP}$ &              $0.053$        &              $0.003$        &              $0.002$        \\

$\ROT_{3}(\pi/2)$ & $\text{MP}$ &              $0.057$        &              $0.004$        &              $0.002$        \\

$\CZ_{0,1}$ & $\text{UMP}$ &              $0.031$        &              $0.005$        &              $0.004$        \\

$\CZ_{1,2}$ & $\text{UMP}$ &              $0.144$        &              $0.029$        &              $0.018$        \\

$\CZ_{2,3}$ & $\text{UMP}$ &              $0.073$        &              $0.008$        &              $0.005$        \\

$\CZ_{0,3}$ & $\text{UMP}$ &              $0.046$        &              $0.005$        &              $0.003$        \\

\hline\hline
\end{tabularx}
}
\end{table}



\begin{table}[H]
\caption[Gate-error quantifiers for $\ROT(\pi/2)$ and $\CZ$ (two-qubit effective Hamiltonian NIGQC model).]{Gate-error quantifiers for $\ROT(\pi/2)$ and $\CZ$ gates. The rows and columns show the same unit-less quantities as \tabref{tab:EMIIMB}. We use the effective Hamiltonian \equref{eq:EHM}, the device parameters listed in \tabref{tab:device_para} and the control pulse parameters listed in \tabaref{tab:CtlSqgIIHB}{tab:CtlTqgIIHB} to obtain the gate-error quantifiers in this table and to model the two-qubit NIGQC illustrated in \figref{fig:device_sketch}(a).\label{tab:EMIIHB}}
{\small
\setlength{\tabcolsep}{4pt}
\begin{tabularx}{\textwidth}{X X X X X X}

\hline\hline

$\text{Gate}$ & $\text{Pulse}$ &               $\DNV$&              $\IFV$&              $\LNV$\\

\hline

$\ROT_{0}(\pi/2)$ & $\text{MP}$ &              $0.0089$       &              $0.0004$       &              $0.0004$       \\

$\ROT_{1}(\pi/2)$ & $\text{MP}$ &              $0.0090$       &              $0.0004$       &              $0.0004$       \\

$\CZ_{0,1}$ & $\text{UMP}$ &              $0.0424$       &              $0.0012$       &              $0.0005$       \\

$\CZ_{0,1}$ & $\text{BMP}$ &              $0.0167$        &              $0.0006$        &              $0.0005$        \\

\hline\hline
\end{tabularx}
}
\end{table}
\begin{table}[H]
\caption[Gate-error quantifiers for $\ROT(\pi/2)$ and $\CZ$ gates (three-qubit effective Hamiltonian NIGQC model).]{Gate-error quantifiers for $\ROT(\pi/2)$ and $\CZ$ gates. The rows and columns show the same unit-less quantities as \tabref{tab:EMIIMB}. We use the effective Hamiltonian \equref{eq:EHM}, the device parameters listed in \tabref{tab:device_para} and the control pulse parameters listed in \tabaref{tab:CtlSqgIIIHB}{tab:CtlTqgIIIHB} to obtain the gate-error quantifiers in this table and to model the three-qubit NIGQC illustrated in \figref{fig:device_sketch}(b).\label{tab:EMIIIHB}}
{\small
\setlength{\tabcolsep}{4pt}
\begin{tabularx}{\textwidth}{X X X X X X}

\hline\hline

$\text{Gate}$ & $\text{Pulse}$ &               $\DNV$&              $\IFV$&              $\LNV$\\

\hline

$\ROT_{0}(\pi/2)$ & $\text{MP}$ &              $0.046$        &              $0.003$        &              $0.002$        \\

$\ROT_{1}(\pi/2)$ & $\text{MP}$ &              $0.040$        &              $0.002$        &              $0.002$        \\

$\ROT_{2}(\pi/2)$ & $\text{MP}$ &              $0.039$        &              $0.002$        &              $0.002$       \\

$\CZ_{0,1}$ & $\text{UMP}$ &              $0.057$        &              $0.003$        &              $0.002$        \\

$\CZ_{0,1}$ & $\text{BMP}$ &              $0.031$        &              $0.004$        &              $0.004$        \\

$\CZ_{1,2}$ & $\text{UMP}$ &              $0.028$        &              $0.006$        &              $0.006$        \\

\hline\hline
\end{tabularx}
}
\end{table}
\begin{table}[H]
\caption[Gate-error quantifiers for $\ROT(\pi/2)$ and $\CZ$ gates (four-qubit effective Hamiltonian NIGQC model).]{Gate-error quantifiers for $\ROT(\pi/2)$ and $\CZ$ gates. The rows and columns show the same unit-less quantities as \tabref{tab:EMIIMB}. We use the effective Hamiltonian \equref{eq:EHM}, the device parameters listed in \tabref{tab:device_para} and the control pulse parameters listed in \tabaref{tab:CtlSqgIVHB}{tab:CtlTqgIVHB} to obtain the gate-error quantifiers in this table and to model the four-qubit NIGQC illustrated in \figref{fig:device_sketch}(c).\label{tab:EMIVHB}}
{\small
\setlength{\tabcolsep}{4pt}
\begin{tabularx}{\textwidth}{X X X X X X}

\hline\hline

$\text{Gate}$ & $\text{Pulse}$ &               $\DNV$&              $\IFV$&              $\LNV$\\

\hline

$\ROT_{0}(\pi/2)$ & $\text{MP}$ &              $0.060$        &              $0.004$        &              $0.003$       \\

$\ROT_{1}(\pi/2)$ & $\text{MP}$ &              $0.056$        &              $0.003$        &              $0.002$       \\

$\ROT_{2}(\pi/2)$ & $\text{MP}$ &              $0.054$        &              $0.003$        &              $0.002$       \\

$\ROT_{3}(\pi/2)$ & $\text{MP}$ &              $0.060$        &              $0.004$        &              $0.002$       \\

$\CZ_{0,1}$ & $\text{UMP}$ &              $0.051$        &              $0.004$        &              $0.003$       \\

$\CZ_{0,1}$ & $\text{BMP}$ &              $0.041$        &              $0.004$        &              $0.004$       \\

$\CZ_{1,2}$ & $\text{UMP}$ &              $0.146$        &              $0.015$        &              $0.005$       \\

$\CZ_{2,3}$ & $\text{UMP}$ &              $0.078$        &              $0.010$        &              $0.007$       \\

$\CZ_{0,3}$ & $\text{UMP}$ &              $0.058$        &              $0.005$        &              $0.003$       \\

\hline\hline
\end{tabularx}
}
\end{table}


\end{subappendices}

%
%
%
%
%

	\end{appendices}
	\hypersetup{linkcolor=black}
	\addcontentsline{toc}{chapter}{List of figures}
	\listoffigures
  	\clearpage
  	\addcontentsline{toc}{chapter}{List of tables}
  	\listoftables
	\clearpage
	\chapter{List of acronyms}
	\begin{acronym}[LAPACK]
		\acro{BASH}{Bourne again shell}
		\acro{BMP}{Bimodal pulse}
		\acro{CUDA}{Compute unified device architecture}
		\acro{ELR}{Energy level repulsion}
		\acro{EMF}{Electromotive forces}
		\acro{FD}{Full diagonalisation}
		\acro{HB}{Harmonic basis}
		\acro{IGQC}{Ideal gate-based quantum computer}
		\acro{JURECA}{Jülich research on exascale cluster architectures}
		\acro{JUWELS}{Jülich wizard for European leadership science}
		\acro{LAPACK}{Linear algebra package}
		\acro{MKL}{Math kernel library}
		\acro{MP}{Microwave pulse}
		\acro{MPI}{Message passing interface}
		\acro{NIGQC}{Non-ideal gate-based quantum computer}
		\acro{OpenMP}{Open multi-processing}
		\acro{PFA}{Product formula algorithm}
		\acro{PGQC}{Prototype gate-based quantum computer}
		\acro{QFT}{Quantum Fourier transformation}
		\acro{SLURM}{Simple Linux utility for resource management}
		\acro{SSH}{Secure shell}
		\acro{TB}{Transmon basis}
		\acro{TDSE}{Time-dependent Schrödinger equation}
		\acro{TLS}{Two-level system}
		\acro{UMP}{Unimodal pulse}
		\acro{VHA}{Variational hybrid algorithm}		
	\end{acronym}
	\clearpage
	
	\chapter{List of publications}
	\begin{enumerate}
	  \item H. Lagemann, D. Willsch, M. Willsch, F. Jin, H. De Raedt, and K. Michielsen, "On the fragility of gate-error metrics in simulation models of flux-tunable transmon quantum computers", submitted to Phys. Rev. A. URL: \href{https://arxiv.org/abs/2211.11011}{Link}
  
	  \item H. Lagemann, D. Willsch, M. Willsch, F. Jin, H. De Raedt, and K. Michielsen, "Numerical analysis of effective models for flux-tunable transmon systems", Phys. Rev. A 106, 022615 (2022). URL: \href{https://journals.aps.org/pra/abstract/10.1103/PhysRevA.106.022615}{Link}
  
	  \item F. Jin, D. Willsch, M. Willsch, H. Lagemann, K. Michielsen and H. De Raedt, "Random State Technology", J. Phys. Soc. Jpn. 90, 012001 (2021). URL: \href{https://journals.jps.jp/doi/10.7566/JPSJ.90.012001}{Link}
  
	  \item D. Willsch, H. Lagemann, M. Willsch, F. Jin, H. De Raedt and K. Michielsen, "Benchmarking Supercomputers with the Jülich Universal Quantum Computer Simulator", NIC Symposium 2020, Publication Series of the John von Neumann Institute for Computing (NIC) NIC Series 50, 255 - 264 (2020). URL: \href{https://juser.fz-juelich.de/record/874419/files/NIC_2020_Michielsen.pdf}{Link}
	\end{enumerate}
	\clearpage

	\chapter{Acknowledgments}
	First of all, I would like to thank Kristel Michielsen and Hans De Raedt for introducing me to the field of computational physics and for providing me with the interesting problem I worked on for the last three years.

	Of course, I also thank the rest of the QIP group for having all these fun group meetings and lunch breaks. Special thanks goes to Dennis Willsch, Madita Willsch and Fengping Jin. Dennis always helped me, with seemingly endless patience, solve large and small problems. Furthermore, with his dissertation he laid the foundation for many of the results presented in this thesis. Madita is probably the best proofreader I know and clearly helped me improve my writing skills. Fengping is a fun office mate, but also a scientific force to be reckoned with!

	Furthermore, I would like to thank David DiVincenzo for co-supervising this dissertation and for providing valuable feedback regarding subjects I was not overly familiar with at the beginning of my time as a PhD student.

	I acknowledge support from the project OpenSuperQ (820363) of the EU Quantum Flagship. Thanks goes to all project partners but special thanks goes to Christopher Warren and Jorge Fernandez-Pendas from Chalmers University of Technology as well as Stefania Lazar from ETH Zurich for answering my numerous questions about the experiments they are working on. Furthermore, I acknowledge the Gauss Centre for Supercomputing e.V. for providing computing time on the GCS Supercomputer JUWELS \cite{JUWELS} at the Jülich Supercomputing Centre (JSC) and the project Jülich UNified Infrastructure for Quantum computing (JUNIQ) that has received funding from the German Federal Ministry of Education and Research (BMBF) and the Ministry of Culture and Science of the State of North Rhine-Westphalia.

	I would also like to thank Klaus Morawetz for giving me a brutal but fruitful crash course in theoretical physics as well as Navina Kleemann and Binh Nguyen for making the years as undergraduate students more enjoyable and sometimes bearable.

	Last but not least, I thank family and friends which have barely seen me for the last three years. Certainly, not everyone grows up in an environment which supports critical thought and autonomy. Furthermore, not all five year old grandsons get a proper answer to the question: how does a nuclear power plant work, paired with a small introduction to nuclear physics.
	\clearpage

}

\addcontentsline{toc}{chapter}{Bibliography}
\printbibheading
\printbibliography[type=book,heading=subbibliography,title={Books}]
\printbibliography[type=misc,heading=subbibliography,title={Electronic references}]
\printbibliography[type=thesis,heading=subbibliography,title={Theses}]
\printbibliography[type=software,heading=subbibliography,title={Software}]
\defbibfilter{papers}{type=article or type=inproceedings or type=report or type=incollection}
\printbibliography[filter=papers,heading=subbibliography,title={Other references}]

@thesis{EMPTY,
  author      = {Nobody}
}

@mastersthesis{Willsch2016Master,
  author      = {Willsch, Dennis},
  institution = {RWTH Aa-chen University},
  title       = {{S}imulation of quantum computer hardware based on superconducting circuits},
  addendum    = {\url{http://hdl.handle.net/2128/21812}},
  type        = {Master's thesis},
  url         = {http://juser.fz-juelich.de/record/819153},
  cid         = {I:(DE-Juel1)JSC-20090406},
  cin         = {JSC},
  pid         = {G:(DE-HGF)POF3-511},
  pnm         = {511 - Computational Science and Mathematical Methods (POF3-511)},
  reportid    = {FZJ-2016-04866},
  school      = {RWTH Aachen University},
  typ         = {PUB:(DE-HGF)19},
  year        = {2016},
}

@mastersthesis{LagemannMSCThesis,
  author       = {Lagemann, Hannes},
  title        = {{D}evelopment and implementation of a gate-based quantum
                  computer simulator for high-dimensional {H}ilbert spaces},
  school       = {RW-TH Aachen},
  type         = {Masterarbeit},
  reportid     = {FZJ-2019-06906},
  pages        = {70},
  year         = {2019},
  note         = {Masterarbeit, RWTH Aachen, 2019},
  cin          = {JSC},
  cid          = {I:(DE-Juel1)JSC-20090406},
  pnm          = {511 - Computational Science and Mathematical Methods
                  (POF3-511)},
  pid          = {G:(DE-HGF)POF3-511},
  typ          = {PUB:(DE-HGF)19},
  url          = {https://juser.fz-juelich.de/record/868377}
}

@phdthesis{Willsch2020,
  author       = {Willsch, Dennis},
  title        = {{S}upercomputer simulations of transmon quantum computers},
  school       = {RWTH Aachen University},
  type         = {Dissertation},
  year         = {2020},
  url          = {https://juser.fz-juelich.de/record/885927},
}

@phdthesis{WillschM2020,
      author       = {Willsch, Madita Franziska},
      othercontributors = {Michielsen, Kristel Francine and DiVincenzo, David},
      title        = {{S}tudy of quantum annealing by simulating the time
                      evolution of flux qubits},
      school       = {RWTH Aachen University},
      type         = {Dissertation},
      year         = {2020},
      url          = {https://publications.rwth-aachen.de/record/795009}
}

@phdthesis{Cottet2002,
  title = {Implementation of a quantum bit in a superconducting circuit},
  author = {Audrey Cottet},
  degree ={PhD},
  year = {2002},
  institution= {Université Paris 6},
  url={http://www.phys.ens.fr/~cottet/ACottetThesis.pdf}
}

@phdthesis{Sanders16,
	author={Yuval Sanders},
	title={Characterizing Errors in Quantum Information Processors},
	publisher={UWSpace},
  school = {University of Waterloo},
  type = {Dissertation},
	year={2016},
	url={http://hdl.handle.net/10012/10467}
}

@phdthesis{Roth20,
	author={ Marco Roth },
	title={{A}nalysis of scalable coupling schemes for superconducting quantum computers},
  school= {RWTH Aachen University},
  type= {Dissertation},
	year={2020},
	url={https://publications.rwth-aachen.de/record/787496}
}

@phdthesis{Ulrich17,
      author       = {Ulrich, Jascha},
      title        = {{L}arge impedances and {M}ajorana bound states in
                      superconducting circuits},
      school       = {RWTH Aachen University},
      type         = {Dissertation},
      year         = {2017},
      url          = {https://publications.rwth-aachen.de/record/684349},
}

@book{QEC13,
  place={Cambridge},
  editor = {Daniel A. Lidar and Todd A. Brun},
  title={Quantum Error Correction},
  DOI={10.1017/CBO9781139034807},
  publisher={Cambridge University Press},
  year={2013},
  url={https://www.cambridge.org/core/books/quantum-error-correction/B51E8333050A0F9A67363254DC1EA15A}
}

@book{Ashcroft76,
  author = {Ashcroft, N. W. and Mermin, N. D.},
  publisher = {Holt-Saunders},
  title = {{S}olid {S}tate {P}hysics},
  year = {1976}
}

@book{AB06,
  author = {Sanjeev Arora and Boaz Barak},
  title = {Computational Complexity: A Modern Approach},
  publisher = {Cambridge University Press},
  year = {2006},
  url = {http://theory.cs.princeton.edu/complexity/}
}

@book{FANO60,
  title={Electromagnetic Fields, Energy and Forces},
  author = {Robert M. Fano and Lan Jen Chu and Richard B. Adler},
  publisher={John Wiley \& Sons},
  year={1960},
  url={https://mitpress.mit.edu/books/electromagnetic-fields-energy-and-forces}
}

@book{Cormen09,
  author = {Cormen, Thomas H. and Leiserson, Charles E. and Rivest, Ronald L. and Stein, Clifford},
  title = {Introduction to Algorithms},
  publisher = {The MIT Press},
  edition = {3},
  year = {2009},
  url = {https://mitpress.mit.edu/books/introduction-algorithms-third-edition}
}

@book{GustafsonSigal2011,
  title ={Mathematical Concepts of Quantum Mechanics},
  author ={Stephen J. Gustafson and Israel Michael Sigal},
  publisher = {Springer},
  year={2011},
  edition={Second},
  url={https://link.springer.com/book/10.1007/978-3-642-21866-8}
}

@book{HoffmanKunze71,
    author = {Hoffman, Kenneth and Kunze, Ray},
    title = {Linear algebra},
    edition= {Second},
    publisher= {Prentice-Hall, Inc., Englewood Cliffs, N.J.},
    year = {1971}
}

@book{GoluVanl96,
  title  = {Matrix Computations},
  author = {Golub, Gene H. and Van Loan, Charles F.},
  publisher= {The Johns Hopkins University Press},
  year= {1996},
  edition = {Third},
  url={https://jhupbooks.press.jhu.edu/title/matrix-computations}
}

@book{CT10,
  address = {New York},
  author = {{Cohen-Tannoudji}, C. and {Grynberg}, G. and {Dupont-Roc}, J.},
  url = {https://www.bibsonomy.org/bibtex/2eb02c7aecb613d60725f7d48f52a82fb/jhigbie},
  publisher = {Wiley},
  title = {Atom-Photon Interactions: Basic Processes and Applications},
  year = {1992}
}

@book{Weinberg2015,
  place={Cambridge},
  edition={2},
  title={Lectures on Quantum Mechanics},
  DOI={10.1017/CBO9781316276105},
  publisher={Cambridge University Press},
  author={Weinberg, Steven},
  year={2015}
}

@book{Watrous2018,
  place={Cambridge},
  title={The Theory of Quantum Information},
  DOI={10.1017/9781316848142},
  publisher={Cambridge University Press},
  author={Watrous, John},
  year={2018},
  url={https://www.cambridge.org/core/books/theory-of-quantum-information/AE4AA5638F808D2CFEB070C55431D897}
}

@book{Maxwell10,
  place={Cambridge},
  series={Cambridge Library Collection - Physical  Sciences},
  title={A Treatise on Electricity and Magnetism},
  volume={2},
  DOI={10.1017/CBO9780511709340},
  publisher={Cambridge University Press},
  author={Maxwell, James Clerk},
  year={2010},
  collection={Cambridge Library Collection - Physical  Sciences}
}

@book{Hager:2010,
  author = {Hager, Georg and Wellein, Gerhard},
  title = {Introduction to High Performance Computing for Scientists and Engineers},
  publisher = {CRC Press, Inc.},
  year = {2010}
}

@book{Nielsen:2011:QCQ:1972505,
 author = {Nielsen, Michael A. and Chuang, Isaac L.},
 title = {Quantum Computation and Quantum Information: 10th Anniversary Edition},
 publisher = {Cambridge University Press},
 year = {2011}
 }

@book{Zangwill13,
     author        = {Zangwill, Andrew},
     title         = {Modern Electrodynamics},
     publisher     = {Cambridge University Press},
     year          = {2013},
     url           = {https://cds.cern.ch/record/1507229}
}

@book{Wendt58,
     author        = {G. Wendt and Ronold W. P. King and F. E. Borgnis and C. H. Papas and H. Bremmer},
     title         = {Elektrische Felder und Wellen / Electric Fields and Waves},
     publisher     = {Springer},
     year          = {1958},
     url           = {https://link.springer.com/book/10.1007/978-3-642-45895-8}
}

@book{Balian91,
     author        = {Roger Balian},
     title         = {From Microphysics to Macrophysics},
     publisher     = {Springer},
     volume        = {1},
     year          = {1991},
     url           = {https://link.springer.com/book/10.1007/978-3-540-45475-5}
}

@book{Balian07,
     author        = {Roger Balian},
     title         = {From Microphysics to Macrophysics},
     publisher     = {Springer},
     volume        = {2},
     year          = {2007},
     url           = {https://link.springer.com/book/10.1007/978-3-540-45480-9?wt_mc=ThirdParty.SpringerLink.3.EPR653.About_eBook}
}

@book{Balanis12,
  title={Advanced Engineering Electromagnetics},
  author={Constantine A. Balanis},
  year={2012},
  publisher={Wiley}
}

@book{Esquinazi98,
     editor        = {Pablo Esquinazi},
     title         = {Tunneling Systems in Amorphous and Crystalline Solids},
     publisher     = {Springer},
     edition       = {1},
     year          = {1998},
     url           = {https://link.springer.com/book/10.1007/978-3-662-03695-2#toc}
}

@book{DiVincenzo13,
      key          = {845776},
      editor       = {DiVincenzo, David},
      title        = {{Q}uantum {I}nformation {P}rocessing},
      volume       = {52},
      address      = {Jülich},
      publisher    = {Forschungszentrum Jülich GmbH Zentralbibliothek, Verlag},
      reportid     = {FZJ-2018-02985},
      isbn         = {978-3-89336-833-4},
      series       = {Schriften des Forschungszentrums Jülich. Reihe
                      Schlüsseltechnologien / Key Technologies},
      pages        = {getr. Zählung},
      year         = {2013},
      organization  = {IFF-Ferienschule,},
      cin          = {PGI-2 / IAS-3},
      cid          = {I:(DE-Juel1)PGI-2-20110106 / I:(DE-Juel1)IAS-3-20090406},
      pnm          = {424 - Exploratory materials and phenomena (POF2-424)},
      pid          = {G:(DE-HGF)POF2-424},
      typ          = {PUB:(DE-HGF)3 / PUB:(DE-HGF)26},
      url          = {https://juser.fz-juelich.de/record/845776}
}

@misc{Mcdonald12voltagedrop,
    author = {Kirk T. McDonald},
    title = {Voltage Drop, Potential Difference and EMF},
    year = {2012},
    url={https://www.hep.princeton.edu/~mcdonald/examples/volt.pdf}
}

@misc{Martinis14,
	author = {John M. Martinis and A. Megrant},
	title = {UCSB final report for the CSQ program: Review of decoherence and materials physics for superconducting qubits},
	year = {2014},
	eprint = {arXiv:1410.5793v1},
}

@misc{farhi2014quantum,
      title={A Quantum Approximate Optimization Algorithm},
      author={Edward Farhi and Jeffrey Goldstone and Sam Gutmann},
      year={2014},
      eprint={1411.4028},
      archivePrefix={arXiv},
      primaryClass={quant-ph}
}

@misc{Naghiloo19,
      title={Introduction to Experimental Quantum Measurement with Superconducting Qubits},
      author={Mahdi Naghiloo},
      year={2019},
      eprint={1904.09291},
      archivePrefix={arXiv},
      primaryClass={quant-ph},
      url={https://arxiv.org/abs/1904.09291}
}

@misc{Uhlig2020,
  doi = {10.48550/ARXIV.2002.01274},
  url = {https://arxiv.org/abs/2002.01274},
  author = {Uhlig, Frank},
  title = {Coalescing Eigenvalues and Crossing Eigencurves of 1-Parameter Matrix Flows},
  publisher = {arXiv},
  year = {2020},
  copyright = {arXiv.org perpetual, non-exclusive license}
}

@misc{Srinivasan2020,
  doi = {10.48550/ARXIV.2006.14254},
  url = {https://arxiv.org/abs/2006.14254},
  author = {Srinivasan, Usha and Kidambi, Rangachari},
  title = {A sorting algorithm for complex eigenvalues},
  publisher = {arXiv},
  year = {2020},
  copyright = {Creative Commons Attribution Non Commercial Share Alike 4.0 International}
}

@article{Burgarth21,
  doi = {10.22331/q-2022-06-14-737},
  url = {https://doi.org/10.22331/q-2022-06-14-737},
  title = {One bound to rule them all: from {A}diabatic to {Z}eno},
  author = {Burgarth, Daniel and Facchi, Paolo and Gramegna, Giovanni and Yuasa, Kazuya},
  journal = {{Quantum}},
  issn = {2521-327X},
  publisher = {{Verein zur F{\"{o}}rderung des Open Access Publizierens in den Quantenwissenschaften}},
  volume = {6},
  pages = {737},
  month = {6},
  year = {2022}
}

@article{Krinner21,
  author={Krinner, Sebastian
  and Lacroix, Nathan
  and Remm, Ants
  and Di Paolo, Agustin
  and Genois, Elie
  and Leroux, Catherine
  and Hellings, Christoph
  and Lazar, Stefania
  and Swiadek, Francois
  and Herrmann, Johannes
  and Norris, Graham J.
  and Andersen, Christian Kraglund
  and M{\"u}ller, Markus
  and Blais, Alexandre
  and Eichler, Christopher
  and Wallraff, Andreas},
  title={Realizing repeated quantum error correction in a distance-three surface code},
  journal={Nature},
  year={2022},
  month={May},
  day={01},
  volume={605},
  number={7911},
  pages={669-674},
  issn={1476-4687},
  doi={10.1038/s41586-022-04566-8},
  url={https://doi.org/10.1038/s41586-022-04566-8}
}

@article{Yan18,
  title = {Tunable Coupling Scheme for Implementing High-Fidelity Two-Qubit Gates},
  author = {Yan, Fei and Krantz, Philip and Sung, Youngkyu and Kjaergaard, Morten and Campbell, Daniel L. and Orlando, Terry P. and Gustavsson, Simon and Oliver, William D.},
  journal = {Phys. Rev. Applied},
  volume = {10},
  issue = {5},
  pages = {054062},
  numpages = {9},
  year = {2018},
  month = {Nov},
  publisher = {American Physical Society},
  doi = {10.1103/PhysRevApplied.10.054062},
  url = {https://link.aps.org/doi/10.1103/PhysRevApplied.10.054062}
}

@article{Werninghaus2021,
  author={Werninghaus, M.
  and Egger, D. J.
  and Roy, F.
  and Machnes, S.
  and Wilhelm, F. K.
  and Filipp, S.},
  title={Leakage reduction in fast superconducting qubit gates via optimal control},
  journal={npj Quantum Information},
  year={2021},
  month={Jan},
  day={29},
  volume={7},
  number={1},
  pages={14},
  issn={2056-6387},
  doi={10.1038/s41534-020-00346-2},
  url={https://doi.org/10.1038/s41534-020-00346-2}
}

@article{Murray2021,
  title = {Material matters in superconducting qubits},
  journal = {Materials Science and Engineering: R: Reports},
  volume = {146},
  pages = {100646},
  year = {2021},
  issn = {0927-796X},
  doi = {https://doi.org/10.1016/j.mser.2021.100646},
  url = {https://www.sciencedirect.com/science/article/pii/S0927796X21000413},
  author = {Conal E. Murray}
}

@article{Gu21,
  title = {Fast Multiqubit Gates through Simultaneous Two-Qubit Gates},
  author = {Gu, Xiu and Fern\'andez-Pend\'as, Jorge and Vikst\aa{}l, Pontus and Abad, Tahereh and Warren, Christopher and Bengtsson, Andreas and Tancredi, Giovanna and Shumeiko, Vitaly and Bylander, Jonas and Johansson, G\"oran and Kockum, Anton Frisk},
  journal = {PRX Quantum},
  volume = {2},
  issue = {4},
  pages = {040348},
  numpages = {28},
  year = {2021},
  month = {Dec},
  publisher = {American Physical Society},
  doi = {10.1103/PRXQuantum.2.040348},
  url = {https://link.aps.org/doi/10.1103/PRXQuantum.2.040348}
}

@article{Burkard04,
  title = {Multilevel quantum description of decoherence in superconducting qubits},
  author = {Burkard, Guido and Koch, Roger H. and DiVincenzo, David P.},
  journal = {Phys. Rev. B},
  volume = {69},
  issue = {6},
  pages = {064503},
  numpages = {20},
  year = {2004},
  month = {Feb},
  publisher = {American Physical Society},
  doi = {10.1103/PhysRevB.69.064503},
  url = {https://link.aps.org/doi/10.1103/PhysRevB.69.064503}
}

@article{Vool17,
  author = {Vool, Uri and Devoret, Michel},
  year = {2017},
  month = {7},
  pages = {},
  title = {Introduction to Quantum Electromagnetic Circuits},
  volume = {45},
  journal = {International Journal of Circuit Theory and Applications},
  doi = {10.1002/cta.2359},
  url={https://onlinelibrary.wiley.com/doi/abs/10.1002/cta.2359}
}

@article{McKay17,
  title = {Efficient $Z$ gates for quantum computing},
  author = {McKay, David C. and Wood, Christopher J. and Sheldon, Sarah and Chow, Jerry M. and Gambetta, Jay M.},
  journal = {Phys. Rev. A},
  volume = {96},
  issue = {2},
  pages = {022330},
  numpages = {8},
  year = {2017},
  month = {Aug},
  publisher = {American Physical Society},
  doi = {10.1103/PhysRevA.96.022330},
  url = {https://link.aps.org/doi/10.1103/PhysRevA.96.022330}
}

@article{Motzoi09,
  title = {Simple Pulses for Elimination of Leakage in Weakly Nonlinear Qubits},
  author = {Motzoi, F. and Gambetta, J. M. and Rebentrost, P. and Wilhelm, F. K.},
  journal = {Phys. Rev. Lett.},
  volume = {103},
  issue = {11},
  pages = {110501},
  numpages = {4},
  year = {2009},
  month = {Sep},
  publisher = {American Physical Society},
  doi = {10.1103/PhysRevLett.103.110501},
  url = {https://link.aps.org/doi/10.1103/PhysRevLett.103.110501}
}

@article{Riwar21,
  author={Riwar, R.-P. and DiVincenzo, D. P.},
  title={Circuit quantization with time-dependent magnetic fields for realistic geometries},
  journal={npj Quantum Information},
  year={2022},
  month={Mar},
  day={25},
  volume={8},
  number={1},
  pages={36},
  issn={2056-6387},
  doi={10.1038/s41534-022-00539-x},
  url={https://doi.org/10.1038/s41534-022-00539-x}
}

@article{Krinner2019,
  author={Krinner, S. and Storz, S. and Kurpiers, P. and Magnard, P. and Heinsoo, J. and Keller, R. and L{\"u}tolf, J. and Eichler, C. and Wallraff, A.},
  title={Engineering cryogenic setups for 100-qubit scale superconducting circuit systems},
  journal={EPJ Quantum Technology},
  year={2019},
  month={May},
  day={28},
  volume={6},
  number={1},
  pages={2},
  issn={2196-0763},
  doi={10.1140/epjqt/s40507-019-0072-0},
  url={https://doi.org/10.1140/epjqt/s40507-019-0072-0}
}

@article{Heinsoo18,
  title = {Rapid High-fidelity Multiplexed Readout of Superconducting Qubits},
  author = {Heinsoo, Johannes and Andersen, Christian Kraglund and Remm, Ants and Krinner, Sebastian and Walter, Theodore and Salath\'e, Yves and Gasparinetti, Simone and Besse, Jean-Claude and Poto\ifmmode \check{c}\else \v{c}\fi{}nik, Anton and Wallraff, Andreas and Eichler, Christopher},
  journal = {Phys. Rev. Applied},
  volume = {10},
  issue = {3},
  pages = {034040},
  numpages = {14},
  year = {2018},
  month = {Sep},
  publisher = {American Physical Society},
  doi = {10.1103/PhysRevApplied.10.034040},
  url = {https://link.aps.org/doi/10.1103/PhysRevApplied.10.034040}
}

@article{McEwen22,
  author={McEwen, Matt
  and Faoro, Lara
  and Arya, Kunal
  and Dunsworth, Andrew
  and Huang, Trent
  and Kim, Seon
  and Burkett, Brian
  and Fowler, Austin
  and Arute, Frank
  and Bardin, Joseph C.
  and Bengtsson, Andreas
  and Bilmes, Alexander
  and Buckley, Bob B.
  and Bushnell, Nicholas
  and Chen, Zijun
  and Collins, Roberto
  and Demura, Sean
  and Derk, Alan R.
  and Erickson, Catherine
  and Giustina, Marissa
  and Harrington, Sean D.
  and Hong, Sabrina
  and Jeffrey, Evan
  and Kelly, Julian
  and Klimov, Paul V.
  and Kostritsa, Fedor
  and Laptev, Pavel
  and Locharla, Aditya
  and Mi, Xiao
  and Miao, Kevin C.
  and Montazeri, Shirin
  and Mutus, Josh
  and Naaman, Ofer
  and Neeley, Matthew
  and Neill, Charles
  and Opremcak, Alex
  and Quintana, Chris
  and Redd, Nicholas
  and Roushan, Pedram
  and Sank, Daniel
  and Satzinger, Kevin J.
  and Shvarts, Vladimir
  and White, Theodore
  and Yao, Z. Jamie
  and Yeh, Ping
  and Yoo, Juhwan
  and Chen, Yu
  and Smelyanskiy, Vadim
  and Martinis, John M.
  and Neven, Hartmut
  and Megrant, Anthony
  and Ioffe, Lev
  and Barends, Rami},
  title={Resolving catastrophic error bursts from cosmic rays in large arrays of superconducting qubits},
  journal={Nature Physics},
  year={2022},
  month={Jan},
  day={01},
  volume={18},
  number={1},
  pages={107-111},
  issn={1745-2481},
  doi={10.1038/s41567-021-01432-8},
  url={https://doi.org/10.1038/s41567-021-01432-8}
}

@article{GKS76,
  author = {Gorini,Vittorio  and Kossakowski,Andrzej  and Sudarshan,E. C. G. },
  title = {Completely positive dynamical semigroups of N‐level systems},
  journal = {Journal of Mathematical Physics},
  volume = {17},
  number = {5},
  pages = {821-825},
  year = {1976},
  doi = {10.1063/1.522979},
  URL = {https://aip.scitation.org/doi/abs/10.1063/1.522979},
  eprint = {https://aip.scitation.org/doi/pdf/10.1063/1.522979}
}

@article{Lindblad76,
  author={Lindblad, G.},
  title={On the generators of quantum dynamical semigroups},
  journal={Communications in Mathematical Physics},
  year={1976},
  month={Jun},
  day={01},
  volume={48},
  number={2},
  pages={119-130},
  issn={1432-0916},
  doi={10.1007/BF01608499},
  url={https://doi.org/10.1007/BF01608499}
}

@article{Turing36,
  author = {Turing, Alan M.},
  journal = {Proceedings of the London Mathematical Society},
  number = {42},
  pages = {230--265},
  title = {On Computable Numbers, with an Application to the {E}ntscheidungsproblem},
  url = {https://londmathsoc.onlinelibrary.wiley.com/doi/abs/10.1112/plms/s2-42.1.230},
  volume = {2},
  year = {1937}
}

@article{Arute19,
  title	= {Quantum Supremacy using a Programmable Superconducting Processor},
  author	= {Frank Arute and Kunal Arya and Ryan Babbush and Dave Bacon and Joseph Bardin and Rami Barends and Rupak Biswas and Sergio Boixo and Fernando Brandao and David Buell and Brian Burkett and Yu Chen and Jimmy Chen and Ben Chiaro and Roberto Collins and William Courtney and Andrew Dunsworth and Edward Farhi and Brooks Foxen and Austin Fowler and Craig Michael Gidney and Marissa Giustina and Rob Graff and Keith Guerin and Steve Habegger and Matthew Harrigan and Michael Hartmann and Alan Ho and Markus Rudolf Hoffmann and Trent Huang and Travis Humble and Sergei Isakov and Evan Jeffrey and Zhang Jiang and Dvir Kafri and Kostyantyn Kechedzhi and Julian Kelly and Paul Klimov and Sergey Knysh and Alexander Korotkov and Fedor Kostritsa and Dave Landhuis and Mike Lindmark and Erik Lucero and Dmitry Lyakh and Salvatore Mandrà and Jarrod Ryan McClean and Matthew McEwen and Anthony Megrant and Xiao Mi and Kristel Michielsen and Masoud Mohseni and Josh Mutus and Ofer Naaman and Matthew Neeley and Charles Neill and Murphy Yuezhen Niu and Eric Ostby and Andre Petukhov and John Platt and Chris Quintana and Eleanor G. Rieffel and Pedram Roushan and Nicholas Rubin and Daniel Sank and Kevin J. Satzinger and Vadim Smelyanskiy and Kevin Jeffery Sung and Matt Trevithick and Amit Vainsencher and Benjamin Villalonga and Ted White and Z. Jamie Yao and Ping Yeh and Adam Zalcman and Hartmut Neven and John Martinis},
  year	= {2019},
  URL	= {https://www.nature.com/articles/s41586-019-1666-5},
  journal	= {Nature},
  pages	= {505–510},
  volume	= {574}
}

@article{Sanders2015,
	doi = {10.1088/1367-2630/18/1/012002},
	url = {https://doi.org/10.1088/1367-2630/18/1/012002},
	year = {2015},
	month = {dec},
	publisher = {{IOP} Publishing},
	volume = {18},
	number = {1},
	pages = {012002},
	author = {Yuval R Sanders and Joel J Wallman and Barry C Sanders},
	title = {Bounding quantum gate error rate based on reported average fidelity},
	journal = {New Journal of Physics}
}

@article{Jin21,
  author = {Jin ,Fengping and Willsch ,Dennis and Willsch ,Madita and Lagemann ,Hannes and Michielsen ,Kristel and De Raedt ,Hans},
  title = {Random State Technology},
  journal = {Journal of the Physical Society of Japan},
  volume = {90},
  number = {1},
  pages = {012001},
  year = {2021},
  doi = {10.7566/JPSJ.90.012001},
  url = {https://doi.org/10.7566/JPSJ.90.012001},
  eprint = {https://doi.org/10.7566/JPSJ.90.012001}
}

@article{Johnston09,
  author = {Johnston, Nathaniel and Kribs, David and Paulsen, Vern},
  year = {2009},
  month = {12},
  pages = {},
  title = {Computing Stabilized Norms for Quantum Operations via the Theory of Completely Bounded Maps},
  volume = {9},
  journal = {Quantum Information and Computation},
  url={https://www.rintonpress.com/journals/qiconline.html#v9n12}
}

@article{DiCarlo2010,
  author={DiCarlo, L. and Reed, M. D. and Sun, L. and Johnson, B. R. and Chow, J. M. and Gambetta, J. M. and Frunzio, L. and Girvin, S. M. and Devoret, M. H. and Schoelkopf, R. J.},
  title={Preparation and measurement of three-qubit entanglement in a superconducting circuit},
  journal={Nature},
  year={2010},
  month={Sep},
  day={01},
  volume={467},
  number={7315},
  pages={574-578},
  issn={1476-4687},
  doi={10.1038/nature09416},
  url={https://doi.org/10.1038/nature09416}
}

@article{Dirac1925,
  title = {The fundamental equations of quantum mechanics},
  author = {Dirac, P.A.M.},
  journal = {Proceedings of the Royal Society A},
  volume = {109},
  issue = {752},
  pages = {642-653},
  year = {1925},
  month = {December},
  doi = {10.1098/rspa.1925.0150},
  url = {https://royalsocietypublishing.org/doi/10.1098/rspa.1925.0150}
}

@article{Ulrich16,
  title = {Dual approach to circuit quantization using loop charges},
  author = {Ulrich, Jascha and Hassler, Fabian},
  journal = {Phys. Rev. B},
  volume = {94},
  issue = {9},
  pages = {094505},
  numpages = {19},
  year = {2016},
  month = {Sep},
  publisher = {American Physical Society},
  doi = {10.1103/PhysRevB.94.094505},
  url = {https://link.aps.org/doi/10.1103/PhysRevB.94.094505}
}

@article{Yurke84,
  title = {Quantum network theory},
  author = {Yurke, Bernard and Denker, John S.},
  journal = {Phys. Rev. A},
  volume = {29},
  issue = {3},
  pages = {1419--1437},
  numpages = {0},
  year = {1984},
  month = {Mar},
  publisher = {American Physical Society},
  doi = {10.1103/PhysRevA.29.1419},
  url = {https://link.aps.org/doi/10.1103/PhysRevA.29.1419}
}

@article{Dawson2006,
 author = {Christopher M. Dawson and Michael A. Nielsen},
 title = {The Solovay-Kitaev Algorithm},
 journal = {Quantum Info. Comput.},
 year = {2006},
 volume = {6},
 pages = {81--95},
 url = {http://dl.acm.org/citation.cfm?id=2011679.2011685}
}

@article{Barenco95,
  title = {Elementary gates for quantum computation},
  author = {Barenco, Adriano and Bennett, Charles H. and Cleve, Richard and DiVincenzo, David P. and Margolus, Norman and Shor, Peter and Sleator, Tycho and Smolin, John A. and Weinfurter, Harald},
  journal = {Phys. Rev. A},
  volume = {52},
  issue = {5},
  pages = {3457--3467},
  numpages = {0},
  year = {1995},
  month = {Nov},
  publisher = {American Physical Society},
  doi = {10.1103/PhysRevA.52.3457},
  url = {https://link.aps.org/doi/10.1103/PhysRevA.52.3457}
}

@article{Suzuki85,
  author = {Suzuki,Masuo },
  title = {Decomposition formulas of exponential operators and Lie exponentials with some applications to quantum mechanics and statistical physics},
  journal = {Journal of Mathematical Physics},
  volume = {26},
  number = {4},
  pages = {601-612},
  year = {1985},
  doi = {10.1063/1.526596},
  url = {https://doi.org/10.1063/1.526596},
  eprint = {https://doi.org/10.1063/1.526596}
}

@article{Suzuki77,
    author = {Suzuki, Masuo and Miyashita, Seiji and Kuroda, Akira},
    title = "{Monte Carlo Simulation of Quantum Spin Systems. I}",
    journal = {Progress of Theoretical Physics},
    volume = {58},
    number = {5},
    pages = {1377-1387},
    year = {1977},
    month = {11},
    issn = {0033-068X},
    doi = {10.1143/PTP.58.1377},
    url = {https://doi.org/10.1143/PTP.58.1377},
    eprint = {https://academic.oup.com/ptp/article-pdf/58/5/1377/5389651/58-5-1377.pdf},
}

@article{DeRaedt83,
  title = {Applications of the generalized Trotter formula},
  author = {De Raedt, Hans and De Raedt, Bart},
  journal = {Phys. Rev. A},
  volume = {28},
  issue = {6},
  pages = {3575--3580},
  numpages = {0},
  year = {1983},
  month = {Dec},
  publisher = {American Physical Society},
  doi = {10.1103/PhysRevA.28.3575},
  url = {https://link.aps.org/doi/10.1103/PhysRevA.28.3575}
}

@article{Michielsen17,
  title = {Benchmarking gate-based quantum computers},
  journal = {Computer Physics Communications},
  volume = {220},
  pages = {44-55},
  year = {2017},
  issn = {0010-4655},
  doi = {https://doi.org/10.1016/j.cpc.2017.06.011},
  url = {https://www.sciencedirect.com/science/article/pii/S0010465517301935},
  author = {Kristel Michielsen and Madita Nocon and Dennis Willsch and Fengping Jin and Thomas Lippert and Hans {De Raedt}}
}

@article{FRAN61,
    author = {Francis, J. G. F.},
    title = "{The QR Transformation A Unitary Analogue to the LR Transformation—Part 1}",
    journal = {The Computer Journal},
    volume = {4},
    number = {3},
    pages = {265-271},
    year = {1961},
    month = {01},
    doi = {10.1093/comjnl/4.3.265},
    url = {https://doi.org/10.1093/comjnl/4.3.265},
    eprint = {https://academic.oup.com/comjnl/article-pdf/4/3/265/1080833/040265.pdf},
}

@article{FRAN62,
    author = {Francis, J. G. F.},
    title = "{The QR Transformation—Part 2}",
    journal = {The Computer Journal},
    volume = {4},
    number = {4},
    pages = {332-345},
    year = {1962},
    month = {01},
    issn = {0010-4620},
    doi = {10.1093/comjnl/4.4.332},
    url = {https://doi.org/10.1093/comjnl/4.4.332},
    eprint = {https://academic.oup.com/comjnl/article-pdf/4/4/332/8201663/040332.pdf},
}

@article{Krinner2020,
  title = {Benchmarking Coherent Errors in Controlled-Phase Gates due to Spectator Qubits},
  author = {Krinner, S. and Lazar, S. and Remm, A. and Andersen, C.K. and Lacroix, N. and Norris, G.J. and Hellings, C. and Gabureac, M. and Eichler, C. and Wallraff, A.},
  journal = {Phys. Rev. Applied},
  volume = {14},
  issue = {2},
  pages = {024042},
  numpages = {9},
  year = {2020},
  month = {Aug},
  publisher = {American Physical Society},
  doi = {10.1103/PhysRevApplied.14.024042},
  url = {https://link.aps.org/doi/10.1103/PhysRevApplied.14.024042}
}

@article{Blais2020circuit,
  title = {Circuit quantum electrodynamics},
  author = {Blais, Alexandre and Grimsmo, Arne L. and Girvin, S. M. and Wallraff, Andreas},
  journal = {Rev. Mod. Phys.},
  volume = {93},
  issue = {2},
  pages = {025005},
  numpages = {72},
  year = {2021},
  month = {May},
  publisher = {American Physical Society},
  doi = {10.1103/RevModPhys.93.025005},
  url = {https://link.aps.org/doi/10.1103/RevModPhys.93.025005}
}

@article{Amin09,
  title = {Consistency of the Adiabatic Theorem},
  author = {Amin, M. H. S.},
  journal = {Phys. Rev. Lett.},
  volume = {102},
  issue = {22},
  pages = {220401},
  numpages = {4},
  year = {2009},
  month = {Jun},
  publisher = {American Physical Society},
  doi = {10.1103/PhysRevLett.102.220401},
  url = {https://link.aps.org/doi/10.1103/PhysRevLett.102.220401}
}

@article{You,
  title = {Circuit quantization in the presence of time-dependent external flux},
  author = {You, Xinyuan and Sauls, J. A. and Koch, Jens},
  journal = {Phys. Rev. B},
  volume = {99},
  issue = {17},
  pages = {174512},
  numpages = {10},
  year = {2019},
  month = {May},
  publisher = {American Physical Society},
  doi = {10.1103/PhysRevB.99.174512},
  url = {https://link.aps.org/doi/10.1103/PhysRevB.99.174512}
}

@article{Didier,
  title = {Analytical modeling of parametrically modulated transmon qubits},
  author = {Didier, Nicolas and Sete, Eyob A. and da Silva, Marcus P. and Rigetti, Chad},
  journal = {Phys. Rev. A},
  volume = {97},
  issue = {2},
  pages = {022330},
  numpages = {13},
  year = {2018},
  month = {Feb},
  publisher = {American Physical Society},
  doi = {10.1103/PhysRevA.97.022330},
  url = {https://link.aps.org/doi/10.1103/PhysRevA.97.022330}
}

@article{Foxen20,
  title = {Demonstrating a Continuous Set of Two-Qubit Gates for Near-Term Quantum Algorithms},
  author = {Foxen, B. and Neill, C. and Dunsworth, A. and Roushan, P. and Chiaro, B. and Megrant, A. and Kelly, J. and Chen, Zijun and Satzinger, K. and Barends, R. and Arute, F. and Arya, K. and Babbush, R. and Bacon, D. and Bardin, J. C. and Boixo, S. and Buell, D. and Burkett, B. and Chen, Yu and Collins, R. and Farhi, E. and Fowler, A. and Gidney, C. and Giustina, M. and Graff, R. and Harrigan, M. and Huang, T. and Isakov, S. V. and Jeffrey, E. and Jiang, Z. and Kafri, D. and Kechedzhi, K. and Klimov, P. and Korotkov, A. and Kostritsa, F. and Landhuis, D. and Lucero, E. and McClean, J. and McEwen, M. and Mi, X. and Mohseni, M. and Mutus, J. Y. and Naaman, O. and Neeley, M. and Niu, M. and Petukhov, A. and Quintana, C. and Rubin, N. and Sank, D. and Smelyanskiy, V. and Vainsencher, A. and White, T. C. and Yao, Z. and Yeh, P. and Zalcman, A. and Neven, H. and Martinis, J. M.},
  collaboration = {Google AI Quantum},
  journal = {Phys. Rev. Lett.},
  volume = {125},
  issue = {12},
  pages = {120504},
  numpages = {6},
  year = {2020},
  month = {Sep},
  publisher = {American Physical Society},
  doi = {10.1103/PhysRevLett.125.120504},
  url = {https://link.aps.org/doi/10.1103/PhysRevLett.125.120504}
}

@article{Koch,
  title = {Charge-insensitive qubit design derived from the Cooper pair box},
  author = {Koch, Jens and Yu, Terri M. and Gambetta, Jay and Houck, A. A. and Schuster, D. I. and Majer, J. and Blais, Alexandre and Devoret, M. H. and Girvin, S. M. and Schoelkopf, R. J.},
  journal = {Phys. Rev. A},
  volume = {76},
  issue = {4},
  pages = {042319},
  numpages = {19},
  year = {2007},
  month = {Oct},
  publisher = {American Physical Society},
  doi = {10.1103/PhysRevA.76.042319},
  url = {https://link.aps.org/doi/10.1103/PhysRevA.76.042319}
}

@article{Huyghebaert90,
  author    = {J. Huyghebaert and H. {De Raedt}},
  title     = {Product formula methods for time-dependent Schr\"{o}dinger problems},
  doi       = {10.1088/0305-4470/23/24/019},
  number    = {24},
  pages     = {5777--5793},
  url       = {https://doi.org/10.1088%2F0305-4470%2F23%2F24%2F019},
  volume    = {23},
  journal   = {J. Phys. A: Math. Gen.},
  month     = {dec},
  publisher = {{IOP} Publishing},
  year      = {1990},
}

@article{Rol19,
  title = {Fast, High-Fidelity Conditional-Phase Gate Exploiting Leakage Interference in Weakly Anharmonic Superconducting Qubits},
  author = {Rol, M. A. and Battistel, F. and Malinowski, F. K. and Bultink, C. C. and Tarasinski, B. M. and Vollmer, R. and Haider, N. and Muthusubramanian, N. and Bruno, A. and Terhal, B. M. and DiCarlo, L.},
  journal = {Phys. Rev. Lett.},
  volume = {123},
  issue = {12},
  pages = {120502},
  numpages = {6},
  year = {2019},
  month = {Sep},
  publisher = {American Physical Society},
  doi = {10.1103/PhysRevLett.123.120502},
  url = {https://link.aps.org/doi/10.1103/PhysRevLett.123.120502}
}

@article{DiCarlo2009,
  author = {DiCarlo, L. and Chow, J. M. and Gambetta, J. M. and Bishop, Lev S. and Johnson, B. R. and Schuster, D. I. and Majer, J. and Blais, A. and Frunzio, L. and Girvin, S. M. and Schoelkopf, R. J.},
  title = {Demonstration of two-qubit algorithms with a superconducting quantum processor},
  journal = {Nature},
  volume = {460},
  issue = {7252},
  pages = {120502},
  numpages = {6},
  year = {2019},
  month = {July},
  publisher = {Nature},
  doi = {10.1038/nature08121},
  url = {https://www.nature.com/articles/nature08121#citeas}
}

@article{McKay16,
  title = {Universal Gate for Fixed-Frequency Qubits via a Tunable Bus},
  author = {McKay, David C. and Filipp, Stefan and Mezzacapo, Antonio and Magesan, Easwar and Chow, Jerry M. and Gambetta, Jay M.},
  journal = {Phys. Rev. Applied},
  volume = {6},
  issue = {6},
  pages = {064007},
  numpages = {10},
  year = {2016},
  month = {Dec},
  publisher = {American Physical Society},
  doi = {10.1103/PhysRevApplied.6.064007},
  url = {https://link.aps.org/doi/10.1103/PhysRevApplied.6.064007}
}

@article{Bengtsson2020,
  title = {Improved Success Probability with Greater Circuit Depth for the Quantum Approximate Optimization Algorithm},
  author = {Bengtsson, Andreas and Vikst\aa{}l, Pontus and Warren, Christopher and Svensson, Marika and Gu, Xiu and Kockum, Anton Frisk and Krantz, Philip and Kri\ifmmode \check{z}\else \v{z}\fi{}an, Christian and Shiri, Daryoush and Svensson, Ida-Maria and Tancredi, Giovanna and Johansson, G\"oran and Delsing, Per and Ferrini, Giulia and Bylander, Jonas},
  journal = {Phys. Rev. Applied},
  volume = {14},
  issue = {3},
  pages = {034010},
  numpages = {9},
  year = {2020},
  month = {Sep},
  publisher = {American Physical Society},
  doi = {10.1103/PhysRevApplied.14.034010},
  url = {https://link.aps.org/doi/10.1103/PhysRevApplied.14.034010}
}

@article{Willsch2020FluxQubitsQuantumAnnealing,
  author    = {Willsch, Madita and Willsch, Dennis and Jin, Fengping and {De Raedt}, Hans and Michielsen, Kristel},
  title     = {Real-time simulation of flux qubits used for quantum annealing},
  doi       = {10.1103/PhysRevA.101.012327},
  issue     = {1},
  pages     = {012327},
  url       = {https://link.aps.org/doi/10.1103/PhysRevA.101.012327},
  volume    = {101},
  journal   = {Phys. Rev. A},
  month     = {Jan},
  numpages  = {16},
  publisher = {American Physical Society},
  year      = {2020},
}

@article{Fenchel1949,
  title={On Conjugate Convex Functions},
  volume={1},
  DOI={10.4153/CJM-1949-007-x},
  number={1},
  journal={Canadian Journal of Mathematics},
  publisher={Cambridge University Press},
  author={Fenchel, W.},
  year={1949},
  pages={73–77}
}

@article{Roth19,
  title = {Analysis of a parametrically driven exchange-type gate and a two-photon excitation gate between superconducting qubits},
  author = {Roth, Marco and Ganzhorn, Marc and Moll, Nikolaj and Filipp, Stefan and Salis, Gian and Schmidt, Sebastian},
  journal = {Phys. Rev. A},
  volume = {96},
  issue = {6},
  pages = {062323},
  numpages = {9},
  year = {2017},
  month = {Dec},
  publisher = {American Physical Society},
  doi = {10.1103/PhysRevA.96.062323},
  url = {https://link.aps.org/doi/10.1103/PhysRevA.96.062323}
}

@article{Wi17,
  author 	= {Dennis Willsch and Madita Nocon and  Fengping Jin and Hans De Raedt and K. Michielsen},
  title	   	= {Gate-error analysis in simulations of quantum computers with transmon qubits},
  journal	= {Physical Review A},
  year	   	= {2017},
  volume	= {96},
  pages	= {062302-1--062302-11},
  url	  	= {https://journals.aps.org/pra/abstract/10.1103/PhysRevA.96.062302}
}

@article{DeRaedt87,
  author 	= {Hans De Raedt},
  title	   	= {Product formula algorithms for solving the time dependent Schroedinger equation},
  journal	= {Computer Physics Reports},
  year	   	= {1987},
  volume	= {7},
  pages	= {1--72},
  url	  	= {https://www.sciencedirect.com/science/article/abs/pii/0167797787900025}
}

@article{Preskill2018,
  author 	= {Preskill, John},
  title 	= {Quantum Computing in the NISQ era and beyond},
  journal 	= {Quantum},
  year 	= {2018},
  volume 	= {2},
  pages 	= {79},
  url 		= {https://doi.org/10.22331/q-2018-08-06-79}
}

@article{DERAEDT201947,
  author 	= {Hans De Raedt and Fengping Jin and Dennis Willsch and Madita Willsch and Naoki Yoshioka and Nobuyasu Ito and Shengjun Yuan and Kristel Michielsen},
  title 		= {Massively parallel quantum computer simulator, eleven years later},
  journal 	= {Computer Physics Communications},
  year 		= {2019},
  pages 	= {47--61},
  volume 	= {237},
  url		= {http://www.sciencedirect.com/science/article/pii/S0010465518303977}
}

@article{Andersen2020,
  author={Andersen, Christian Kraglund
  and Remm, Ants
  and Lazar, Stefania
  and Krinner, Sebastian
  and Lacroix, Nathan
  and Norris, Graham J.
  and Gabureac, Mihai
  and Eichler, Christopher
  and Wallraff, Andreas},
  title={Repeated quantum error detection in a surface code},
  journal={Nature Physics},
  year={2020},
  month={Aug},
  day={01},
  volume={16},
  number={8},
  pages={875-880},
  issn={1745-2481},
  doi={10.1038/s41567-020-0920-y},
  url={https://doi.org/10.1038/s41567-020-0920-y}
}

@article{Lacroix2020,
  title = {Improving the Performance of Deep Quantum Optimization Algorithms with Continuous Gate Sets},
  author = {Lacroix, Nathan and Hellings, Christoph and Andersen, Christian Kraglund and Di Paolo, Agustin and Remm, Ants and Lazar, Stefania and Krinner, Sebastian and Norris, Graham J. and Gabureac, Mihai and Heinsoo, Johannes and Blais, Alexandre and Eichler, Christopher and Wallraff, Andreas},
  journal = {PRX Quantum},
  volume = {1},
  issue = {2},
  pages = {110304},
  numpages = {16},
  year = {2020},
  month = {Oct},
  publisher = {American Physical Society},
  doi = {10.1103/PRXQuantum.1.020304},
  url = {https://link.aps.org/doi/10.1103/PRXQuantum.1.020304}
}

@article{Ganzhorn20,
  title = {Benchmarking the noise sensitivity of different parametric two-qubit gates in a single superconducting quantum computing platform},
  author = {Ganzhorn, M. and Salis, G. and Egger, D. J. and Fuhrer, A. and Mergenthaler, M. and M\"uller, C. and M\"uller, P. and Paredes, S. and Pechal, M. and Werninghaus, M. and Filipp, S.},
  journal = {Phys. Rev. Research},
  volume = {2},
  issue = {3},
  pages = {033447},
  numpages = {18},
  year = {2020},
  month = {Sep},
  publisher = {American Physical Society},
  doi = {10.1103/PhysRevResearch.2.033447},
  url = {https://link.aps.org/doi/10.1103/PhysRevResearch.2.033447}
}

@article{shor1997algorithm,
  author    = {Peter W. Shor},
  title     = {Polynomial-Time Algorithms for Prime Factorization and Discrete Logarithms on a Quantum Computer},
  doi       = {10.1137/S0097539795293172},
  number    = {5},
  pages     = {1484-1509},
  url       = {https://doi.org/10.1137/S0097539795293172},
  volume    = {26},
  journal   = {SIAM J. Comput.},
  year      = {1997},
}

@article{Mueller2019,
	doi = {10.1088/1361-6633/ab3a7e},
	url = {https://doi.org/10.1088/1361-6633/ab3a7e},
	year = 2019,
	month = {oct},
	publisher = {{IOP} Publishing},
	volume = {82},
	number = {12},
	pages = {124501},
	author = {Clemens Müller and Jared H Cole and Jürgen Lisenfeld},
	title = {Towards understanding two-level-systems in amorphous solids: insights from quantum circuits},
	journal = {Reports on Progress in Physics},
  url= {https://iopscience.iop.org/article/10.1088/1361-6633/ab3a7e}
}

@article{Burnett2019,
  author={Burnett, Jonathan J. and Bengtsson, Andreas and Scigliuzzo, Marco and Niepce, David and Kudra, Marina and Delsing, Per and Bylander, Jonas},
  title={Decoherence benchmarking of superconducting qubits},
  journal={npj Quantum Information},
  year={2019},
  month={Jun},
  day={26},
  volume={5},
  number={1},
  pages={54},
  issn={2056-6387},
  doi={10.1038/s41534-019-0168-5},
  url={https://doi.org/10.1038/s41534-019-0168-5}
}

@article{JUWELS,
  author    = {{J\"{u}lich Supercomputing Centre}},
  title     = {{JUWELS: Modular Tier-0/1 Supercomputer at the J\"{u}lich Supercomputing Centre}},
  journal   = {Journal of large-scale research facilities},
  year      = {2019},
  volume    = {5},
  pages     = {A135},
  doi       = {10.17815/jlsrf-5-171},
  url       = {http://dx.doi.org/10.17815/jlsrf-5-171},
}

@article{JURECA,
  author    = {{J\"{u}lich Supercomputing Centre}},
  title     = {{JURECA: Modular supercomputer at J\"{u}lich Supercomputing Centre}},
  journal   = {Journal of large-scale research facilities},
  year      = {2018},
  volume    = {4},
  pages     = {A132},
  doi       = {10.17815/jlsrf-4-121-1},
  url       = {http://dx.doi.org/10.17815/jlsrf-4-121-1},
}

@article{Baker22,
  author = {Baker,Aneirin J.  and Huber,Gerhard B. P.  and Glaser,Niklas J.  and Roy,Federico  and Tsitsilin,Ivan  and Filipp,Stefan  and Hartmann,Michael J. },
  title = {Single shot i-Toffoli gate in dispersively coupled superconducting qubits},
  journal = {Applied Physics Letters},
  volume = {120},
  number = {5},
  pages = {054002},
  year = {2022},
  doi = {10.1063/5.0077443},
  url = {https://doi.org/10.1063/5.0077443},
  eprint = {https://doi.org/10.1063/5.0077443}
}

@article{Berke21,
  author={Berke, Christoph
  and Varvelis, Evangelos
  and Trebst, Simon
  and Altland, Alexander
  and DiVincenzo, David P.},
  title={Transmon platform for quantum computing challenged by chaotic fluctuations},
  journal={Nature Communications},
  year={2022},
  month={May},
  day={06},
  volume={13},
  number={1},
  pages={2495},
  issn={2041-1723},
  doi={10.1038/s41467-022-29940-y},
  url={https://doi.org/10.1038/s41467-022-29940-y}
}

@article{Roth22,
  author={Roth, Thomas and Ma, Ruichao and Chew, Weng Cho},
  journal={IEEE Antennas and Propagation Magazine},
  title={The Transmon Qubit for Electromagnetics Engineers: An Introduction.},
  year={2022},
  volume={},
  number={},
  pages={2-14},
  doi={10.1109/MAP.2022.3176593}
}

@article{Wittler21,
  title = {Integrated Tool Set for Control, Calibration, and Characterization of Quantum Devices Applied to Superconducting Qubits},
  author = {Wittler, Nicolas and Roy, Federico and Pack, Kevin and Werninghaus, Max and Roy, Anurag Saha and Egger, Daniel J. and Filipp, Stefan and Wilhelm, Frank K. and Machnes, Shai},
  journal = {Phys. Rev. Applied},
  volume = {15},
  issue = {3},
  pages = {034080},
  numpages = {20},
  year = {2021},
  month = {Mar},
  publisher = {American Physical Society},
  doi = {10.1103/PhysRevApplied.15.034080},
  url = {https://link.aps.org/doi/10.1103/PhysRevApplied.15.034080}
}

@article{Hund1927,
  author={Hund, F.},
  title={Zur Deutung der Molekelspektren. I},
  journal={Zeitschrift f{\"u}r Physik},
  year={1927},
  month={Oct},
  day={01},
  volume={40},
  number={10},
  pages={742-764},
  issn={0044-3328},
  doi={10.1007/BF01400234},
  url={https://doi.org/10.1007/BF01400234}
}

@incollection{HTCN06,
  author      = {Hans De Raedt and Kristel Michielsen},
  title       = {Computational Methods for Simulating Quantum Computers},
  editor      = {Michael Rieth and Wolfram Schommers},
  booktitle   = {Handbook of Theoretical and Computational Nanotechnology},
  publisher   = {American Scientific Publishers},
  year        = {2006}
}

@inProceedings{DV97,
  author    = {M. H. Devoret},
  booktitle = {Fluctuations quantiques : Les Houches, Session LXIII},
  title     = {Quantum fluctuations in electrical circuits},
  pages     = {351--386},
  publisher = {Elsevier},
  booktitle = {Fluctuations quantiques: Les Houches, session LXIII, 27 Juin - 28 Juillet 1995 = Quantum fluctuations},
  editor    = {Serge Reynaud and Elisabeth Giacobino and Jean Zinn-Justin},
  isbn      = {0-444-82593-2},
  year      = {1997},
  url       = {https://boulderschool.yale.edu/sites/default/files/files/devoret_quantum_fluct_les_houches.pdf}
}

@inProceedings{Willsch20NIC,
      author       = {Willsch, Dennis and Lagemann, Hannes and Willsch, Madita and Jin, Fengping and Michielsen, Kristel and De Raedt, Hans},
      title        = {{B}enchmarking {S}upercomputers with the {J}ülich
                      {U}niversal {Q}uantum {C}omputer {S}imulator},
      volume       = {50},
      address      = {Jülich},
      publisher    = {Forsch-ungszentrum Jülich GmbH Zentralbibliothek, Verlag},
      reportid     = {FZJ-2020-01429},
      series       = {Publication Series of the John von Neumann Institute for
                      Computing (NIC) NIC Series},
      pages        = {255 - 264},
      year         = {2020},
      comment      = {NIC Symposium 2020},
      booktitle     = {NIC Symposium 2020},
      month         = {Feb},
      url          = {https://juser.fz-juelich.de/record/874419},
}

@inProceedings{shor1994factoring,
  Title                    = {Algorithms for quantum computation: discrete logarithms and factoring},
  Author                   = {Peter W. Shor},
  Booktitle                = {Proceedings 35th Annual Symposium on Foundations of Computer Science},
  Year                     = {1994},
  Month                    = {Nov},
  Pages                    = {124-134},
  Doi                      = {10.1109/SFCS.1994.365700}
}

@Inbook{vonNeumann1993,
  author={von Neumann, J.
  and Wigner, E. P.},
  editor={Wightman, Arthur S.},
  title={{\"U}ber merkw{\"u}rdige diskrete Eigenwerte},
  bookTitle={The Collected Works of Eugene Paul Wigner: Part A: The Scientific Papers},
  year={1993},
  publisher={Springer Berlin Heidelberg},
  address={Berlin, Heidelberg},
  pages={291-293},
  isbn={978-3-662-02781-3},
  doi={10.1007/978-3-662-02781-3_19},
  url={https://doi.org/10.1007/978-3-662-02781-3_19}
}

@software{NLopt,
  author = {Steven G. Johnson},
  title = {The NLopt nonlinear-optimization package},
  url = {http://github.com/stevengj/nlopt}
}

@software{PACK99,
  title= {LAPACK:Linear Algebra PACKage},
  author= {E. Anderson and Z. Bai and C. Bischof and S. Blackford and J. Demmel J. Dongarra and J. Du Croz and A. Greenbaum and S. Hammarling and A. McKenney and D. Sorensen},
  url = {http://www.netlib.org/lapack/}

}

@software{MKL09,
  title={MKL:Math Kernel Library},
  author=  {Intel Corporation},
  url={https://software.intel.com/content/www/us/en/develop/tools/oneapi/components/onemkl.html#gs.axn2um}
}

@software{JugitJUSQUACE,
  title = {JUSQUACE:JUelich Superconducting QUAntum Computer Emulator},
  author ={Hannes Lagemann},
  url={https://jugit.fz-juelich.de/qip/jusquace},
  year = {2020}
}

\chapter*{Eidesstattliche Erkl\"arung}
	Ich, Hannes Alfred Lagemann, erkl\"are hiermit, dass diese Dissertation und die darin dargelegten Inhalte die eigenen sind und selbstst\"andig, als Ergebnis der eigenen origin\"aren Forschung, generiert wurden.\\
	
	\noindent Hiermit erkl\"are ich an Eides statt
	\begin{enumerate}
	\item Diese Arbeit wurde vollständig oder gr\"oßtenteils in der Phase als Doktorand dieser Fakult\"at und Universit\"at angefertigt;

	\item Sofern irgendein Bestandteil dieser Dissertation zuvor für einen akademischen Abschluss oder eine andere Qualifikation an dieser oder einer anderen Institution verwendet wurde, wurde dies klar angezeigt;

	\item Wenn immer andere eigene- oder Ver\"offentlichungen Dritter herangezogen wurden, wurden diese klar benannt;

	\item Wenn aus anderen eigenen- oder Ver\"offentlichungen Dritter zitiert wurde, wurde stets die Quelle hierfür angegeben. Diese Dissertation ist vollst\"andig meine eigene Arbeit, mit der Ausnahme solcher Zitate;
	
	\item Alle wesentlichen Quellen von unterstützung wurden benannt;
	
	\item Wenn immer ein Teil dieser Dissertation auf der Zusammenarbeit mit anderen basiert, wurde von mir klar gekennzeichnet, was von anderen und was von mir selbst erarbeitet wurde;
	
	\item Teile dieser Arbeit wurden zuvor ver\"offentlicht und zwar in: H. Lagemann, D. Willsch, M. Willsch, F. Jin, H. De Raedt, and K. Michielsen, "Numerical analysis of effective models for flux-tunable transmon systems", Phys. Rev. A 106, 022615 (2022). 
	\end{enumerate}
	\vspace{1.0 cm}
	\noindent\begin{minipage}[l]{8.0cm}
        \begin{flushleft}\hrulefill \\
        Datum und Unterschrift
		\end{flushleft}
    \end{minipage}
	\clearpage
\end{document}